%% file: thesis.tex
\documentclass[reqno,12pt,oneside]{report} 

\usepackage{rac}         
\usepackage{aas_macros}  
\usepackage[intlimits]{amsmath} 
\usepackage{amsxtra}     
\usepackage{amsthm}
\usepackage{amssymb}
\usepackage{amsfonts}
\usepackage{bmpsize}
\usepackage[dvipdfmx]{graphicx}    
\usepackage{bmpsize}
\usepackage{rotating}
\usepackage{color}
\usepackage{subfigure}   
\usepackage{verbatim}
\usepackage[printonlyused]{acronym} 
\usepackage{setspace}  
\usepackage{dsfont}
\usepackage[margin=1in]{geometry}

\usepackage{tikz}
\usetikzlibrary{arrows,shapes}
\usetikzlibrary{trees}
\usetikzlibrary{patterns}
\usetikzlibrary{snakes}
\usetikzlibrary{matrix,arrows} 				
\usetikzlibrary{positioning}				
\usetikzlibrary{calc,through}				
\usetikzlibrary{decorations.pathreplacing}  
\usepackage{pgffor}							

\usetikzlibrary{decorations.pathmorphing}	
\usetikzlibrary{decorations.markings}
\tikzset{
    vector/.style={decorate, decoration={snake}, draw},
	provector/.style={decorate, decoration={snake,amplitude=2.5pt}, draw},
	antivector/.style={decorate, decoration={snake,amplitude=-2.5pt}, draw},
        smallvector/.style={decorate, decoration={snake,amplitude=1.5pt,post length=0.5mm}, draw},
    fermion/.style={draw=black, postaction={decorate},
        decoration={markings,mark=at position .55 with {\arrow[draw=black]{>}}}},
    fermionbar/.style={draw=black, postaction={decorate},
        decoration={markings,mark=at position .55 with {\arrow[draw=black]{<}}}},
    fermionnoarrow/.style={draw=black},
    gluon/.style={decorate, draw=black,
        decoration={coil,amplitude=4pt, segment length=5pt}},
    scalar/.style={dashed,draw=black, postaction={decorate},
        decoration={markings,mark=at position .55 with {\arrow[draw=black]{>}}}},
    scalarbar/.style={dashed,draw=black, postaction={decorate},
        decoration={markings,mark=at position .55 with {\arrow[draw=black]{<}}}},
    scalarnoarrow/.style={dashed,draw=black},
    electron/.style={draw=black, postaction={decorate},
        decoration={markings,mark=at position .55 with {\arrow[draw=black]{>}}}},
    bigvector/.style={decorate, decoration={snake,amplitude=4pt}, draw},
    arrow/.style={draw=black, postaction={decorate},
        decoration={markings,mark=at position 1 with {\arrow[draw=black]{>}}}},
}

\tikzstyle{block} = [draw, rectangle, 
    minimum height=3em, minimum width=6earticlem]

\newcommand\fft[2]{\frac{#1}{#2}}
\newcommand\ft[2]{{\textstyle\frac{#1}{#2}}}
\newcommand\ffd[2]{{\displaystyle\frac{#1}{#2}}}
\newcommand\nn{\nonumber}

\theoremstyle{plain}

\theoremstyle{definition}

\theoremstyle{remark}

\numberwithin{theorem}{chapter}     

\makeatletter
\def\cleardoublepage{\clearpage\if@twoside \ifodd\c@page\else
\hbox{}
\thispagestyle{empty}
\newpage
\if@twocolumn\hbox{}\newpage\fi\fi\fi}
\makeatother 

\begin{document}

\bibliographystyle{unsrt}    

\titlepage{Holography, Supergravity, and the Weak Gravity Conjecture}{Brian McPeak}{Doctor of Philosophy}
{Physics}{2020}
{Professor James T. Liu, Chair\\
 Professor Roy Clarke\\
 Professor Henriette Elvang\\
 Professor Kayhan Gultekin \\
 Professor Leopoldo Pando Zayas}

\initializefrontsections


\copyrightpage{Brian McPeak}

\makeatletter
\if@twoside \setcounter{page}{4} \else \setcounter{page}{1} \fi
\makeatother
 
\dedicationpage{To Lexi}

\startacknowledgementspage
\input{Intro/Acknowledgements}
\label{Acknowledgements}


\tableofcontents     

\startabstractpage
{Holography, Supergravity, and the Weak Gravity Conjecture}{Brian McPeak}{Chair: James Liu}
\input{Abstract/Abstract}
\label{Abstract}

\startthechapters 

 \chapter{Introduction}
 \label{chap:Intro}
 \input{Intro/Intro}

 \chapter{Holography and the Weyl Anomaly}
 \label{chap:Weyl}

\input{Chap2/chap2}

 \chapter{Consistent Truncations on the Lunin-Maldacena Background}
 \label{chap:Truncations}
 \input{Chap3/chap3}
 
  \chapter{The Weak Gravity Conjecture and Black Hole Entropy}
 \label{chap:WGC}
 \input{Chap4/chap4}

\startappendices
\addtocontents{toc}{\protect\setcounter{tocdepth}{0}}

 \appendix{Heat kernel for spins up to two}
 \label{app:A_b6fields}
 \input{Appendices/App1}

  \appendix{Heat kernel for general spins}
 \label{app:B_generalheatkernel}
 \input{Appendices/App2}


\appendix{EFT Basis and On-Shell Matrix Elements}
 \label{app:D_EFT Basis}
 \input{Appendices/App4}

 \appendix{Corrections to the Maxwell equation}
 \label{app:E_Maxwell Corrections}
 \input{Appendices/App5}

 \appendix{Variations of Four-Derivative Operators with respect to the Metric}
 \label{app:F_Lagrangian Corrections}
 \input{Appendices/App6}

 \appendix{Proof of Convexity of the Extremality Surface}
 \label{app:G_Convexity}
 \input{Appendices/App7}

  \appendix{Entropy Shifts from the On-Shell Action}
 \label{app:H_Entropy}
 \input{Appendices/App8}

 \addtocontents{toc}{\protect\setcounter{tocdepth}{2}}

\startbibliography 
 \begin{singlespace} 
  \bibliography{References}   
 \end{singlespace}


\end{document}

%% file: Intro/Acknowledgements.tex
First and foremost, I would like to thank Jim Liu, who is a wonderful advisor and a true lover of physics. His advice, support, and insight are a large part of what made the last five years so exciting and fulfilling. The review sections of this thesis are filled with physics that he showed me on his blackboard, often more than once. 

I am also grateful to Henriette Elvang for being such a great teacher, mentor, and role model, and for helping me so much during my career. Thanks to Leo Pando Zayas, for years of teaching and an almost-collaboration. Thanks to Finn Larsen, for his advice and a number of educational conversations. The atmosphere and intellectual environment on the third floor of Randall is really special and I hope I can bring a part of it with me wherever I go. 

I want to thank my collaborators Jim, Callum, and Sera, for bringing their energy and expertise to our shared projects, which were a great pleasure. I hope our academic paths cross again.

I am thankful to Jim Gates, Tom Cohen, and William Linch, without whom I may have never learned to love theory, and definitely would have never made it to Michigan.

To my fellow students Scott, Callum, Shruti, Noah, Josh, Vimal, Anthony, Wenli, and Marina, for the hours we spent talking about physics in order to procrastinate from doing other physics.  

To my Mom and Dad, for answering my questions during long car rides when I was little.

%% file: Abstract/Abstract.tex
This dissertation represents work on three different subjects relating to quantum gravity and the AdS/CFT correspondence. 

First, we review a holographic computation of the one-loop corrections to the Weyl anomaly on Ricci flat backgrounds in six dimensions. This allows us to determine the correction to one linear combination of the anomaly coefficients. Then, we will show that these corrections may be obtained from the six-dimensional superconformal index.

The second section will cover consistent truncations on the Lunin-Maldacena (LM) background \cite{Lunin:2005jy}. We show how to restore minimal supersymmetry to the model of \cite{Lunin:2005jy} by determining the reduction ansatz which includes the graviton and a gauge field, which comprise the graviton multiplet of $\mathcal{N} = 2$ supergravity in five dimensions. Then we discuss our (previously unpublished) attempt to construct a truncation which includes a scalar field corresponding to the $\beta$-deformation parameter of the dual field theory. We show that if such a solution exists, it must differ somewhat drastically from the LM background. 

Finally, we discuss higher-derivative corrections to black hole solutions and the weak gravity conjecture (WGC) in a few settings. First we compute higher-derivative corrections to the extremality bounds for black holes which are charged under an arbitrary number of $U(1)$ gauge fields, and determine the constraints implied by the WGC. Next we consider the higher-derivative corrections to thermodynamic quantities for charged black holes in Anti-de Sitter space. We confirm and clarify a previously noted relationship between the shift to the extremality bound and the shift to the Wald entropy. We also show that if the shift in the Wald entropy is assumed to be positive, then the coefficient of the $R_{\mu \nu \rho \sigma} R^{\mu \nu \rho \sigma}$ term in the effective Lagrangian must be positive as well.

%% file: Intro/Intro.tex
The story of theoretical physics in the last 100 years has been dominated by two main characters: quantum field theory and gravity. Quantum field theory is, unsurprisingly, a theory of fields. Fundamental particles are excitations of these fields-- from the electromagnetic fields to fields of electrons, protons, pions, the Higgs boson, and any other particle either known or imagined. And it is a quantum theory. This means that nature is not just described by a single possible history-- the electron is spin up, or the photon went through slit A. Instead, different states can be added or subtracted, and interfere-- the electron is spin up plus spin down, the photon went through both slits. Symmetry acts as an organizing principle of these quantum fields, allowing us to classify them by their quantum numbers like spin and charge. The successes of quantum field theory are numerous; models have been written down for almost every phenomenon in particle physics, and many beyond, including cosmological processes like inflation, and condensed matter phenomena like superconductivity. 

Among the most important successes of quantum field theory, and all of physics in the 20th century, is the Standard Model, which describes three of the four fundamental forces-- electromagnetism, the weak force, and strong forces-- to extreme precision. One observable, the anomalous magnetic dipole moment of the electron, has been measured and found to match with the prediction of the Standard Model to within one part in one billion. That the theory is so successful is actually remarkable-- all known theories in physics have some range of validity outside of which a different or more fundamental theory is required. Newtonian physics works fine unless you are moving too fast or gravity is too strong-- in those cases, special relativity or general relativity are required. Electromagnetism might be great for describing energy and charge flowing through circuits, but if those circuits are too small then the quantum mechanical nature of the electrons that make up the currents becomes relevant. Chiral perturbation theory does a good job of describing pions and nucleons and other bound states of quarks, but fails to describe the quarks' interactions in regimes of high energy or high density. All over physics, we have examples where our models break down and are better described by other theories that are more fundamental or more useful.

But so far, we have not been able to push the Standard Model past its regime of validity. And this success, the remarkable accuracy of the theory, is also a source of frustration. Probing whatever lies beyond it using colliders will require access to higher energy experiments than we currently have access to. And we know that something must lie beyond it. There are issues with the internal logic of the model, such as the hierarchy problem, which is that the Higgs mass is so low (compared to its natural scale) that it appears to be fine-tuned. And there is a variety of observed phenomena not explained by the Standard Model, including the neutrino masses, the asymmetry of matter and anti-matter, and the nature of dark matter. And of course, there is a giant hole, already suggested above where we said ``three of the four fundamental forces". 

The remaining force is gravity. Another major success story of theoretical physics is the discovery and development of General Relativity. General Relativity gives a beautiful geometric description of gravity. The presence of mass bends spacetime, and objects follow paths of least distance (geodesics) in the curved geometry. This is encoded by the fundamental relation of General Relativity, the Einstein equation: 
\begin{align}
    R_{\mu \nu} - \frac{1}{2} \, R \,  g_{\mu \nu} + \Lambda \, g_{\mu \nu} \ = \ T_{\mu \nu} \, .
    \label{eq:Einstein}
\end{align}
On the left-hand-side, we have the Ricci tensor $R_{\mu \nu}$, the Ricci scalar $R$, the metric $g_{\mu \nu}$, and the cosmological constant $\Lambda$. These variables describe the curvature of the spacetime at any point inside it. On the right-hand-side, we have the stress tensor $T_{\mu \nu}$. This represents the matter living inside the spacetime. Equation (\ref{eq:Einstein}) says that the shape of spacetime is determined by the matter we put in it, and at the same time, the motion of the matter is determined by the shape of spacetime. The solutions of this equation describe a broad array physical phenomena, from black holes mergers and expanding universes to the clockwork motion of the planets revolving around the sun.

However, like the Standard Model, General Relativity cannot be the end of the story. First of all, it is problematic to include quantum matter on the right-hand-side of equation (\ref{eq:Einstein}) but keep classical gravity on the left-hand-side. This can be argued by thought experiments similar to Schr{\"o}dinger's cat, but instead of a random quantum decay deciding whether to kill the cat, a random quantum decay decides whether to destroy the world. If the world can be in a superposition of exploded and normal, then it seems the gravitational fields it generates must be in superpositions as well.

Another hint at gravity beyond General Relativity comes from black holes. One of the simplest solutions admitted by the Einstein equation is a single stationary, spherically symmetric black hole. The metric of this solution is given by
\begin{align}
    ds^2 = - \left( 1 - \frac{2 M}{r} \right) dt^2 + \left( 1 - \frac{2 M}{r} \right)^{-1} dr^2 + r^2 d \Omega^2
    \label{Schwarzschild}
\end{align}
This metric has issues at $r = 0$: it blows up to infinity! (It also has issues where $r = 2 M$ but it turns out that that point is not a real problem because the curvature is finite there). So if we want to describe what happens near $r = 0$, which is a point inside the black hole, we will need a description beyond General Relativity, because General Relativity gives a nonsensical answer. 

This black hole solution gives another, subtler hint about quantum gravity. It was realized in the 1970s that black holes behave thermally. This means, basically, that they have temperature and entropy, and that they behave according to the laws of thermodynamics. The temperature $T$ and entropy $S$ are related to surface gravity $\kappa$ and surface area $S$ by\footnote{Here we are using ``natural units," which means that $\hbar = G_N = c = 1$.}
\begin{align}
   T = \frac{\kappa}{2 \pi} \ , \qquad  S = \frac{A}{4} \, .
\end{align}
With these replacements in mind, the laws of black hole thermodynamics state are the following:
\begin{enumerate}
    \setcounter{enumi}{-1}
    \item The surface gravity $\kappa$ is constant over its event horizon
    \item The conservation of energy (including angular velocity $\Omega$, angular momentum $J$, potential $\Phi$ and charge $Q$) in the form
        $$ d M = \frac{1}{8 \pi} \kappa d A + \Omega d J + \Phi d Q $$
    \item The surface area of a black hole can never decrease
    \item The entropy of a black hole goes to a constant as the temperature goes to zero \footnote{The stronger form of the third law, that the entropy goes to zero as temperature goes to zero, is not true for some black holes.}
\end{enumerate}
It is, in many ways, shocking to find that the laws of thermodynamics apply to black holes. And it begs a further question: if black holes have entropy, then what are their microstates? After all, we have known since the 19th century that thermodynamics has statistical underpinnings, whereby macroscopic variables such as temperature, pressure, and volume emerge from the interactions of the particles or phonons or whatever microscopic objects make up the thermal system. According to this viewpoint, the entropy is given by
\begin{align}
    S = k_b \log W \, ,
\end{align}
where $W$ is the number of microstates of the system (at given values of the other macroscopic variables). So the thermodynamic description of black holes leads us to ask: what is their micrscopic description? What underlying dynamics lead to the degeneracy of states implied by their (potentially very large) surface areas?

It was not known at the outset, but the theory of quantized strings that was being developed in parallel to the theory of black hole thermodynamics addresses these questions. String theory began as an attempt to describe the strong force, but these efforts were abandoned following the discovery and success of quantum chromodynamics. It was realized in the 1970s, however, that the spectrum of the closed string contained a massless spin-two excitation-- the defining properties of a graviton, the quantum particle carrying the gravitational interaction. String theory does give a microscopic description of the gravitational dynamics underlying black holes. In some cases, this is enough to directly compute the black hole entropy by counting microstates-- a major success in this direction was \cite{Strominger:1996sh}, where the horizon area of a class of supersymmetric black holes in five dimensions was shown to be the same as log of the number of supersymmetry-preserving brane configurations.

Nonetheless, the project of understanding quantum gravity through string theory is nowhere near complete. The low-energy dynamics are well understood in the form of supergravity, but very high-energy processes require an understanding of the role played by black holes and other non-perturbative objects. We currently lack such an understanding. 

\section{AdS/CFT}

A major breakthrough for both quantum gravity and quantum field theory came in 1997 with the discovery of the AdS/CFT correspondence \cite{Maldacena:1997re}. On one side of the correspondence lies Anti-de Sitter space (AdS), which is a spacetime-- a solution of Einstein equations-- with constant negative curvature. On the other side is conformal field theory (CFT), which is a subset of quantum field theories that contain extra symmetries, notably including scale invariance. AdS/CFT states that the dynamics of quantum gravity in AdS can be be exactly described by a CFT living on the boundary of the spacetime. This discovery provided a bridge between the ideas of quantum field theory and quantum gravity. It also gave a concrete realization of the idea, inherent in the story about black hole entropy, that information about a region of spacetime can be represented on the region's boundary. A number of other different strains of research also anticipated aspects of the correspondence. One such project was the understanding of asymptotic symmetries in general relativity, which culminated in the discovery \cite{brown1986} that the asymptotic symmetry group of AdS${}_3$ matches the symmetry group of CFTs in two dimensions. Other work \cite{tHooft:1993dmi, Susskind:1994vu} anticipated the philosophy that bulk dynamics may be described by dynamics at the boundary, but did not have a specific demonstration of this idea. 

An important insight of \cite{Maldacena:1997re} was a concrete realization of the holographic principle. Consider $N$ parallel $D3$ branes, which are $3+1$-dimensional objects of IIB string theory in $9+1$ dimensions. The degrees of freedom of this system include the fluctuations of the branes and the stringy fluctuations in the space outside the branes (the ``bulk"). At the low-energies, the brane dynamics are described by $\mathcal{N} = 4$ super-Yang Mills theory and the string dynamics in the bulk are described by supergravity in flat space.

From another point of view, the $D3$ branes may be considered as massive charged sources in 10-dimensional type IIB supergravity. A solution of type IIB supergravity describing these branes was given in \cite{Horowitz:1991cd}. A key feature of this solution is that near the branes (the ``near-horizon region"), the spacetime looks like $AdS_5 \times S^5$. Beyond this (infinitesimally small) region, the supergravity dynamics is described by supergravity in flat space as well.

A key step of \cite{Maldacena:1997re}, which is essentially what allows the duality to work at all, is to take the string tension to infinity at the same time as the distance between branes is taken to zero. Then the mass of strings stretched between branes is kept fixed. On one hand, in this limit, the brane degrees of freedom completely decouple from the bulk supergravity degrees of freedom. From the other point of view, the excitations of the near-horizon geometry completely decouple from those of the flat-space region. The decoupling of the flat-space supergravity in both pictures leads to the natural conjecture that, at low energies, the gauge theory describing the brane dynamics is the same as supergravity in the $AdS_5 \times S^5$ near-horizon region. A further conjecture relates the full, UV-complete theories (rather than just the low-energy limits): $\mathcal{N} = 4$ super-Yang Mills is equivalent to type IIB string theory in $AdS_5 \times S^5$.

In the 23 years since \cite{Maldacena:1997re}, the correspondence has been found to go far beyond its initial stringy realizations. Most of the details of this set-up seem to be dispensable. We do know a few of the features that seem to be required for a CFT to have an AdS gravity dual. These include large central charge, which essentially gives the CFT enough degrees of freedom to make up for the fact that it has fewer dimensions, and a large gap in the conformal dimension of single-trace operators with spin larger than two, which is required for bulk physics to be local on scales below the AdS radius \cite{Heemskerk:2009pn}. But the correspondence has been applied in different numbers of dimensions and has succeeded in providing a dual descriptions of a vast array of phenomena in quantum field theory and gravity, including black hole evaporation, renormalization group flows, and models of superconductivity.

The AdS/CFT correspondence is one of our best and most important tools for studying quantum gravity because it gives a precise formulation of how the theory is to be formulated in one specific type of spacetime. It is also extremely useful in going in the other direction-- that is, in using gravity to study quantum field theory. This has been fruitful primarily to study quantum field theories at strong coupling, because this is precisely the region of parameter space where the bulk theory of gravity is described by the fairly well-understood General Relativity in AdS. In this thesis, I will review how my work over the past four years has addressed some of the major problems in quantum field theory and gravity, including supersymmetry, supergravity, AdS/CFT, and the physics of black holes.

\section{Overview and Summary of the Publications Discussed in this Dissertation}

The rest of this dissertation will be divided into three parts, which will cover the three main topics I have worked on during my PhD at the University of Michigan. A more extensive introduction to these topics will be given in each chapter. 

\begin{itemize}
    \item \textbf{The Weyl anomaly and the superconformal index} \\
    AdS/CFT is a duality at large $N$, where $N$ is typically the rank of the gauge group or another measure of the degrees of freedom. The difficulty in performing computations beyond the leading order in $N$ is currently a major obstacle to a more complete and detailed understanding of the correspondence. In chapter II, we discuss our work on a one-loop (subleading in $N$) calculation of the Weyl anomaly. Anomalies are important objects in field theory because they are often unchanged under renormalization group flows. The fact that they match at weak and strong coupling makes them useful observables for studying the AdS/CFT correspondence. The Weyl anomaly, in particular, is of central importance because of its role in classifying the CFT and its degrees of freedom. Our calculation provides an interesting example of a holographic one-loop calculation, and allow us to compute corrections to the Weyl anomalies in cases where they were previously unknown.  
    
        \begin{itemize}
            \item In the paper listed \cite{Liu:2017ruz}, we use a previous conjecture for the one-loop corrections to the holographic Weyl anomaly to compute the corrections for 6-dimensional conformal field theories. The method applies to theories on Ricci-flat backgrounds and for SUSY multiplets with a highest spin of two.
            \item In ref. \cite{Liu:2018eml}, we use the above results for the one-loop correction to the anomaly to derive a differential operator that gives the anomaly coefficients when acting on the superconformal index.
        \end{itemize}
        
    \item \textbf{Consistent truncations on the Lunin-Maldacena background} \\
    We only experience four spacetime dimensions, yet string theory is most naturally defined in higher numbers of dimensions. Therefore in order to understand if string theory describes our world, we must understand how higher-dimensional theories can appear as lower-dimensional ones. Consistent truncation is one answer to this-- it basically amounts to the conditions under which the modes propagating in extra dimensions may be removed from the theory. Currently, a full systematic treatment of allowed consistent truncations does not exist. In chapter III, we describe our work on constructing truncations on a particular background of type IIB supergravity. We show how this background can be upgraded to include supersymmetry, and we give some evidence that it may also admit a new class of truncation. However, we are unable to fully construct the latter. 
    
        \begin{itemize}
            \item Ref. \cite{Liu:2019cea} gives a reduction ansatz for minimal gauged supergravity on the Lunin-Maldacena background. This requires adding a gauge field to fill out the gravity multiplet, and the specific form of the ansatz bears some similarity with constructions of gauged supergravity on Sasaki-Einstein manifolds.
            \item In an unpublished effort with James Liu, we try to find a solution to type IIB supergravity that includes an extra scalar field $\gamma$ that is dual to one of the exactly marginal deformation of $\mathcal{N} = 4$ super-Yang Mills. Working perturbatively in the field $\gamma$, we find the first order solution but found an obstruction at second order. This indicates that the solution we are after either does not exist, or requires more extensive modifications of the LM solution.
        \end{itemize}
        
    \item \textbf{Higher-derivative corrections to black holes and the weak gravity conjecture}\\ 
    Absent a full description of our universe using string theory, it would be nice to understand if there are features that are expected in all possible models. One such feature is the subject of the weak gravity conjecture (WGC). This posits that in theories of quantum gravity with long-range forces, there must be particles that are self-repulsive. The statement may be motivated by black hole decay and it appears to be true in all known examples from string theory. However, so far it is unproven, and its relationship to a number of other conjectures about quantum gravity remains mysterious. In chapter IV, we study several new aspects of this statement, including theories with multiple long-range forces, and theories in Anti-de Sitter space.
        \begin{itemize}
            \item Ref. \cite{{Jones:2019nev}} extended previous work on the black hole weak gravity conjecture to theories with arbitrary numbers of electric and magnetic charges in four asymptotically flat dimensions. We find that requiring that all charged black holes can decay places constraints on the signs of the EFT coefficients.
            
            \item Ref. \cite{Cremonini:2019wdk} considers the higher-derivative corrections to Reissner-Nordstr{\"o}m black holes in Anti-de Sitter space. We show that the four-derivative corrections to the extremality bound are related to the corrections to the Wald entropy, verifying the claim \cite{Goon:2019faz} for this specific case. We also show that if the entropy shift from higher-derivative corrections is always positive, then the coefficient of $R_{\mu \nu \rho \sigma} R^{\mu \nu \rho \sigma}$ must be positive. 
        \end{itemize}
\end{itemize}

%% file: Chap2/chap2.tex
\section{Review: Superconformal Field Theories}
\label{chap2:Overview}

Understanding quantum field theories at strong coupling is one of the most challenging and important problems in theoretical physics. Superconformal field theories are a restricted class where we may obtain a number of answers that are not available for general quantum field theories. This is because their rich symmetry structure of these allows for special tools to study them. In particular, the AdS/CFT correspondence and supersymmetric localization have led to a number of insights into the structure of these theories. 

The basic symmetries of flat space are translations, generated by $P_\mu$ and Lorentz transformations, generated by $M_{\mu \nu}$. For superconformal field theories, the group of spacetime symmetries is enlarged by adding (1) scale transformations $D$ and special conformal transformations $K_{\mu}$, which together generate the conformal group, (2) fermionic generators of supersymmetry transformations $Q$ and $S$, and (3) generators of $R$-symmetry $T$, which act on the SUSY generators, and whose form depends on the amount of supersymmetry

\subsection{Weyl Anomaly}

Curved backgrounds usually break the spacetime symmetries of a theory. Nonetheless it is often possible to discuss versions those symmetries. For conformal symmetry, we introduce the notion of Weyl invariance, which means that the metric is unchanged by transformations of the form
\begin{align}
     g_{\mu \nu} \rightarrow e^{-2 \sigma(x)} g_{\mu \nu} \, .
\end{align}
This is different from a conformal transformation, which acts on the coordinates and the metric. Weyl transformations are typically what we have in mind when we say that conformal field theories are ``the same at all scales". Classically, the stress tensor encodes the changed in the action due to a change in the metric
\begin{align}
    \delta S = \int T^{\mu \nu} \delta g_{\mu \nu} \, .
\end{align}
Under an infinitesimal Weyl transformation, where $\delta g_{\mu \nu} = - 2  \sigma(x) g_{\mu \nu}$, then the change in the action is 
\begin{align}
    \delta S = - 2 \int T_{\mu}{}^{\mu} \sigma(x) \, .
\end{align}
Since this must hold for all functions $\sigma(x)$, we conclude that conformal invariance requires that $ T_{\mu}{}^{\mu} = 0$, or that the stress tensor is traceless. Like all physical symmetries, Weyl invariance may be broken by quantum corrections. This breaking is measured by the expectation value of the trace of the stress tensor. In two dimensions, this takes the form
\begin{align}
    \langle T \rangle \ = \ - \frac{c}{12} R \, ,
\end{align}
where $c$ is the central charge and $R$ is the Ricci scalar of the background. In this way, the Weyl anomaly is similar to a `t Hooft anomaly, in that it becomes measurable when the theory is coupled to a non-trivial background. 

The central charge $c$ of a CFT is an important number for describing the theory. In two dimensions, Cardy's formula \cite{Cardy:1986ie} demonstrates that the central charge $c$ is a reliable measure of the degrees of freedom.  Furthermore, its physical implication can be seen from the Zamolodchikov $c$-theorem \cite{Zamolodchikov:1986gt} which states that an effective $c$ function can be defined that is monotonically decreasing along renormalization group flows to the infrared.  While the picture is perhaps the clearest in two dimensions, recent work extending these results to higher-dimensional CFTs has further emphasized the importance of Weyl anomalies in more general situations.

\subsection{Holographic Weyl Anomaly}

The AdS/CFT correspondence provides an ideal framework for investigating various anomalies, as they may often be reliably computed on both sides of the strong/weak coupling duality. Such calculations can provide a test of the AdS/CFT correspondence and can also provide additional insights on strongly coupled CFTs. The Weyl anomaly was first discussed in the context of holography in \cite{Henningson:1998gx}. The partition function of the boundary theory should be the same as that of the gravitational bulk theory. Therefore the Weyl anomaly may be measured ``holographically" by looking at the effect on the bulk partition function of a transformation that scales the boundary metric. In particular, for a partition function given by
\begin{align}
    Z = \int \mathcal{D} \phi \exp{(-S[\phi])},
\end{align}
we define the anomaly $\mathcal{A} = \langle T \rangle$ by
\begin{align}
    \delta\log Z = - \int d^d x \sqrt{\det g}\, \delta \sigma \mathcal A.
\end{align}
From the holographic point of view, the leading order partition function may obtained from on-shell action. The leading-order computation requires expanding the action in terms of an IR cutoff $\epsilon$ and functions of the curvature invariants $a_{(i)}$ in the following way:
\begin{align}
    S = \int \sqrt{\det h} \left( \epsilon^{-d / 2} a_{(0)} +  \epsilon^{-d / 2 + 1} a_{(2)} + ... + \epsilon^{-1} a_{(d - 2)} + \log \epsilon \, a_{(d)} + \mathcal{L}_{fin}\right) \, ,
\end{align}
Here $h$ is the boundary metric. The entire action is invariant under $\delta h = - 2 \sigma h, \ \delta \epsilon = -2 \sigma \epsilon$. The negative powers of $\epsilon$ are all divergent before and after the transformation, so they are irrelevant. However, the log-divergent term picks up a finite shift as $ \epsilon \rightarrow \epsilon -2 \sigma \epsilon$. This must be cancelled since the entire action is invariant; therefore the finite piece of the Lagrangian must be shifted by $-\log (1 - 2 \sigma) a_{(d)}$. 

The demanding part of the calculation is to compute the functions $a_{(i)}$, which may be accomplished by expanding the bulk metric in powers of the radial coordinate $r$ near the boundary at $r = 0$. In the AdS$_5$/CFT$_4$ case, the anomaly takes the form
\begin{align}
    \mathcal{A} = \frac{1}{16\pi^2} \left( a \, E_{(4)} + c \, I_{(4)} \right) \, ,
\end{align}
where the Euler density $E_{(4)}$ and the conformal invariant $I_{(4)}$ are made by contracting the curvature invariants. In this case, the holographic computation of the Weyl anomaly gives us the familiar result
\begin{equation}
    c=a=\frac{N^2}4\frac{\pi^3}{\mathrm{vol}(\Sigma^5)},
    \label{eq:aclead}
\end{equation}
where IIB supergravity has been compactified on AdS$_5\times\Sigma^5$.  Additional corrections to the leading order expression may arise from higher derivative modifications to the supergravity action as well as from quantum (i.e. loop) effects.

The Weyl anomaly goes beyond the leading order in the $1 / N$ expansion. Holographically, the log-divergent part of the one-loop effective action provides an $\mathcal O(1)$ correction to the Weyl anomaly coefficients $a$ and $c$.  This was initially computed for the case of AdS$_5\times S^5$ in \cite{Bilal:1999ph,Mansfield:1999kk,Mansfield:2000zw,Mansfield:2002pa,Mansfield:2003gs}, where it was observed that the leading order result (\ref{eq:aclead}) is shifted according to $N^2\to N^2-1$, in agreement with expectations for $SU(N)$ gauge symmetry.  More recently, the one-loop computation in AdS$_5$ has been extended to holographic field theories with reduced or even no supersymmetry \cite{Ardehali:2013gra,Ardehali:2013xya,Ardehali:2013xla,Beccaria:2014xda}.

The one-loop holographic computation is essentially a sum over contributions from all states in the spectrum of single-trace operators.  Curiously, when arranged in terms of 4-dimensional $\mathcal N=1$ superconformal multiplets, the contribution from long multiplets vanish identically.  As a result, only short representations contribute to the $\mathcal O(1)$ shift in $a$ and $c$.  This allows for a close connection between the central charges and the superconformal index, which also encodes knowledge of the shortened spectrum \cite{Ardehali:2014zba,Ardehali:2014esa} (see also \cite{DiPietro:2014bca}).

\subsubsection{Weyl anomaly in Six Dimensions}

6-dimensional superconformal field theories are noteworthy because six is the highest possible dimension for superconformal invariance. Furthermore, such theories can be reduced on Riemann surfaces to give a large class of theories in four dimensions.  However much less is known about 6-dimensional superconformal field theories, and the situation is complicated by the fact that we have to consider four central charges, $\{a, c_1, c_2, c_3\}$. In general, the anomaly takes the following form
\begin{align}
 (4 \pi)^3 \mathcal A= -a E_6 + (c_1 I_1 + c_2 I_2 + c_3 I_3) + D_\mu J^\mu,
\end{align}
where the coefficients $E_6$ and $I_i$ are defined by the curvature of the background geometry:
\begin{align}
\begin{split}
    E_6 &= \epsilon^{abcdef} \epsilon^{ghijkl} R_{ab gh} R_{cd ij} R_{efkl}  \\
    I_1&=C^a{}_{mn}{}^bC^m{}_{pq}{}^nC^p{}_{ab}{}^q,\\
    I_2&=C^{ab}{}_{mn}C^{mn}{}_{pq}C^{pq}{}_{ab},\\
    I_3&=C^{mnpq}\square C_{mnpq}+\cdots.  
\end{split}
\end{align}

Supersymmetry reduces the number of independent coefficients by imposing relations between them. For convenience, we will define
\begin{equation}
c=\fft{c_2-c_3}{32},\qquad c'=\fft{c_1-4c_2}{192},\qquad c''=\fft{c_1-2c_2+6c_3}{192}.
\label{eq:cc'c''}
\end{equation}
For all superconformal theories, the combination $c''$ will vanish. Furthermore, for extended $\mathcal{N} = (2,0)$ supersymmtery, $c'$ vanishes as well, and we are left with only two coefficients.  

We have defined $c$ so that we may use the combination $c - a$, which is familiar from the analogous combination that appears in four dimensions. This is demonstrated by Einstein gravity on AdS$_7$, where we find relations of the form \cite{Henningson:1998gx,Bastianelli:2000hi}
\begin{equation}
    c_1=4c_2=-12c_3=96 c = 96a\sim\mathcal O(N^3),
\end{equation}
which is the 6-dimensional analog of (\ref{eq:aclead}). The relations between the $c_i$ coefficients given above arise naturally in the holographic computation, and are consistent with 6-dimensional $(2,0)$ superconformal invariance.

We would like to go beyond the leading order for the 6-dimensional SCFTs. The most extensively studied $(2,0)$ theory of relevance is that of $N$ coincident M5-branes, which is dual to supergravity on AdS$_7\times S^4$.  Here the conjectured expression for the central charges are \cite{Tseytlin:2000sf,Mansfield:2003bg,Beccaria:2014qea}
\begin{equation}
    a=-\frac{1}{288}(4N^3-\ft94N-\ft74),\qquad c=-\frac{1}{288}(4N^3-3N-1).
    \label{eq:M5ac}
\end{equation}
The $\mathcal O(N)$ terms arise from $R^4$ corrections \cite{Tseytlin:2000sf}, while the $\mathcal O(1)$ terms arise at one-loop \cite{Mansfield:2003bg,Beccaria:2014qea}.  The $\mathcal O(1)$ shift $\delta a=7/1152$ was computed in \cite{Beccaria:2014qea} by evaluating the one-loop partition function on global (Euclidean) AdS$_7$ with $S^6$ boundary.  However, the conjectured $\delta c=1/288$ has not yet been directly computed, as the most straightforward computation of one-loop determinants involve highly symmetric spaces with conformally flat boundaries.  In such cases, the Weyl invariants vanish, so no information is provided about the $c_i$ coefficients.

An alternative approach to the computation of $\delta a$ and $\delta c$ was developed in \cite{Mansfield:1999kk,Mansfield:2000zw,Mansfield:2002pa,Mansfield:2003gs} based on a functional Schr\"odinger approach.  In this approach, the contribution of each state to the $\mathcal O(1)$ shift in the Weyl anomaly takes the form
\begin{equation}
\delta\mathcal A=-\fft12\left(\Delta-\fft{d}2\right)b_d,
\label{eq:MN}
\end{equation}
where $\Delta$ is the conformal dimension and $b_d$ is the heat kernel coefficient for the corresponding AdS$_{d+1}$ field when restricted to the $d$-dimensional boundary.  In principle, since the 6-dimensional $b_6$ coefficient may be computed on a general curved background, this allows for a full determination of not just the $a$ coefficient but the $c_i$'s as well.

It has been argued in \cite{Beccaria:2014xda}, however, that the expression (\ref{eq:MN}) cannot in general be valid, as the contribution for a single field should have a more complicated dependence on the conformal dimension $\Delta$.  This can be seen explicitly in comparison with the expression for $\delta a$ obtained directly from the one-loop determinant on global AdS.  Curiously, however, when (\ref{eq:MN}) is summed over the states of a complete supermultiplet, the resulting expression appears to be valid on Ricci-flat backgrounds as it passes all consistency checks and has the expected connection to the index \cite{Beccaria:2014xda,Ardehali:2014zba}.

Another line of reasoning has been developed to determine the anomaly coefficients directly from the appropriate conformal higher spin operators on the boundary \cite{Beccaria:2017dmw,Beccaria:2017lcz}. In particular, it is argued that AdS fields with higher dimensions $\Delta$ correspond to boundary fields whose kinetic operators are greater than second order in derivatives. The factorization of these operators on Ricci-flat backgrounds may serve as a justification of the functional Schr\"odinger method presented in \cite{Mansfield:1999kk,Mansfield:2000zw,Mansfield:2002pa,Mansfield:2003gs}.

\subsection{Superconformal Index}

Another important part of this chapter will be the superconformal index\footnote{This brief introduction to the index largely follows that of \cite{Romelsberger:2007ec} and \cite{Ardehali:2016kza}}. A supersymmetric theory is one that has symmetry generators, which we shall call $Q$ and $Q^{\dagger}$, that satisfy anti-commutation relations rather than commutation relations. In the simplest case, with zero spatial dimensions,
\begin{align}
    \left\{ Q \, , \, Q^{\dagger} \right\} \ = \ H \ , \qquad Q^2 = Q^{\dagger 2} = 0
\end{align}
The existence of such an operator implies that the states $|s \rangle$ must come in pairs that have the same energy. To see this, consider a state $|b \rangle$ such that $ H |b \rangle = E |b \rangle$. Then define $| f \rangle = (Q + Q^{\dagger})  |b \rangle$. Then the commutation relation implies that $ H |f \rangle = E |f \rangle$: 
\begin{align}
    H | f \rangle = H (Q + Q^{\dagger})  |b \rangle = ( Q Q^{\dagger} Q + Q^{\dagger} Q Q^{\dagger} ) |b \rangle =  (Q + Q^{\dagger}) H |b \rangle = E | f \rangle
\end{align}
There is an exception to this argument: states with $E = 0$ do \textit{not} need to be paired because $H | b \rangle = 0 \implies (Q + Q^{\dagger}) | b \rangle = 0$. Recall that energies are always positive for supersymmetric theories:
\begin{align}
    \langle s | H | s \rangle =  \langle s | Q Q^{\dagger} | s \rangle  + \langle s | Q^{\dagger} Q | s \rangle = 2 \langle s | Q^{\dagger} Q | s \rangle = \big| Q |s \rangle \big|^2 > 0
\end{align}
So in principle it is possible that supersymmetric theories have undpaired ground states. These states may be counted by the Witten index:
\begin{align}
    \mathcal{I} = \sum_s (-1)^F e^{-\beta E_s} = \#_b - \#_f
\end{align}
Here $F$ is the fermion number, which equals 1 for fermionic ($f$) states and 0 for bosonic ($b$) states. This index essentially counts the difference in the number of bosonic and fermionic states. 

Now let us consider superconformal field theories in four dimensions. Then with minimal $\mathcal{N} = 1$ supersymmetry, we have four supercharges, $\{ Q_1, Q_2, Q_1^{\dagger}, Q_2^{\dagger} \}$, (and a corresponding set of conformal supercharges $S$). We need to pick a pair of charges to define the index. Let us choose $Q_1$ and $Q_1^{\dagger}$, which have commutation relations
\begin{align}
    \left\{ Q_1 \, , \, Q_1^{\dagger} \right\} \ = \ H - 2 J_3 - \frac{3}{2} R : = \delta
\end{align}
where $H$ is the Hamiltonian in radial quantization, $J$ is angular momentum, and $R$ is the r-charge. We may construct an object analogous to the Witten index:
\begin{align}
    \mathcal{I} = \sum_s (-1)^F e^{-\beta \delta}
\end{align}
In this case, only the unpaired states that satisfy $E - 2 j_3 - r = 0$ contribute; such states will be said to be part of short multiplets or short representations. 

If we like, we may further \textit{refine} the index by terms that commute with $Q_1$. For example, 
\begin{align}
    \mathcal{I}(\beta_0) = \sum_s (-1)^F e^{-\beta_0 (E - r/2)} e^{- \beta \delta}
        \label{ind1}
\end{align}
includes an extra regulating factor that makes the index finite for a number of SCFTs. More generally, we could include charges $C_i$ and their fugacities, $\mu_i$, which would give an index 
\begin{align}
    \mathcal{I}(\mu_i) = \sum_s (-1)^F \prod_i \mu_i^{C_i}  e^{- \beta \delta}
    \label{ind2}
\end{align}
The expressions (\ref{ind1}) and (\ref{ind2}) are typically called the \textit{superconformal index} \cite{Kinney:2005ej, Romelsberger:2005eg}. The superconformal index counts the short representations of the superconformal group. As we shall see below, the $\mathcal{O}(1)$ corrections to the Weyl anomaly are also zero for long representations. This allows for a rich interplay between the anomaly and the superconformal index.

\subsection{Overview}

In this chapter, we will study the one-loop contribution to $\delta(c-a)$ using holography and the superconformal index.

In the first section, we use (\ref{eq:MN}) to compute the $\mathcal O(1)$ contribution to the holographic Weyl anomaly of $\mathcal N=(1,0)$ theories from maximum spin-2 multiplets in the bulk.  Since we consider Ricci-flat backgrounds, we only obtain information on $\delta(c-a)$, and are unable to probe $c'$, which may be non-zero in the case of $\mathcal N=(1,0)$ supersymmetry. This is similar to the AdS$_5$/CFT$_4$ case, where $b_4\sim\delta(c-a)R_{\mu\nu\rho\sigma}^2$ on Ricci-flat backgrounds.  As a consistency check, we find that $\delta(c-a)$ vanishes for long representations of $\mathcal N=(1,0)$ supersymmetry, as expected.

In the second section, we extend the relation of holographic central charges to the superconformal index in the case of AdS$_7$/CFT$_6$. Using the results for $\delta a$ and $\delta(c-a)$ for $(1,0)$ theories, we demonstrate below how they may be obtained from the large-$N$ single-trace index.  In particular, we construct differential operators that extract $\delta a$ and $\delta(c-a)$ from the index in the high-temperature limit.  The expression for $\delta a$ is fully constrained, while that for $\delta(c-a)$ has one undetermined coefficient related to our lack of knowledge of the $\mathcal O(1)$ holographic Weyl anomaly beyond spin two.

\section{The $\mathcal O(1)$ contribution to the holographic Weyl anomaly} 

As indicated above, the anomaly is given by the central charges and the curvature invariants in the following form:
\begin{equation}
    (4 \pi)^3 \mathcal A=-aE_6+(c_1I_1+c_2I_2+c_3I_3)+D_\mu J^\mu,
    \label{eq:Asix}
\end{equation}

The procedure we use to obtain the $\mathcal O(1)$ shift in the anomaly for $\mathcal N=(1,0)$ theories is to sum the expression (\ref{eq:MN}) over complete representations of the corresponding $OSp(8^*|2)$ supergroup.  However, we first start with states in the bosonic subgroup $OSp(8^*|2)\supset SO(2)\times SU(4)\times SU(2)_R$ labeled by $D(\Delta,j_1,j_2,j_3)$ along with $R$-symmetry representation $r$.  We thus have
\begin{equation}
    \delta\mathcal A(\mathrm{rep})=-\fft12\sum_{\mathrm{rep}}(\Delta-3)b_6(j_1,j_2,j_3).
    \label{eq:deltaA}
\end{equation}
In the following, we first work out the heat kernel coefficients $b_6(j_1,j_2,j_3)$ on a Ricci-flat background, and then perform the sum over complete supermultiplets with maximum spin two.

\subsection{Heat kernel coefficients}

For an operator $\Delta=-\nabla^2-E$ where $E$ is some endomorphism, the 6-dimensional Seeley-DeWitt coefficient $b_{6}(\Delta)$ takes the form \cite{Gilkey:1975iq,Bastianelli:2000hi}
\begin{align}
    b_{6}(\Delta) &= \frac{1}{(4 \pi)^3 7!} \mathrm{Tr}\biggl[18A_1 + 17 A_2 - 2A_3 - 4 A_4 + 9 A_5 + 28 A_6 - 8 A_7 + 24 A_8 + 12 A_9 \nn\\
    &\quad+ \frac{35}{9} A_{10} -\frac{14}{3} A_{11} + \frac{14}{3} A_{12} - \frac{206}{9} A_{13} +\frac{64}{3} A_{14} -\frac{16}{3} A_{15} + \frac{44}{9} A_{16} +\frac{80}{9} A_{17}\nn \\
    &\quad+ 14 \Bigl( 8 V_1 + 2V_2 + 12V_3 -12 V_4 + 6V_5 -4V_6 +5V_7 +6V_8 + 60V_9 +30 V_{10} \nn\\
    &\kern2em+ 60 V_{11} + 30 V_{12} + 10 V_{13} + 4 V_{14} + 12 V_{15} + 30 V_{16} + 12V_{17} + 5 V_{18} -2 V_{19} + 2 V_{20}  \Bigl) \biggl].
\label{eq:b6coef}
\end{align}
Here the $A_a$'s form a basis of curvature invariants \cite{parker1987,Bastianelli:2000hi}, and the $V_a$'s are built from the endomorphism $E$ and the curvature $F_{ij}$ of the connection \cite{Bastianelli:2000hi}. In particular, while the coefficients of the $A_a$'s are universal, the $V_a$ terms are specific to the representation.

We follow the conventions spelled out in Appendix~A of \cite{Bastianelli:2000hi}, which also give explicit expressions for the $A_a$'s and $V_a$'s.  However, we are concerned with only the combinations that are non-vanishing on Ricci-flat backgrounds.  These are
\begin{equation}
A_5 = (\nabla_i R_{abcd})^2,\quad A_9 = R_{abcd} \nabla^2 R^{abcd},\quad
A_{16} = R_{ab}{}^{cd}R_{cd}{}^{ef}R_{ef}{}^{ab},\quad A_{17} = R_{aibj}R^{manb}R^i{}_m{}^j{}_n.
\end{equation}
The full list of $A_a$'s, and expressions for the $V_a$'s are given in Appendix~\ref{app:A_b6fields}.

The invariants $E_6$ and $I_1$, $I_2$, and $I_3$ may be written in terms of the basis $A_a$ functions. On a Ricci-flat background, they become
\begin{align}
    E_6 = 32 A_{16} - 64 A_{17} , \quad I_1 = - A_{17}  , \quad I_2 = A_{16}  , \quad    I_3 = 3A_5 + 6 A_9 + 2 A_{16} + 8 A_{17}.
    \label{eq:E6Ii}
\end{align}
As these quantities are not all independent, we will be unable to determine the individual central charges $\{a,c_i\}$ using only a Ricci-flat background.  Note that we may construct two combinations that are total derivatives
\begin{align}
    D_1&=\nabla_a(R_{mnij}\nabla_a R_{mnij})=A_5+A_9,\nn\\
    D_2&=2\nabla_a(R_{mnij}\nabla_m R_{anij})=-A_5+A_{16}+4A_{17}.
\end{align}
This allows us to rewrite (\ref{eq:E6Ii}) in terms of the two invariants $A_{16}$ and $A_{17}$
\begin{equation}
    E_6 =  32 A_{16} - 64 A_{17} , \quad I_1 = - A_{17}  , \quad I_2 = A_{16}  , \quad    I_3 = -A_{16} -4 A_{17}+6D_1-3D_2.
\end{equation}
On a Ricci-flat background, we have the relations $E_6= 32(2I_1+I_2)$ and $I_3=4I_1-I_2$ up to a total derivative.  As a result, the 6-dimensional anomaly, (\ref{eq:Asix}), takes the form
\begin{equation}
   (4 \pi)^3 \mathcal A=32(c-a)A_{16}-64(c-a+3c'')A_{17}+D_\mu J^\mu,
    \label{eq:Aca}
\end{equation}
on Ricci-flat backgrounds.  The implication of this expression is that we will only be able to obtain information on the $\mathcal O(1)$ contribution to $c-a$ and to $c''$.  Since the latter must vanish for superconformal theories, it will serve as a consistency check of our approach.  This leaves us with a holographic determination of $\delta(c-a)$, which may be combined with the result of \cite{Beccaria:2014qea} for the $\delta a$ coefficient to extract both $\delta c$ and $\delta a$.  This, in principle, provides a complete determination of the $\mathcal O(1)$ shift in the holographic Weyl anomaly of $\mathcal N=(2,0)$ theories.  Unfortunately the additional anomaly coefficient $\delta c'$ for $\mathcal N=(1,0)$ theories cannot be determined in this manner on Ricci-flat backgrounds.

Ideally, we would like to have an expression for the heat kernel coefficient $b_6(\Delta)$ for fields transforming in an arbitrary $(j_1,j_2,j_3)$ representation of the 6-dimensional $SU(4)$ Euclidean rotation group.  However, this requires understanding of arbitrary higher-spin Laplacians, which currently eludes us.  There is also some potential ambiguity in relating `on-shell' states in AdS$_7$ to their corresponding boundary Laplacians in the functional Schrodinger approach of \cite{Mansfield:1999kk}.  We thus restrict to spins up to two.  The relevant $b_6$ coefficients evaluated on a Ricci-flat background are summarized in Table~\ref{tbl:b6coef}.  The coefficients for $\phi$, $\psi$, $A_\mu$ and $B_{\mu\nu}$ were computed in \cite{Bastianelli:2000hi}, while the remaining ones are worked out in Appendix~\ref{app:B_generalheatkernel}.


\begin{table}[t]
\begin{center}
  \begin{tabular}{| c | c | c | c | c | c ||c|c|}
    \hline
    Field & $SU(4)$ Rep & $c_5$ & $c_9$ & $c_{16}$ & $c_{17}$&$\gamma_{16}$&$\gamma_{17}$ \\ \hline \hline
    $\phi$ & $(0,0,0)=\mathbf1$ & $9$ & $12$ & ${44}/{9}$ & ${80}/{9}$  &$17/9$&$-28/9$ \\ \hline
    $\psi$ & $(1,0,0)=\mathbf4$ & $-20$ & $-36$ & ${-202}/{9}$ & ${-436}/{9}$&$-58/9$&$140/9$  \\ \hline
    $A_{\mu}$ & $(0,1,0)=\mathbf6$ &$-58$ & $-96$ & ${-164}/{3}$ & ${-344}/{3}$&$-50/3$&$112/3$  \\ \hline
    $C^{+}_{\mu \nu \rho}$ & $(2,0,0)=\mathbf{10}$ & $174$ & $456$ & ${-5608}/{9}$ & ${26504}/{9}$&$-8146/9$&$16352/9$  \\ \hline
    $\Psi_{\mu}$ & $(1,1,0)=\mathbf{\overline{20}}$ & $292$ & $828$ & ${3526}/{9}$ & ${22012}/{9}$ &$-1298/9$&$2716/9$ \\ \hline
    $B_{\mu \nu}$ & $(1,0,1)=\mathbf{15}$ & $107$ & $348$ &${2992}/{3}$ & ${-1616}/{3}$ &$2269/3$&$-4508/3$ \\ \hline
    $G_{\mu \nu}$ & $(0,2,0)=\mathbf{20'}$ & $544$ & $1416$ & ${-1388}/{9}$ & ${49984}/{9}$ &$-9236/9$&$18592/9$ \\ \hline
  \end{tabular}
\end{center}
\caption{Heat kernel coefficients $(4\pi)^37!b_6=c_5A_5+c_9A_9+c_{16}A_{16}+c_{17}A_{17}$ for fields of spins up to two on a Ricci-flat background.  In the last two columns, we tabulate $\gamma_{16}$ and $\gamma_{17}$, where $(4\pi)^37!b_6=\gamma_{16}A_{16}+\gamma_{17}A_{17}+D_\mu J^\mu$.
\label{tbl:b6coef}}
\end{table}

\subsection{$\mathcal N=(1,0)$ Theory}

We now turn to the superconformal theories, starting with the $\mathcal N=(1,0)$ theory. We expect that the anomaly vanishes when summed over long representations, and we will see that this is indeed the case.  The $\mathcal N=(1,0)$ superconformal algebra is $OSp(8^*|2)$, with bosonic subgroup $SO(2,6)\times SU(2)_R$.  Here $SO(2,6)$ is either the isometry group of AdS$_7$ or the 6-dimensional conformal group.  We label representations of $OSp(8^*|2)\supset SO(2,6)\times SU(2)_R\supset SO(2)\times SU(4)\times SU(2)_R$ by conformal dimension $\Delta$, $SU(4)$ Dynkin labels $(j_1,j_2,j_3)$ and $SU(2)_R$ Dynkin label $k$ (so that $SU(2)$ `spin' is given by $k/2$).

Unitary irreducible representations of the $\mathcal N=(1,0)$ theory have been studied and explicitly constructed in \cite{Minwalla:1997ka,Dobrev:2002dt,Bhattacharya:2008zy,Buican:2016hpb,Cordova:2016emh}.  The theory has one regular and three isolated short representations, given generically by
\begin{equation}
\begin{tabular}{ll}
$A[j_1,j_2,j_3;k]$:\qquad\hbox{}&$\Delta=\ft12(j_1+2j_2+3j_3)+2k+6$,\\
$B[j_1,j_2,0;k]$:&$\Delta=\ft12(j_1+2j_2)+2k+4$,\\
$C[j_1,0,0;k]$:&$\Delta=\ft12j_1+2k+2$,\\
$D[0,0,0;k]$:&$\Delta=2k$.
\end{tabular}
\label{eq:1,0short}
\end{equation}
For maximum spin two, however, we must restrict to $j_1=j_2=j_3=0$.  In this case, it is a simple exercise to perform the sum (\ref{eq:deltaA}) over the multiplet using the values of $\gamma_{16}$ and $\gamma_{17}$ given in Table~\ref{tbl:b6coef}.  Comparison with (\ref{eq:Aca}) then allows us to extract $\delta(c-a)$ and $\delta c''$.  The results are summarized in Table~\ref{tbl:(1,0)}.

\begin{table}[t]
\resizebox{\columnwidth}{!}{
\begin{tabular}{| l | c | c | c | c | c | c |}
 \hline
 &&$D[0,0,0;k]$&$C[0,0,0;k]$&$B[0,0,0;k]$&$A[0,0,0;k]$&$L[0,0,0;k]$\\
 Level & $SU(4)$ & $\Delta=2k$ & $\Delta = 2k + 2$ &$\Delta = 2k+4$ & $\Delta = 2k + 6$ & $\Delta > 2k+ 6$ \\ \hline
 $\Delta$ & $\mathbf1$ & $\scriptstyle k$ &$\scriptstyle k$ & $\scriptstyle k$ & $\scriptstyle k$ & $\scriptstyle k$ \\
 \hline
 $\Delta + \frac{1}{2}$ & $\mathbf4$ & $\scriptstyle k-1$ & $\scriptstyle k-1, k+1$ & $\scriptstyle k-1, k+1$ & $\scriptstyle k-1, k+1$ & $\scriptstyle k-1, k+1$ \\
 \hline
 $\Delta + 1$ & $\mathbf{10}$ &  & $\scriptstyle k$ & $\scriptstyle k$ & $\scriptstyle k$ & $\scriptstyle k$ \\
         & $\mathbf6$ & $\scriptstyle k-2$ & $\scriptstyle k-2, k$ & $\scriptstyle k-2, k, k+2$ & $\scriptstyle k-2, k, k+2$ & $\scriptstyle k-2, k, k+2$ \\
 \hline
 $\Delta + \frac{3}{2}$ & $\mathbf{\overline{20}}$ &  & $\scriptstyle k-1$ & $\scriptstyle k-1, k+1$ & $\scriptstyle k-1, k+1$ & $\scriptstyle k-1, k+1$ \\
         & $\mathbf{\overline4}$ & $\scriptstyle k-3$ & $\scriptstyle k-3, k-1$ & $\scriptstyle k-3, k-1, k+1$ & $\scriptstyle k-3, k-1, k+1, k+3$ & $\scriptstyle k-3, k-1, k+1, k+3$ \\
 \hline
 $\Delta + 2$ & $\mathbf{20'}$ & & & $\scriptstyle k$ & $\scriptstyle k$ & $\scriptstyle k$ \\
         & $\mathbf{15}$  & & $\scriptstyle k-2$ & $\scriptstyle k-2, k$ & $\scriptstyle k-2, k, k+2$ & $\scriptstyle k-2, k, k+2$ \\
         & $\mathbf1$   & $\scriptstyle k-4$ & $\scriptstyle k-4, k-2$ & $\scriptstyle k-4, k-2, k$ & $\scriptstyle k-4, k-2, k, k+2$ & $\scriptstyle k-4, k-2, k, k+2, k+4$ \\
         \hline
$\Delta + \frac{5}{2}$ & $\mathbf{20}$ &  &  & $\scriptstyle k-1$ & $\scriptstyle k-1, k+1$ & $\scriptstyle k-1, k+1$ \\
         & $\mathbf4$ & & $\scriptstyle k-3$ & $\scriptstyle k-3, k-1$ & $\scriptstyle k-3, k-1, k+1$ & $\scriptstyle k-3, k-1, k+1, k+3$ \\
         \hline
$\Delta + 3$ & $\mathbf{\overline{10}}$ &  &  &  &$\scriptstyle k$ & $\scriptstyle k$ \\
         & $\mathbf6$ &  &  & $\scriptstyle k-2$ & $\scriptstyle k-2, k$ & $\scriptstyle k-2, k, k+2$\\
 \hline
$\Delta + \frac{7}{2}$ & $\mathbf{\overline4}$ & & & & $\scriptstyle k-1$ & $\scriptstyle k-1, k+1$ \\
\hline
$\Delta + 4$ & $\mathbf1$ &  &  &  &  & $\scriptstyle k$\\
\hline
\hline
Anomaly&$2^5\cdot6!\delta(c-a)$ & $1$ &$57+180k$&$303+180k$& $-1$ & $0$  \\
&$\delta c''$     & $0$ & $0$ & $0$ & $0$ & $0$ \\        
\hline      
\end{tabular}}
\caption{The $\mathcal N=(1,0)$ multiplets with maximum spin two, and corresponding holographic Weyl anomaly coefficients $\delta(c-a)$ and $\delta c''$.  Here $k$ is the $SU(2)_R$ Dynkin label (with spin~$=k/2$).  The shortening conditions correspond to those of (\ref{eq:1,0short}), while the last column is the maximum spin-two long representation.}
\label{tbl:(1,0)}
\end{table}

As a consistency check, we note that the anomaly coefficient $c''$ vanishes identically after summation over a complete multiplet.  This is a requirement of supersymmetry, but is not manifest from the individual $b_6$ coefficients in Table~\ref{tbl:b6coef}.  We also see that the anomaly vanishes for the long representation, which agrees with expectations from the AdS$_5$ case \cite{Ardehali:2013gra,Beccaria:2014xda}.  As for the non-vanishing contributions, note that $\delta(c-a)$ for the $A$ and $D$ type multiplets are equal and opposite. This must be the case, as $A[0,0,0;k]$ and $D[0,0,0;k+2]$ are ``mirror shorts" that sum to become a long multiplet.

Finally, recall that the $\mathcal N=(1,0)$ theory admits three independent anomaly coefficients, which we have parametrized as $a$, $c$ and $c'$.  Since we only consider Ricci-flat backgrounds, we have only been able to determine the difference $\delta(c-a)$.  This may be combined with the holographic $\delta a$ coefficient obtained in \cite{Beccaria:2014qea} to separate out the contributions to $\delta a$ and $\delta c$.  These results are presented in Table~\ref{tbl:10ac}.  However, we are unable to determine $\delta c'$ unless we can move away from Ricci-flat backgrounds.

\begin{table}[t]
\resizebox{\columnwidth}{!}{
\begin{tabular}{| c | c | c | c |}
    \hline
    Multiplet & $\Delta$ & $2^5\cdot6!\delta a$ & $2^5\cdot6!\delta(c-a)$\\
    \hline
    $L[0,0,0;k]$ & $>2k+6$ & $0$ & $0$  \\
    \hline
    $A[0,0,0;k]$ & $2k+6$ & $10\Delta^2(\Delta^2-2)+\fft{11}3$ & $-1$ \\
    \hline
    $B[0,0,0;k]$ & $2k+4$ & $-10(\Delta-\fft23)^2(3(\Delta-\fft23)^2-14)-\fft{530}9(\Delta-\fft23)-\fft{419}9 $ & $90(\Delta-\fft23)+3$\\
    \hline
    $C[0,0,0;k]$ & $2k+2$ & $10(\Delta-\fft43)^2(3(\Delta-\fft43)^2-14)-\fft{530}9(\Delta-\fft43)+\fft{419}9$ & $90(\Delta-\fft43)-3$ \\
    \hline
    $D[0,0,0;k]$ & $2k$ & $-10(\Delta-2)^2((\Delta-2)^2-2)-\fft{11}3$ & $1$ \\
    \hline
\end{tabular}}

\caption{Contribution to the Weyl anomaly coefficients $\delta a$ and $\delta c$ from maximum spin two multiplets for the $\mathcal N=(1,0)$ theory. Here $c$ is related to the conventional anomaly coefficients $c_i$ according to (\ref{eq:cc'c''}).  The $\delta a$ coefficient is computed using the results of \cite{Beccaria:2014qea}.}
\label{tbl:10ac}
\end{table}

\subsection{$\mathcal N=(2,0)$ Theory}

We may perform the same computation for the $\mathcal N=(2,0)$ theory, noting however that only the 1/2-BPS multiplets have spins less than or equal to two.  In this case, the superconformal algebra decomposes as $OSp(8^*|4)\supset SO(2,6)\times Sp(4)_R\supset SO(2)\times SU(4)\times Sp(4)_R$.  The shortening conditions follow the same pattern as (\ref{eq:1,0short}), however with extended $R$-symmetry \cite{Minwalla:1997ka,Dobrev:2002dt,Bhattacharya:2008zy,Buican:2016hpb,Cordova:2016emh}
\begin{equation}
\begin{tabular}{ll}
$A[j_1,j_2,j_3;k_1,k_2]$:\qquad\hbox{}&$\Delta=\ft12(j_1+2j_2+3j_3)+2(k_1+k_2)+6$,\\
$B[j_1,j_2,0;k_1,k_2]$:&$\Delta=\ft12(j_1+2j_2)+2(k_1+k_2)+4$,\\
$C[j_1,0,0;k_1,k_2]$:&$\Delta=\ft12j_1+2(k_1+k_2)+2$,\\
$D[0,0,0;k_1,k_2]$:&$\Delta=2(k_1+k_2)$.
\end{tabular}
\end{equation}
Here $(k_1,k_2)$ are Dynkin labels for $Sp(4)$, with $(1,0)$ denoting the $\mathbf4$ and $(0,1)$ denoting the $\mathbf5$.  For maximum spin two, we restrict to the 1/2-BPS multiplets $D[0,0,0;0,k]$ with $\Delta=2k$.  (The case $k=1$ is the free tensor multiplet, while $k=2$ is the stress tensor multiplet.)

\begin{table}[t]
\centering
\begin{tabular}{| c | c | c | c | c |}
    \hline
    $\Delta$ & $SU(4)$ & $D[0,0,0;0,2]$ & $D[0,0,0;0,3]$ &  $D[0,0,0;0,k\ge4]$\\
    \hline
    \hline
    $2k$ & $\mathbf1$ & $(0,2)$ & $(0,3)$ & $(0, k)$ \\
    \hline
    $2k + \frac{1}{2}$ & $\mathbf4$ & $(1,1)$ & $(1,2)$ & $(1, k-1)$\\
    \hline
    $2k + 1$ & $\mathbf6$ & $(2,0)$ & $(2,1)$ & $(2, k-2)$\\
            & $\mathbf{10}$ & $(0,1)$ & $(0,2)$ & $(0, k-1)$\\
    \hline
    $2k + \frac{3}{2}$ & $\mathbf{\overline{20}}$ & $(1,0)$ & $(1,1)$ & $(1,k-2)$\\
                    & $\mathbf{\overline4}$ & & $(3,0)$ & $(3,k-3)$\\
    \hline
    $2k + 2$ & $\mathbf{20'}$ & $(0,0)$ & $(0,1)$ & $(0,k-2)$ \\
            & $\mathbf{15}$ & & $(2,0)$ & $(2,k-3)$\\
            & $\mathbf1$ & & & $(4,k-4)$\\
    \hline
    $2k+\frac{5}{2}$& $\mathbf{20}$ & & $(1,0)$ & $(1,k-3)$ \\
                & $\mathbf4$ & & & $(3, k-4)$\\
    \hline
    $2k+3$ & $\mathbf{\overline{10}}$ & & $(0,0)$ & $(0,k-3)$\\
        & $\mathbf6$ & & & $(2,k-4)$ \\
    \hline
    $2k+\frac{7}{2}$& $\mathbf{\overline4}$ & & & $(1,k-4)$\\
    \hline
    $2k+4$ & $\mathbf1$ & & & $(0,k-4)$\\
    \hline
    \hline
    Anomaly & $384\delta(c-a)$ & $13$ & $37$ & $6k(k-1)+1$\\ 
            &$\delta c''$&$0$&$0$&$0$ \\
    \hline
\end{tabular}

\caption{The $\mathcal N=(2,0)$ 1/2-BPS (maximum spin two) representation $D[0,0,0;0,k]$ and corresponding holographic Weyl anomaly coefficients $\delta(c-a)$ and $\delta c''$.  Entries are $Sp(4)_R$ representations specified by Dynkin labels $(k_1,k_2)$.}
\label{tbl:(2,0)}
\end{table}

The holographic computation of $\delta(c-a)$ and $\delta c''$ for the $D[0,0,0;0,k]$ multiplets are shown in Table~\ref{tbl:(2,0)}.  The case $k\ge4$ is generic, and we do not include $k=1$, which is a supersingleton and would not appear in a holographic computation.  The special case $k=3$ fits into the generic pattern. In fact so does $k=2$, although it requires separate treatment because of the presence of massless modes.  For $k=2$, the states in $D[0,0,0;0,2]$ are
\begin{align}
    &D(4;0,0,0)_{\mathbf{14}}+D(4\ft12;1,0,0)_{\mathbf{16}}+D(5;0,1,0)_{\mathbf{10}}\nonumber\\
    &+D(5;2,0,0)_{\mathbf5}+D(5\ft12;1,1,0)_{\mathbf4}+D(6;0,2,0)_{\mathbf1},
\end{align}
where $D(\Delta;j_1,j_2,j_3)$ labels the $SO(2,6)$ representation and the subscript labels the $Sp(4)_R$ representation.  The massless vector, gravitino and graviton representations can be obtained from the corresponding massive representations by subtracting out null states according to
\begin{align}
    D(5;0,1,0)&=D(5+\epsilon;0,1,0)-D(6;0,0,0),\nn\\
    D(5\ft12;1,1,0)&=D(5\ft12+\epsilon;1,1,0)-D(6\ft12;1,0,0),\nn\\
    D(6;0,2,0)&=D(6+\epsilon;0,2,0)-D(7;0,1,0).
\end{align}
(Note that the three-form, $D(5;2,0,0)$, is massive, so no subtraction is required.) Taking these null states into account then gives the result $\delta(c-a)=13/384$ for $k=2$ shown in Table~\ref{tbl:(2,0)}.

Although $k=2$ and $k=3$ are special cases, the holographic anomaly coefficient $\delta(c-a)=(1/384)(6k(k-1)+1)$ is in fact universal.  Combining this with $\delta a=-(7/1152)(6k(k-1)+1)$ obtained in \cite{Beccaria:2014qea} then allows us to separate out the individual coefficients
\begin{equation}
    D[0,0,0;0,k\ge2]:\qquad\delta a=-\fft1{288}\cdot\fft74\left(6k(k-1)+1\right),\qquad\delta c=-\fft1{288}\left(6k(k-1)+1\right).
\end{equation}

As an application, consider the $\mathcal N=(2,0)$ theory obtained by compactifying 11-dimensional supergravity on AdS$_7\times S^4$.  The Kaluza-Klein spectrum is simply
\begin{equation}
\oplus_{k\ge2}D[0,0,0;0,k],
\end{equation}
where $k=2$ corresponds to the `massless' supergravity sector.
The anomaly coefficients $\delta a$ and $\delta c$ may be computed by summing over the Kaluza-Klein levels
\begin{equation}
     \delta a=-\fft1{288}\cdot\fft74\sum_{k=2}^\infty(6k(k-1)+1),\qquad
     \delta c=-\fft1{288}\sum_{k=2}^\infty(6k(k-1)+1).
\end{equation}
Following \cite{Beccaria:2014qea}, we regulate the sums using a hard cutoff. This amounts to setting $\sum_{k = 1}^{\infty} k^n = 0$ for any $n \geq0$. This implies $\sum_{k = 2}^\infty f(k)=-f(1)$, where $f(k)$ is polynomial in $k$.  As a result, the regulated anomaly for AdS$_7\times S^4$ is
\begin{equation}
     \delta a=\fft1{288}\cdot\fft74,\qquad\delta c=\fft1{288}.
\end{equation}
This is equal and opposite to the result for the conformal anomaly of the free tensor multiplet computed in \cite{Tseytlin:2000sf}, and agrees with the $\mathcal O(1)$ contributions in (\ref{eq:M5ac}).

\section{Central Charges from the The Superconformal Index}

We now turn to the second major theme of this chapter, which is how to obtain these corrections using differential operators acting on the superconformal index.

\subsection{The Superconformal Index for the $(1,0)$ Theory}

The 4-dimensional superconformal index was introduced in \cite{Romelsberger:2005eg,Kinney:2005ej} and generalized to additional dimensions in \cite{Bhattacharya:2008zy}.  Before discussing the index, we first briefly review the $\mathcal N=(1,0)$ theory.  Six dimensions is the highest dimension that admits superconformal symmetry, and $(1,0)$ supersymmetry is minimal.  The superconformal algebra decomposes as $OSp(8^*|2)\supset SO(2,6)\times SU(2)_R\supset U(1)_\Delta\times SU(4)\times SU(2)_R$.  Unitary representations may be labeled by conformal dimension $\Delta$, $SU(4)$ Dynkin labels $(j_1,j_2,j_3)$ and the $SU(2)_R$ label $k$ (with `spin' $k/2$).

Long representations of $(1,0)$ have $\Delta>\fft12(j_1+2j_2+3j_3)+2k+6$, while short representations fall into four categories, comprising one regular and three isolated short multiplets.  The shortening conditions are given by \cite{Bhattacharya:2008zy,Buican:2016hpb,Cordova:2016emh}
\begin{align}
    A[j_1,j_2,j_3;k]:&&&\Delta=\ft12(j_1+2j_2+3j_3)+2k+6,\nn\\
    B[j_1,j_2,0;k]:&&&\Delta=\ft12(j_1+2j_2)+2k+4,\nn\\
    C[j_1,0,0;k]:&&&\Delta=\ft12j_1+2k+2,\nn\\
    D[0,0,0;k]:&&&\Delta=2k.
    \label{eq:10short}
\end{align}
Long representations are generated by the action of all 16 real supercharges and have states with dimensions ranging from $\Delta$ to $\Delta+4$, while the successive shortened representations generically have dimensions going up to $\Delta+7/2$, $\Delta+3$, $\Delta+5/2$ and $\Delta+2$, respectively.  The latter $D$ multiplets are generated by eight supercharges and are half-BPS.

We now turn to the 6-dimensional $(1,0)$ index, which was introduced in \cite{Bhattacharya:2008zy} as
\begin{align}
    \mathcal{I}(p,q,s) = \mathrm{Tr}_{\mathcal{H}}(-1)^{j_1 + j_3}e^{- \beta \delta} q^{\Delta - \frac{1}{2}k}s^{j_1}p^{j_2} ,
    \label{eq:index}
\end{align}
where $\delta = \Delta - 2k - \frac{1}{2}(j_1 + 2 j_2 + 3 j_3)$.  Recall here that $(j_1,j_2,j_3)$ labels the $SU(4)$ Lorentz representation, and that $j_1+j+3$ represents the fermion number.  In particular, the index is a Witten index refined by fugacities $q$, $s$ and $p$ associated with the charges $\Delta-k/2$, $j_1$ and $j_2$ that commute with the supercharge $\mathcal Q$ used to define the index.  While the trace is {\it a priori} over all states in the spectrum, only those satisfying $\delta=0$ will contribute.  Thus the index is actually independent of $\beta$, and only receives contributions from shortened multiplets.

Since we are motivated by the holographic dual, our main interest is on the single-trace index, which corresponds to the single particle spectrum.  In this case, the expression (\ref{eq:index}) has a particularly simple form.  To see this, we first note that the charges $j_1$ and $j_2$ in (\ref{eq:index}) are $SU(3)$ weights corresponding to the breaking of $SU(4)$ by the defining supercharge $\mathcal Q$.  As a result, the index can be decomposed as a sum over $SU(3)$ characters $\chi_{(j_1,j_2)}(s,p)$ given by the Weyl character formula:
\begin{equation}
\chi_{(j_1,j_2)}(s,p)= \frac{\begin{pmatrix}{s^{j_1+1}p^{j_2+1}-s^{-j_2-1}p^{-j_1-1}  
    + s^{j_2+1}p^{-j_1-j_2-2}-s^{j_1+j_2+2}p^{-j_2-1} }\\{
    +s^{-j_1-j_2-2}p^{j_1+1} 
    -s^{-j_1-1}p^{j_1+j_2+2}} \end{pmatrix}}{\left(\sqrt{sp }-\ffd{1}{\sqrt{ sp}}\right)\left(\ffd{s}{\sqrt{p}}-\ffd{\sqrt{p}}{s}\right)\left(\ffd{p}{\sqrt{s}}-\ffd{\sqrt{s}}{p}\right) 
   }
\end{equation}
Moreover, for a given representation, the index receives contributions from both superconformal primaries and their descendants.  The contributions from the latter are captured by the denominator factor
\begin{equation}
    \fft1{\mathcal D(p,q,s)}=\fft1{(1-qs^{-1})(1-qp)(1-qs/p)}=1+q\chi_{(0,1)}(s,p)+q^2\chi_{(0,2)}(s,p)+\cdots.
\end{equation}
As a result, the single-trace index for a given short representation takes the form
\begin{equation}
    \mathcal I(p,q,s)\sim q^{\Delta-\fft{k}2}\fft{\chi(s,p)}{\mathcal D(p,q,s)},
\end{equation}
for some appropriate $SU(3)$ character $\chi(s,p)$.  The indices were worked out on a representation by representation basis in \cite{Buican:2016hpb}, and we summarize the results in Table~\ref{tb:index}.

\begin{table}[t]
\begin{center}
  \begin{tabular}{| l | l | l |}
    \hline
    \ Multiplet & \ Shortening Condition & \ $\mathcal D(p,q,s)\mathcal{I}_R(p,q,s)$ \ \\
    \hline
    \hline
    \ $A[j_1,j_2,j_3;k]$ \ & \ $\Delta=\ft12(j_1+2j_2+3j_3)+2k+6$ \ & \ $(-1)^{j_1 + j_3+1} \, q^{\Delta-\fft12k} \, \chi_{(j_1, j_2)}(s,p)$ \ \\ \hline
    \ $B[j_1,j_2,0;k]$ \ & \ $\Delta=\ft12(j_1+2j_2)+2k+4$ & \ $(-1)^{j_1} \, q^{\Delta-\fft12k} \, \chi_{(j_1, j_2+1)}(s,p)$ \ \\ \hline
    \ $C[j_1,0,0;k]$ \ & \ $ \Delta=\ft12j_1+2k+2$ & \ $(-1)^{j_1+1} \, q^{\Delta-\fft12k} \, \chi_{(j_1+1, 0)}(s,p)$ \ \\ \hline
    \ $D[0,0,0;k]$ \ & \ $\Delta=2k$ & \ $ \,  q^{\Delta-\fft12k} \, \chi_{(0, 0)}(s,p)$ \ \\ \hline
  \end{tabular}
\end{center}
\caption{Contribution to the single-trace index for $(1,0)$ short multiplets with Dynkin labels $(j_1, j_2, j_3)$, conformal weight $\Delta$, and $R$-charge $k$.  Here we are taking generic values for $j_1,j_2,j_3$ and $k$; some special cases arise at small values of the quantum numbers.
\label{tb:index}}
\end{table}


\subsection{Central Charges from the Index}

For $(1,0)$ theories with a large-$N$ dual, we generally expect the central charges to scale as $\mathcal O(N^3)$.  Holographically, the leading contribution comes from the tree-level bulk action \cite{Henningson:1998gx}. Sub-leading terms of $\mathcal O(N)$ arise from $\alpha'^3R^4$ corrections and terms of $\mathcal O(1)$ from the one-loop determinant.  It is the latter terms that we focus on.

\subsubsection{A Differential Operator for $\delta a$}

We first examine the $\mathcal O(1)$ contribution $\delta a$ to the $a$ central charge.  This was evaluated in \cite{Beccaria:2014qea} for an arbitrary representation of the $SO(2,6)$ conformal group labeled by $D(\Delta,j_1,j_2,j_3)$ by computing the heat kernel group theoretically on global AdS$_7$. The result can be expressed as
\begin{align}
    \delta a(\Delta,j_1,j_2,j_3)&=\fft{(-1)^{j_1+j_3}(\Delta-3)}{2^5\cdot6!}\biggl[\fft1{21}(\Delta-3)^6d(j_1,j_2,j_3)\nn\\
    &\kern4em-(\Delta-3)^4\left(I_2(j_1,j_2,j_3)+\fft13d(j_1,j_2,j_3)\right)\nn\\
    &\kern4em+(\Delta-3)^2\biggl(\fft{70}{51}I_4(j_1,j_2,j_3)+\fft{75}{17}\fft{I_2(j_1,j_2,j_3)^2}{d(j_1,j_2,j_3)}+\fft{50}{17}I_2(j_1,j_2,j_3)\nn\\
    &\kern9em+\fft49d(j_1,j_2,j_3)\biggr)
    -\fft{75}4\fft{I_3(j_1,j_2,j_3)}{d(j_1,j_2,j_3)}\biggr],
\label{eq:btda}
\end{align}
where the sign factor $(-1)^F=(-1)^{j_1+j_3}$ distinguishes between bosons and fermions.  Here we have rewritten the expression of \cite{Beccaria:2014qea} in terms of $SU(4)$ invariants where
\begin{equation}
    d(j_1,j_2,j_3)=\fft1{12}(j_1+1)(j_2+1)(j_3+1)(j_1+j_2+2)(j_2+j_3+2)(j_1+j_2+j_3+3),
\end{equation}
is the dimension of the representation and the $I_a$'s are indices
\begin{align}
    I_2(j_1,j_2,j_3)&=\fft1{60}d(j_1,j_2,j_3)[3j_1^2+12j_1+4j_1j_2+2j_1j_3+4j_2^2+4j_2j_3+16j_2+3j_3^2+12j_3],\nn\\
    I_3(j_1,j_2,j_3)&=\fft1{60}d(j_1,j_2,j_3)(j_1-j_3)(j_1+j_3+2)(j_1+2j_2+j_3+4),\nn\\
    I_4(j_1,j_2,j_3)&=\fft1{420}d(j_1,j_2,j_3)[3 j_1^4 + 8 j_1^3 j_2 + 2 j_1^2 j_2^2 - 12 j_1 j_2^3 - 6 j_2^4 + 4 j_1^3 j_3 + 2 j_1^2 j_2 j_3\nn\\
    &\kern4em- 18 j_1 j_2^2 j_3 - 12 j_2^3 j_3 - 4 j_1^2 j_3^2 + 2 j_1 j_2 j_3^2 + 2 j_2^2 j_3^2 + 4 j_1 j_3^3 + 8 j_2 j_3^3 + 3 j_3^4\nn\\
    &\kern4em+24 j_1^3 + 30 j_1^2 j_2 - 50 j_1 j_2^2 - 48 j_2^3 + 6 j_1^2 j_3 - 28 j_1 j_2 j_3 - 50 j_2^2 j_3 + 6 j_1 j_3^2\nn\\
    &\kern4em+ 30 j_2 j_3^2 + 24 j_3^3
    +54 j_1^2 - 34 j_1 j_2 - 122 j_2^2 - 2 j_1 j_3 - 34 j_2 j_3 +  54 j_3^2\nn\\
    &\kern4em+24 j_1 - 104 j_2 + 24 j_3
 ],
\end{align}
normalized to unity for the fundamental $(1,0,0)$ representation.

For the $(1,0)$ superconformal case, we compute the shift $\delta a$ for each supermultiplet by summing (\ref{eq:btda}) over the individual states comprising the representation.  The multiplet structure has been worked out explicitly in \cite{Buican:2016hpb,Cordova:2016emh}, and using those results, we may obtain $\delta a$ for each type of shortened multiplet given in (\ref{eq:10short}):
\begin{equation}
    \delta a=\begin{cases}
    (-1)^{j_1+j_3+1}\mathcal A(j_1,j_2,\Delta-\ft12k),&A[j_1,j_2,j_3;k];\nn\\
    (-1)^{j_1}\mathcal A(j_1,j_2+1,\Delta-\ft12k),&B[j_1,j_2,0;k];\nn\\
    (-1)^{j_1+1}\mathcal A(j_1+1,0,\Delta-\ft12k),&C[j_1,0,0;k];\nn\\
   \mathcal A(0,0,\Delta-\ft12k),&D[0,0,0;k].
    \end{cases}
    \label{eq:hda}
\end{equation}
Here $\mathcal A(j_1,j_2,\hat\Delta)$ has the universal form
\begin{align}
    2^5\cdot6!\mathcal A(j_1,j_2,\hat\Delta)&=-10\left(\fft43\hat\Delta-2\right)^4 d(j_1,j_2)+20\left(\fft43\hat\Delta-2\right)^2[4I_2(j_1,j_2)+d(j_1,j_2)]\nn\\
    &\quad+\fft{530}9\left(\fft43\hat\Delta-2\right)I_3(j_1,j_2)-\fft{80}9[I_{2,2}(j_1,j_2)+3I_2(j_1,j_2)]-\fft{11}3d(j_1,j_2),
    \label{eq:calA}
\end{align}
where
\begin{equation}
    d(j_1,j_2)=\ft12(j_1+1)(j_2+1)(j_1+j_2+2),
\end{equation}
is the dimension of the $SU(3)$ representation and the $I_a$'s are indices
\begin{align}
    I_2(j_1,j_2)&=\fft1{12}d(j_1,j_2)[j_1^2+3j_1+j_1j_2+j_2^2+3j_2],\nn\\
    I_3(j_1,j_2)&=\fft1{60}d(j_1,j_2)(j_1-j_2)(j_1+2j_2+3)(2j_1+j_2+3),\nn\\
    I_{2,2}(j_1,j_2)&=\fft35I_2(j_1,j_2)\left(8\fft{I_2(j_1,j_2)}{d(j_1,j_2)}-1\right),
    \label{eq:SU3index}
\end{align}
normalized to unity for the fundamental $(1,0)$ representation.  Since $SU(3)$ has rank two, it only has two independent Casimir invariants, with corresponding indices $I_2$ and $I_3$.  Therefore the fourth order index $I_{2,2}$ is not independent, but can be decomposed in terms of $I_2$ as indicated above.

It is now apparent that the structure of the holographic $\delta a$ in (\ref{eq:hda}) closely resembles that of the single-trace index as shown in Table~\ref{tb:index}.  This connection can be made precise by associating the factor $q^{\Delta-\fft12k}\chi_{(j_1,j_2)}(s,p)$ in the index with the anomaly function $\mathcal A(j_1,j_2,\Delta-\fft12)$.  This is easily done once we realize that the indices can be obtained from the $SU(3)$ character $\chi_{(j_1,j_2)}(s,p)$.  The relation is not unique, but one possibility is to take
\begin{align}
    d(j_1,j_2)&=\chi_{(j_1,j_2)}(s,p)\big|_{s=p=1},\nn\\
    I_2(j_1,j_2)&=\ft{1}{2}
    (s \partial_s)^2 \chi_{(j_1, j_2)}(s,p)\big|_{s=p=1},\nn \\
    I_3(j_1,j_2)&= (p \partial_p)(s \partial_s)^2 \chi_{(j_1, j_2)}(s,p)\big|_{s=p=1},\nn\\
    I_{2,2}(j_1,j_2)&=\ft{1}{2}(s\partial_s)^4 \chi_{(j_1, j_2)}(s,p)\big|_{s=p=1}.
    \label{eq:Ichi}
\end{align}
The reason we have left $I_{2,2}$ in the $\delta a$ expression (\ref{eq:calA}) is now apparent, as it can be obtained directly from the character as opposed to the square of $I_2$.

Combining the above observations, we are now led to the final expression relating $\delta a$ to the single-trace index
\begin{align}
    \delta a&=\fft1{2^5\cdot6!}\biggl[-10\left(\fft43q\partial_q-2\right)^4+20\left(\fft43q\partial_q-2\right)^2(4\hat I_2+1)+\fft{530}9\left(\fft43q\partial_q-2\right)\hat I_3\nn\\
    &\kern4.3em-\fft{80}9(\hat I_{2,2}+3\hat I_2)-\fft{11}3\biggr]\mathcal D(p,q,s)\mathcal I(p,q,s)\bigg|_{p=q=s=1}.
    \label{eq:afromI}
\end{align}
Here the $\hat I_a$'s correspond to the differential operators used in (\ref{eq:Ichi}) to obtain the indices from the group character.

\subsubsection{A Differential Operator for $\delta (c-a)$}

We now turn to consideration of holographic $\delta(c-a)$.  So far, this has only been worked out for maximum spin-two multiplets, so the information is necessarily incomplete.  Nevertheless, there is still a useful connection to be made, and the data is shown in Table~\ref{tb:c-aindex}.  Noting that the relevant $SU(3)$ representations are the singlet, triplet and anti-triplet, and that the indices, (\ref{eq:SU3index}), are normalized to unity for the triplet, we obtain the expression
\begin{equation}
    \delta(c-a)=\fft1{2^5\cdot6!}\left[-90\left(\fft43q\partial_q-2\right)\hat I_3+1+\lambda(\hat I_{2,2}-\hat I_2)\right]\mathcal D(p,q,s)\mathcal I(p,q,s)\bigg|_{p=q=s=1},
    \label{eq:c-afromI}
\end{equation}
where $\lambda$ is an undetermined constant.  This ambiguity arises because the combination $I_{2,2}-I_2$ vanishes for the singlet and (anti-)triplet representations.

\begin{table}[t]
\begin{center}
  \begin{tabular}{| l | l | l | l |}
    \hline
    \ Multiplet & \ Shortening Condition\ & \ $\mathcal D(p,q,s)\mathcal{I}_R(p,q,s)$\ &\ $2^5\cdot6!\delta(c-a)$ \\
    \hline
    \hline
    \ $A[0,0,0;k]$ \ & \ $\Delta=2k+6$ \ & \ $-q^{\hat\Delta} \, \chi_{(0,0)}(s,p)$ \ &$-1$\\ \hline
    \ $B[0,0,0;k]$ \ & \ $\Delta=2k+4$ & \ $q^{\hat\Delta} \, \chi_{(0,1)}(s,p)$ \ &\ $3+90(\fft43\hat\Delta-2)$\ \\ \hline
    \ $C[0,0,0;k]$ \ & \ $ \Delta=2k+2$ & \ $- q^{\hat\Delta} \, \chi_{(1, 0)}(s,p)$ \ &\ $-3+90(\fft43\hat\Delta-2)$\ \\ \hline
    \ $D[0,0,0;k]$ \ & \ $\Delta=2k$ & \ $q^{\hat\Delta} \, \chi_{(0, 0)}(s,p)$ \ &\ $1$\ \\ \hline
  \end{tabular}
\end{center}
\caption{The single-trace index and holographic $\delta(c-a)$ for maximum spin-two $(1,0)$ short multiplets.  The $\delta(c-a)$ results are taken from \cite{Liu:2017ruz}, but are given here in terms of $\hat\Delta\equiv\Delta-\fft12k$.
\label{tb:c-aindex}}
\end{table}

\section{Remarks}

In this chapter, we have computed the one-loop correction to $(c - a)$, and we have shown how to obtain this quantity and the corrections to $a$ from the superconformal index. A few comments are now in order. 

\subsection{Applicability of our Prescription}

We have used the functional Schr\"odinger method of  \cite{Mansfield:1999kk,Mansfield:2000zw,Mansfield:2002pa,Mansfield:2003gs}. It is reasonable to question whether the use of (\ref{eq:MN}) is valid, as it disagrees with the direct computation of $\delta a$ performed in \cite{Beccaria:2014qea,Beccaria:2014xda}.  A quick way to see this is to note that $\delta a$ in Table~\ref{tbl:10ac} is a fourth order polynomial in $\Delta$, while the result of summing (\ref{eq:MN}) over a supermultiplet can be at most quadratic in $\Delta$.  (One power comes directly from (\ref{eq:MN}), while another can arise from the dimension of the shortened representation.)  If $\delta(c-a)$ was expected to be cubic or higher in $\Delta$, then our result, as shown in the last column of Table~\ref{tbl:10ac}, cannot possibly be correct. 

However, $c-a$ can be at most linear in $\Delta$, which is consistent with application of (\ref{eq:MN}). To see this, recall that, in superconformal field theories, the stress tensor is contained in a multiplet of currents, so that there is a corresponding multiplet of anomalies. For $\mathcal N=(1,0)$ theory, the `t~Hooft anomalies are characterized by the anomaly polynomial
\begin{equation}
    \mathcal I_8=\fft1{4!}[\alpha c_2(R)^2+\beta c_2(R)p_1(T)+\gamma p_1(T)^2+\delta p_2(T)],
\end{equation}
and the relation to the Weyl anomaly coefficients has recently been worked out \cite{Cordova:2015fha,Beccaria:2015ypa,Yankielowicz:2017xkf,Beccaria:2017dmw}
\begin{equation}
    a=-\fft1{72}(\alpha-\beta+\gamma+\ft38\delta),\qquad c-a=-\fft\delta{192},\qquad c'=\fft1{432}(\beta-2\gamma+\ft12\delta).
\end{equation}
Since $\alpha$ is the coefficient of the $[SU(2)_R]^4$ anomaly, it can be at most fifth power in $\Delta$, where the extra power comes from the dimension of the representation.  Similarly, $\beta$ can be at most cubic in $\Delta$, while $\gamma$ and $\delta$ can be at most linear in $\Delta$.  This in turn demonstrates that $a$ will be at most fifth power in $\Delta$, $c'$ will be at most cubic and $c-a$ will be at most linear.  Thus the functional Schr\"odinger method is indeed compatible with $\delta(c-a)$.  However, we also see this approach cannot be used to compute either $\delta a$ alone or $\delta c'$. 

\subsection{Possibility of Higher Spin}

While we have focused on short multiplets with spins $\le2$, it would be desirable to work more generally with higher-spin multiplets.  To do so, we would need knowledge of the $b_6$ coefficients for arbitrary spin fields.  This in turn depends on the form of the higher-spin Laplacian.  In general, this depends on the bulk dynamics of the higher-spin field and the further restriction to the boundary following from the procedure of \cite{Mansfield:1999kk,Mansfield:2000zw,Mansfield:2002pa,Mansfield:2003gs}.  For higher-spin bosons, it is natural to take a bulk Laplacian of the form $\Delta=-\Box-E$ with the endomorphism $E=\Sigma_{ab}R^{abcd}\Sigma_{cd}$, where $\Sigma_{ab}$ are $SU(4)$ generators in the appropriate bosonic higher-spin representation.  However, the situation is less clear for fermions.  The natural generalization would be to simply take $\Sigma_{ab}$ to be in a fermionic higher-spin representation.  However, this does not agree with the square of the Dirac operator for ordinary spin-1/2 fermions.  Nevertheless, it is possible that the use of a universal endomorphism term for bosons and fermions would be appropriate when tracing over supermultiplets.  Along these lines, we have computed the $b_6$ coefficient for general higher-spin representations in Appendix~\ref{app:B_generalheatkernel}.

\subsection{High Temperature Limit of the Index}

In holographic 6d SCFTs, the leading order behavior of the central charges scales as $N^3$, and the first subleading corrections arise at $\mathcal O(N)$.  So in practice the $\mathcal O(1)$ terms that we have identified from the single-trace index are rather small corrections.  Nevertheless, their structure can provide a hint at a more complete relationship between the full index and central charges.  The full index, of course, differs from the single-trace index, but can be related through the plethystic exponential.  As in the AdS$_5$/CFT$_4$ case considered previously \cite{Ardehali:2014zba,Ardehali:2014esa}, we expect that the connection of $\delta a$ and $\delta(c-a)$ to the single-trace index generalizes in terms of the high-temperature structure of the full index \cite{DiPietro:2014bca,Ardehali:2015hya,Assel:2015nca,Ardehali:2015bla,DiPietro:2016ond}.

What we mean here by the high-temperature limit comes from the connection between the superconformal index and the supersymmetric partition function on $S^n\times S^1$ \cite{Lorenzen:2014pna,Assel:2015nca}:
\begin{equation}
    \mathcal I(\beta)=e^{\beta E_{\mathrm{susy}}}Z_{S^n\times S^1_\beta},
\end{equation}
where $E_{\mathrm{susy}}$ is the supersymmetric Casimir energy and the inverse temperature $\beta$ is associated with the radius of $S^1$.  As highlighted in \cite{DiPietro:2014bca,Assel:2014paa,Ardehali:2015hya}, the 4-dimensional index has a high-temperature expansion of the form
\begin{equation}
    \log\mathcal I(\beta)\sim\fft{16\pi^2(c-a)'}{3\beta}-4(2a-c)\log\left(\fft\beta{2\pi}\right)+\fft{4(3c+a)\beta}{27}+\cdots,
\end{equation}
where the prime denotes a possible shift related to the displacement of the minimum of the effective potential away from the origin \cite{Ardehali:2015bla,DiPietro:2016ond}.  The linear term in $\beta$ is the 4-dimensional supersymmetric Casimir energy, and when generalized to the squashed sphere is connected to the holographic one-loop computation of $\delta a$ and $\delta c$ \cite{Ardehali:2015hya}.

In six dimensions, the high-temperature expansion of the index instead takes the form \cite{DiPietro:2014bca}
\begin{equation}
    \log\mathcal I(\beta)\sim\fft{8\pi^4}{9\beta^3}C_0+\fft{\pi^2}{6\beta}C_1+\cdots+\beta E_{\mathrm{susy}}+\ldots,
\end{equation}
where it was suggested that the factors $C_0$ and $C_1$ are related to the 't~Hooft anomaly coefficients
\begin{equation}
    \mathcal I_8=\fft1{4!}[\alpha c_2(R)^2+\beta c_2(R)p_1(T)+\gamma p_1(T)^2+\delta p_2(T)],
\end{equation}
by
\begin{equation}
    C_0=\gamma+\ft14\delta,\qquad C_1=\ft92\beta-8\gamma+\delta.
\end{equation}
While the holographic $\delta a$ and $\delta(c-a)$ are related to $E_{\mathrm{susy}}$, and therefore do not constrain $C_0$ and $C_1$, one may hope that aspects of the holographic dual can nevertheless refine our understanding of these terms.  In any case, we note that, while $C_0$ receives non-vanishing contributions from free $(1,0)$ scalar and tensor multiplets \cite{Benvenuti:2016dcs,Bak:2016vpi,Gustavsson:2018sgi}, it nevertheless vanishes in the $(2,0)$ theory \cite{Bhattacharya:2008zy,Kim:2012qf,Kim:2013nva,DiPietro:2014bca,Kim:2016usy}.  This leaves us with the question of whether any additional meaning can be attributed to $C_0$.  One way to distinguish $(1,0)$ from $(2,0)$ theories is the vanishing of the $c'$ central charge in the latter.  However, the relation \cite{Cordova:2015fha,Beccaria:2015ypa,Yankielowicz:2017xkf,Beccaria:2017dmw}
\begin{equation}
    a=-\fft1{72}(\alpha-\beta+\gamma+\ft38\delta),\qquad c-a=-\fft\delta{192},\qquad c'=\fft1{432}(\beta-2\gamma+\ft12\delta),
\end{equation}
demonstrates that this cannot be the complete story.  Likewise, the relation between $C_1$ and the central charges is not clear either.  These issues merit further study, as their resolution will lead to a deeper understanding of 6-dimensional SCFTs.

%% file: Chap3/chap3.tex
\section{Review: Type IIB Supergravity}

So far in this dissertation, we have studied quantum field theory on fixed backgrounds, and we have considered supersymmetry to be a global transformation. If we allow for local supersymmetry transformations, then we are forced to allow the background manifold to fluctuate. The result is a theory of gravity. This is roughly because the commutation relations of supersymmetry transformation include spacetime symmetries, so local supersymmetry forces the spacetime symmetries to be local as well. Invariance under local spacetime symmetries requires that the metric transform as a fluctuating gauge field just as invariance under a local $U(1)$ symmetry requires a field $A_{\mu}$. For these reasons, the theories with local supersymmetry transformations are called theories of \textit{supergravity}. 

\subsection{Type IIB Supergravity}

There are a number of supergravity theories, depending on the number of dimensions and supersymmetries, and the type of matter fields that are included. Here we will review one example that is particularly interesting due to its role in the AdS/CFT correspondence: type IIB supergravity. This 10-dimensional theory has two chiral supercharges (unlike its partner type IIA, which has a left-handed and a right-handed supercharge). The minimal spinor representations in 9+1 dimensions are 16-dimensional, so type IIB supergravity is a theory with maximal supersymmetry-- that is, it has 32 supercharges. 

The theory can be obtained as the low-energy limit of type IIB string theory. The fermionic part includes two gravitinos and two dilatinos. The bosonic matter content is broken into two sectors based on the periodicity of fields defining them in the string theory. The NS-NS fields are the metric $g_{\mu\nu}$, a two-form $B_{\mu\nu}$ and the dilaton $\phi$. The R-R fields are the axion $\chi\equiv C_0$, another two-form $C_2$ and a four-form $C_4$ with a self-dual field strength. The field strengths obtained from these potentials are defined by:
\begin{align}
    F_1 &= d \chi, \qquad  \qquad H_3 = d B_2, \qquad  \qquad F_3 = dC_2 \, - \chi H_3,\nn \\
    F_5 &= d C_4 - \frac{1}{2} \, ( \, C_2 \wedge \, H_3 \, - B_2 \, \wedge \, d C_2 \, ),
\end{align}
where $F_5=*F_5$.  The Bianchi identities then follow:
\begin{align}
    d F_1 &= 0, \qquad \qquad d F_3 - H_3 \wedge F_1 = 0,\nn \\
    d H_3 &= 0, \qquad \qquad d F_5 - H_3 \wedge F_3 = 0 \, .
\end{align}

The type IIB supergravity equations of motion cannot be derived from a covariant action because the self-duality of $F_5$ implies that its kinetic term vanishes.  However, the equations are known.  The form-field equations are
\begin{align}
    d( e^{2 \phi} * F_1) &=- e^{\phi} H_3 \wedge * F_3,\nn\\
    d*d\phi &= e^{2 \phi} F_1 \wedge * F_1 - \ft12 e^{- \phi} H_3 \wedge * H_3+\ft12 e^{\phi} F_3 \wedge * F_3,\nn \\
    d( e^{- \phi} * H_3) &=e^{\phi} F_1 \wedge * F_3 +F_3 \wedge F_5,\nn\\
    d( e^{\phi} * F_3) &=-H_3 \wedge F_5,
\end{align}
and the Einstein equation in Ricci form is
\begin{align}
    R_{\mu\nu}&=\fft12\partial_\mu\phi\partial_\nu\phi+\fft12e^{2\phi}\partial_\mu\chi\partial_\nu\chi+\fft14e^{-\phi}\left(H_{\mu\rho\sigma}H_\nu{}^{\rho\sigma}-\fft1{12}g_{\mu\nu}H_{\lambda\rho\sigma}H^{\lambda\rho\sigma}\right)\nn\\
    &\qquad+\fft14e^\phi\left(F_{\mu\rho\sigma}F_\nu{}^{\rho\sigma}-\fft1{12}g_{\mu\nu}F_{\lambda\rho\sigma}F^{\lambda\rho\sigma}\right)+\fft1{4\cdot4!}F_{\mu\lambda\rho\sigma\tau}F_\nu{}^{\lambda\rho\sigma\tau}.
\end{align}
Note that we did not include the equations of motion for the fermions. This is because for our purposes (and almost all purposes) the equations of motion are of interest because we are interested in their solutions. These solutions make up the classical backgrounds of the theory. Classical backgrounds always have vanishing fermionic fields. 

\subsection{Kalua-Klein Reduction and Consistent Truncations}

Type IIB string theory (and its low energy limit, supergravity) is dual to $\mathcal{N} = 4$ SYM in four dimensions. But AdS/CFT is often cited as a duality between gravitational theories in $d+1$ dimensions and conformal field theories in $d$ dimensions. The resolution to this apparent tension lies in KK-reduction, whereby a field on a non-compact times a compact manifold are represented as an infinite tower of fields on only the non-compact manifold. For the simplest example of this, consider a massless scalar field in 5 dimensions, $\{x_0, ..., x_4\}$.
\begin{align}
    \Phi = \Phi(x_0, x_1, x_2, x_3, x_4) \, .
\end{align}
But now imagine that we \textit{compactify} the last dimension, which we will now call $y$, on a circle of radius $R$. Then the dependence of the function in the $y$-direction becomes periodic, and we express it as a Fourier series:
\begin{align}
    \Phi = \sum_{n = 0} \phi_n (x_0, x_1, x_2, x_3) \, e^{2 \pi i n y / R } \, .
    \label{KKred}
\end{align}
We have now replaced the 5-dimensional field $\Phi$ with an infinite number of four-dimensional fields $\phi_n$. The 5-dimensional Klein-Gordon equation reads
\begin{align}
    \begin{split}
        & \Box_5 \Phi = (\Box_4 + \partial_y \partial^y) \sum_n \phi_n (x_0, x_1, x_2, x_3) \, e^{2 \pi i n y / R } = 0 \, , \\   
        &   \qquad \implies   \sum_n   \left( \Box_4 - (2 \pi n / R)^2 \right) \phi_n (x_0, x_1, x_2, x_3) \, e^{2 \pi i n y / R } = 0 \, .
    \end{split}
\end{align}
Since these Fourier modes are all independent, this implies that all of these terms are separately zero, so we get a Klein-Gordon equation for each mode:
\begin{align}
    \left( \Box_4 - (2 \pi n / R)^2 \right) \phi_n = 0 \, .
\end{align}
From the equations of motion of our tower of fields, we find that our 5-dimensional field $\Phi$ has broken down into a single massless scalar $\phi_0$ plus an infinite number of scalars $\phi_n$ with mass $m = 2 \pi n / R$. This is the idea of Kaluza-Klein (KK) analysis. The same may be performed on other compact manifolds-- however, the Fourier analysis part may be very difficult, as the spectrum of the Laplacian is required. For higher-dimensional spheres, however, the generalization is fairly straightforward, and the internal part of the KK modes are the spherical harmonics. 

Now we will introduce a few more terms. \textit{Consistent truncation} refers to throwing out fields from a field theory in such a way that the remaining fields do not source the removed fields. For an extremely elementary example, consider a theory with two massless fields
\begin{align}
    \mathcal{L} = -(\partial a)^2 -(\partial b)^2 + a^2 b
\end{align}
The equations of motion are 
\begin{align}
    \Box a  &= 2 a b \\
    \Box b &= a^2
\end{align}
In this example, we can consistently truncate $a$ because if we set it to zero, then $a$ has no source. But we can not consistently truncate $b$ because it is sourced by $a$. This is equivalent to saying: if we set $a = 0$ from the outset, it will stay off, because it will have no source. But $b$ will not stay equal to zero as long as $a$ remains. 

Finally, we introduce the idea of \textit{dimensional reduction}. This is when we consistently truncate all but the massless mode in the KK analysis. Getting rid of the higher modes might be well justified if we are working at energies far beneath the energy of the first excited mode. This goes like $1 / R$ in our simple circle example, and generally it is inversely proportional to the size of the compact manifold. Keeping only the massless mode truly does reduce the number of dimensions: from the (\ref{KKred}), it is easy to see that, because $n = 0$, the massless mode does not include any dependence on $y$. Only the higher modes fluctuate in the $y$-direction. Dimensional reductions are a subset of consistent truncations, which may or may not reduce the number of dimensions. 

In practice, dimensional reductions are constructed using a reduction ansatz, which is a solution for the higher-dimensional fields in terms of the lower-dimensional ones. Consistency is checked by ensuring that the lower-dimensional equations of motion imply to the  higher-dimensional ones. In our simple example above, we could say that the reduction ansatz is $a = 0$, $b =b'$, where $b'$ is a massless scalar. Then it is clear that any solution of $\Box b' = 0$ is also a solution of the untruncated equations. Doing the same for $a = a'$, $b =0$ does not result in a solution to the original equations.

\subsection{$\beta$-deformations and the Lunin-Maldacena background}

Let us now turn to the dual theory of IIB string theory / supergravity:  $\mathcal{N} = 4$ super Yang-Mills. As we have mentioned before, $\mathcal{N} = 4$ SYM is a conformal field theory. One natural question about conformal field theories is whether they come in continuous families (perhaps parameterized by continuous parameters) or if they are isolated points in the space of theories. For some conformal theories, the continuous families of theories are described by extra operators that can be added to the Lagrangian. Such operators are called \textit{marginal deformations}, and the their couplings parameterize the space of conformal theories. 

Recall that for the class of $\mathcal{N} = 1$ theories (of which $\mathcal{N} = 4$ SYM is a member), the interactions can be organized using the \textit{superpotential} $W$, which is a function of the \textit{superfields} $\Phi$:
\begin{align}
    \mathcal{L} \supset \int \, d \theta^2 \, W(\Phi) \, ,
\end{align}
where $\theta$ are the superspace coordinates. Now, $\mathcal{N} = 4$ supersymmetry is a very strong constraint. In fact, there is only one superpotential that preserves this much supersymmetry: 
\begin{align}
    W_{\mathcal{N} = 4} = h \, \mathrm{Tr} \, \left( \Phi_1 \Phi_2 \Phi_3 - \Phi_1 \Phi_3 \Phi_2 \right)
\end{align}

It was shown in \cite{Leigh:1995ep} that $\mathcal{N} = 4$ SYM has two marginal deformations that preserve $\mathcal{N} = 1$ supersymmetry. The marginal deformations of \cite{Leigh:1995ep}, also called the $\beta$-deformations, enter in the Lagrangian through $\mathrm{Tr} \, \left( \Phi_1 \Phi_2 \Phi_3 + \Phi_1 \Phi_3 \Phi_2 \right) $ and $ \mathrm{Tr} \, \left( \Phi_1^3 + \Phi_2^3 + \Phi_3^3 \right) $ terms in the superpotential. 

A longstanding puzzle of the AdS/CFT correspondence is to determine the bulk duals of these theories. On general grounds, the duals of the $\beta$-deformed theories are expected to be type IIB string theory on $AdS_5 \times X^5$, where $X^5$ is a 5-dimensional manifold that should be able to be continuously deformed to a sphere. An answer was given for the first deformation \cite{Lunin:2005jy} using a solution-generating technique designed to preserve the correct symmetries. This method makes use of the fact that the first deformation has an additional pair of $U(1)$ symmetries acting on the superfields $\Phi_i$. The gravity dual of the second deformation is still unknown beyond the second-order result of \cite{Aharony:2002hx}.

\subsection{Overview}

The Lunin-Maldacena (LM) background \cite{Lunin:2005jy} preserves $\mathcal{N} = 2$ supersymmetry in five dimensions, so it is natural to expect that it can be extended to a full consistent truncation of IIB supergravity on $AdS_5$ times a deformed $S^5$. The result is 5-dimensional $\mathcal{N} = 2$ gauged supergravity. This would be in accord with the conjecture that any supersymmetric vacuum solution of KK form can be extended to a full non-linear KK reduction with the full set of corresponding supergravity fields \cite{Duff:1985jd,Gauntlett:2007ma}.

The goal of this chapter is the construct the truncation from type IIB supergravity to $\mathcal{N} = 2$ supergravity in 5 dimensions. This requires that we consistently turn on an $\mathcal N=2$ graviphoton in this background. At the linearized level, there is an obvious procedure for doing so by gauging the $U(1)_R$ isometry of the metric.  However, the non-linear reduction is not as straightforward.  Guided by the consistent truncation of IIB supergravity on a Sasaki-Einstein manifold \cite{Berenstein:2002ke,Buchel:2006gb}, we construct a full non-linear KK reduction to gauged $\mathcal N=2$ supergravity in the Lunin-Maldacena background.  While the Gauntlett-Varela conjecture \cite{Gauntlett:2007ma} has been verified for general AdS$_5$ solutions of M-theory \cite{Gauntlett:2004zh,Gauntlett:2006ai}, the present construction yields a non-trivial example starting directly from a IIB supergravity point of view. This will comprise the first section of the chapter.

We are particularly interested in the Lunin-Maldacena case because its starting point can be viewed as $\mathrm{AdS}_5\times S^5$ deformed by turning on a field in the `massive' KK tower.  Although the $\beta$ deformation is non-dynamical here, its presence nevertheless creates some tension between having non-trivial excitations in the KK tower and a consistent truncation that aims to remove such fields. In the second chapter, we  will present our progress in constructing a full non-linear reduction ansatz that includes the deformation parameter $\gamma$ as a dynamical field. We find that the most naive extensions of the LM background retaining such a scalar are not consistent, and we comment on what such a solution, if it exists, might look like. 

\section{Reduction to $\mathcal N=2$ gauged supergravity}

We begin with the Lunin-Maldacena background, which, following the notation of \cite{Lunin:2005jy}, takes the form
\begin{align}
    ds^2&= G^{- 1/4} \left[ ds^2_{\mathrm{AdS}_5} + \sum_i( d\mu_i^2 + G \mu_i^2 d\phi_i^2)  + 9(\gamma^2+\sigma^2) G\mu_1^2 \mu_2^2 \mu_3^2 d \psi^2 \right], \nn\\
    e^{-\phi} &= G^{-1/2} H^{-1}, \qquad\chi =\gamma \sigma g_{0,E} H^{-1},\nn \\
    B_2 &= \ \gamma G w_2 - 12 \sigma w_1 \wedge d \psi , \qquad
    C_2 = \ -\sigma G w_2 - 12 \gamma w_1 \wedge d \psi ,\nn \\
    F_5 &=4(1+*)\omega_{\mathrm{AdS}_5}= 4(\omega_{\mathrm{AdS}_5}+Gdw_1\wedge d\phi_1\wedge d\phi_2\wedge d\phi_3),
\label{eq:LM}
\end{align}
where
\begin{align}
    G^{-1} & =  1 + (\gamma^2 + \sigma^2) g_{0,E}, \qquad H = 1 + \sigma^2 g_{0,E},\qquad g_{0,E} =\mu_1^2 \mu_2^2 + \mu_2^2 \mu_3^2 + \mu_3^2 \mu_1^2,\nn\\
    dw_1 &=\mu_1\mu_2\mu_3*_21, \qquad
    w_2 = \mu_1^2 \mu_2^2 d\phi_1\wedge d\phi_2 + \mu_2^2 \mu_3^2 d\phi_2\wedge d\phi_3 + \mu_3^2 \mu_1^2 d\phi_3\wedge d\phi_1.
\label{eq:GHg}
\end{align}
Here, we have written the five-sphere as a $T^3$ fibration over $S^2$, with $\{\phi_i\}$ as the torus coordinates and $\{\mu_i\}$ the `direction cosines' satisfying $\sum_i\mu_i^2=1$.  In addition, $\psi=(\phi_1+\phi_2+\phi_3)/3$ is the diagonal combination that defines the isometry direction dual to $U(1)_R$.  This solution is parametrized by two real constants, $\gamma$ and $\sigma$, which can be combined into a complex deformation parameter $\beta=\gamma-i\sigma$.  At linearized order, this deformation turns on the two-form potentials $B_2$ and $C_2$, which then backreact on the other fields in a manner that is consistent with \cite{Aharony:2002hx}.  Note that here we have chosen the initial IIB axi-dilaton to be $\tau=i$ prior to the $\beta$ deformation.

The one-form $w_1$ was introduced in \cite{Lunin:2005jy} as a potential, and is implicitly defined by its exterior derivative $dw_1$, where $*_21$ is the volume form on $S^2$.  In particular, for constant $\gamma$ and $\sigma$, only $dw_1$ shows up in the field strengths $H_3$ and $F_3$.  However, if the $\beta$ deformation were to be made spacetime dependent, then either $w_1$ would enter directly in the field strengths or some modification would be needed.  Although we do not pursue this approach here, we will nevertheless demonstrate below that including a dynamical graviphoton is sufficient to make a particular choice of $w_1$ physical.

\subsection{The reduction ansatz}

Although there is as yet no fully systematic treatment of consistent truncations, the starting point is clear as we can gain much insight from the linearized KK spectrum.  Since the deformed background in (\ref{eq:LM}) preserves $\mathcal N=2$ supersymmetry, our aim is to obtain a truncation to the bosonic sector of $\mathcal N=2$ supergravity.  In particular, this involves the generalization of the AdS$_5$ background to an arbitrary 5-dimensional space with metric $g_{\mu\nu}$ along with the addition of a graviphoton $A$ with field strength $F=dA$. 

To do this we will take advantage of the natural Sasaki-Einstein structure of $S^5$. Recall that Sasaki-Einstein manifolds are those that are both Sasaki and Einstein, and that a Riemannian manifold $\mathcal{S}$ is Sasaki if and only if its metric cone ($C = \mathbb{R}_{>0} \times \mathcal{S}, \  ds^2(C) = dr^2 + r^2 ds^2(S)$) is K{\"a}hler. The simplest example in five dimensions (and the one that is relevant for us) is the sphere, which has metric cone $\mathbb{C}^3 \backslash \{ 0 \}$. We will write the solution AdS$_5 \times S^5$ as a general Sasaki-Einstein compactification that retains the graviphoton, and then we will transform to the $\beta$-deformed theory.

Sasakian manifolds admit a Killing vector field known as the Reeb vector. When this vector fields' orbit closes, as is the case for the sphere, they define a foliation of $SE_5$. Then $SE_5$ may be written as a circle bundled over a 4-dimensional K{\"a}hler base as:
\begin{equation}
    ds^2(SE_5) = ds^2(B) + \eta^2, 
\label{SE metric}
\end{equation}
with $d\eta=2J$ where $J$ is the K\"ahler form on the base.  In the case where $SE_5 = S^5$, the K{\"a}hler base is $CP^2$.

Since the graviphoton gauges the $U(1)_R$ isometry generated by $\partial/\partial\psi$, the metric ansatz is obtained by the replacement $d\phi_i\to d\phi_i+A$.  However, this is not yet complete, as the five-form field strength also gains graviphoton contributions in a Freund-Rubin setup.  In the absence of the $\beta$ deformation, a consistent Sasaki-Einstein truncation takes the form \cite{Berenstein:2002ke, Buchel:2006gb}
\begin{align}
    ds^2&=g_{\mu\nu}dx^\mu dx^\nu+ds^2(B)+(\eta+A)^2,\nn\\
    F_5&=(1+*)(4*_51-*_5F\wedge J)\nn\\
    &=4*_51+2J\wedge J\wedge(\eta+A)-*_5F\wedge J+F\wedge J\wedge(\eta+A),
\label{eq:SE}
\end{align}
where $*_5$ is the Hodge dual with respect to the 5-dimensional metric $g_{\mu\nu}$. 

With (\ref{eq:SE}) as a starting point, we can turn on the Lunin-Maldacena deformation, which also brings the IIB axi-dilaton and two-form potentials into play.  The resulting ansatz takes the form
\begin{align}
    ds^2 &= G^{- 1/4} \left[g_{\mu\nu}dx^\mu dx^\nu + \sum_i(d\mu_i^2 + G\mu_i^2 (d\phi_i+A)^2)  + 9( \gamma^2+\sigma^2)G \mu_1^2 \mu_2^2 \mu_3^2 (d \psi + A)^2 \right],\nn \\ 
    \quad e^{-\phi}&= G^{-1/2} H^{-1}, \qquad \chi =  \gamma \sigma g_{0,E} H^{-1},\nn \\
    B_2  &= \gamma G w_2 - 12 \sigma w_1 \wedge \left( d \psi  + A \right),\qquad
    C_2  =  -\sigma G w_2 - 12 \gamma w_1\wedge \left( d \psi  + A \right),\nn \\
    F_5 &= 4*_51 + 4G d w_1\wedge(d\phi_1+A)\wedge(d\phi_2+A)\wedge (d\phi_3+A)\nn\\
    &\qquad- *_5F \wedge J + F  \wedge J \wedge(\eta+A) + 12 G \left( \gamma^2 + \sigma^2 \right) F \wedge w_1 \wedge w_2.
    \label{eq:GVansatz}
\end{align}
The scalar functions $G$, $H$ and $g_{0,E}$ are unchanged from (\ref{eq:GHg}), while $w_2$ now takes the form
\begin{equation}
    w_2=\mu_1^2\mu_2^2(d\phi_1+A)\wedge(d\phi_2+A)+\mu_2^2\mu_3^2(d\phi_2+A)\wedge(d\phi_3+A)+\mu_3^2\mu_1^2(d\phi_3+A)\wedge(d\phi_1+A).
\end{equation}
In addition, the forms pertaining to the Sasaki-Einstein structure can be expressed in terms of the $S^5$ quantities as
\begin{align}
    \eta+A &= \sum_i\mu_i^2(d\phi_i+A)=A + \sum_i\mu_i^2 d\phi_i,\nn\\
    2 J &= \sum_i d \mu_i^2 \wedge (d\phi_i+A)=\sum_id\mu_i^2\wedge d\phi_i.
\end{align}
Here we have made use of the constraint $\sum_i\mu_i^2=1$.

Note that the final term in the five-form ansatz in (\ref{eq:GVansatz}) is required by self-duality, as it is obtained by expanding out the 10-dimensional self-dual expression $F_5=(1+*)(4*_51-*_5F\wedge J)$ in the Lunin-Maldacena background.  It is interesting that the one-form $w_1$ appears directly, and not as a potential, in this term.  This is also the case for the three-form field strengths
\begin{align}
     H_3 &= \gamma \, G \, d w_2 - 12 \sigma\, dw_1\wedge(d\psi+A)+ 12 \sigma F \wedge w_1 - \gamma\left( \gamma^2 + \sigma^2 \right) \, G^2 \, d g_{0, E} \wedge w_2 ,\nn\\
    F_3 &= -\sigma \, H^{-1} \, d w_2 - 12 \gamma \, H^{-1}\,dw_1 \wedge(d\psi+A) + 12 \gamma  \, H^{-1}F \wedge w_1\nn \\
    & \kern4em + \sigma \left( \gamma^2 + \sigma^2 \right)  G \, H^{-1} \, d g_{0, E} \wedge w_2.
    \label{eq:3form}
\end{align}
As a result, turning on the graviphoton selects a preferred $w_1$ given as
\begin{equation}
    w_1 = -\frac{1}{12}\bigg[ (\mu_2^2 - \mu_3^2) \mu_1 d \mu_1 + (\mu_3^2 - \mu_1^2) \mu_2 d \mu_2 + (\mu_1^2 - \mu_2^2) \mu_3 d \mu_3 \bigg].
\end{equation}
It follows that
\begin{align}
    dw_1&=\fft13\left[\mu_1\mu_2d\mu_1\wedge d\mu_2+\mu_2\mu_3d\mu_2\wedge d\mu_3+\mu_3\mu_1d\mu_3\wedge d\mu_1\right]\nn\\
    &=\mu_1\mu_2\mu_3*_21,
\end{align}
where we have chosen an orientation such that
\begin{equation}
    *_2d\mu_i=\epsilon_{ijk}\mu_jd\mu_k,\qquad d\mu_i\wedge d\mu_j=\epsilon_{ijk}\mu_k*_21.
\end{equation}
From this point of view, $w_1$ is in fact physical, and can be expressed more compactly as
\begin{equation}
    w_1=\fft1{12}*_2d(\mu_1\mu_2\mu_3).
\end{equation}

\subsection{Verification of the ansatz}

We have verified that the above ansatz satisfies the IIB axi-dilaton and form field equations of motion.  Although we did not fully verify the IIB Einstein equation, we expect it to work as well.  The IIB equations of motion are satisfied provided the metric $g_{\mu\nu}$ and graviphoton $A_\mu$ obey the corresponding equations obtained from the bosonic Lagrangian of 5-dimensional $\mathcal N=2$ gauged supergravity
\begin{equation}
    e^{-1}\mathcal L_5=R*_51+12*_51-\fft32F\wedge*_5F+F\wedge F\wedge A.
    \label{eq:5dim}
\end{equation}
The graviphoton kinetic term can be made canonical by the rescaling $A\to A/\sqrt3$.

In order to verify the ansatz, we had to compute the 10-dimensional Hodge dual of the field strengths.  This was done by splitting the 10-dimensional space into a warped product of 5-dimensional spacetime, the $S^2$ base and the $T^3$ fiber
\begin{equation}
    ds^2=G^{-1/4}\left[g_{\mu\nu}dx^\mu dx^\nu+\sum_id\mu_i^2+G\left(\sum_ie_i^2+(\gamma^2+\sigma^2)\mu_1^2\mu_2^2\mu_3^2\Bigl(\sum_i\fft{e_i}{\mu_i}\Bigr)^2\right)\right],
\end{equation}
where $e_i=\mu_i(d\phi_i+A)$.  We use $*_5$, $*_2$ and $*_3$ to denote the Hodge duals within these three subspaces, respectively (without the overall $G^{-1/4}$ factor), and $*$ without subscript to denote the Hodge dual taken in the full 10-dimensional IIB metric (including $G^{-1/4}$).  In this case, we have the useful identities
\begin{align}
    *_31&=Ge_1\wedge e_2\wedge e_3,\nn\\
    *_3e_1 &= e_2 \wedge e_3 - G (\gamma^2 + \sigma^2) \mu_2 \mu_3 w_2,\nn \\
    *_3(e_1\wedge e_2)&=e_3+(\gamma^2+\sigma^2)\mu_1^2\mu_2^2\mu_3^{\vphantom2}\sum_i\fft{e_i}{\mu_i},\nn\\
    *_3(e_1\wedge e_2\wedge e_3)&=G^{-1},
\end{align}
along with cyclic permutations.  From these, we can obtain
\begin{equation}
    *_3\Bigl(\sum_i\fft{e_i}{\mu_i}\Bigr)=\fft{G}{\mu_1\mu_2\mu_3}w_2,\qquad*_3w_2=\mu_1\mu_2\mu_3G^{-1}\sum_i\fft{e_i}{\mu_i}.
\end{equation}

Verification of the form field equations of motion is straightforward although somewhat tedious.  Here we present some of the expressions that were useful in performing this check.  The IIB dilaton and RR scalar are naturally combined into the complex axi-dilaton
\begin{equation}
    \tau=\chi+ie^{-\phi}=(\gamma\sigma g_{0,E}+iG^{-1/2})H^{-1},
\end{equation}
with corresponding one-form field strength
\begin{equation}
    d\tau=-\ft12i(\sigma+i\gamma G^{1/2})^2H^{-2}G^{-1/2}dg_{0,E}.
\end{equation}
The three-form field strengths were given above in (\ref{eq:3form}), and can be combined into the complex three-form
\begin{align}
    G_3=F_3-ie^{-\phi}H_3&=(\sigma+i\gamma G^{1/2})H^{-1}\Bigl[-dw_2+4iG^{1/2}*_21\wedge*_3w_2-12iG^{-1/2}F\wedge w_1\nn\\
    &\kern9em+(\gamma^2+\sigma^2)Gdg_{0,E}\wedge w_2\Bigr],
\end{align}
with 10-dimensional Hodge dual
\begin{align}
    *G_3&=(\sigma+i\gamma G^{1/2})H^{-1}\Bigl[-G^{-1/2}*_{10}dw_2+4i*_51\wedge w_2+12i*_5F\wedge*_2w_1\wedge e_1\wedge e_2\wedge e_3\nn\\
    &\kern9em-(\gamma^2+\sigma^2)G^{1/2}*_51\wedge*_2dg_{0,E}\wedge*_3w_2\Bigr].
\end{align}
The axi-dilaton equation is then satisfied identically, while the three-form and five-form equations of motion are satisfied so long as the graviphoton satisfies the 5-dimensional equation of motion $d*_5F=F\wedge F$ originating from (\ref{eq:5dim}).

\section{Further Truncations}

Above we have extended the Lunin-Maldacena solution into a full consistent truncation of IIB supergravity to the bosonic sector of pure 5-dimensional $\mathcal N=2$ gauged supergravity.  It is of course interesting to ask if further consistent truncations generalizing the Lunin-Maldacena solution are possible. 

One interesting place to explore this idea is in the scalars that are dual to the exactly marginal deformations. The deformations are given by operators with $\Delta = 4$ on the boundary, so their duals should be the exactly massless dynamical fields in the bulk. As an example, first consider the $\mathcal{N} = 4$-preserving deformation-- this corresponds to the operator that is equal to the Lagrangian itself. Moving around in the space of $\mathcal{N} = 4$-preserving conformal field theories amounts to changing the coupling of the theory, $\tau_{YM} = \theta / 2 \pi  + 4 \pi i / g^2_{YM} $. The dual of this deformation is the axidilaton $\tau_s$. 

We find that it is possible to retain a dynamical 5-dimensional axi-dilaton $\tau_s=\tau_{1s}+i\tau_{2s}$ in the Lunin-Maldacena solution.  In fact, it can be shown that the solution of \cite{Lunin:2005jy} remains valid without modification, even for a dynamical $\tau_s$.  To demonstrate this, it is convenient to express the fields as
\begin{align}
    ds^2&= G^{- 1/4} \left[g_{\mu\nu}dx^\mu dx^\nu + \sum_i( d\mu_i^2 + G \mu_i^2 d\phi_i^2)  + 9\fft{|\beta|^2}{\tau_{2s}}G\mu_1^2 \mu_2^2 \mu_3^2 d \psi^2 \right], \nn\\
    e^{-\phi} &= \tau_{2s}G^{-1/2} H^{-1}, \qquad\chi =\tau_{1s}-\beta_1\beta_2 g_{0,E} H^{-1},\nn \\
    B_2 &= \fft{\beta_1}{\tau_{2s}}G w_2 - 12 \sigma w_1 \wedge d \psi , \qquad
    C_2 = \left(\beta_2+\fft{\tau_{1s}}{\tau_{2s}}\beta_1\right) G w_2 - 12 \gamma w_1 \wedge d \psi ,\nn \\
    F_5 &= 4(*_51+Gdw_1\wedge d\phi_1\wedge d\phi_2\wedge d\phi_3),
\end{align}
where
\begin{equation}
    G^{-1} = 1 + \fft{|\beta|^2}{\tau_{2s}} g_{0,E},\qquad
    H = 1 + \fft{\beta_2^2}{\tau_{2s}}g_{0,E},
\end{equation}
and we have introduced the shifted $\beta$-deformation parameter
\begin{equation}
    \beta=\beta_1+i\beta_2=\gamma-\tau_s\sigma.
    \label{eq:beta}
\end{equation}

A dynamical $\tau_s$ modifies the 10-dimensional one-form field strength
\begin{align}
    d\tau&=\fft{i}2(\beta_1+i\beta_2G^{-1/2})^2G^{1/2}H^{-2}dg_{0,E}\nn\\
    &\quad+\left(1+\fft{i\beta_1\beta_2}{\tau_{2s}}g_{0,E}G^{1/2}\right)H^{-1}\left(d\tau_{1s}+iG^{-1/2}H^{-1}d\tau_{2s}\right)+\fft{i}2\fft{\beta_2^2-\beta_1^2}{\tau_{2s}}g_{0,E}G^{1/2}H^{-1}d\tau_{2s},
\end{align}
as well as the complex three-form field strength
\begin{align}
    G_3&=\fft{\beta_1+i\beta_2G^{-1/2}}H\biggl[-iG^{1/2}dw_2-4G*_21\wedge*_3w_2+i\fft{|\beta^2|}{\tau_{2s}}G^{3/2}dg_{0,E}\wedge w_2\nn\\
    &\kern7.5em+\fft{G}{\tau_{2s}}\left((d\tau_{1s}-iG^{1/2}d\tau_{2s})+2i\fft{G^{1/2}H}{\beta_1+i\beta_2G^{-1/2}}(\beta_1d\tau_{2s}-\beta_2d\tau_{1s})\right)\wedge w_2\biggr].
\end{align}
The resulting equations of motion are then consistent with the 5-dimensional Lagrangian
\begin{equation}
    e^{-1}\mathcal L_5=R*_51+12*_51-\fft1{2\tau_{2s}^2}d\tau_s\wedge*d\bar\tau_s.
\end{equation}
Although we have not included the graviphoton here, we expect that a full consistent truncation can be obtained that retains the complete set of fields of the generic squashed Sasaki-Einstein reduction.

\subsection{Fluctuating $\gamma(x)$}

For a much less trivial example than $\tau_s$, we can consider the fields dual to the $\mathcal{N} = 1$-preserving deformation. Since $\gamma$ and $\sigma$ of the LM solution are the bulk parameters that characterize the strength of the deformation, they are the fields dual to the boundary deformation. So we would like to know if they may be consistently made dynamical. There is a crucial difference between this and dynamical $\tau_s$, in that $\gamma$ and $\sigma$ are part of the first excited KK level and moreover carry non-trivial dependence on the internal coordinates.  Stated differently, while it is always possible to obtain a consistent truncation by restricting to singlets under an internal symmetry group, in this case there is no such obvious subgroup that will retain $\gamma$ and $\sigma$ while removing the rest of the KK tower.

At the same time, however, the Lunin-Maldacena solution itself allows us to move continuously along the exactly marginal deformation parametrized by $\gamma$ and $\sigma$.  This raises the possibility that they may couple to higher states in the KK tower in a controlled manner.  After all, the truncation is consistent when these fields are set to constants, corresponding to turning on constant sources for the dual operators.  Additional motivation for a possible consistent truncation arises by noting that the shifted deformation parameter $\beta$ in (\ref{eq:beta}) can be spatially varying when $\tau_s$ is made dynamical.  This hints that an independent dynamical $\beta$ may be obtained using the dynamical $\tau_s$ solution as a starting point.  Nonetheless, we have found that a straightforward promotion of $\beta$ to an independently varying field does not lead to a consistent solution of the IIB equations of motion, so further study of the system will be required to see if such a truncation is possible.

The most obvious thing to is to simply make the $\gamma$ that appears in the potentials a function of spacetime. Then the equations of motion acquire terms proportional to $\partial \gamma$ and $(\partial \gamma)^2$ (and $\partial^2 \gamma$ terms, but these are set to zero according to the equation of motion for $\gamma$, which is massless). It is straightforward to check that this does not yield a solution to the 10-dimensional equations of motion. So if there is a solution, modifications must be made to the reduction ansatz.

\subsubsection{Zeroth- and First-Order Solution} 

In the absence of an inspired guess at the solution, we take a systematic approach of determining the solution order-by-order in the deformation $\gamma$\footnote{We set $\sigma=0$ for simplicity. In the case of a constant $\gamma$ and $\sigma$, the solution exists even if one of them is turned off-- we have no reason to expect that this not to be the case for fluctuating $\gamma$ and $\sigma$. In fact, the 5-dimensional action of the $\gamma$ and $\sigma$ are given in (3.30) of \cite{Lunin:2005jy}, and we can check explicitly that they are decoupled.}. The zeroth order solution is the LM solution with no $\gamma$ or $\sigma$, which reduces to the Freund-Rubin solution:
\begin{align}
    ds^2&= ds^2_{\mathrm{AdS}_5} + \sum_i( d\mu_i^2 + G \mu_i^2 d\phi_i^2) \, , \nn\\
    & \phi = \phi_0, \qquad F_5 = 4(1 + \star_0) w_{AdS_5}\, ,
\end{align}
where $\phi_0$ is a constant and $\star_0$ is the 10-dimensional Hodge star using the uncorrected metric.

A first-order solution requires modification of the three-forms field strengths. It is perhaps rather surprising, but the solution given in (\ref{eq:LM}) does not work at first order. The linear KK reduction ansatz for a number of the low-lying fields were worked out long ago \cite{Kim:1985ez}, and we find that a modification is needed of $C_2$ to satisfy the IIB equations of motion to even leading order. The resulting three-forms are 
\begin{align}
    H^{(3)} = \gamma d w_2 + d \gamma \wedge w_2 \, , \qquad F^{(3)} = - 4 \gamma \star_2 1 \wedge \star_3 w_2 + \frac{d \gamma \wedge \star_5 d w_2}{4}
\end{align}
The $\star_5$, $\star_2$, and $\star_3$ are the Hodge stars within the spacetime coordinates $x^{\mu}$, the $(\alpha, \, \theta)$ coordinates on the two-sphere, and the three-torus coordinates $(\phi_1, \phi_2, \phi_3)$, respectively. The modification makes the $C_2$ much more complicated, but actually simplifies the expression for $F_3$ and makes it more symmetric with $H_3$. This modification differs from the original LM solution by a term that is pure gauge if $\gamma$ is non-dynamical. 

One might ask if this is the unique form for the first-order solution. Computing the spherical harmonics on $S^5$ shows that these are the only harmonics. We have assumed that the solution follows a pattern where $\phi$, $F_5$, and $g$ contain even powers in $\gamma$, and $H_3$ and $F_3$ contain odd powers. This seems likely, especially given that it is the patterned followed by the LM solution for constant $\gamma$.

\subsubsection{Second-Order Solution: Dilaton}

We have attempted to find a second-order solution as well, which requires modification of $\phi$, $F_5$, and $g$. Consider first the dilaton, whose equation of motion is:
\begin{align}
    \Delta \phi + \frac{1}{2} e^{-\phi} |H^{(3)}|^2 - \frac{1}{2} e^{\phi} |F^{(3)}|^2 = 0
\end{align}
The second-order modifications to the dilaton decouple from the modifications to $F_5$ and $g$. It is straightforward to solve this: the dilaton modifications are sourced by the first-order modifications to $H_3$ and $F_3$. We first compute these terms:
\begin{align}
    & \frac{1}{2} e^{-\phi} |H^{(3)}|^2 - \frac{1}{2} e^{\phi} |F^{(3)}|^2 = \gamma^2  4 \, G^2 \,  \mu_1 \mu_2 \mu_3 \, (-1 + 4 g_{0E}) \, w_{\text{AdS}} \wedge \star_2 1 \wedge d \phi_1 d \phi_2 d \phi_3
    \\
    & \qquad   \frac{1}{4}  \, G^2 \,  \mu_1 \mu_2 \mu_3 \, (-1 + 4 g_{0E}) \, d \gamma \wedge \star d \gamma \wedge \star_2 1 \wedge d \phi_1 d \phi_2 d \phi_3
\end{align}
Now we can solve this by taking its integral, and setting that equal to $\star_{10} d \phi$. The solution to this requires the following modification of $\phi$:
\begin{align}
    e^{-2 \phi} = 1 + \gamma^2 g_{0E} \rightarrow 1 + \gamma^2 \left( - \frac{1}{4} + g_{0E} \right) \, .
\end{align}
This seemingly random change may be understood by taking a full truncation including both $\gamma(x)$ and $\tau_2(x)$. In that case, we have
\begin{align}
    e^{-2 \phi} = \tau_2^2 +  \tau_2 \gamma^2 g_{0E}
\end{align}
The equation of motion is consistent if you require the equations of motion that arise from the action in (3.30) of \cite{Lunin:2005jy}. Alternative, this may be viewed as a modification $\tau_2 \rightarrow 1 + \gamma^2 /8 + ...$, where the ellipses stand for terms higher order in $\gamma$. This is simply the solution of the $\tau_2$ EOM in terms of $\gamma$.

\subsubsection{Second-Order Solution: Five-Form}

Next we consider the five-form. This is trickier because second-order change in the metric can show up in the five-form equation of motion through the Hodge star. The EOM is:
\begin{align}
    d F^{(5)} = H^{(3)} \wedge F^{(3)}
\end{align}
This equation is solved by adding a term to $F^{(5)}$ (which is the term proportional to $d \gamma$) and then modifying the metric so that the entire five-form is self dual. Basically this amounts to finding a term that is an integral of the left-over pieces of the original equation. These are given by:
\begin{align}
    d F^{(5)}_{LM} - H^{(3)} \wedge F^{(3)} =  2 \, \mu_1 \mu_2 \mu_3 \left( 1 - 4 g_{0E} \right) \, \gamma \, d \gamma \wedge \star_2 1 \wedge d \phi_1 d \phi_2 d \phi_3
\end{align}
The integral of this term is the additional piece:
\begin{align}
    d F^{(5)}_{LM} - H^{(3)} \wedge F^{(3)} = - d \delta F^{(5)} = d \left( -\frac{1}{4} \, \mu_1 \mu_2 \mu_3 \, \gamma \, d \gamma \wedge \star_2 d g_{0E}  \wedge d \phi_1 d \phi_2 d \phi_3 \right)
\end{align}
Therefore the term in parentheses on the RHS is modification needed for the five form. The integral is not unique, but we have chosen it to be proportional to $d \gamma$ so that the LM solution is restored when $d \gamma \rightarrow 0$. Another modification of $F_5$ is needed to ensure that the five-form is self dual. This is just the Hodge star of $\delta F_5$:
\begin{align}
     \delta F_5 \ & = \ -\frac{1}{4} \, \mu_1 \mu_2 \mu_3  \, \gamma \, d \gamma \wedge \star_2 \,  d g_{0E}  \wedge d \phi_1 d \phi_2 d \phi_3 \\
    \star_{10} \  \delta F_5 \ &  = \ \frac{1}{4} \,  \gamma \,  (\star_5 \, d \gamma )  \wedge d g_{0E}
\end{align}
Instead of adding a $\star_{10} \delta F_5$ term, we could instead try to modify the metric so that $F_5 + \delta F_5$ is self-dual. This approach works as well, but neither one allows the Einstein equation to be solved.

\subsubsection{Second-Order Solution: Einstein Equation}

We find an obstruction to a second-order solution in the Einstein equation. Recall the Einstein equation takes the form
\begin{align}
    R_{\mu\nu}&=\fft12\partial_\mu\phi\partial_\nu\phi+\fft12e^{2\phi}\partial_\mu\chi\partial_\nu\chi+\fft14e^{-\phi}\left(H_{\mu\rho\sigma}H_\nu{}^{\rho\sigma}-\fft1{12}g_{\mu\nu}H_{\lambda\rho\sigma}H^{\lambda\rho\sigma}\right)\nn\\
    &\qquad+\fft14e^\phi\left(F_{\mu\rho\sigma}F_\nu{}^{\rho\sigma}-\fft1{12}g_{\mu\nu}F_{\lambda\rho\sigma}F^{\lambda\rho\sigma}\right)+\fft1{4\cdot4!}F_{\mu\lambda\rho\sigma\tau}F_\nu{}^{\lambda\rho\sigma\tau}.
\end{align}
Note that the dilaton does not contribute to the right-hand side because the leading-order is constant and the first correction is second-order in $\gamma$, so the first correction to the RHS goes like $\gamma^4$. We assume that $\chi$ does not contribute either, because it has no part that is constant in $\gamma$. It is not clear how to contract the index on the derivative if it were first-order in $\partial_{\mu} \gamma$, and if it is second order, then its contribution to the Einstein equation will be fourth order, like that of $\phi$.

So we are left with the three-forms and the five-form on the RHS. We have computed the LHS $-$ RHS of this equation in Mathematica. The Ricci-tensor was computed with the unmodified part of the Einstein equation, so the LHS $-$ RHS must be cancelled out with modifications to the metric or further modifications to the five-form. The result is
\begin{align}
    R_{\mu \nu} = \begin{pmatrix} A & 0 \\ 0 & B \end{pmatrix}. 
\end{align}
The spacetime part $A$ is given by
\begin{align}
    \resizebox{\columnwidth}{!}{$
    \begin{pmatrix}
    \frac{\gamma_0^2}{4} + \frac{ (\partial \gamma)^2}{16}(4 g_{0E} - 1) & \frac{\gamma_0 \gamma_1}{4} & \frac{\gamma_0 \gamma_2}{4} & \frac{\gamma_0 \gamma_3}{4} & \frac{\gamma_0 \gamma_4}{4}   \\
    \frac{\gamma_1 \gamma_0}{4} & \frac{\gamma_1^2}{4} + \frac{ (\partial \gamma)^2}{16}(4 g_{0E} - 1) & \frac{\gamma_1 \gamma_2}{4} & \frac{\gamma_1 \gamma_3}{4} & \frac{\gamma_1 \gamma_4}{4}  \\
    \frac{\gamma_2 \gamma_0}{4} & \frac{\gamma_2 \gamma_1}{4} & \frac{\gamma_2^2}{4}  + \frac{ (\partial \gamma)^2}{16}(4 g_{0E} - 1)& \frac{\gamma_2 \gamma_3}{4}  & \frac{\gamma_2 \gamma_4}{4}  \\
    \frac{\gamma_3 \gamma_0}{4} & \frac{\gamma_3 \gamma_1}{4} & \frac{\gamma_3 \gamma_2}{4} & \frac{\gamma_3^2}{4} + \frac{ (\partial \gamma)^2}{16}(4 g_{0E} - 1) & \frac{\gamma_3 \gamma_4}{4}  \\
    \frac{\gamma_4 \gamma_0}{4} & \frac{\gamma_4 \gamma_1}{4} & \frac{\gamma_4 \gamma_2}{4} & \frac{\gamma_4 \gamma_3}{4} & \frac{\gamma_4^2}{4} + \frac{ (\partial \gamma)^2}{16}(4 g_{0E} - 1)
    \end{pmatrix}
    $}
\end{align}
where we have used the shorthand $\gamma_{\mu} = \partial_{\mu} \gamma$. This block is okay: many of these terms reduce to the 5-dimensional stress tensor 
\begin{align}
    T_{\mu \nu} = \partial_{\mu} \gamma  \, \partial_{\nu} \gamma - \frac{1}{2}g_{\mu \nu}  (\partial \gamma)^2
\end{align}
The diagonal parts may be cancelled by various modifications to the metric or further modifications to $F_5$. The internal block $B$ is given by
\begin{align}
    \resizebox{\columnwidth}{!}{$
    \begin{pmatrix}
    \frac{(\partial \gamma)^2}{16} \left( 1 + 4 g_{0E} - \frac{8 \mu_2^2 \mu_3^2}{\mu_2^2  + \mu_3^2} \right) & \frac{(\partial \gamma)^2}{4} \frac{\mu_1 \mu_2 \mu_3 (\mu_3^2 - \mu_2^2)}{\sqrt{ \mu_2^2 + \mu_3^2}} & 0 & 0 & 0 \\
    \frac{(\partial \gamma)^2}{4} \frac{\mu_1 \mu_2 \mu_3 (\mu_3^2 - \mu_2^2)}{\sqrt{ \mu_2^2 + \mu_3^2}} & \frac{(\partial \gamma)^2}{16} \left( (\mu_2^2 + \mu_3^2) (1 - 4 g_{0E}) + 8 \mu_2^2 \mu_3^2 \right)   & 0 & 0 & 0 \\
    0 & 0 & - \frac{(\partial \gamma)^2}{16}   \mu _1^2 \left(4 g_{0E}+2 \mu
   _1^2-1\right) & \frac{(\partial \gamma)^2}{8}   \mu _1^2 \mu
   _2^2  & \frac{(\partial \gamma)^2}{8}  \mu _1^2 \mu
   _3^2 \\
    0 & 0 & \frac{(\partial \gamma)^2}{8}  \mu _1^2 \mu
   _2^2 &  - \frac{(\partial \gamma)^2 }{16}  \mu _2^2 \left(4 g_{0E}+2 \mu
   _2^2-1\right) & \frac{(\partial \gamma)^2}{8}   \mu _2^2 \mu
   _3^2\\
    0 & 0 & \frac{(\partial \gamma)^2}{8}   \mu _1^2 \mu
   _3^2 & \frac{(\partial \gamma)^2}{8}  \mu _2^2 \mu
   _3^2 &  - \frac{(\partial \gamma)^2}{16}   \mu _3^2 \left(4 g_{0E}+2 \mu
   _3^2-1\right) \\
    \end{pmatrix}
    $}
\end{align}
This block is where the problem arises. The issue is that off-diagonal components on the lower-right. There is no way to cancel out the $(3,4)$, $(3,5)$, and $(4,5)$ components. The five-form can only contribute to the diagonal components (this is not obvious, but it is because the zeroth order solution is the antisymmetric tensor). And the metric can either (a) be proportional to $\gamma^2$, which breaks the Lunin-Maldacena solution at leading order, or (b) be proportional to $(\partial \gamma)^2$, in which case the Ricci tensor gets uncancelled terms with four derivatives, like $\partial_\mu \partial_\nu \gamma \, \partial^\mu \partial^\nu \gamma$. 

Given the assumptions we have made, it is not possible to solve the Einstein equation at second order. The primary assumptions are that the fields are split into even and odd powers of $\gamma$, and that the fields have no explicit dependence on the internal dimensions $\phi_1$, $\phi_2$, and $\phi_3$. Removing this constraint, in particular, allows for much more general solutions. However, it is reasonable to guess that the $U(1)$ isometries along the $\phi$-directions are maintained, as these were required for the technique that generated the LM solution in the first place. It may be that there exists a solution to the Einstein equation that relaxes these assumptions, but we have been unable to guess it.

%% file: Chap4/chap4.tex
\section{Review: The Weak Gravity Conjecture}
\label{chap4:intro}

As we have mentioned, string theory is widely believed to provide a UV complete description of quantum gravity. There is a problem though: the theory allows for an astronomical number of vacua, which manifest at low energies as effective field theories (EFTs). This set of consistent string vacua is known as the \textit{Landscape}. Due to the large number of low-energy descriptions, it may be difficult or impossible to find a vacuum that describes our world. Recently a different approach has proven useful: rather than searching through vacua, we should study the general conditions under which an EFT admits a UV completion that includes quantum gravity. Theories that admit no such completion are said to be in the \textit{Swampland} \cite{Vafa:2005ui}. A number of Swampland criteria have been put forward (for a review of the program, see \cite{Brennan:2017rbf, Palti:2019pca}). 

One candidate for a general principle constraining consistent string vacua is the weak gravity conjecture (WGC) \cite{ArkaniHamed:2006dz}. Various forms of the conjecture have been proposed, but roughly it states that EFTs that arise as low energy descriptions of theories of quantum gravity must have a state with a greater charge than mass-- i.e. for which ``gravity is the weakest" force. Were this not the case, extremal or near-extremal black holes would unable to decay because emitting a sub-extremal state would cause the left-over black hole to be superextremal, violating cosmic censorship. This, in turn is problematic because it leads to the existence of an arbitrarily large number of stable states, which is believed to be pathological \cite{Vafa:2005ui}. We now review these arguments in more detail.

\subsection{WGC in Flat Space}
\label{subsec:review}

The original WGC was formulated as a Swampland criterion \cite{ArkaniHamed:2006dz}: \textit{in a UV complete model of quantum gravity, there should not exist an infinite tower of exactly stable states in a fixed direction in charge space.} Such an infinite tower might lead to a species problem or remnant issues \cite{Susskind:1995da, Banks:2006mm}. No proof of this statement has been given, but it is consistent with all known explicit examples of string compactifications and is conceptually consistent with a number of other conjectures about quantum gravity, such as the finiteness principle and the absence of global symmetries \cite{Vafa:2005ui}.

The conjecture can be interpreted as a statement about the (in-)stability of nearly extremal black holes. Consider the context of a single gauge field in 4-dimensional flat space\footnote{Keep in mind that many of the considerations here will change when we consider AdS.}. The low energy description is Einstein-Maxwell theory, whose action is
\begin{align}
    S = \int \text{d}^4 x \sqrt{-g}\left[\frac{M_{\text{Pl}}^2}{4}R - \frac{1}{4}F_{\mu\nu}F^{\mu\nu} \right] \, ,
\end{align}
The spectrum of large black holes of this theory corresponds to the familiar Kerr-Newman solutions, which are characterized by their mass, angular momentum, and charge under the gauge field $F_{\mu \nu}$. There do not exist black holes for just any combination of these parameters however; the \textit{extremality bound} for black holes gives a lower bound on the mass given the charge and angular momentum. Violation of the extremality bound results in existence of a naked singularity in the spacetime.

From here on, we will restrict to zero angular momentum\footnote{One might wonder if there is a version of the weak gravity conjecture for angular momentum-- that might constrain the ratio of mass to spin. However, rotating black holes can decrease their angular momentum by emitting scalar particles with orbital angular momentum, which is an important difference between the spinning and charged black holes.}. In this case, the extremality bound for the Einstein-Maxwell theory becomes the requirement 
\begin{align}
    Q^2 < M^2/M_{\text{Pl}}^2
    \label{eq:EMext}
\end{align}
If it is forbidden to have an infinite tower of stable states, then near-extremal ($Q \sim M$) black holes above some critical charge must be able to decay. Whether this is kinematically possible (i.e. consistent with conservation of mass and charge) depends on the spectrum of charged states with masses lighter than the black hole. It is easy to see from the inequality (\ref{eq:EMext}): if a black holes with $Q \sim M$ emits a state with a mass larger than its charge, than the leftover black hole will have charge larger than its mass-- in violation of the extremality bound. Therefore for these black holes, which have charge very near their mass, to decay, the theory must contain a state that is \textit{self-repulsive}, meaning $q_i^2 \geq m_i^2/M_{\text{Pl}}^2$ (regardless of whether we include higher-derivative corrections). If there are no self-repulsive states then such a decay is impossible and an infinite tower of extremal black holes are exactly stable, violating the Swampland criterion. This leads to the common formulation of the WGC:\\

\textbf{Weak Gravity Conjecture (Single Charge):} \textit{In a UV complete model of quantum gravity there must exist some state with} $Q^2 \geq M^2/M_{\text{Pl}}^2$.\\

In the context of a specific model, to show that the WGC is violated requires complete knowledge of the spectrum of charged states. To show that it is satisfied however, requires only the existence of a single self-repulsive state. It is useful to separate charged states into three regimes according to their masses:

\begin{enumerate}
	\item \textbf{Particle regime} ($M\ll M_{\text{Pl}}$): States in this regime are well-described by ordinary quantum field theory on a fixed spacetime background. 
	\item \textbf{Stringy regime} ($M \lesssim M_{\text{Pl}}$): States in this regime are intrinsically related to the UV completion. They can usually only be calculated from a detailed understanding of the UV physics such as an explicit string compactification.
	\item \textbf{Black hole regime} ($M \gg M_{\text{Pl}}$): States in this regime are well-described by classical black hole solutions in the relevant low-energy model of gravity. 
\end{enumerate}

\subsection{The Black Hole WGC}

One interesting proposal is that the self-repulsive states required by the WGC are the black holes \cite{Kats:2006xp}. Naively, it would seem impossible for a charged black hole to be self-repulsive since this would violate the extremality bound. However, a theory of quantum gravity may not exactly be Einstein-Maxwell theory at low energies; it may have other states at higher energies. These states will manifest at lower energies as higher-derivative corrections, and these corrections will shift the extremality bound. For large black holes, with $Q^2\gg 1$, these corrections can be calculated perturbatively in $1/Q^2$, with the leading corrections corresponding to four-derivative effective operators. The authors of \cite{Kats:2006xp} analyzed electrically charged solutions to the following effective action
\begin{equation}
\label{4dersingle}
S = \int \text{d}^4 x \sqrt{-g}\left[\frac{M_{\text{Pl}}^2}{4}R - \frac{1}{4}F_{\mu\nu}F^{\mu\nu} + \alpha \left(F_{\mu\nu} F^{\mu\nu}\right)^2  + \beta  \left(F_{\mu\nu} \tilde{F}^{\mu\nu}\right)^2 +\gamma F_{\mu\nu}F_{\rho\sigma}W^{\mu\nu\rho\sigma}\right] \, ,
\end{equation}
where $W^{\mu \nu \rho \sigma}$ is the Weyl tensor. To leading-order, the corrected extremality bound is 
\begin{equation}
    \frac{M_{\text{Pl}}^2Q^2}{M^2} \leq 1+\frac{4}{5Q^2}(2\alpha -\gamma)+\mathcal{O}\left(\frac{1}{Q^4}\right).
\end{equation}
The $\mathcal{O}\left(1/Q^4\right)$ contributions correspond to next-to-leading-order in the four-derivative operators and leading-order in six-derivative operators. If the corrected extremality bound is positive
\begin{equation}\label{singleext}
    2\alpha-\gamma>0,
\end{equation}
then extremal black holes with finite charge are self-repulsive and the WGC is satisfied in the black hole regime. Conversely, if the corrected extremality bound is negative
\begin{equation}
    2\alpha-\gamma<0,
\end{equation}
then the decay of asymptotically large extremal black holes into extremal black holes with large but finite charge is kinematically impossible. This does not necessarily mean that the WGC is violated, but rather that if it is valid then there must exist a self-repulsive state in either the stringy or particle regimes. 

Various arguments have been given that (\ref{singleext}) should always be true, even from a low-energy perspective. These include arguments from unitarity, causality \cite{Hamada:2018dde}, positivity of the S-matrix \cite{Bellazzini:2019xts}, shifts to entropy bounds \cite{Cheung:2018cwt}, and renormalization group running \cite{Charles:2019qqt}. 


\subsection{The Entropy-Extremality Relation}

One intriguing proof of the WGC in flat space relates the extremality shift to the shift in the Wald entropy. The entropy for black holes in higher-derivative theories is given by the Wald entropy \cite{Wald:1993nt}:
\begin{align}
    S = - 2 \pi \int_{\Sigma} \frac{\delta \mathcal{L}}{\delta R_{\mu \nu \rho \sigma}} \epsilon_{\mu \nu} \epsilon_{\rho \sigma} \, .
    \label{Wald_Entropy}
\end{align}
This integral is performed over the horizon $\Sigma$. $\mathcal{L}$ is the Lagrangian for the effective theory. The higher-derivative corrections will shift Wald entropy by their appearence in the Lagrangian, and by shifting the location of the horizon that is integrated over. For the theory describe by (\ref{4dersingle}), the corrections to the Wald entropy in the near-extremal limit are \cite{Cheung:2018cwt}
\begin{align}
    \Delta S |_{Q, M} = -\frac{2}{5 T_0} (2 \alpha - \gamma) \, ,
\end{align}
where $T_0$ refers to the unshifted black hole temperature. The same combination $(2 \alpha - \gamma)$ appears in the shifted extremality bound (\ref{singleext}) so the black hole WGC will be satisfied as long as the entropy shift is positive. The authors of \cite{Cheung:2018cwt} present an argument that the higher-derivative corrections should increase the entropy, thereby proving the Black Hole WGC. This motivates the definition of the following conjecture:\\

\textbf{Entropy Shift Conjecture}: \textit{The higher-derivative corrections to the Wald entropy for a solution at fixed charge and mass are always positive in theories with a UV completions that include quantum gravity}\\

The argument for the entropy shift positivity is not expected to be fully general; it applies to higher-derivative corrections that arise from integrating out massive particles at tree-level. However it is not clear if there is a counterexample for UV complete theories (see the appendix of \cite{Hamada:2018dde} for a theory with a negative entropy shift), so the status of the entropy shift conjecture is unknown. The relation between the entropy shift and extremality, however, appears to be very robust. A purely thermodynamic proof in \cite{Goon:2019faz}, where no assumptions were made about the particulars of the background. Another derivation, which prevents an infinite black hole entropy for very-near-extremal black holes, was given in \cite{Hamada:2018dde}.


\subsection{Overview}
\label{subsec:results}

This chapter will review work on two main subjects. The first on generalizing the discussion of the black hole WGC to theories that consist of a graviton plus $N$ $U(1)$ gauge fields. We consider black hole solutions with general electric and magnetic charges. 

The two-derivative approximation to the EFT has many accidental symmetries, including an $O(N)$ global flavor symmetry, parity and $U(N)$ electromagnetic duality symmetry. We do not assume that any of these symmetries, and instead analyze the most general possible set of three and four-derivative operators. This leads to the Lagrangian
\begin{align}
  \begin{split}
    S &= \int \mathrm{d}^4x \sqrt{- g} \Big[ \frac{M_{\text{Pl}}^2}{4}R - \frac{1}{4} F^i_{\mu \nu} F^{i \, \mu \nu} 
    + a_{ijk}F^i_{\mu\nu}F^{j\nu\rho}{F^k_{\rho}}^\mu+ b_{ijk}F^i_{\mu\nu}F^{j\nu\rho}{\tilde{F}^k_{\rho}} {}^\mu \\
    &\hspace{25mm}+  \alpha_{i j k l} \, F^i_{\mu \nu} F^{j \, \mu \nu} F^k_{\rho \sigma} F^{l \, \rho \sigma}
    + \beta_{i j k l} \,  F^i_{\mu \nu} \tilde{F}^{j \, \mu \nu} F^k_{\rho \sigma} \tilde{F}^{l \, \rho \sigma} 
    \\
    & \qquad \qquad \qquad + \gamma_{i j} \, F^i_{\mu \nu} F^j_{\sigma \rho} W^{\mu \nu \sigma \rho}
    + \chi_{i j k l} \, \tilde{F}^i_{\mu \nu} F^{j \, \mu \nu} F^k_{\rho \sigma} F^{l \, \rho \sigma} 
    + \omega_{i j} \, F^i_{\mu \nu} \tilde{F}^j_{\sigma \rho} W^{\mu \nu \sigma \rho} \Big].
  \end{split}
  \label{ActionJM}
\end{align}

In section \ref{sec:extshift}, we calculate the leading-order corrections to dyonic, non-rotating, extremal black hole solutions corresponding to the effective action (\ref{ActionJM}); various technical details are given in appendices \ref{app:E_Maxwell Corrections} and \ref{app:F_Lagrangian Corrections}. The corrected extremality bound is inferred by demanding the existence of a horizon (\ref{extremality result}) and is found to depend on all five of the four-derivative operators, including parity-violating operators when magnetic charges are present. It is shown that the three-derivative operators do not give corrections to spherically symmetric solutions at any order in the perturbative expansion. 

Next we would like to analyze the decay of these black holes. \cite{Cheung:2014vva} discussed the necessary condition on the particle spectrum for a black hole with multiple charges to decay. The spectrum of light states is assumed to consist of a set of particles with masses $m_i$ and electric and magnetic charges $\vec{q}_i$ and $\vec{p}_i$. Then the condition that the decay of asymptotically large extremal black holes be allowed is given by the \textit{convex hull condition}\cite{Cheung:2014vva}:\\

\textbf{Weak Gravity Conjecture (Multiple Charges):} \textit{In a UV complete model of quantum gravity, the convex hull of the set of charge-to-mass vectors}
\begin{equation}
    \vec{z}_i \equiv \frac{M_{\text{Pl}}}{m_i}
        \begin{pmatrix}
            \vec{q}_i\\
            \vec{p}_i
        \end{pmatrix},
\end{equation}
 \textit{for every charged state in the spectrum, with mass $m$, electric charges $\vec{q}=(q^1,q^2...)$ and magnetic charges $\vec{p}=(p^1,p^2,...)$, must enclose the unit ball $|\vec{z}|^2\leq 1$.}\\

In section \ref{sec:decay}, we analyze the necessary kinematic conditions under which multiply-charged black holes can decay into smaller charge black holes. First we describe the natural generalization of the convex hull condition to the black hole regime, and then we argue (with a proof relegated to appendix \ref{app:G_Convexity}) that in the large black hole regime, when the perturbative expansion in $1/Q^2$ is justified, the extremality surface is always convex. The black hole WGC is then shown to reduce to the condition that a quartic form (\ref{extremalityform}) is everywhere positive. This amounts to a conditions on the Wilson coefficients $\{a_{ijk},b_{ijk},\alpha_{ijkl},\beta_{ijkl},\gamma_{ij},\chi_{ijkl},\omega_{ij}\}$ under which the convex hull condition is satisfied by contributions from the \textit{black hole regime}. The condition is analyzed in detail in two examples; first we consider the black hole that is charged under two electric charges $q_1$ and $q_2$, and second we consider the black hole that has both an electric charge $q$ and a magnetic charge $p$ under a single $U(1)$ gauge field. 

The second half of this chapter will be devoted to a similar set of calculations in AdS in a general number of dimensions, and we will restrict the low energy spectrum to a graviton and a single vector field. As we will see, many parts of the WGC story do not apply in AdS for an obvious reason: the relationship between mass and charge of an extremal black holes in AdS is \textit{already non-linear at the two-derivative level}\footnote{By ``extremal," we mean that the temperature is zero. This is not the same as the BPS limit in AdS.}.
Therefore it is not at all clear what is gained by studying the higher-derivative corrections to the extremal mass-to-charge ratio%
\footnote{Other aspects of the WGC have been discussed in AdS.  See e.g.~\cite{Nakayama:2015hga, Harlow:2015lma, Montero:2016tif, Montero:2018fns}.}.
Furthermore, massive particles emitted from a black hole cannot fly off to infinity in AdS as they can in flat space, so if the WGC allows for the instability of black holes in AdS, it must be through a completely different mechanism. 

Regardless, the entropy-extremality relationship is expected to hold in AdS as it does in flat space (and indeed, an example in AdS${}_4$ was given in \cite{Goon:2019faz}). The remainder of this chapter is devoted to analyzing the entropy shift. In section \ref{sec:AdSRNcor}, we compute the first order corrections to the AdS Reissner-Nordstr{\"o}m (RN) black hole solutions, and we use the solution to compute the shifts to extremality and the Wald entropy. We verify that the relationship \cite{Cheung:2018cwt, Goon:2019faz} between the shift to mass and shift to entropy is valid for AdS-RN black holes and discuss a slight extension whereby these quantities are both proportional to the charge shift as well. 

In section \ref{sec:EuclideanAction}, we reproduce these results from a thermodynamic point of view. It was shown \cite{Reall:2019sah} that the first-order corrections to the solutions are not needed to compute the first order corrections to thermodynamic quantities. In this section, we verify that this is the case for AdS-RN backgrounds by computing the four-derivative corrections to the renormalized on-shell action. From this we compute the free energy and other thermodynamic quantities. We find that the thermodynamic calculation using the method of \cite{Reall:2019sah} match with the results obtained using the corrected solutions. Specifically, we find an exact match in even dimensions, while in odd dimensions the free energy and associated thermodynamic quantities are renormalization-scheme dependent, and agree with the geometric calculation in a physically motivated \textit{zero Casimir} scheme.

In section \ref{sec:EntropyConstraints}, we review the argument \cite{Cheung:2018cwt} for the positivity of the entropy shift, and comment on a potential issue with applying it to AdS. The positivity of the entropy shift requires that the black hole solutions are local minima of the path integral, so we compute the specific heat and electrical permittivity to determine the regions of parameter space where the black holes will be stable. Finally, we determine the constraints placed on the EFT coefficients by assuming that the entropy shift is positive for all stable black holes. The constraints include the requirement that the coefficient of Riemann-squared is positive. As this coefficient is proportional to the difference $c - a$ between the central charges of the dual CFT, we conclude that the positivity of the entropy shift will be violated in theories where $c - a < 0$. Some of the details about. We relegate to appendix \ref{app:H_Entropy} the specific form of the entropy shifts and bounds on the EFT coefficients for AdS${}_5$ through AdS${}_7$. 

\section{Extremality Shift with multiple $U(1)$s}
\label{sec:extshift}

In this section we will determine the effect of higher-derivative operators on the extremality bound using the method developed in \cite{Kats:2006xp}. In the case of multiple charges, this amounts to delineating the space of allowed charge combinations $Q = \sqrt{ q_1^2 + p_1^2 + ...}$ for a given mass $m$. We use the presence of a naked singularity, or absence of an event horizon, to rule out charge configurations at a given mass; such combinations of charge and mass will be called \textit{superextremal}. 

In pure Einstein-Maxwell theory, the superextremal black holes have $Q/m > 1$. We refer to such an inequality as the \textit{extremality bound}. This requirement derives from the positivity of the discriminant of the function $1/ g_{rr}$, which itself comes from the requirement that that function should have a zero (i.e. the event horizon).  We will see that the higher-derivative corrections have the effect of shifting the right-hand side of this bound by factors proportional to the Wilson coefficients and suppressed by factors of $1 / Q$. Generically, $n$-derivative operators will contribute a term in the extremality bound that is proportional to $1 / Q^{n-2}$.




This approach is necessarily first-order in the EFT coefficients; if we were to compute the shift to second-order in the four-derivative coefficients, we would need also to consider the first-order effect of six-derivative operators, as these contribute at the same order in $1/Q$. This means that at each step we eliminate all terms that are beyond leading-order in the four-derivative coefficients. 

\subsection{No Correction from Three-Derivative Operators}

When $N\geq 3$ the leading effective interactions are given by three-derivative operators:
\begin{equation}
    S_{3} = \int \text{d}^4 x \sqrt{-g}\left[\frac{M_{\text{Pl}}^2}{4}R - \frac{1}{4} F^i_{\mu \nu} F^{i \, \mu \nu} +a_{ijk}F^i_{\mu\nu}F^{j\nu\rho}{F^k_{\rho}}^\mu+ b_{ijk}F^i_{\mu\nu}F^{j\nu\rho}{\tilde{F}^k_{\rho}}{}^\mu\right],
\end{equation}
where the dual field strength tensor is defined as
\begin{equation}
    \tilde{F}^{i\mu\nu} = \frac{1}{2}\epsilon^{\mu\nu\rho\sigma}F^i_{\rho\sigma} \, .
\end{equation}
From the index structure of the three-derivative operators (alternatively from the structure of the corresponding local matrix elements given in appendix \ref{app:D_EFT Basis}) one can show that both $a_{ijk}$ and $b_{ijk}$ are totally antisymmetric. 

We analyze solutions to the equations of motion:
\begin{align}
    \label{3dereom}
    \nabla_\mu F^{i\mu\nu} &= -6a_{ijk} \nabla_\mu \left(F^{j\nu\rho}{F^k_\rho}^\mu\right) -6b_{ijk} \nabla_\mu \left({F^j_\alpha}^\nu \tilde{F}^{k\mu\alpha}\right),\nonumber\\
    R_{\mu\nu}-\frac{1}{2}Rg_{\mu\nu} &= \frac{2}{M_{\text{Pl}}^2}\left[F^i_{\mu\rho}{F^i_\nu}^\rho -\frac{1}{4}g_{\mu\nu}F^i_{\rho\sigma}F^{i\rho\sigma}\right. \nonumber\\
    & \hspace{5mm}  + 2\left. a_{ijk}\left[F^i_{\alpha\mu}F_\nu^{j\rho}F_\rho^{k\alpha}-\frac{1}{2}g_{\mu\nu}F_{\rho\sigma}^i F^{j\sigma\alpha}F_\alpha^{k\rho}\right] +2 b_{ijk}F^i_{\mu\rho}F^j_{\nu\sigma}\tilde{F}^{k\rho\sigma}\right].
\end{align}
By an elementary spurion analysis it is clear that there can be no modification of the extremality bound at $\mathcal{O}(a,b)$. Promoting $a_{ijk}$ and $b_{ijk}$ to background fields transforming as totally anti-symmetric tensors of the (explicitly broken) flavor symmetry group $O(N)$, at leading order the extremality shift can depend only on invariants of the form $a_{ijk} q^i q^j q^k$ or $a_{ijk} q^i q^j p^k$, which vanish. At next-to-leading order there could be contributions of the form $a_{ijk}a_{klm}q^ip^jq^lp^m$, which do not obviously vanish for similarly trivial reasons. If present such contributions would appear at the same order, $\mathcal{O}\left(1/Q^2\right)$ as the leading-order contributions from the four-derivative operators. 

Interestingly these $\mathcal{O}(a^2,ab,b^2)$ corrections also vanish. To show this, we evaluate the right-hand-side of (\ref{3dereom}) on a spherically symmetric ansatz, 
\begin{align}
\begin{split}
        ds^2 \ = \ & g_{tt}(r) \, dt^2 + g_{rr}(r) \, r^2 dr^2 + d \Omega^2 , \qquad F^{i \, tr}(r), \qquad   F^{i \, \theta \phi}(r), 
\end{split}
\end{align}
with the remaining components of the field strength tensors set to zero. The higher-derivative terms are seen to vanish due to the structure of the index contractions. The equations of motion for the non-zero components $g_{tt},\; g_{rr},\; F^{i t r}, \; F^{i \theta \phi}$ are \textit{identical} to the equations of motion of two-derivative Einstein-Maxwell. The Reissner--Nordstr{\"o}m black hole remains the unique spherically symmetric solution to the higher-derivative equations of motion with a given charge and mass.

It is interesting to note that the above argument fails if the solution is only axisymmetric, as in the general Kerr-Newman solution. For spinning, dyonic black holes, the three-derivative operators might give $\mathcal{O}\left(1/Q^2\right)$ corrections to the extremality bounds. We leave the analysis of this case to future work.

\subsection{Four-Derivative Operators}

The three-derivative operators have no contribution on spherically symmetric backgrounds. Thus, the leading shift to the extremality bound comes from four-derivative operators. We consider the action
\begin{align}
  \begin{split}
    S_4 &= \int \mathrm{d}^4x \sqrt{- g} \Big( \frac{R}{4} - \frac{1}{4} F^i_{\mu \nu} F^{i \, \mu \nu} 
      + \alpha_{i j k l} \, F^i_{\mu \nu} F^{j \, \mu \nu} F^k_{\rho \sigma} F^{l \, \rho \sigma}
    + \beta_{i j k l} \,  F^i_{\mu \nu} \tilde{F}^{j \, \mu \nu} F^k_{\rho \sigma} \tilde{F}^{l \, \rho \sigma} 
    \\
    & \qquad \qquad \qquad + \gamma_{i j} \, F^i_{\mu \nu} F^j_{\sigma \rho} W^{\mu \nu \sigma \rho}
    + \chi_{i j k l} \, \tilde{F}^i_{\mu \nu} F^{j \, \mu \nu} F^k_{\rho \sigma} F^{l \, \rho \sigma} 
    + \omega_{i j} \, F^i_{\mu \nu} \tilde{F}^j_{\sigma \rho} W^{\mu \nu \sigma \rho} \Big).
  \end{split}
  \label{Action4}
\end{align}
Here the Latin indices run from $1$ to the number of gauge fields $N$. This is the most general possible set of four-derivative operators for Einstein-Maxwell theory in 4 dimensions. For a thorough discussion on how these operators comprise a complete basis, see appendix \ref{app:D_EFT Basis}. We will see that the parity-odd operators can contribute if we allow for magnetic charges. Our calculation is identical to the one performed in \cite{Kats:2006xp} if we set $N \rightarrow 1$ and turn on only electric charges. We have chosen units with $ M_{\text{Pl}}= 1$ for convenience, though they may be restored via dimensional analysis. 

\subsubsection{Black Hole Background}

First consider the uncorrected theory, which is gravity with $N$ $U(1)$ gauge fields. This theory admits solutions that are black holes with up to $N$ electric and magnetic charges. These solutions take the form: %
\begin{align}
\label{zerosol}
\begin{split}
        ds^2 \ = \ & g_{tt} \, dt^2 + g_{rr} \, dr^2 + r^2 d \Omega^2 , \qquad F^{i \, tr} \ = \ \frac{q^i}{r^2}, \qquad  F^{i \, \theta \phi} \ = \ \frac{p^i}{r^4 \, \sin \theta} \, , \\[5pt]
        & \qquad -g_{tt} \ = \ g^{rr} \ = \ 1 - \frac{2 M}{r} + \frac{Q^2}{r^2} \, .
\end{split}
\end{align}
Here $Q^2 = q^i q^i + p^i p^i$. These backgrounds are spherically symmetric, so we will impose this as a requirement on the shifted background\footnote{Spherical symmetry ensures that $1/g_{rr} = g^{rr}$, even for the corrected solutions. However, $g_{tt}$ and $1/g_{rr}$ will generally receive different corrections, which is why we do not denote these functions with one symbol such as $f(r)$.}. In the case of spherical symmetry, one may rearrange the Einstein equation and integrate to find \cite{Kats:2006xp}
\begin{align}
    \label{T00Int}
    g^{rr} = 1 - \frac{2 M}{r} - \frac{2}{r} \int_r^\infty dr r^2 T_t{}^t  \, .
\end{align}
For the uncorrected theory, the stress tensor is 
\begin{align}
    T_{\mu \nu} = F^i_{\mu \alpha} F^i_{\nu}{}^{\alpha} - \frac{1}{4} F^i_{\alpha \beta} F^{i \alpha \beta} g_{\mu \nu} \, .
\end{align}
In this case, it is easy to see that the effect of the stress tensor is to add the $\frac{q^2 + p^2}{r^2}$ term to $g^{rr}$.

\subsubsection{Corrections to the Background}

Now consider the effect of the four-derivative terms. To compute their effect on the geometry, we must compute their contributions to the stress tensor. We will expand the stress tensor as a power series in the Wilson coefficients as 
\begin{align}
    T = T^{(0)} + T^{(1)}_{\text{Max}} + T^{(1)}_{\text{Lag}} + ...
\end{align}
Here we have written two terms that are proportional to the first power of the Wilson coefficients $( \alpha_{ijkl}, \beta_{ijkl}, ... )$, because there are two different sources of first-order corrections.

The first change $T^{(1)}_{\text{Max}}$ comes from the effect of these operators on solutions to the Maxwell equations, which changes the values of $F^i_{\mu \alpha} F^i_{\nu}{}^{\alpha} - \frac{1}{4} F^i_{\alpha \beta} F^{i \alpha \beta} g_{\mu \nu} $. Thus, $T^{(1)}_{\text{Max}}$ essentially comes from evaluating the zeroth-order stress tensor on the first-order solution of the $F^i$ equations of motion.

The second change $T^{(1)}_{\text{Lag}}$ derives from varying the higher-derivative operators with respect to the metric. Thus, this term is essentially the first-order stress tensor, and we will evaluate it on the zeroth-order solutions to the Einstein and Maxwell equations. The remainder of this section will be devoted to computing each of these contributions. 

\subsubsection{Maxwell Corrections}

The first source of corrections to the stress tensor derives from including the corrections to the value of $F$. The corrected gauge field equation of motion is
\begin{align}
    \begin{split}
    \nabla_{\mu} F^{i \mu \nu} =& \, \nabla_{\mu} \Big( 8 \, \alpha_{ijkl}  F^{j \mu \nu} F^k_{\alpha \beta} F^{l \alpha \beta} + 8 \, \beta_{ijkl}  \tilde{F}^{j \mu \nu} F^k_{\alpha \beta} \tilde{F}^{l \alpha \beta} + 4 \, \gamma_{ij}  F^j_{\alpha \beta} W^{\mu \nu \alpha \beta}  \\
    &  \qquad \qquad + 4 \, \left( \chi_{ijkl} \tilde{F}^{j \mu \nu} F^k_{\alpha \beta} F^{l \alpha \beta} + \chi_{klij} F^{j \mu \nu} \tilde{F}^k_{\alpha \beta} F^{l \alpha \beta} \right) +  4 \, \omega_{ij}  \tilde{F}^j_{\alpha \beta} W^{\mu \nu \alpha \beta} \Big).
    \label{Maxwell}
    \end{split}
\end{align}
We denote the right-hand side of this equation by $\nabla_{\mu} G^{\mu \nu}$. The first-order solution to the Maxwell equation leads to corrections that equal (see appendix \ref{app:E_Maxwell Corrections})
\begin{align}
    \begin{split}
        & (T^{(1)}_{\text{Max}})_t{}^t \ = \ - \left[ \sqrt{-g} G^{i t r} \right]^{(1)} \left[  \sqrt{-g} F^{i t r} \right]^{(0)} / (g_{\theta \theta } g_{\phi \phi}) \, .
        \label{max_form}
    \end{split}
\end{align}
By plugging in the zeroth-order values of the fields into this expression, we compute the corrections to the stress tensor through the Maxwell equation:
\begin{align}
    \begin{split}
        \label{Maxwell Corrections}
        (T^{(1)}_{\text{Max}})_t{}^t  \ & = \ \frac{8}{r^8} \, \Big( 2 \alpha_{ijkl} \, q^i q^j (q^k q^l - p^k p^l) + 4 \,  \beta_{ijkl} \, q^i p^j  q^k p^l  +  2 \gamma_{ij} \, q^i q^j \, (Q^2  - Mr) \\
        & \qquad \quad + \chi_{ijkl} \, \left( q^i p^j( q^k q^l - p^k p^l)  + 2 q^i q^j q^k p^l \right) +  \, 2 \omega_{ij}\, q^i p^j \, (Q^2 - Mr)  \Big) \, .
    \end{split}
\end{align}
The details of this derivation may be found in appendix \ref{app:E_Maxwell Corrections}, but we should comment on a few interesting points. First, note the only $G^{i t r}$ arises in the result. This is due to the Bianchi identity, which does not allow $G^{i \theta \phi}$ to contribute. The Bianchi identity requires that $\partial_r F_{\theta \phi} = 0$, so in fact $F^i_{\theta \phi}$ can get no corrections at any order. 

A subtlety arises from the fact that the metric appears in the expression for the stress tensor. Therefore, it might appear that the first-order corrections to $T_t{}^t$ involve contributions from the first-order value of $F$ and the first-order value of $g$. This would be problematic because the first-order value of $g$ is what we use the stress tensor to compute in the first place. In fact, this is not an issue; only the zeroth-order metric shows up in (\ref{max_form}). This decoupling relies on cancellation between various factors of metric components, as well as spherical symmetry. Without this, the perturbative procedure we use to compute the shift to the metric would not work. We do not expect this decoupling between corrections to the stress tensor and corrections to the metric to happen for general backgrounds. It would be interesting to understand the general circumstances under which it occurs.

\subsubsection{Lagrangian Corrections}

The second source of corrections is comparatively straightforward and comes from considering the higher-derivative terms in the Lagrangian as ``matter" and varying them with respect to the metric. The variations of each term are given in appendix \ref{app:F_Lagrangian Corrections}. The result is 
\begin{align}
    \begin{split}
        (T^{(1)}_{\text{Lag}})_t{}^t \quad &= \quad \frac{1}{r^8} \, \Big( 4 \, \alpha_{ijkl} \, (p^i p^j p^k p^l + 2 q^i q^j p^k p^l - 3 q^i q^j q^k q^l) - 4 \, \beta_{ijkl} \, q^i p^j  q^k p^l\\
        &  -\frac{4}{3} \, \gamma_{ij} \left( q^i q^j (6Q^2 - 2 M r - 3 r^2) + p^i p^j (6Q^2 - 10 Mr - 3 r^2) \right) \\
        & - 16\,  \chi_{ijkl} \, q^i p^j q^k q^l - \frac{8}{3} \, \omega_{ij} q^i p^j(4 Mr - 3 r^2) \Big) \, .
    \end{split}
\end{align}

In both cases, we have simplified the expressions by using the symmetries of the tensor appearing in the higher-derivative terms (e.g. $\alpha_{ijkl} = \alpha_{jikl} = \alpha_{klij}$). 

\subsection{Leading Shift to Extremality Bound}

By adding together both sources of corrections and computing the integral in (\ref{T00Int}), we compute the shift to the radial function $g^{rr}$ defined as, 
\begin{align}
    g^{rr} = 1 - \frac{2 M}{r} + \frac{q^2 + p^2}{r^2} + \Delta g^{rr}.
\end{align}
Then the shift is given by
\begin{align}
    \begin{split}
        \label{metricshift}
        \Delta g^{rr} &= -\frac{4}{15 r^6} \Big( 6 \, \alpha_{ijkl} \, (q^i q^j - p^i p^j)(q^k q^l - p^k p^l) \,
         + \, 24 \beta_{ijkl} q^i p^j q^k p^l \\
        & +\gamma _{ij} \, \left( q^i q^j - p^i p^j\right) \left( 12 Q^2 -25 M r + 10 r^2 \right)  \\
        &  + 12 \,  \chi_{ijkl} \, q^i p^j \, \left(q^k q^l-p^k p^l\right) + 2 \,  \omega _{ij} \,   q^i p^j \left( 12 Q^2 -25 M r + 10 r^2 \right) \Big).
    \end{split}
\end{align}

To find the shift to extremality that results from this, we examine when the new radial function $g^{rr}(r, M, Q)$ has zeros \footnote{Equivalently we could examine the zeros of $g_{tt}$. This must give identical results since the consistency of the metric signature requires that $g_{tt}$ and $g^{rr}$ have the same set of zeros.   }. This equation is sixth order in $r$, but we are only interested in the first-order shift to the solution. We Taylor-expand near the extremal solution where $r = M$ and $Q = M$, and keep only terms that are first-order in Wilson coefficients:
\begin{align}
    \begin{split}
        g^{rr}(r, M, Q) & \, = \, g^{rr}(M, M, M) + (Q - M) \, \partial_Q g^{rr} |_{(M, M, M)} + (r - M) \,  \partial_r g^{rr} |_{(M, M, M)} \\
        & \, = \, \Delta g^{rr}(M, M, M) + (Q - M) \, \partial_Q g^{rr} |_{(M, M, M)}. \,
    \end{split}
\end{align}
We have kept $M$ fixed.  In going from the first to the second line, we have used that the uncorrected metric vanishes at $(M, M, M)$ so $g^{rr}(M, M, M) = \Delta g^{rr}(M, M, M)$. We also used that the uncorrected metric also has vanishing $r-$derivative at $(M, M, M)$, so the last term on the first line may be removed because it is second-order in Wilson coefficients. The requirement that $g^{rr}$ leads to the condition:
\begin{equation}
    g^{rr}(r, M, Q) = 0 \implies  Q - M =  - \frac{\Delta g^{rr}(M, M, M)}{ \partial_Q  g^{rr}(M, M, M)}.
\end{equation}
Now we evaluate this expression and divide by $m$ to find the result for the extremality bound $|\vec{z}|^2 = Q^2/M^2 $
\begin{align} 
    \begin{split}
        |\vec{z}| & \leq 1 +  \frac{2}{5 (Q^2)^3} \Big( 2 \, \alpha_{ijkl} \, (q^i q^j - p^i p^j)(q^k q^l - p^k p^l) \,
         + \, 8 \beta_{ijkl} q^i p^j q^k p^l - \gamma _{ij} \, \left( q^i q^j - p^i p^j\right) Q^2 \\
        &  \hspace{25mm} + 4 \,  \chi_{ijkl} \, q^i p^j \, \left(q^k q^l-p^k p^l\right) - 2 \,  \omega _{ij} \,   q^i p^j Q^2  \Big) + \mathcal{O}\left(\frac{1}{(Q^2)^2}\right).
    \label{extremality result}
    \end{split}
\end{align}
This is the main technical result of \cite{Jones:2019nev}. In the next section, we comment on the constraints that black hole decay might place on these coefficients, and we analyze this expression for the case of black holes with two electric charges, and the case of black holes with a single electric and single magnetic charge.

\section{Black Hole Decay with multiple $U(1)$s}
\label{sec:decay}

As described by \cite{Cheung:2014vva} and reviewed in section \ref{subsec:review}, a state with charge-to-mass vector $\vec{z}$ and total charge $Q^2\equiv \sum_i((q^i)^2+(p^i)^2)$ is kinematically allowed to decay to a general multiparticle state only if $\vec{z}$ lies in the convex hull of the light charged states. In the case of asymptotically large extremal black holes decaying to finite charge black holes, the spectrum of light states corresponds to the region compatible with the extremality bound. This bound describes a surface in $z$-space of the form
\begin{equation}
    |\vec{z}|=1+T(\vec{z},Q^2),
\end{equation}
where $T\rightarrow 0$ as $Q^2\rightarrow \infty$. The convex hull condition \cite{Cheung:2014vva} has a natural generalization to the sector of extremal black hole states:\\

\textbf{Black Hole Convex Hull Condition:} \textit{It is kinematically possible for asymptotically large extremal black holes to decay into smaller finite $Q^2$ black holes only if the convex hull of the extremality surface encloses the unit ball $|\vec{z}|\leq 1$}.\\

\begin{figure}
\includegraphics[scale=0.8]{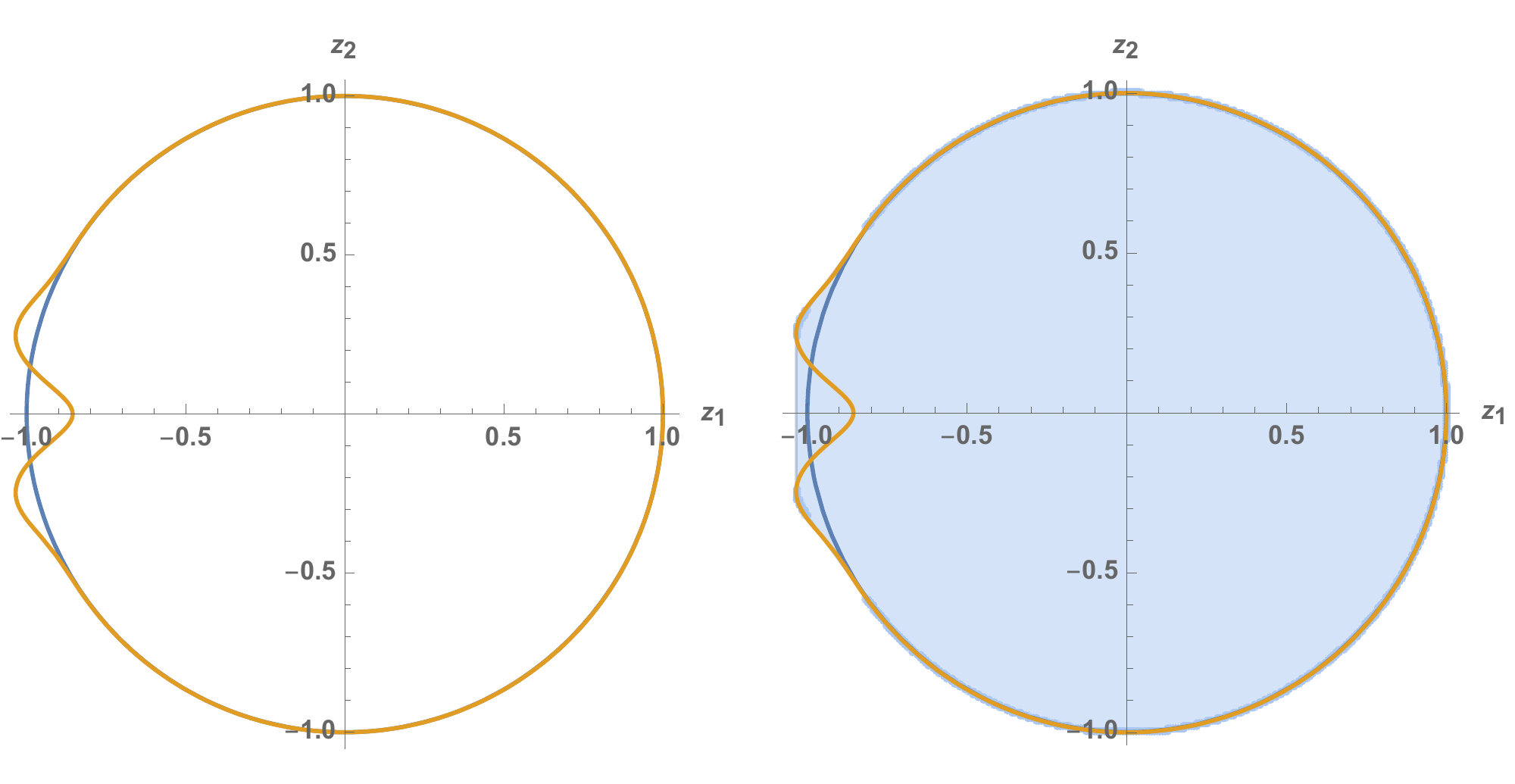}
\begin{flushleft}
\caption{(Left): an extremality curve that naively violates the WGC as it does not enclose the unit circle. (Right): the convex completion of the extremality curve \textit{does} enclose the unit circle, hence the WGC is satisfied. For this to be possible the extremality surface must be somewhere locally non-convex, which is shown in appendix \ref{app:G_Convexity} to be impossible in the perturbative regime.
\label{convex hull}}
\end{flushleft}
\end{figure}

This means that to determine if the decay of a large black hole is kinematically allowed, we must first determine the convex hull of a complicated surface, a task that may only be tractable numerically. As illustrated in figure \ref{convex hull}, it is possible for the convex hull of the extremality surface to enclose the unit ball even if the surface itself does not. Furthermore, the extremality surface may be non-convex even if the magnitude of the corrections is arbitrarily small. 

The condition simplifies somewhat in the $Q^2 \gg 1$ regime, where the corrections to the unit circle derive from the four-derivative terms and are small as a result. In appendix \ref{app:G_Convexity} we prove that if $T(\vec{z},Q^2)$ is a quartic form, as it is in the explicit result (\ref{extremality result}), then the smallness of the deviation \textit{does} imply convexity. In this regime, the convex hull condition is simplified in the sense that the extremality surface always bounds a convex region. At a given $Q^2 \gg 1$, and $\vec{z}$, the black hole extremality bound describes a surface in $z$-space of the form
\begin{equation}
    |\vec{z}|=1+\frac{1}{(Q^2)^3} T_{ijkl}z^i z^j z^k z^l + \mathcal{O}\left(\frac{1}{(Q^2)^2}\right).
\end{equation}
The condition for the multi-charge weak gravity conjecture to be satisfied in the perturbative regime degenerates to the more tractable condition:\\

\textbf{(Perturbative) Black Hole Weak Gravity Conjecture:} \textit{
It is kinematically possible for asymptotically large extremal black holes to decay into smaller finite $Q^2$ extremal black holes if the quartic extremality form} 
\begin{equation}
\label{extremalityform}
    T(q^i,p^i) = T_{ijkl}z^i z^j z^k z^l,
\end{equation}
 \textit{is everywhere non-negative. Using the parametrization of the effective action (\ref{ActionJM}), this bound takes the form}
    \begin{align}
        T(q^i,p^i)\;\;= \;\;&2  \alpha_{ijkl}  (q^i q^j - p^i p^j)(q^k q^l - p^k p^l) +  8 \beta_{ijkl} q^i p^j q^k p^l - \gamma _{ij} \, Q^2\left( q^i q^j - p^i p^j\right)\nonumber \\
        &  \hspace{5mm} + 4 \,  \chi_{ijkl} \, q^i p^j \, \left(q^k q^l-p^k p^l\right) - 2 \,  \omega _{ij} \,  Q^2 q^i p^j \geq 0 \, ,
        \label{extremality bound}
\end{align}
\textit{which follows directly from (\ref{extremality result}).}

\subsection{Examples}

According to the previous section, we can determine whether black holes are stable by checking if the extremality form is anywhere negative. In this section we demonstrate this with a few basic examples. 

\subsubsection{Black Hole With Two Electric Charges}

A black hole that is electrically charged under two $U(1)$ groups provides one simple example. In this case, the extremality bound simplifies to 
\begin{align}
    (2 \alpha_{ijkl} - \gamma_{ij} \delta_{kl}) q^i q^j q^k q^l  > 0.
\end{align}
As the $q$ factors project to the completely symmetric part of this tensor, it is convenient to define $ T_{ijkl} = 2 \alpha_{ \{ijkl\} } - \gamma_{ \{ij} \delta_{kl \} }$, where we have symmetrized the indices with weight one. Expanding the constraint in components leads to 
\begin{align}
    \begin{split}
        T_{1111} \, q_1^4 + T_{1112} \, q_1^3 \, q_2 + T_{1122} \, q_1^2 \, q_2^2 + T_{1222} \, q_1 \, q_2^3  + T_{2222} \, q_2^4 > 0.
        \label{two electric polynomial}
    \end{split}
\end{align}
This polynomial must be positive for all possible combinations of $q_1$ and $q_2$. We use the fact that the polynomial in (\ref{two electric polynomial}) is homogenous, and divide by $(q_2)^4$. Redefining $q_1 / q_2 = x$ simplifies the left-hand-side of the inequality to a polynomial of one variable:
\begin{align}
    \begin{split}
        T_{1111} \, x^4 +  T_{1112} \, x^3 +  T_{1122} \, x^2 + T_{1222} \, x  + T_{2222} > 0.
    \end{split}
    \label{electric one var polynomial}
\end{align}
This polynomial is quartic so one may solve this by studying the explicit expressions for the roots and demanding that they are not real. However the positivity conditions for fourth order polynomials are much simpler and lead to a set of relations among the components of $T_{ijkl}$ (see, for instance, \cite{Wang2005CommentsO}). This allows the problem to be solved entirely in the case of two charges; for $N>2$ one must analyze multivariate polynomials. 

For an example of a theory that may be in the Swampland, consider the following four-derivative terms:
\begin{align}
    \begin{split}
        \mathcal{L}_4 = \alpha_{1111} \, F^1_{\mu \nu} F^{1 \, \mu \nu} F^1_{\rho \sigma} F^{1 \, \rho \sigma} + \alpha_{1122} \, F^1_{\mu \nu} F^{1 \, \mu \nu} F^2_{\rho \sigma} F^{2 \, \rho \sigma} + \alpha_{2222} \, F^2_{\mu \nu} F^{2 \, \mu \nu} F^2_{\rho \sigma} F^{2 \, \rho \sigma},
    \end{split}
\end{align}
where $\alpha_{1111} = 2$, $\alpha_{1122} = -8$, and $\alpha_{2222} = 3$. Then the extremality shift becomes
\begin{align}
    2 \, q_1^4 - 8 \, q_1^2 \, q_2^2 + 3 \, q_2^4 > 0.
\end{align}
The inequality is satisfied when $q_1 = 0$ or $q_2 = 0$, but at $q_1 = q_2$, the extremality shift is negative. Therefore, a black hole with $q_1 = q_2$ in this theory would not be able to decay to smaller black holes. This model requires the existence of self-repulsive states in the spectrum in either the particle or stringy regimes to evade the Swampland.

\subsubsection{Dyonic Black Hole}

Another simple case occurs when there is only a single gauge field but the black hole has both electric and magnetic charge. Then the extremality bound is obtained by removing all indices from (\ref{extremality result}):
\begin{align} 
    2 \alpha \, (q^2 - p^2)^2 + 8 \beta \, q^2 p^2 - \gamma \, (q^2 - p^2) (q^2 + p^2) + 4 \chi \, q p (q^2 - p^2)  - 2 \omega \, q p (q^2 + p^2) \ > \  0.
    \label{onedyonicext}
\end{align}
We recover the results of \cite{Kats:2006xp} when the magnetic charge is set to zero. A single electric charge shifts the extremality as 
\begin{equation}
    \label{singleelectric}
    |z_q| = 1 + \frac{2}{5 |Q|^2} ( 2 \alpha - \gamma ).
\end{equation}
However, a single magnetic charge has the opposite sign for $\gamma$:
\begin{equation}
    \label{singlemagnetic}
    |z_{p}| = 1 + \frac{2}{5 |Q|^2} ( 2 \alpha + \gamma ).
\end{equation}
Requiring that both types of black holes be able to decay places a stronger constraint on $\alpha$ and $\gamma$:
\begin{align}
    \begin{split}
        \label{combined}
        2 \alpha > |\gamma|.
    \end{split}
\end{align}
If we assume that both $p$ and $q$ are non-zero, we can again divide by $p^4$ as we did in the previous section, and again find a polynomial of a single variable:
\begin{align}
     (2 \alpha - \gamma) \, y^4 + (4 \chi - 2 \omega ) \, y^3 + (-4 \alpha  + 8 \beta) \, y^2 + (- 4 \chi - 2 \omega) \, y + (2 \alpha + \gamma) \ > \ 0.
     \label{dyon polynomial}
\end{align}

The generalized bound (\ref{dyon polynomial}) coincides exactly with the (regularized forward-limit) scattering positivity bounds derived in \cite{Bellazzini:2019xts} for arbitrary linear combinations of external states. It is interesting that the requirement that dyonic black holes are unstable gives a new physical motivation for these generalized scattering bounds.

For the case of a single gauge field, a very physical example comes to mind: the Euler-Heisenberg Lagrangian \cite{Heisenberg1936}, in which integrating out electron loops induces a four-point interaction among the gauge fields.\footnote{The electron should also contribute to the $W F F$-type operators as well, but this contribution is suppressed by a factor of $1/z$. The electron is extraordinarily superextremal ($z = 2 \times 10^{21}$) so we can safely ignore these terms for our example.} This model has four derivative terms given by
\begin{align}
    \mathcal{L}_4 = \alpha (F_{\mu \nu} F^{\mu \nu})^2 + \beta (F_{\mu \nu} \tilde{F}^{\mu \nu})^2,
\end{align}
with $\alpha = 4$, $\beta = 7$ (up to overall constants that do not effect the problem). The inequality that must be satisfied is the following:
\begin{align}
    4 y^4 + 40 y^2 + 8 > 0.
\end{align}
Clearly this holds for all values of $y$. Thus, we have found that the Euler-Heisenberg theory is not in the Swampland. This does not require that we know anything about the spectrum, or that the higher-derivative operators came from integrating out a particle at all. Only the four-derivative couplings are needed to learn that this theory allows nearly extremal black holes to decay.

The condition (\ref{onedyonicext}) exhibits an interesting simplification when $\alpha = \beta$ and the remaining coefficients are set to zero. In this case, the condition on the quartic form then reads
\begin{equation}
    \alpha (q^2+p^2)^2 > 0. 
\end{equation}
In this special case the extremality surface becomes invariant under orthogonal rotations in charge-space. In fact, it is simple to verify that this is the only choice of coefficients with this feature. The enhanced symmetry is a consequence of the electromagnetic duality invariance of the equations of motion for this choice of coefficients. In the effective action, the necessary condition for duality invariance is the \textit{Noether-Gaillard-Zumino condition} \cite{Gaillard:1981rj}
\begin{equation}
    F_{\mu\nu}\tilde{F}^{\mu\nu}+G_{\mu\nu}\tilde{G}^{\mu\nu} = 0, \hspace{5mm} \text{where} \hspace{5mm} \tilde{G}_{\mu\nu} \equiv 2 \frac{\delta S}{\delta F^{\mu\nu}}.
\end{equation}
One can verify that this is satisfied if we $\alpha = \beta, \ \gamma = \chi = \omega = 0$ as above, at least to fourth order in derivatives. To make this equation hold to sixth order would require the addition of sixth-derivative operators to the Lagrangian, and so on. For a general analysis of electric-magnetic duality invariant theories, see \cite{Gibbons:1995cv}.

\begin{figure}
\includegraphics[scale=0.8]{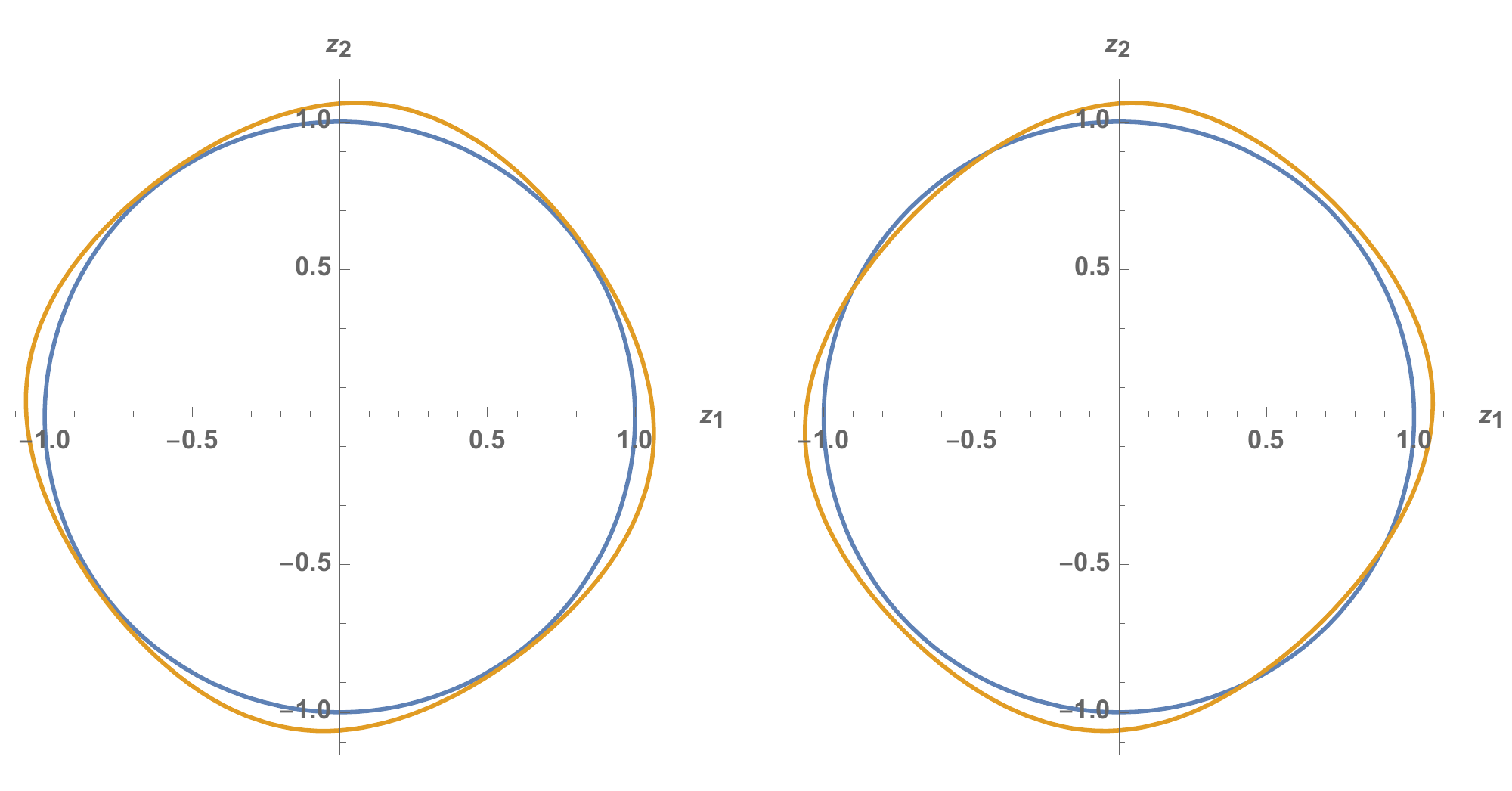}
\begin{flushleft}
\caption{
(Left): the corrections to the extremality curve are everywhere positive, hence the WGC is satisfied. (Right): the corrections to the extremality curve are \textit{not} everywhere positive; large extremal black holes cannot always decay to intermediate mass black holes, whether or not the WGC is satisfied cannot be decided in the low-energy EFT.
\label{positive corrections}}
\end{flushleft}
\end{figure}

\subsection{Unitarity and Causality}

Infrared consistency conditions on the low energy effective theory have been used to bound the coefficients of higher-derivative operators. Such constraints were first considered in the context of the weak gravity conjecture in \cite{Cheung:2014ega}, and were extended to the case of multiple gauge fields in \cite{Andriolo:2018lvp}. Further arguments based on unitarity and causality were given in \cite{Hamada:2018dde}. Here we review these arguments and present a few generalizations.

\subsubsection{Integrating Out Massive Particles}

One source of higher derivative corrections derives from integrating out states in the particle regime. By this we mean states that are well described by ordinary QFT on a fixed spacetime background. Such states necessarily have masses smaller than some cutoff scale $\Lambda_{QFT}$, which is the string scale or whatever scale new physics invalidates the QFT description. We have already seen a simple example of this in the Euler-Heisenberg Lagrangian above.

At tree-level, only neutral particles contribute to the four-point interactions. Consider, for example, a dilaton that couples to the field strengths. The Lagrangian for the scalar theory is
\begin{align}
    \mathcal{L} = \frac{R}{4} - \frac{1}{2} ( \partial \phi )^2 - \frac{m_{\phi}^2}{2} \phi^2 - \frac{1}{4} F^i_{\mu \nu} F^{i \, \mu \nu} + \mu_{i j} \phi F^i_{\mu \nu} F^{j \, \mu \nu}.
\end{align}
We integrate out the scalar to find the effective four-derivative coupling by matching to the low-energy EFT at the scale $\Lambda_{\text{UV}}\lesssim m_\phi$
\begin{align}
    \mathcal{L}_4 \supset \frac{M_{\text{Pl}}^4}{m_{\phi}^2}  ( \mu_{ij} \mu_{kl} + \mu_{ik} \mu_{jl} + \mu_{il} \mu_{jk} ) F^i_{\mu \nu} F^{j \, \mu \nu} F^k_{\rho \sigma} F^{l \, \rho \sigma}.
\end{align}
Therefore, in this simple setup, the coefficient $\alpha_{ijkl}$ takes the form
\begin{align}
    \alpha_{ijkl} =  \frac{1}{m_{\phi}^2}  ( \mu_{ij} \mu_{kl} + \mu_{ik} \mu_{jl} + \mu_{il} \mu_{jk} ).
\end{align}

For a single gauge field $\alpha =  \frac{3 \mu^2}{m_{\phi}^2} $. Unitarity requires that $\mu$ is real, which implies that $\alpha$ is positive \cite{Hamada:2018dde}. It is easy to see that this is still the case when there are more gauge fields. The extremality form for this theory is 
\begin{align}
    \alpha_{ijkl} q^i q^j q^k q^l = \frac{3}{m_{\phi}^2} (\mu_{ij} q^i q^j)^2,
\end{align}
which must be positive.\footnote{Note that unlike the case of single gauge field, unitarity does not bound all the coefficients separately. For instance, in the two charge case, $\mu_{11} = 1$, $\mu_{22} = -1$, and $\mu_{12} = 0$ would lead to $\alpha_{1122} = -1/m_{\phi}^2$.} The same reasoning shows that integrating out an axion, which couples to $F^i \tilde{F}^j$, generates $\beta_{ijlk}$, and that its contribution to the extremality form is also positive.

Light charged particles cannot contribute at tree-level so their leading contributions are at loop-level. The diagrams that contribute in this case are:
\begin{center}
	\begin{tikzpicture}[scale=0.9, line width=1 pt]
	\begin{scope}
\draw[vector] (0,0)--(-1,1);
\draw[scalar] (0,0)--(2,0);
\draw[scalarbar] (0,0)--(0,-2);
\draw[scalar] (2,0)--(2,-2);
\draw[scalarbar] (0,-2)--(2,-2);
\draw[vector] (2,0)--(3,1);
\draw[vector] (0,-2)--(-1,-3);
\draw[vector] (2,-2)--(3,-3);
	\node [above left] at (-1,1) {$\gamma_i$};
	\node [below left] at (-1,-3) {$\gamma_j$};
	\node [above right] at (3,1) {$\gamma_l$};
	\node [below right] at (3,-3) {$\gamma_k$};
	\node at (1,-4) {(a)};
	\end{scope}
	\begin{scope}[shift={(5,-1)}]
	\draw [vector] (2,2)--(3,1);
	\draw [scalar] (3,1)--(4,0);    
	\draw [vector] (2,-2)--(3,-1);	
	\draw [scalarbar] (3,1)--(3,-1);
	\draw [scalarbar] (3,-1)--(4,0);    
	\draw [vector] (4,0.05)--(5,0.05);	
	\draw [vector] (4,-0.05)--(5,-0.05);	
	\draw [vector] (5,0)--(6,1);
	\draw [vector] (5,0)--(6,-1);
	\node [above left] at (2,2) {$\gamma_i$};
	\node [below left] at (2,-2) {$\gamma_j$};
	\node [above right] at (6,1) {$\gamma_l$};
	\node [below right] at (6,-1) {$\gamma_k$};
	\node at (4,-3) {(b)};
	\end{scope} 
	\end{tikzpicture}
\end{center}
\begin{center}
    \begin{tikzpicture}[scale=0.9, line width=1 pt]
    \begin{scope}[shift={(0,0)}]
	\draw [vector] (-1,1)--(0,0);
	\draw [vector] (-1,-1)--(0,0);
	\draw [vector] (0,0.05)--(1.1,0.05);
	\draw [vector] (0,-0.05)--(1.1,-0.05);	
	\draw [scalar] (3,0) arc (0:180:1);
	\draw [scalarbar] (3,0) arc (0:-180:1); 
	\draw [vector] (3,0.05)--(4,0.05);	
	\draw [vector] (3,-0.05)--(4,-0.05);	
	\draw [vector] (4,0)--(5,1);
	\draw [vector] (4,0)--(5,-1);
	\node [above left] at (-1,1) {$\gamma_i$};
	\node [below left] at (-1,-1) {$\gamma_j$};
	\node [above right] at (5,1) {$\gamma_l$};
	\node [below right] at (5,-1) {$\gamma_k$};
	\node at (2,-3) {(c)};
	\end{scope} 
	\begin{scope}[shift={(7,0)}]
	\draw [vector] (2,2)--(3,1);
	\draw [scalar] (3,1)--(4.5,0);    
	\draw [vector] (2,-2)--(3,-1);	
	\draw [scalarbar] (3,1)--(3,-1);
	\draw [scalarbar] (3,-1)--(4.5,0);    
	\draw [vector] (4.5,0.05)--(6,0.05);	
	\draw [vector] (4.5,-0.05)--(6,-0.05);	
	\node [above left] at (2,2) {$\gamma_i$};
	\node [below left] at (2,-2) {$\gamma_j$};
	\node [right] at (6,0) {$h$};
	\node at (4,-3) {(d)};
	\end{scope}
    \end{tikzpicture}
\end{center}
These contribute at the same order except they have relative factors of $z_{\phi}$, the particle's charge-to-mass ratio, coming from counting couplings and propagators. Diagram (a) goes like $z_{\phi}^4$, (b) like $z_{\phi}^2$, (c) like $z_{\phi}^0$; diagram (d) contributes at order $z_{\phi}^2$.  The field-strength four-point interaction is generated by the first three diagrams. In the limit where $z_{\phi} \gg 1$, diagram (a) dominates all the others (as we noted above in the Euler-Heisenberg example) and the extremality form becomes
\begin{align}
    T_{ijlk} q^i q^j q^k q^l= \alpha_{ijkl} q^i q^j q^k q^l = (z_{\phi}^i q^i)^4,
\end{align}
Again, we find a manifestly positive contribution. For $z_{\phi} $ near or less than one, both $\alpha_{ijkl}$ and $\gamma_{ij}$ are generated by diagrams that are order $z_{\phi}^0$. In that case this scaling argument does not apply, and the order one constants need to be included in the analysis. These arguments are schematic and largely review what was already considered in \cite{Andriolo:2018lvp}.

One might wonder whether this analysis is relevant to the parity-odd operators. Interestingly, \cite{Colwell:2015wna} has shown how to generalize the Euler-Heisenberg Lagrangian by integrating out a monopole or dyonic charge. The effective Lagrangian was derived in that paper (and earlier in \cite{Kovalevich:1997de}) to be
 \begin{align}
     \begin{split}
     \mathcal{L}_4 =&  \big( 4 (\hat{q}^2 - \hat{p}^2)^2 + 28\hat{q}^2 \hat{p}^2 \big) (F^2)^2
    + \big( 7 (\hat{q}^2 - \hat{p}^2)^2 +  16 \hat{q}^2 \hat{p}^2 \big) (F \tilde{F} )^2 - 12 \hat{q} \hat{p}  (\hat{q}^2 - \hat{p}^2) F^2 (F \tilde{F}).
    \label{monopole}
     \end{split}
 \end{align}
where the $\hat{q}$ and $\hat{p}$ refer to the electric and magnetic charges of the dyon that is integrated out (not the charges of the black hole). This procedure generates the parity-violating four-photon coupling as well as the two parity-even ones. This is not surprising given that magnetic charges violate parity in their interactions with the gauge field. What is more interesting is that this term is \textit{not} a square, unlike every other term appearing in the effective Lagrangian. The sign of the generated term depends on the sign of the product of the electric and magnetic charges of the particle. In terms of the polynomial derived in (\ref{dyon polynomial}), the condition that must be met to satisfy the WGC is:
\begin{align}
    \begin{split}
        &  \big( \hat{q}^4 + 5 \hat{q}^2 \hat{p}^2 + \hat{p}^4 \big)\, x^4
        + \, 3 \, \big( \hat{q}^3  \hat{p} - \hat{q}  \hat{p}^3 \big) \, x^3
        + \,  \big( 5 \hat{q}^4 -8 \hat{q}^2 \hat{p}^2 + 5 \hat{p}^4  \big) \, x^2 \\
        & \qquad \qquad  + \, 3  \, \big( \hat{q}^3  \hat{p} - \hat{q}  \hat{p}^3 \big) \, x 
        \,+ \,  \big( \hat{q}^4 + 5 \hat{q}^2 \hat{p}^2 + \hat{p}^4 \big)  \, > \, 0.
    \end{split}
\end{align}
This polynomial is always positive, so the Lagrangian given in (\ref{monopole}) does not allow for stable black holes and satisfies the WGC.
 
\subsubsection{Causality Constraints}

Another set of arguments for bounds on the EFT coefficients rely on causality. These were first considered in \cite{Cheung:2014ega} and generalized to multiple gauge fields in \cite{Hamada:2018dde}. Two methods were used, and they were shown to give the same result. The first is to consider the propagation of photons on a photon gas background. Requiring that photons travel do not travel superluminally constrains the four-photon interaction. The second method uses analyticity and unitarity to relate the EFT coefficients to an integral over the imaginary part of the amplitude, which is manifestly positive. The bounds obtained this way for multiple gauge fields are
\begin{align}
    \label{causality}
    \sum_{ij} \left( \alpha_{ \{i j \} \{k l \} } + \beta_{ \{i j \} \{k l \} }  \right)  u^i v^j u^k v^l \geq 0.
\end{align}
This inequality must hold for any vectors $\vec{u}$ and $\vec{v}$. This bound is independent from the bounds that we have derived in (\ref{extremality bound}), so it is not enough to imply the WGC on its own. 

So far these arguments have only bounded the four-photon interactions. Another causality-based argument was made in \cite{Hamada:2018dde} that bounds the photon-photon-graviton interaction parameterized by $\gamma$. They argued that the addition of this four-derivative term introduces causality violation at a scale $E \sim M_{\text{Pl}} / \gamma^{1/2}$ (a fact noticed in \cite{Camanho:2014apa}). Therefore new physics must arise at scale $\Lambda_{QFT} \lesssim M_{\text{Pl}} / \gamma^{1/2}$, which means $\gamma \lesssim (M_{\text{Pl}} / \Lambda_{QFT})^2$. This argument suggests that perhaps the $W F F$ four-derivative terms are generically bounded by causality to be much smaller than a number of possible contributions to the $F^4$ terms. It would be interesting to extend the analysis of \cite{Camanho:2014apa} to the more general set of operators used here, but this is beyond the scope of this dissertaion. 

\subsection{Renormalization of Four-Derivative Operators}
\label{subsec:renorm}

The Wilson coefficients that appear in the extremality shift (\ref{extremality result}) are determined by UV degrees-of-freedom integrated out of the low-energy effective field theory. In section \ref{sec:decay} we gave explicit examples of contributions to the Wilson coefficients from integrating out massive particle states, both at tree- and loop-level. To consistently calculate the correction to the extremality bound for a black hole with total charge $Q^2$, we must first calculate the renormalization group evolution from the \textit{matching scale} $\mu^2 \sim \Lambda_{\text{UV}}^2$ to the \textit{horizon scale} $\mu^2 \sim M_{\text{Pl}}^2/Q^2$. For black holes with $Q^2\gg 1$ these scales can be arbitrarily separated and the effects of the logarithmic running of the Wilson coefficients can be dramatic. 

In the single $U(1)$ case it was recently argued \cite{Charles:2019qqt} that as we RG flow towards the deep IR, $Q^2\rightarrow \infty$, the logarithmic running of a particular combination of Wilson coefficients dominates the extremality shift, independent of the values of the coefficients at the matching scale. Explicitly, the extremality bound takes the form
\begin{equation}
\frac{Q^2}{M^2}\leq 1 + \frac{4}{5Q^2}\left(\frac{c}{16\pi^2}\text{log}\left(\frac{\Lambda_{\text{UV}}^2 Q^2}{M_{\text{Pl}}^2}\right)+2\alpha_{\text{UV}}-\gamma_{\text{UV}}\right).
\end{equation}
If $c>0$ then at some finite value of the charge $Q^2$ extremal black holes must be self-repulsive. This was shown to be the case in \cite{Charles:2019qqt} for various explicit theories, including the single $U(1)$ model (\ref{4dersingle}). Since the renormalization group coefficient $c$ depends only on the massless degrees of freedom, this analysis depends only on the universality class of the model. For those classes in which this conclusion holds, the WGC is always satisfied independently of the details of the UV completion, and in that sense is no longer a useful Swampland criterion.

This argument generalizes to an arbitrary number of $U(1)$ gauge fields. Since there are many more four-derivative operators, we must make use of a non-renormalization theorem that arises as a consequence of the accidental $U(N)$ electromagnetic duality symmetry of the two-derivative approximation. The theorem we require is \\

\textbf{Non-Renormalization of Duality Violating Operators}: \textit{In Einstein-Maxwell with $N$ $U(1)$ gauge fields, a four-derivative operator $\mathcal{O}_i$ is renormalized at one-loop only if it generates an on-shell local matrix element that is an invariant tensor of the maximal compact electromagnetic duality group $U(N)$.}\\

This result was first noted long-ago following a detailed calculation of the UV divergence \cite{Deser:1974cz,Deser:1974xq}, and recently generalized (including massless scalars) to the full non-compact duality group $Sp(2N)$ in \cite{Charles:2017dbr}. A novel proof using on-shell methods was given in \cite{Jones:2019nev}. This proof does not require a detailed calculation; only an analysis of the possible divergences is needed.

By simple dimensional analysis we know that the counter-terms to one-loop divergences in Einstein-Maxwell are four-derivative operators. In \cite{Jones:2019nev} we give a complete classification of local matrix elements corresponding to four-derivative operators, so together with the above non-renormalization theorem, we know that most general \textit{local} UV divergence is given by
\begin{equation}
\label{form}
    \left[\mathcal{A}_4^{\text{1-loop}}\left(1_{\gamma,i}^+,2_{\gamma, j}^+,3_\gamma^{-,k},4_{\gamma}^{-,l}\right)\right]_{\text{UV}} = \frac{c}{16\pi^2\epsilon}\left({\delta_i}^k{\delta_j}^l+{\delta_i}^l{\delta_j}^k\right)[12]^2\langle 34\rangle^2.
\end{equation}

At one-loop, the divergence fixes the dependence of the scattering amplitude on the renormalization group scale $\mu^2$. After adding a counterterm with coefficient $\alpha (\mu) $ to remove the UV divergence, the \textit{physical} scattering amplitude should be independent of $\mu^2$
\begin{equation}
    \mathcal{A}_4^{\text{1-loop}}\left(1_{\gamma,i}^+,2_{\gamma, j}^+,3_\gamma^{-,k},4_{\gamma}^{-,l}\right) = \left[\alpha(\mu^2)+\frac{c}{8\pi^2}\log(\mu^2)\right]\left({\delta_i}^k{\delta_j}^l+{\delta_i}^l{\delta_j}^k\right)[12]^2\langle 34\rangle^2+\mathcal{O}\left(\epsilon^0\right),
\end{equation}
which gives the logarithmic running of the Wilson coefficient
\begin{equation}
    \alpha(\mu^2) = -\frac{c}{8\pi^2}\log\left(\frac{\mu^2}{\Lambda_{\text{UV}}^2}\right),
\end{equation}
where $\Lambda_{\text{UV}}$ is some UV matching scale, assumed to be arbitrarily larger than the horizon scale. The ultraviolet divergence in Einstein-Maxwell coupled to $N$ $U(1)$ gauge fields was first calculated long-ago \cite{Deser:1974cz,Deser:1974xq}, and then recalculated using unitarity methods \cite{Dunbar:1995ed,Norridge:1996he} 
\begin{equation}
    \label{expUV}
    \left[\mathcal{A}_4^{\text{1-loop}}\left(1_{\gamma,i}^+,2_{\gamma, j}^+,3_\gamma^{-,k},4_{\gamma}^{-,l}\right)\right]_{\text{UV}} = \frac{1}{16\pi^2\epsilon}\left(\frac{137    }{120}+\frac{N-1}{20}\right)\left({\delta_i}^k{\delta_j}^l+{\delta_i}^l{\delta_j}^k\right)[12]^2\langle 34\rangle^2.
\end{equation}
This gives the RG coefficient in (\ref{form}) as 
\begin{equation}
    c = \frac{137    }{120}+\frac{N-1}{20}.
\end{equation}
From this matrix element we can reverse engineer the corresponding four-derivative operator
\begin{equation}
    S \supset \alpha(\mu^2)\left(\delta_{ik}\delta_{jl}+\delta_{il}\delta_{jk}\right)\int \text{d}^4 x \sqrt{-g}\left[ \left(F^i_{\mu \nu} F^{j \, \mu \nu} F^k_{\rho \sigma} F^{l \, \rho \sigma} +  F^i_{\mu \nu} \tilde{F}^{j \, \mu \nu} F^k_{\rho \sigma} \tilde{F}^{l \, \rho \sigma}\right)\right].
\end{equation}
Note that we have lost manifest duality invariance when passing from on-shell scattering amplitudes to the effective action and so have made the replacement ${\delta_i}^j\rightarrow \delta_{ij}$. As an important cross-check, the effect of such an operator on the perturbed metric at leading order in $\alpha$ is given by (\ref{metricshift}) to be
\begin{equation}
    \Delta g^{rr} = -\frac{24\alpha(\mu^2)}{15r^6}\sum_{i=1}^N\left(q_i^2+p_i^2\right),
\end{equation}
which manifests the expected electromagnetic duality symmetry, further enhanced to $O(2N)$. 

When evaluating the extremality form, $\mu$ should be taken to be the horizon scale $\mu^2 \sim M_{\text{Pl}}^4/M^2 \sim M_{\text{Pl}}^2/Q^2 $. Since $c>0$, as $Q^2\rightarrow \infty$ the logarithmic term becomes \textit{large} and \textit{positive}. With the logarithmic running included the extremality form at the horizon scale is given by
\begin{align} 
\label{formlog}
    T(q^i,p^i) &= \frac{1}{8\pi^2}\left(\frac{137    }{120}+\frac{N-1}{20}\right)(Q^2)^2\log\left(\frac{\Lambda_{\text{UV}}^2 Q^2}{M_{\text{Pl}}^2}\right)+\alpha^{\text{UV}}_{ijkl} \, (q^i q^j - p^i p^j)(q^k q^l - p^k p^l) \nonumber\\
    &  \hspace{15mm} + \, 8 \beta^{\text{UV}}_{ijkl} q^i p^j q^k p^l - \gamma^{\text{UV}}_{ij} \, \left( q^i q^j - p^i p^j\right)Q^2+ 4 \,  \chi^{\text{UV}}_{ijkl} \, q^i p^j \, \left(q^k q^l-p^k p^l\right)\nonumber\\
    & \hspace{15mm} - 2 \,  \omega^{\text{UV}}_{ij} \,   q^i p^j Q^2,
\end{align}
where $Q^2 =\sum_i(q_i^2+p_i^2)$. In this expression $\alpha^{\text{UV}}$, $\beta^{\text{UV}}$, $\gamma^{\text{UV}}$, $\chi^{\text{UV}}$, and $\omega^{\text{UV}}$ refers to the values of the Wilson coefficients at the matching scale $\Lambda_{\text{UV}}$. Importantly, the logarithmic term is $O(2N)$ invariant and therefore gives an isotropic contribution to the extremality form. This contribution scales like $Q^4 \log Q$, while the rest of the terms scale like $Q^4$. Therefore it dominates over all other contributions. We conclude that for sufficiently large $Q^2$, the extremality form is positive, independent of the values of the Wilson coefficients at the matching scale $\Lambda_{\text{UV}}$, and consequently the multi-charge WGC is always satisfied in the black hole regime.

Here the full $U(N)$ duality invariance of the UV divergence (enhanced to $O(2N)$ in the quartic form) was essential to the argument. It would not have been enough that some Wilson coefficients had a positive logarithmic running, to prove the multi-charge WGC we require positivity in all directions, which as we have shown follows from a generalized non-renormalization theorem as a consequence of tree-level $U(N)$ duality symmetry of Einstein-Maxwell.

It is interesting to note that we can \textit{almost} reach this same conclusion without knowing the explicit form of the UV divergence (\ref{expUV}). In \cite{Hamada:2018dde} the causality bound (\ref{causality}) was applied to the Wilson coefficients at the UV matching scale $\Lambda_{\text{UV}}$ and consequently to constrain the properties of the states integrated out. But this bound must remain valid even deeper in the IR where, as we have seen, the logarithmic running dominates. If the RG coefficient $c$ had been negative, then the bound (\ref{causality}) is eventually violated, indicating the presence of superluminal propagation at very low energies. Since we expect that Einstein-Maxwell is not inconsistent in the deep IR, it must be the case that $c \geq 0$ even without doing a detailed one-loop calculation. This argument has nothing to say about the possibility that $c=0$. Only an explicit calculation is sufficient to demonstrate the existence of a non-vanishing one-loop divergence.

\subsection{Potential for Future Work}

The argument we have given above requires that electromagnetic duality invariance is not broken at two-derivative order. It would be interesting to study generalizations where the duality is broken at leading order, such as when a dilaton couples to the field strength. Moreover, this argument depends in an essential way on a symmetry of Einstein-Maxwell that is only present in four-dimensions. In $d\neq 4$ there is no reason to expect that such a non-renormalization theorem should be valid and so it is not clear if the weak gravity conjecture is similarly trivialized by non-trivial RG running.  

Considering scalar fields might also offer the opportunity to check whether the conditions on the EFT coefficients are satisfied in specific models. One such example is the 4-dimensional STU model \cite{Cvetic:1999xp}, which retains four Abelian gauge fields and three dilatonic scalar fields. More generally, the photon and graviton are often accompanied by light scalar moduli in UV complete models from string compactifications. This means that a full understand of the relationship between the weak gravity conjecture and higher-derivative corrections requires studying the role played by scalar fields. We leave these and other generalizations to future work.


\section{Corrections to the AdS-RN Geometry}

We wil now focus our attention on Anti-de Sitter space. Consider Einstein-Maxwell theory in a $(d+1)$-dimensional AdS spacetime of size $l$.
The first non-trivial terms in the derivative expansion of the effective action arise at the four-derivative level, and by appropriate field redefinitions we may choose a complete basis of dimension-independent operators:
\begin{align}
    \begin{split}
        I &= -\frac{1}{16 \pi } \int d^{d+1} x \sqrt{-g} \Bigg[  \frac{d (d-1)}{l^2} + R -\frac{1}{4} F^2 \\
        & \qquad \qquad \qquad \qquad +l^2  \epsilon \Big( c_1 R_{abcd} R^{abcd} +  c_2 R_{abcd} F^{ab} F^{cd} + c_3 (F^2)^2 + c_4 F^4 \Big) \Bigg].
    \end{split}
    \label{ActionCJLM}
\end{align}
Note that additional CP-odd terms can arise in specific dimensions, but will not contribute to the static, stationary spherically symmetric black holes that we are considering here.  This basis parallels that of \cite{Myers:2009ij}, which used the same set of dimensionless Wilson coefficients, but focused on the $(4+1)$-dimensional case.  Depending on the origin of the AdS length scale $l$, one may expect these coefficients to be parametrically small, of the form $c_i\sim(\Lambda l)^{-2}$, where $\Lambda$ denotes the scale at which the EFT breaks down. In particular, this will be the case in order for the action (\ref{ActionCJLM}) to be under perturbative control. We have also introduced the small bookkeeping parameter $\epsilon$, which will allow us to keep track of which terms are first order in the $c_i$ coefficients.

\subsection{The Zeroth Order Solution}

At the two-derivative level, this action admits a family of AdS-RN black holes parametrized by uncorrected mass $m$ and charge $q$,
\begin{align}
    \begin{split}
        &  ds^2 = -f(r) dt^2 + g(r)^{-1} dr^2 + r^2 d \Omega_{d - 1,k}^2 \, , \\
        &  f(r)  = g(r) = k - \frac{m}{r^{d-2}} + \frac{q^2}{4 r^{2d-4}} + \frac{r^2}{l^2}, \\
        & A = \left( - \frac{1}{c} \frac{q}{r^{d-2}} + \Phi \right) dt,  \quad \Phi = \frac{1}{c} \frac{q}{r_h^{d-2}}\, , \qquad c = \sqrt{ \frac{ 2(d-2)}{(d-1)} }\,  .
        \label{radial function}
    \end{split}
\end{align}
Here $r_h$ is the outer horizon radius, and the parameter $k=0,\pm1$ specifies the horizon geometry, with $k=1$ corresponding to the unit sphere.  The constant $\Phi$ is chosen so that the $A_t$ component of the gauge field vanishes on the horizon, and represents the potential difference between the asymptotic boundary and the horizon. 

Typically, we will consider lower case letters $(m, q, ...)$ to be parameters in the theory, while upper case letters $(M, Q, S, T, ...)$ will denote physical quantities that may or may not receive corrections. We will add a subscript zero (e.g. $M_0$) to denote the uncorrected contribution to quantities that do receive order $c_i$ corrections. The shifts, which are equal to the corrected quantities minus the uncorrected ones, will be denoted by the $\epsilon$ derivative. However, we will sometimes use $\Delta$ when it is convenient, with subscripts indicating quantities held fixed, for example, we have 
\begin{align}
    (\Delta M)_T \equiv \lim_{\epsilon \rightarrow 0}\left( M(T,\epsilon) - M_0(T)\right)  \equiv \lim_{\epsilon \rightarrow 0}\left( \frac{ \partial M}{\partial \epsilon} \right)_T. 
\end{align}
Finally, in sections IV and V we will use dimensionless quantities $(\nu, \xi)$ for convenience. These are defined by $\nu = (r_h)_0 / l$ and $Q = (1 - \xi) Q_{\text{ext}} $.

\subsection{The First Order Solution}

We now turn to the first order solution in terms of the Wilson coefficients $c_i$.  We follow the procedure outlined in Ref.~\cite{Kats:2006xp}, but work in an AdS$_{d+1}$ background.  While general $(d+1)$-dimensional results may be worked out analytically, we took a shortcut of working with explicit dimensions four through eight and then fitting the coefficients to extract results for arbitrary dimension.  Since the four-derivative terms are built from tensors with eight indices and hence four metric contractions, the resulting expressions will scale at most as $d^4$.  Hence the coefficients are fully determined by the results in five different dimensions.

Following \cite{Kats:2006xp}, we start with the \textit{effective} stress tensor, where corrections come from two sources. The first is from substituting in the corrected Maxwell field to the zeroth order electromagnetic stress tensor, and the second is from the explicit four-derivative corrections to the stress tensor evaluated on the zeroth order solution.  The result of computing both of these contributions to the time-time component of the stress tensor is
\begin{align}
    \begin{split}
        T_t{}^t&= -\frac{(d-1) (d-2) \, q^2}{4 \, r^{2d - 2}} 
        + \frac{d (d-1)}{ l^2} \\
        & + c_1 \Bigg( \frac{ (d - 2) (8 d^3 - 24 d^2 + 15 d + 3) \, q^4 l^2}{ 8 r^{4d - 4}}
        - \frac{ (d - 1) (d - 2) (4 d^2 - 9 d + 3 ) \, m q^2 l^2 }{r^{3d - 2}} 
        \\
        & \qquad \qquad \qquad + k \frac{4  d (d-1) (d-2)^2 \, l^2 q^2}{r^{2d}}
        - \frac{ d (d-1) (d-2) (d-3) \, l^2 m^2}{r^{2d}} \\
        & \qquad \qquad \qquad + \frac{(d- 2) (2d - 3)(2 d^2 - 5 d + 1) \, q^2 }{r^{2 d-2}} + \frac{2 d (d - 3) }{l^2} \Bigg)
        \\
        & + c_2 \Bigg( \frac{ (d-1)^3 (d-2) \, q^4 l^2}{r^{4d-4}}
        - \frac{(d-1)^2 (3d^2 - 8 d + 4) \, q^2 m l^2}{r^{3d-2}} 
        + k \frac{2 d (d-1)^2 (d-2) \, q^2 l^2}{r^{2d}} \\
        & \qquad \qquad \qquad + \frac{2 (d-1)^3 (d-2) \, q^2}{r^{2d-2}} \Bigg)
         + \left( 2 c_3 + c_4 \right) \Bigg( \frac{(d-1)^2 (d-2)^2 q^4 l^2}{2 \, r^{4 d-4}} \Bigg) \, .
         \label{StressTensor}
    \end{split}
\end{align}
The shift to the geometry may be obtained from the corrections to the stress tensor \cite{Kats:2006xp}, 
\begin{align}
    \Delta g = \frac{1}{(d - 1) r^{d - 2}} \int \, dr \,  r^{d-1}\Delta T_t{}^t \, ,
\end{align}
and after integrating the $\mathcal{O}(c_i)$ terms in (\ref{StressTensor}), we find 
\begin{align}
    \begin{split}
        \Delta g (r) &= c_1 
        \Bigg( -\frac{ (d - 2) (8 d^3 - 24 d^2 + 15 d + 3) \, q^4 l^2}{ 8 (d - 1) (3 d - 4) r^{4d - 6}}
        + \frac{(d - 2) (4 d^2 - 9 d + 3 ) \, m q^2 l^2 }{2 (d - 1) r^{3d - 4}} 
        \\
        & \qquad \qquad  - k \frac{4  (d-2)^2 \, l^2 q^2}{r^{2d-2}}
        + \frac{  (d-2) (d-3) \, l^2 m^2}{r^{2d-2}} \\
        & \qquad \qquad 
        - \frac{ (2d - 3)(2 d^2 - 5 d + 1) \, q^2 }{(d - 1)r^{2 d-4}}
        + \frac{2 (d - 3) r^2}{(d - 1)l^2} \Bigg)
        \\
        & + c_2 \Bigg( - \frac{ (d-1)^2 (d-2) \, q^4 l^2}{(3d - 4)r^{4d - 6}}
        + \frac{ (3d^2 - 8 d + 4) \, q^2 m l^2}{2 r^{3d-4}} \\
        & \qquad \qquad 
        - k \frac{2 (d-1) (d-2) \, q^2 l^2}{r^{2d-2}} 
        - \frac{2 (d-1)^2 \, q^2}{r^{2d-4}} \Bigg) \\
        & + \left( 2 c_3 + c_4 \right) \Bigg(- \frac{(d - 1) (d - 2)^2 q^4 l^2 }{ (6d - 8) r^{4d - 6}} \Bigg) \, .
    \end{split}
    \label{eq:Deltag}
\end{align}
The time component of the metric can then be obtained using the relation \cite{Kats:2006xp} 
\begin{align}
    f(r) = (1 + \gamma(r)) g(r),
\end{align}
where $\gamma(r)$ is defined by\footnote{
We note that the definition of $\gamma$ implies that it is positive provided that the null energy condition holds. 
}
\begin{align}
    \gamma(r) =    -\frac{1}{(d - 2)} \int dr r \left( T_t{}^t - T_r{}^r \right) .
    \label{eq:gamma}
\end{align}
For our particular case we find:
\begin{align}
    \gamma(r) =\left( c_1 \frac{(d - 2)(2 d^2 - 5 d + 1) }{(d - 1) } + c_2 d (d - 2)\right)\fft{q^2l^2}{r^{2d-2}}.
    \label{eq: gammaVal}
\end{align}
Finally, we have
\begin{align}
    \begin{split}
        F_{tr} &= \sqrt{\frac{(d - 2)(d - 1) }{2}} \Bigg[(1-8c_2) \frac{q}{r^{d - 1}}+4c_2(d-1)(d-2)\fft{qml^2}{r^{2d-1}}\\
        &\kern4em+\left( \fft{c_1}2\frac{(2 d^2 - 5 d + 1)}{(d - 1)} -\fft{ c_2}2 (7d - 12) - 4\left( 2 c_3 + c_4 \right)(d - 1) \right) (d-2)\fft{q^3l^2}{r^{3d-3}}
        \Bigg] \, ,
    \end{split}
    \label{eq:Ftr}
\end{align}
which we note is independent of the geometry parameter $k$, as was the case in \cite{Cremonini:2009ih}.

\subsubsection{Asymptotic Conditions and Conserved Quantities}

The first order solution can be summarized as
\begin{align}
    &ds^2=-\left(1+\gamma(r)\right)g(r)dt^2+g(r)^{-1}dr^2+r^2d\Omega_{d-1,k}^2 \, ,
\end{align}
where
\begin{equation}
    g(r)=k-\fft{m}{r^{d-2}}+\fft{q^2}{4r^{2d-4}}+\fft{r^2}{l^2}+\Delta g.
\end{equation}
The corrected metric functions, $\Delta g$ and $\gamma(r)$, are given in (\ref{eq:Deltag}) and (\ref{eq:gamma}), respectively.  In addition, the full electric field is given in (\ref{eq:Ftr}).  For a given zeroth order AdS radius $l$, this solution is specified by two parameters, $m$ and $q$, which correspond to the mass and charge of the uncorrected black hole.
At the same time, the corrected solution includes a number of integration constants, two of which we have implicitly set to zero in the integral expressions for $\Delta g$ and $\gamma(r)$.  The constant related to $\Delta g$ can be absorbed by a shift in $m$, and a third constant from the corrected Maxwell equation can be absorbed by a shift in $q$.  The constant related to $\gamma(r)$ can be absorbed at the linearized level by a rescaling of the time coordinate, and hence can be thought of as a redshift factor.

In order to make the correspondence between the parameters of the solution, $m$ and $q$, and the physical mass and charge of the black hole more precise, consider the part of $\Delta g$ that is leading in $r$. We can see that there is a term that goes like $c_1 \frac{r^2}{l^2}$ that dominates over all other terms in the correction. Therefore, for large values of $r$, the solution takes the form
\begin{align}
    f(r)\approx g(r)&=k-\fft{m}{r^{d-2}}+\left(1+c_1\fft{2(d-3)}{d-1}\right)\fft{r^2}{l^2}+\cdots,\nn\\
    F_{tr}&=\sqrt{\fft{(d-2)(d-1)}2}(1-8c_2)\fft{q}{r^{d-1}}+\cdots.
\end{align}
Our first observation is that the AdS radius gets modified because the Riemann-squared term is non-vanishing on the original uncorrected background.  This suggests that we define an effective AdS radius
\begin{align}
    l^2 = \lambda^2 l^2_{\text{eff}} , \qquad \qquad \lambda^2 = \left( 1 + c_1 \frac{2 (d - 3)}{(d - 1)} \right).
    \label{eq:lamdef}
\end{align}
This shift by $\lambda$ is unavoidable when turning on the $c_1$ Wilson coefficient.  However, in principle we still have a choice of whether we hold $l$ or $l_{\text{eff}}$ fixed when turning on the four-derivative corrections.

In what follows, we always choose to keep $l$ fixed.  Then, since the effective AdS radius is shifted, the asymptotic form of the metric is necessarily modified as well.  From a holographic point of view, this leads to a modification of the boundary metric
\begin{equation}
    ds^2\sim r^2\left(\fft{dt^2}{l^2}+d\Omega_{d-1,k}^2\right)\quad\longrightarrow\quad ds^2\sim r^2\left(\fft{dt^2}{l_{\text{eff}}^2}+d\Omega_{d-1,k}^2\right).
\end{equation}
This is generally undesirable, as we would like to compare thermodynamic quantities in a framework where we hold the boundary metric fixed while turning on the Wilson coefficients.  One way to avoid this shift in the boundary metric is to introduce a `redshift' factor
\begin{equation}
    t=\bar t/\lambda,
\end{equation}
to compensate for the shift in $l_{\text{eff}}$.  In terms of the time $\bar t$, the solution now takes the form
\begin{align}
    ds^2&=-\bar f(r)\, d\bar t^2+g(r)^{-1}dr^2+r^2d\Omega_{d-1,k}^2,\nn\\
    F_{\bar tr}&=\lambda^{-1}F_{tr}=\sqrt{\fft{(d-2)(d-1)}2}(1-8c_2)\fft{q/\lambda}{r^{d-1}}+\cdots,
\end{align}
where
\begin{align}
    \bar f(r)&=\lambda^{-2}(1+\gamma(r))g(r)=k/\lambda^2-\fft{m/\lambda^2}{r^{d-2}}+\fft{r^2}{l^2}+\cdots,\nn\\ g(r)&=k-\fft{m}{r^{d-2}}+\fft{r^2}{l_{\text{eff}}^2}+\cdots.
\end{align}

We now turn to the charge and mass of the solution measured with respect to the redshifted $\bar t$ time.  For the charge $Q$, we take the conserved Noether charge
\begin{equation}
    Q=\fft1{16\pi}\int_{\Sigma_{d-1}}*\mathcal F,
\end{equation}
where $\mathcal F$ is the effective electric field
\begin{equation}
    \mathcal F_{\mu\nu}=F_{\mu\nu}+l^2\left(-4c_2R_{\mu\nu\rho\sigma}F^{\rho\sigma}-8c_3F_{\mu\nu}(F^2)-8c_4F_{\nu\rho}F^{\rho\sigma}F_{\sigma\mu}\right).
\end{equation}
The result is
\begin{equation}
    Q=\left.\fft{1+8c_2}{16\pi}\omega_{d-1}\lambda r^{d-1}F_{\bar t r}\right|_{r\to\infty}=\sqrt{\fft{(d-2)(d-1)}2}\fft{\omega_{d-1}}{16\pi}q,
    \label{eq:Qdef}
\end{equation}
where $\omega_{d-1}$ is the volume of the unit $S^{d-1}$.  The $1/16\pi$ factor arises from the prefactor in the action (\ref{ActionCJLM}) where we have set Newton's constant $G=1$.

Unlike in the asymptotically Minkowski case, some care needs to be taken in obtaining the mass of the black hole.  With an eye towards holography, we choose to define the mass from the boundary stress tensor \cite{Balasubramanian:1999re}.  The standard approach to holographic renormalization involves the addition of appropriate local boundary counterterms so as to render the action finite.  This was performed in \cite{Cremonini:2009ih} for $R^2$-corrected bulk actions, and since only the $c_1R_{abcd}R^{abcd}$ term in (\ref{ActionCJLM}) leads to an additional divergence, we can directly use the result of \cite{Cremonini:2009ih}.  The result is
\begin{equation}
    M=\fft{\omega_{d-1}}{16\pi}(1+4c_1(d-3))\fft{(d-1)m}\lambda,
\end{equation}
where we have taken into account the scaling of the mass by the redshift factor $\lambda$.  Substituting in $\lambda$ from (\ref{eq:lamdef}) then gives
\begin{equation}
    M=\fft{\omega_{d-1}}{16\pi}(d-1)(1+\rho)m,
    \label{eq:massdef}
\end{equation}
where
\begin{align}
    \rho = c_1 \frac{(d - 3) (4 d - 5)}{d - 1}.
    \label{eq:rhodef}
\end{align}
Note that we are taking the mass here to \textit{exclude} the Casimir energy that is normally part of the boundary stress tensor.  This will be important when comparing with the thermodynamic quantities extracted from the regulated on-shell action in section~IV.  Working in the setup of holographic renormalization ensures that the mass $M$ and charge $Q$ defined in (\ref{eq:massdef}) and (\ref{eq:Qdef}), respectively, yield a consistent framework for black hole thermodynamics.  

\label{sec:AdSRNcor}

\subsection{Mass, Charge, and Entropy from the AdS-RN Geometry Shift}

 Given the first-order solution, we can calculate shifts to the mass, $\Delta M$, and entropy, $\Delta S$, of the black hole induced by the four-derivative corrections. In these computations it is important to keep in mind what is being held fixed as we turn on the Wilson coefficients $c_i$.  The main parameters we consider here are the mass $M$ and charge $Q$, which are related to the two parameters, $m$ and $q$, of the solution by (\ref{eq:massdef}) and (\ref{eq:Qdef}), respectively. In addition we consider the thermodynamic quantities $T$ (temperature) and $S$ (entropy), although they are not all independent.  Note that we always consider the AdS radius $l$ to be fixed, although interesting results have been obtained by mapping it to thermodynamic pressure.

Singly-charged, non-rotating black holes may be described by any two of mass $M$, charge $Q$ and the horizon radius $r_h$.  Of course, any number of other parameters may be used as well, such as the temperature $T$ or an extremality parameter, such as was used in \cite{Cheung:2018cwt}. If we further impose the extremality condition $T=0$ on the solution, then only a single parameter is needed. Clearly this is only true for non-rotating black holes with a single gauge field, as more general solutions may have additional charges or angular momenta. It is important to keep in mind what is being held fixed when we turn on the higher-derivative corrections, as the results will depend on this choice.  For example, we will see below that the shift to $M/Q$ depends on whether the mass, charge or horizon radius is held fixed when comparing the corrected with uncorrected quantities. 

Recall that, in our first-order solution, the geometry is essentially given by the radial function
\begin{align}
\label{gcorr}
    g^{rr}=g(r) = k - \frac{m}{r^{d-2}} + \frac{q^2}{4 r^{2d-4}} + \frac{r^2}{l^2} + \, \Delta g \, ,
\end{align}
where $\Delta g$ denotes the contributions of the higher-derivative corrections to the geometry, and $\epsilon$ is a small parameter we use to keep track of where $\mathcal{O}(c_i)$ corrections come in. Using the fact that both $g(r_h)$ and $g'(r_h)$ vanish at extremality, 
we may express the extremal mass and charge as a function of the horizon radius,
\begin{align}
    \begin{split}
        M_{\text{ext}} &= 2 V (d - 1) r_h^{d-2} \left( \left(k +  \frac{d - 1}{d - 2 } \frac{r_h^2}{l^2} \right) \left(1 + \epsilon \rho \right) + \epsilon \, \Delta g + \frac{r_h}{2(d - 2)} \epsilon \, \Delta g' \right) \, , \\
        Q_{\text{ext}}^2 & = 2 V^2 (d - 1)(d - 2) r_h^{2(d-2)} \left( k + \frac{d }{d - 2 } \frac{r_h^2}{l^2}  + \epsilon \, \Delta g + \frac{r_h}{d - 2} \epsilon \, \Delta g' \right) \, ,
        \label{extremal_mq}
    \end{split}
\end{align}
where $M$ and $Q$ are the asymptotic quantities defined in (\ref{eq:massdef}) and (\ref{eq:Qdef}), and we have defined $V=\omega_{d-1}/16\pi$. Though we have expressed $M$ and $Q$ as functions of $r_h$, these expressions are valid regardless of which of the three quantities is being held fixed. For example, if we work at fixed charge, then $Q$ gets no $\mathcal{O}(\epsilon)$ corrections, in which case $M$ and $r_h$ will both receive corrections.  

\subsubsection{Extremality at Leading Order}

 Before discussing the extremality and entropy shifts, we consider the leading order relations between $M_0$, $Q_0$ and $(r_h)_0$ for extremal black holes. We will repress the $0$ subscripts in this subsection, but we mean the uncorrected quantities. Setting $\epsilon=0$ in (\ref{extremal_mq}) immediately gives the relations
\begin{align}
    \begin{split}
        M_{\text{ext}} &= 2 V (d - 1) r_h^{d-2} \left(k +  \frac{d - 1}{d - 2 } \frac{r_h^2}{l^2} \right)\, ,\\
        Q_{\text{ext}}^2 &= 2 V^2 (d - 1)(d - 2) r_h^{2(d-2)} \left( k + \frac{d }{d - 2 } \frac{r_h^2}{l^2} \right) \, .
        \label{leading_extremal_mq}
    \end{split}
\end{align}
In principle, we can eliminate $r_h$ from these equations to obtain the relation between mass and charge for extremal AdS black holes.  However, for general dimension $d$, there is no simple expression that directly encodes this relation.  Nevertheless, we can consider the limit of small and large black holes.

For small black holes ($r_h \ll l$), we take $k=1$ (ie a spherical horizon) and find
\begin{align}
    M_{\text{ext}} \sim Q_{\text{ext}} \sim r_h^{d - 2} \, ,
\end{align}
so one recovers the simple $M \sim Q$ scaling that appears in flat space.  (Note that asymptotically Minkowski black holes necessarily have spherical horizons.)  For large black holes ($r_h \gg l$), on the other hand, the scaling is very different from that of flat space,
\begin{align}
    M_{\text{ext}} \sim r_h^{d} \, , \qquad Q_{\text{ext}} \sim r_h^{d - 1}\qquad\Rightarrow\qquad M_{\text{ext}} \sim \left( Q_{\text{ext}} \right)^{\frac{d}{d - 1}} \, .
\end{align}
In fact, this is precisely the scaling behavior expected based on the relationship between minimal scaling dimension and charge for boundary operators with large global charges \cite{Loukas:2018zjh}.

\subsubsection{Mass Shift at Fixed Charge}

 Now we consider the effect of four-derivative corrections. If we hold the charge fixed, then the shift to extremality is entirely due to the change in the mass.  This may computed from the expression (\ref{extremal_mq}) for the mass by taking a derivative with respect to $\epsilon$, which parametrizes the higher-derivative corrections, leading to
\begin{align}
    \begin{split}
     & \left( \frac{ \partial M }{\partial \epsilon} \right)_{Q, T = 0} =  V (d - 1) r_h^{d - 2} \Bigg( 2 \Delta g + \frac{1}{d - 2} r_h \Delta g' \\
     & \quad + 2 \rho \left( k + \frac{d - 1}{d - 2} \frac{r_h^2}{l^2} \right) + \frac{2}{(d - 2) r_h} \left( (d - 2)^2 k + d (d - 1) \frac{r_h^2}{l^2}  \right) \left( \frac{ \partial r_h } {\partial \epsilon} \right)  \Bigg) \, ,
    \end{split}
\end{align}
where we have taken into account the fact that when the charge is fixed, we must allow the horizon radius $r_h$ to vary with $\epsilon$. To compute the shift $\partial r_h/\partial\epsilon$, we use the fact that we are holding $Q$ fixed. Then we use the expression for $Q_{\text{ext}}$ in (\ref{extremal_mq}) and demand that $\left({\partial Q}/{\partial \epsilon} \right)_{T = 0}= 0$ to obtain an equation for $\partial r_h/\partial\epsilon$. This procedure leads to the rather simple result
\begin{align}
    \begin{split}
     & \left( \frac{ \partial M } {\partial \epsilon} \right)_{Q, T = 0} = V (d - 1)  r_h^{d - 2} \left( \Delta g + 2 \rho \left( k + \frac{d - 1}{d - 2} \frac{r_h^2}{l^2} \right) \right) \, .
     \label{mass_shift}
    \end{split}
\end{align}
Note that the dependence on $\Delta g'$ has vanished. From the geometric point of view, this non-trivial cancellation is crucial for the extremality-entropy relation to hold. 

\subsubsection{Charge Shift at Fixed Mass}

 If we instead hold the mass fixed, the entire shift in the extremality is due to the shift in charge. Following the same procedure as in the fixed charge case, but this time demanding $\partial M_{\text{ext}}/\partial\epsilon=0$, we find the relation:
\begin{align}
    \left( \frac{ \partial Q^2 } {\partial \epsilon} \right)_{M, T = 0} = - 2 V^2 (d - 1) (d - 2) r_h^{2d - 4} \left( \Delta g + 2 \rho  \left( k + \frac{d - 1}{d - 2} \frac{r_h^2}{l^2}\right) \right) \, .
\end{align}
Here we also find a cancellation of all $\Delta g'$ terms. Moreover, this shift is proportional to the mass shift at fixed charge
\begin{equation}
     \left( \frac{ \partial Q^2 } {\partial \epsilon} \right)_{M, T = 0}=-2V(d-2)r_h^{d-2}\left( \frac{ d M } {d \epsilon} \right)_{Q, T = 0}\,.
\end{equation}
This relationship more clear when we write this as the shift of $Q$ rather than $Q^2$. Using $\Delta Q^2 = 2 Q \Delta Q$, we find
\begin{align}
        Q \left( \frac{ \partial Q } {\partial \epsilon} \right)_{M, T = 0}=- V(d-2)r_h^{d-2}\left( \frac{ \partial M } {\partial \epsilon} \right)_{Q, T = 0}\,.
\end{align}
Finally, we use $\Phi = Q / (d - 2)  V  r^{d - 2}$ to write:
\begin{align}
        \left( \frac{ \partial M } {\partial \epsilon} \right)_{Q, T = 0} = 
        - \Phi \left( \frac{ \partial Q } {\partial \epsilon} \right)_{M, T = 0} \,.
\end{align}
So we see that the mass shift is related to the charge shift times the potential. In Appendix A, we derive this statement for a general thermodynamic system and show that it holds for any extensive charge and its conjugate. 

One physical consequence of this fact is that the entropy-extremality relationship (with a different proportionality factor) will hold regardless of whether the mass or charge is held fixed. As far as we know, this has not been noticed before in the literature.

\subsubsection{Summary of Extremality Shifts}

 The shifts to extremality may be obtained from these mass and charge shifts. For completeness, we also present calculation at fixed horizon radius, as this extremality shift has previously been considered in the literature as well \cite{Myers:2009ij, Cremonini:2009ih},
\begin{align}
    \begin{split}
        \left( \frac{M}{Q} \right)_{Q, T = 0} \quad &= \quad  \left( \frac{M}{Q}  \right)_0  \left( 1 + \rho +  \Delta g \ \frac{1}{2\left( k + \frac{d - 1}{d - 2} \frac{r_h^2}{l^2} \right)} \right), \\
        \left( \frac{M}{Q} \right)_{M, T = 0} \quad &= \quad  \left( \frac{M}{Q}  \right)_0  \left( 1 + \rho \ \frac{k + \frac{d - 1}{d - 2} \frac{r_h^2}{l^2}}{k + \frac{d}{d - 2} \frac{r_h^2}{l^2}} +  \Delta g \  \frac{1}{2\left( k + \frac{d}{d - 2} \frac{r_h^2}{l^2} \right)}   \right), \\ 
        \left( \frac{M}{Q} \right)_{r_h, T = 0} \quad &= \quad \left( \frac{M}{Q}  \right)_0 \left(  1 + \rho + \frac{ \Delta g \left( k + \frac{d + 1}{d - 2} \frac{r_h^2}{l^2}  \right) + r_h \Delta g' \frac{1}{(d - 2)^2} \frac{r_h^2}{l^2}} {2 \left( k + \frac{d - 1}{d - 2} \frac{r_h^2}{l^2}  \right) \left( k + \frac{d}{d - 2} \frac{r_h^2}{l^2}  \right) } \right),
    \end{split}
    \label{eq:MQ3}
\end{align}
where the corrections are encoded in $\rho$ and $\Delta g$ given in (\ref{eq:rhodef}) and (\ref{eq:Deltag}), respectively (and $\Delta g'$ as well for the fixed $r_h$ case).  For these final results, we have set $\epsilon=1$.  However, the expressions are only valid to first order in the Wilson coefficients $c_i$.  Here the uncorrected charge to mass ratio may be obtained from (\ref{leading_extremal_mq}), and takes the form
\begin{align}
    \left( \frac{M}{Q}  \right)_0 \quad = \quad   \sqrt{\frac{2 (d - 1)}{d - 2}} \frac{ k +  \frac{d - 1}{d - 2 } \frac{r_h^2}{l^2} }{  \sqrt{k + \frac{d }{d - 2 } \frac{r_h^2}{l^2}} } \, .
    \label{eq:MQ0}
\end{align}
Note that, in (\ref{eq:MQ3}), the horizon radius $r_h$ may be taken to be the uncorrected radius, and can be obtained from either $M$ or $Q$ using the leading order expressions (\ref{leading_extremal_mq}).  In (\ref{eq:MQ0}), the leading order expression for $r_h$ should be used.  Finally, note that $\Delta g$ depends on the parameters $m$ and $q$ as well as the radius $r$.  The $m$ and $q$ parameters are directly obtained from $M$ and $Q$ using (\ref{eq:massdef}) and (\ref{eq:Qdef}), and again the leading order horizon radius can be used in $\Delta g$.

\subsection{Wald Entropy}

 We now compare the shift in mass at fixed charge and temperature to the shift in entropy at fixed mass and charge. The entropy for black holes in higher-derivative theories is given by the Wald entropy \cite{Wald:1993nt}:
\begin{align}
    S = - 2 \pi \int_{\Sigma} \frac{\delta \mathcal{L}}{\delta R_{\mu \nu \rho \sigma}} \epsilon_{\mu \nu} \epsilon_{\rho \sigma} \, .
    \label{Wald_Entropy1}
\end{align}
For spherically symmetric backgrounds, the integral over the horizon $\Sigma$ gives a factor of the area $A$. The two-derivative contribution to the entropy is simply $S^{(2)}=A/4$, while the four-derivative terms yield
\begin{align}
    S^{(4)} =\left. -2\pi A \frac{\delta \Delta \mathcal{L}}{\delta R_{\mu\nu\rho\sigma}} \epsilon_{\mu\nu}\epsilon_{\mu\nu}\right|_{\partial^4} = - \frac{A}{4} l^2 (4 c_1 R_{trtr} + 2 c_2 F_{tr}F_{tr}) \, .
\end{align}
The total entropy is the sum of these terms,
\begin{align}
    \begin{split}
        S =\left. \frac{A}{4} \left(1 - \epsilon  \left( 4 c_1 l^2 R_{trtr} + 2 c_2 l^2 F_{tr} F_{tr} \right) \right)\right|_{r_h} \,,
    \end{split}
    \label{eq:Wald}
\end{align}
where we once again introduced $\epsilon$ to parametrize the expansion.  Here the horizon area is given by $A=\omega_{d-1}r_h^{d-1}$, where $r_h$ is the corrected horizon radius.  On the other hand, the $R_{trtr}$ and $F_{tr}F_{tr}$ terms need only be computed on the zeroth-order background,
\begin{align}
    \begin{split}
        R_{trtr} &= \frac{1}{l^2} + \frac{(2 d - 3)(Q/V)^2}{2 (d - 1) r^{2 d - 2}} - \frac{(d - 2) M/V}{2 r^{d}} \, , \\
        F_{tr}F_{tr} &= \frac{(Q/V)^2}{r^{2 d - 2}} \, .    
    \end{split}
\end{align}
It does not matter whether we use the corrected or uncorrected quantities here because they already show up in a term that is order $\epsilon$. Note also that, while the expression for the Wald entropy (\ref{eq:Wald}) is given in terms of $M$, $Q$ and $r_h$ of the fully corrected solution, only two of these quantities are independent.

We now examine the entropy shift for a given solution at fixed mass $M$ and charge $Q$.  For the moment, we work at arbitrary $M$ and $Q$, and not necessarily at extremality.  The general expression for the entropy shift is then
\begin{equation}
    \left(  \frac{\partial S}{\partial \epsilon} \right)_{Q, M} = \frac{A}{4} \left( (d - 1)\left( \frac{1}{r_h} \frac{ \partial r_h }{\partial \epsilon}\right)_{Q,M}  -  \left( 4 c_1 l^2 R_{trtr} + 2 c_2 l^2 F_{tr} F_{tr} \right) \right) \, ,
    \label{eq:Sshift}
\end{equation}
where the first term was obtained by
\begin{align}
    \frac{1}{A} \frac{ \partial A }{\partial \epsilon} = (d - 1) \frac{1}{r_h} \frac{ \partial r_h }{\partial \epsilon} \, .
\end{align}
Here it is important to note that the horizon radius $r_h$ receives a correction when working at fixed $M$ and $Q$.  If, on the other hand, we were to keep the horizon radius fixed (as is done in \cite{Myers:2009ij}), we would find only the second (interaction) term in (\ref{eq:Sshift}), and the entropy shift would be independent of $c_3$ and $c_4$.

To compute ${\partial r_h}/{\partial \epsilon}$, we start with the horizon condition $g(r_h)=0$ where $g(r)$ is given by (\ref{gcorr}) with $m$ and $q$ rewritten in terms of $M$ and $Q$.  Taking a derivative and solving for $\partial r_h/\partial\epsilon$ then gives
\begin{align}
    \fft1{r_h}\frac{ \partial r_h }{\partial \epsilon} =  - \frac{\rho M+V(d-1)r_h^{d-2}\Delta g} {(d-2)(M-(M_{\text{ext}})_0)} \, .
    \label{eq:drhde}
\end{align}
where $(M_{\text{ext}})_0$ is the leading order extremal mass given in (\ref{leading_extremal_mq}).  As we can see, this expression diverges if the leading order solution is extremal.  This is in fact not a surprise, as leading order extremality implies a double root at the horizon.  The higher order corrections will lift this double root and hence cannot be parametrized as a linear shift in $\epsilon$.

In order to avoid the divergence, we can instead consider a leading order solution taken slightly away from extremality.  As long as we are sufficiently close to extremality, the first term in (\ref{eq:Sshift}) will dominate the entropy shift.  Noting further that, at extremality, the numerator of (\ref{eq:drhde}) becomes proportional to the mass shift (\ref{mass_shift}) at fixed charge, we can rewrite (\ref{eq:Sshift}) as
\begin{equation}
    \left(\fft{\partial S}{\partial \epsilon}\right)_{Q,M}\!\!=-\fft{A}4\left(\fft{d-1}{(d-2)(M-(M_{\text{ext}})_0)}\left(\fft{\partial M}{\partial \epsilon}\right)_{Q,T=0}\!\!+\fft{d-1}{d-2}\rho+4c_1l^2R_{trtr}+2c_2l^2F_{tr}F_{tr}\right).
\end{equation}
The deviation away from extremality can be written in terms of the leading order temperature,
\begin{align}
    4 \pi T_0 =|g'((r_h)_0)|_{\epsilon = 0} = \fft{(d-2)(M-(M_{\text{ext}})_0)}{V(d-1)(r_h)_0^{d-1}}\, .
\end{align}
The total shift to the entropy is then given by 
\begin{align}
    \left(\fft{\partial S}{\partial\epsilon}\right)_{Q,M}=-\fft1{T_0}\left(\fft{\partial M}{\partial \epsilon}\right)_{Q,T=0}-\fft{A}4\left(\fft{d-1}{d-2}\rho+4c_1l^2R_{trtr}+2c_2l^2F_{tr}F_{tr}\right)\,.
\end{align}
Finally, as $T_0 \rightarrow 0$ we reproduce the relation
\cite{Cheung:2018cwt, Goon:2019faz}
\begin{align}
    \left(  \frac{ \partial M}{\partial \epsilon} \right)_{Q, T = 0} =- T_0 \left(  \frac{ \partial S }{\partial \epsilon} \right)_{Q, M} \, .
\end{align}
Note that this relation was obtained using only the general feature that the corrected geometry may be written in terms of a shift $\Delta g$ to the radial function $g(r)$.  In particular, we never had to use the explicit form of $\Delta g$ given in (\ref{eq:Deltag}).

\subsection{Explicit Results for the Entropy Shifts}

 In order to compare with the next section, we include some explicit results for the mass shifts. In section V, we will see what constraints may be placed on the EFT coefficients by imposing that entropy shift is positive. We will use the mass shift here, to remove the factor of $T_0$. The entropy shift is positive when the mass shift at constant charge is negative. It is easy to see that the shifts here are positive when all the coefficients are positive. 

 For AdS${}_4$, we find:
\begin{align}
\label{TdS4}
    \begin{split}
            T_0 \Delta S = \frac{1}{5 r_h l^2}\Big( 4 c_1 (l^2 + 3 r_h^2)^2  + 2 c_2  (l^2 + 3 r_h^2) (l^2 + 18 r_h^2) + 8 (2 c_3 + c_4 ) (l^2 + 3 r_h^2)^2  \Big) \, .
    \end{split}
\end{align}
For AdS${}_5$, we get:
\begin{align}
\label{TdS5}
    \begin{split}
            T_0 \Delta S =  \frac{\pi}{16l^2}  \Big( & c_1 (31 l^4 + 128 l^2 r_h^2 + 138 r_h^4)\\
    & \qquad + c_2 24 (l^2 + 2 r_h^2) (l^2 + 6 r_h^2)+ (2 c_3 + c_4 ) 72 (l^2 + 2 r_h^2)^2 \Big) \, .
    \end{split}
\end{align}
AdS${}_6$:
\begin{align}
    \begin{split}
            T_0 \Delta S = \frac{2 \pi}{99 l^2}  \Big( & c_1 r_h (369 l^4 + 1263 l^2 r_h^2 + 1124 r_h^4)\\
    & \qquad  + c_2 4 r_h (3 l^2 + 5 r_h^2) (27 l^2 + 100 r_h^2)+ (2 c_3 + c4 ) 96 r_h (3 l^2 + 5 r_h^2)^2 \Big) \, .
    \end{split}
\end{align}
AdS${}_7$:
\begin{align}
    \begin{split}
            T_0 \Delta S =  \frac{\pi^2}{224 l^2}  \Big( & c_1 \, (1384 l^4 r_h^2 + 4236 l^2 r_h^4 + 3345 r_h^6) \\
    & \qquad  + c_2 \, 40 (2 l^2 + 3 r_h^2) (16 l^2 + 45 r_h^2) + (2 c_3 + c4 ) \, 800 (2l^2 + 3 r_h^2)^2 \Big) \, .
    \end{split}
\end{align}

\section{Thermodynamics from the On-Shell Euclidean Action}
\label{sec:EuclideanAction}

 The ultimate goal of our analysis is to determine the leading higher-derivative corrections to relations between certain global properties of black hole solutions. These relations are of a thermodynamic nature, and arise by taking various derivatives of the free-energy corresponding to the appropriate ensemble. As is well-known \cite{Hawking:1979ig}, the classical free-energy of a black hole can be calculated using the saddle-point approximation of the Euclidean path integral with appropriate boundary conditions. In the \textit{Gibbs} or \textit{grand canonical ensemble}, the appropriate quantity is the Gibbs free-energy, which may be calculated from the on-shell Euclidean action
\begin{equation}
  \beta G(T,\Phi) = I_E[g^E_{\mu\nu}\left(T,\Phi\right),A^E_\mu(T,\Phi)],
\end{equation}
where $\beta = T^{-1}$, and  $g^E_{\mu\nu}\left(T,\Phi\right)$ and $A^E_\mu(T,\Phi)$ are Euclideanized solutions to the classical equations of motion with temperature $T$ and potential $\Phi$. Similarly in the \textit{canonical ensemble} the corresponding quantity is the Helmholtz free-energy, given by
\begin{equation}
  \beta F(T,Q) = I_E[g^E_{\mu\nu}\left(T,Q\right),A^E_\mu(T,Q)],
\end{equation}
where $g^E_{\mu\nu}\left(T,Q\right)$ and $A^E_\mu(T,Q)$ are Euclideanized solutions with temperature $T$ and electric charge $Q$. In both expressions, $I_E$ is the \textit{renormalized} Euclidean on-shell action.

The Euclidean action with cosmological constant is IR divergent when evaluated on a solution.  However, it may be given a satisfactory finite definition by first regularizing the integral with a radial cutoff $R$. To render the variation principle well-defined on a spacetime with boundary we must add an appropriate set of Gibbons-Hawking-York (GHY) \cite{York:1972sj,Gibbons:1976ue} (in the case of the canonical ensemble, also Hawking-Ross \cite{Hawking:1995ap}) terms \textit{in addition to} a set of boundary counterterms. The complete on-shell action then consists of three contributions
\begin{equation}
    I_{E} = I_{\text{bulk}} + I_{\text{GHY}} + I_{\text{CT}}.
\end{equation}
If the counterterms are chosen correctly, they will cancel the divergence of the bulk and Gibbons-Hawking-York terms, rendering the results finite as $R\rightarrow \infty$. In AdS there is a systematic approach to generating such counterterms via the method of \textit{holographic renormalization} \cite{Henningson:1998gx,Balasubramanian:1999re,Emparan:1999pm}; since the logic of this approach is well-described in detail elsewhere (see e.g.~\cite{Skenderis:2002wp}) we will not review it further, but simply make use of known results. Explicit expressions for the needed GHY and counterterms (including the four-derivative corrections used in this dissertation) valid in $\text{AdS}_d$, $d=4,5,6$ can be found in \cite{Liu:2008zf,Cremonini:2009ih}.

Once the free-energy is calculated, the remaining thermodynamic quantities can be determined straightforwardly by using the definitions of the free-energies and the first-law of black hole thermodynamics
\begin{equation}
  F = E-TS,\hspace{5mm} G= E-TS-\Phi Q, \hspace{5mm} dE = TdS+\Phi dQ.
\end{equation}
The expressions calculated using these Euclidean methods should agree with the Lorentzian or geometric calculations in the previous section.  Note, however, that there is a bit of a subtlety with the notion of black hole mass here, as the thermodynamic relations are for the energy $E$ of the system. In holographic renormalization, there is always an ambiguity in the addition of finite counterterms that shift the value of the on-shell action. The standard approach is to fix the ambiguity by demanding that even-dimensional global AdS has zero vacuum energy while odd-dimensional global AdS has non-zero vacuum energy that is interpreted as a Casimir energy in the dual field theory.  In this case the thermodynamic energy is the sum of the black hole mass and the Casimir energy
\begin{equation}
    E=M+E_c,
\end{equation}
and the mass $M$ of the black hole is only obtained after subtracting out the Casimir energy contribution, as we did in section~II.

The purpose of introducing this alternative approach is not just to give a cross-check on the results of the previous section, but also to verify a recent general claim by Reall and Santos \cite{Reall:2019sah}. The $\mathcal{O}(\epsilon)$ corrections we are considering can be calculated by first evaluating the free-energy or on-shell action at the same order. Naively, this would require evaluating three contributions
\begin{align}
  I_E[g^E_{\mu\nu},A^E_\mu] &= I^{(2)}_E[g^{(2)E}_{\mu\nu},A^{(2)E}_\mu] + \epsilon \left(\frac{\partial}{\partial \epsilon} I^{(2)}_E[g^{(2)E}_{\mu\nu}+\epsilon g^{(4)E}_{\mu\nu} ,A^{(2)E}_\mu+\epsilon A^{(4)E}_\mu] \right)\biggr\vert_{\epsilon=0} \nonumber\\
                            &\hspace{5mm} +\epsilon I^{(4)}_E[g^{(2)E}_{\mu\nu},A^{(2)E}_\mu] + \mathcal{O}\left(\epsilon^2\right),
\end{align}
where $(2)$ and $(4)$ denote two and four derivative terms in the action and their corresponding perturbative contributions to the solution.  The central claim in \cite{Reall:2019sah} is that the first term at $\mathcal{O}(\epsilon)$ is actually \textit{zero}, and that therefore we do not need to explicitly calculate the $\mathcal{O}(\epsilon)$ corrections to the equations of motion. For black hole solutions of the type considered in this chapter, we can evaluate the leading corrections without much difficulty, but for more general situations with less symmetry this may not be possible. In such a case the Euclidean method is more powerful, as has recently been demonstrated with calculation of corrections involving angular momentum \cite{Cheung:2019cwi} or dilaton couplings \cite{Loges:2019jzs}. 

Although the result of \cite{Reall:2019sah} was demonstrated in the grand canonical ensemble, it is straightforward to see that it implies an identical claim about the leading corrections in the canonical ensemble. While the quantities of interest can be extracted from either, the explicit expressions encountered in the latter are usually far simpler and therefore more convenient. Recall that we can change ensemble by a Legendre transform of the free-energy
\begin{equation}
  F(T,Q) = G(T,\Phi(Q)) + \Phi(Q) Q, \hspace{5mm} Q = -\left(\frac{\partial G}{\partial \Phi}\right)_{T},
\end{equation}
where the right-hand-side is defined in terms of the implicit inverse function $\Phi(Q)$. At fixed $T$ and $Q$, the potential $\Phi$ receives corrections from the higher-derivative interactions, and so, expanding the right-hand-side to $\mathcal{O}(\epsilon)$, we have
\begin{align}
  F(T,Q) &= G^{(2)}(T,\Phi^{(2)}(Q)) + \epsilon\left(\frac{\partial}{\partial \epsilon}G^{(2)}(T,\Phi^{(2)}(Q)+\epsilon\Phi^{(4)}(Q))\right)\biggr\vert_{\epsilon=0} \nonumber\\
&\hspace{5mm}+ \epsilon G^{(4)}(T,\Phi^{(2)}(Q)) + \Phi^{(2)}(Q)Q +\epsilon \Phi^{(4)}(Q)Q +\mathcal{O}\left(\epsilon^2\right).
\end{align} 
Recognizing that 
\begin{equation}
  \left(\frac{\partial}{\partial \epsilon}G^{(2)}(T,\Phi^{(2)}(Q)+\epsilon\Phi^{(4)}(Q))\right)\biggr\vert_{\epsilon=0} = \Phi^{(4)}(Q)\left(\frac{\partial G^{(2)}}{\partial \Phi}\right)_T\biggr\vert_{\Phi=\Phi^{(2)}(Q)} = -\Phi^{(4)}(Q)Q,
\end{equation}
we see that the leading correction to the Helmholtz free energy is simply given by 
\begin{equation}
   F(T, Q) = F^{(2)}(T, Q) +\epsilon G^{(4)}(T,\Phi^{(2)}(Q)) + \mathcal{O}\left(\epsilon^2\right).
\end{equation}
In terms of the on-shell Euclidean action, using the result of Reall and Santos, this is then equivalent to
\begin{equation}
\label{F4}
  F^{(4)}(T,Q) = \frac{1}{\beta}I^{(4)}_E\left(g^{(2)E}_{\mu\nu}\left(T,Q\right),A^{(2)E}_\mu(T,Q)\right),
\end{equation}
where here $I_E^{(4)}$ denotes the contribution of the four-derivative terms to the renormalized on-shell action. Note that this includes potential four-derivative Gibbons-Hawking-York terms, but as this argument makes clear, will not include any additional Hawking-Ross terms. This expression is the analogue of the Reall-Santos result, but in the canonical ensemble. It says that the leading correction to the Helmholtz free-energy is given by evaluating the four-derivative part of the renormalized on-shell action on a solution to the two-derivative equations of motion with temperature $T$ and charge $Q$. 

Below we will give a brief review of the well-known thermodynamic relations at two-derivative order, and then using the above result we will calculate the leading corrections and verify explicitly that they agree with the results of the previous section.

\subsection{Two-Derivative Thermodynamics}

 As described above, the regularized on-shell action has a bulk as well as various boundary contributions. At two-derivative order and in $d$-dimensions these have the explicit form
\begin{align}
  I^{(2)}_{\text{bulk}} &= -\frac{1}{16 \pi } \int d^{d+1} x \sqrt{g} \Big(  \frac{d (d-1)}{l^2} + R -\frac{1}{4} F^2 \Big), \nonumber\\
  I^{(2)}_{\text{GHY}} &= -\frac{1}{8 \pi } \int d^{d} x \sqrt{h} K, \nonumber\\
  I^{(2)}_{\text{CT}} &= \frac{1}{8 \pi } \int d^{d} x \sqrt{h} \left( \frac{d - 1}{l} + \frac{l}{2 (d - 2)} \mathcal{R}\right),
\end{align}
where $h_{ab}$ and $\mathcal{R}_{ab}$ are the metric and Ricci tensor of the induced geometry on the boundary at $r = R$. Note that in $I^{(2)}_{\text{CT}}$ we have included the minimal set of counterterms necessary to cancel the IR divergence in $d=3$ and $d=4$. For $d>4$, additional counterterms beginning at quadratic order in the boundary Riemann tensor are necessary to cancel further divergences. 

The regularized bulk action has a well-defined variational principle provided that $\delta A_{a} = 0$ at $r = R$. This amounts to holding $\Phi$ fixed, and thus it corresponds to boundary conditions compatible with the grand canonical ensemble. For many applications, we will want to hold the charge fixed. From a thermodynamic point of view, we want to use the extensive quantity $Q$ instead of the intensive $\Phi$, so we must compute the Helmholtz free energy instead of the Gibbs free energy. Holding $Q$ fixed requires different boundary conditions, and in particular
the further addition of a Hawking-Ross boundary term \cite{Hawking:1995ap} 
\begin{align}
    I^{(2)}_{\text{HR}} =  \frac{1}{16 \pi} \int d^{d} x \sqrt{h} n_\mu F^{\mu b} A_b \, ,
\end{align}
where $n_\mu$ is the normal vector on the boundary and $A_a$ is the pull-back of the gauge potential. To summarize, the total two-derivative on-shell action
\begin{equation}
  I^{(2)}_E = I^{(2)}_{\text{bulk}}+I^{(2)}_{\text{GHY}}+I^{(2)}_{\text{HR}}+I^{(2)}_{\text{CT}},
\end{equation}
evaluated on the Euclideanized solution to the two-derivative equations of motion
\begin{align}
    \begin{split}
        ds_E^2 =& f(r) d\tau^2 + g(r)^{-1} dr^2 + r^2 d \Omega_{d-1}^2 \, , \qquad f(r) = g(r) = 1 - \frac{m}{r^{d-2}} + \frac{q^2}{4 r^{2d-4}} + \frac{r^2}{l^2}, \\
        & \quad A_E = i\left( - \frac{1}{c} \frac{q}{r^{d-2}} + \Phi \right) d\tau,  \qquad c = \sqrt{ \frac{ 2(d-2)}{(d-1)} }, \qquad \Phi = \frac{1}{c} \frac{q}{l^{d-2}\nu^{d-2}}\, ,
    \end{split}
\end{align}
is equal to $\beta F^{(2)}(T,Q)$, where $F^{(2)}$ is the two-derivative contribution to the Helmholtz free-energy. In the above we have introduced the dimensionless variable $\nu\equiv (r_h)_0/l$, where $(r_h)_0$ is the location of the outer-horizon of the two-derivative solution with temperature $T$ and charge $Q$. Note also that here, and for the remainder of this section, we will consider only spherical $k=1$ black holes. Since $\nu$ satisfies $f(\nu)=0$, we can solve for the parameter $m$ as
\begin{equation}
  m = \nu^{d-2}+\frac{q^2}{4\nu^{d-2}}+\frac{\nu^d}{l^2}.
\end{equation}
In the Euclidean approach to calculating the leading corrections to the thermodynamics, it will prove natural to continue to use $\nu$ and $q$ to parametrize the space of black hole solutions, even when the four-derivative corrections are included. This means that it is also natural to write all thermodynamic quantities in these variables, which requires the use of standard thermodynamic derivative identities to rewrite derivatives. Recall that the parameter $q$ and the physical charge $Q$ are not the same, but are related by an overall constant given in (\ref{eq:Qdef}).  Therefore holding $Q$ fixed is the same as holding $q$ fixed. Explicitly, the two-derivative free-energy calculated in this way in $\text{AdS}_{4}$ is given by
\begin{equation}
\label{F2d3}
 F_{d=3}^{(2)}(q,\nu) = -\frac{l \nu ^3}{4}+\frac{l \nu }{4}+\frac{3 q^2}{16 l \nu } ,
\end{equation}
and in $\text{AdS}_5$ by
\begin{equation}
\label{F2d4}
 F_{d=4}^{(2)}(q,\nu) = -\frac{1}{8} \pi  l^2 \nu ^4+\frac{1}{8} \pi  l^2 \nu ^2+\frac{5 \pi  q^2}{32 l^2 \nu
   ^2}+\frac{3 \pi  l^2}{32} .
\end{equation}
Once the free-energy is calculated, the entropy and energy are given by
\begin{align}
S = - \left( \frac{\partial F}{ \partial T} \right)_Q, \hspace{5mm} E = F+TS.
\end{align}
In terms of our natural variables, we can reexpress the entropy as
\begin{equation}
  S(q,\nu) = \left(\frac{\partial F}{\partial \nu}\right)_q \left[\left(\frac{\partial T}{\partial \nu}\right)_q\right]^{-1},
\end{equation}
where the temperature is given by
\begin{equation}
  T(q,\nu) = \frac{(d-2)q ^2 l^{1-d}\nu^{1-d}}{4\pi}+ \frac{(d-1)\nu^2+d-2}{4\pi l}.
\end{equation}
Note that this expression is exact, meaning it does not receive corrections when we include the four-derivative interactions. It is therefore useful to introduce the function
\begin{equation}
\label{qext}
    q^2_{\text{ext}}(\nu) = -\frac{2 \left(d \nu ^2+d-\nu ^2-2\right) (l \nu )^{d-2}}{(d-2) },
\end{equation}
such that taking the limit $q^2\rightarrow  q^2_{\text{ext}}(\nu)$ is equivalent to taking the extremal limit $T\rightarrow 0$. 

If we extract the energy $E=F+TS$ from the expressions (\ref{F2d3}) and (\ref{F2d4}), we find that it agrees with the mass, (\ref{eq:massdef}), for $\text{AdS}_4$ but not $\text{AdS}_5$. This is not surprising as the thermodynamic energy $E$ and mass $M$ of the black hole in AdS$_5$ differ by a Casimir energy contribution that is independent of $q$ and $\nu$.  We can, of course remove the Casimir energy by the addition of \textit{finite boundary counterterms}, or equivalently by a change in holographic renormalization scheme. The expression (\ref{F2d4}) is calculated in a \textit{minimal subtraction} scheme, in which the possible finite counterterms are zero and the Casimir energy is present.

Physically, it is useful work in a scheme in which the energy $E$ coincides with the mass $M$ of the black hole, without a Casimir contribution. In such a \textit{zero Casimir} scheme, the energy of pure $\text{AdS}_5$ is defined to be zero. Calculating the free-energy from the on-shell action of pure $\text{AdS}_5$ with generically parametrized four-derivative counterterms we find that this scheme requires the following modification from the minimal subtraction counterterms 
\begin{equation}
    I_{\text{CT}}^{(2)} \longrightarrow I_{\text{CT}}^{(2)}+\frac{1}{8 \pi } \int d^{4} x \sqrt{h} \left( -\frac{l^3 }{96}\right) \mathcal{R}^2.
\end{equation}
The free energy calculated with this modified on-shell action agrees exactly with the expectation using (\ref{eq:massdef}). Note that the entropy, since it is given by a derivative of the free-energy, is independent of the choice of scheme. The zero Casimir scheme is a physically motivated choice, but certainly not unique. 

\subsection{Four-Derivative Corrections to Thermodynamics}

 To evaluate the four-derivative corrections we make use of the result (\ref{F4}). As in the two-derivative contribution, the on-shell action is properly defined by a regularization and renormalization procedure. For the operators in (\ref{ActionCJLM}) with Wilson coefficients $c_2$, $c_3$ and $c_4$ the required $I_{\text{bulk}}^{(4)}$ contribution is actually finite, while for the term in (\ref{ActionCJLM}) proportional to $c_1$, we must again regularize and renormalize by adding infinite boundary counterterms. The required explicit expressions, as well as the complete set of four-derivative GHY terms, can be found in \cite{Liu:2008zf,Cremonini:2009ih}. 
The calculation is otherwise identical to the two-derivative contribution described above, and in $\text{AdS}_4$ we find
\begin{align}
    F^{(4)}_{d=3}(q,\nu) = &c_1 \left(-\frac{ \left(20 l^4 \nu ^4-5 l^2 \nu ^2 q^2+q^4\right)}{20 l^5 \nu
   ^5}-\frac{3 \nu   }{l}\right)+\frac{c_2 q^2  \left(l^2 \left(20 l^2
   \nu ^2-7 q^2\right)-60 l^4 \nu ^4\right)}{80 l^7 \nu ^5}\nonumber\\
    &    -\frac{c_3 q^4 }{5
   l^5 \nu ^5}-\frac{c_4 q^4 }{10 l^5 \nu ^5}.
\end{align}
The complete free-energy, up to $\mathcal{O}(\epsilon^2)$ contributions, is then given by
\begin{equation}
    F_{d=3}(q,\nu) = F_{d=3}^{(2)}(q,\nu)+\epsilon  F_{d=3}^{(4)}(q,\nu) + \mathcal{O}(\epsilon^2).
\end{equation}
From this explicit expression we can then calculate the entropy
\begin{align}
    S_{d=3}&=\pi  l^2 \nu ^2-\frac{4 \pi  c_1 \epsilon  \left(4 l^4 \nu ^4 \left(1-3 \nu ^2\right)-3 l^2 \nu ^2
   q^2+q^4\right)}{4 l^2 \left(3 \nu ^2-1\right) \nu ^4+3 \nu ^2 q^2}-\frac{\pi  c_2 q^2
   \epsilon  \left(12 l^2 \nu ^2 \left(\nu ^2-1\right)+7 q^2\right)}{4 l^2 \left(3 \nu
   ^2-1\right) \nu ^4+3 \nu ^2 q^2}\nonumber\\
   & \hspace{5mm}   -\frac{16 \pi  c_3 q^4 \epsilon }{4 l^2 \left(3 \nu
   ^2-1\right) \nu ^4+3 \nu ^2 q^2}-\frac{8 \pi  c_4 q^4 \epsilon }{4 l^2 \left(3 \nu
   ^2-1\right) \nu ^4+3 \nu ^2 q^2}+\mathcal{O}(\epsilon^2),
\end{align}
and mass (which coincides with the thermal energy)
\begin{align}
   M_{d=3} &=\frac{1}{2} l \left(\nu ^3+\nu \right)+\frac{q^2}{8 l
   \nu }+\frac{c_1 q^4 \epsilon  \left(q^2-4 l^2 \nu ^2 \left(9 \nu ^2+2\right)\right)}{40 l^5
   \left(3 \nu ^2-1\right) \nu ^7+30 l^3 \nu ^5 q^2}\nonumber\\
   &\hspace{5mm}+\frac{c_2 q^2 \epsilon  \left(80 l^4
   \nu ^4 \left(-9 \nu ^4+6 \nu ^2+1\right)-8 l^2 \nu ^2 \left(39 \nu ^2+7\right) q^2+7
   q^4\right)}{40 l^3 \nu ^5 \left(4 l^2 \nu ^2 \left(3 \nu ^2-1\right)+3 q^2\right)}\nonumber\\
   &\hspace{5mm}+\frac{2
   c_3 q^4 \epsilon  \left(q^2-4 l^2 \nu ^2 \left(9 \nu ^2+2\right)\right)}{5 l^3 \nu ^5
   \left(4 l^2 \nu ^2 \left(3 \nu ^2-1\right)+3 q^2\right)}+\frac{c_4 q^4 \epsilon 
   \left(q^2-4 l^2 \nu ^2 \left(9 \nu ^2+2\right)\right)}{5 l^3 \nu ^5 \left(4 l^2 \nu ^2
   \left(3 \nu ^2-1\right)+3 q^2\right)}+\mathcal{O}(\epsilon^2).
\end{align}
Taking the extremal limit we find the following expression for the mass shift
\begin{align}
    \label{massshift}
    (\Delta M_{d=3})_{Q,T=0} = &-\frac{4 c_1 l \left(3 \nu ^2+1\right)^2}{5 \nu }-\frac{2 c_2 l (3 \nu ^2+1)(18 \nu ^2+1)}{5 \nu }\nonumber\\
    &   -\frac{16 c_3 l \left(3 \nu ^2+1\right)^2}{5 \nu }-\frac{8
   c_4 l \left(3 \nu ^2+1\right)^2}{5 \nu },
\end{align}
which agrees exactly with the results we have derived using the shifted solution. Strictly, the two expressions are parameterized in terms of different variables ($\nu$ the uncorrected horizon vs. $r_h$ the corrected horizon), but these differ by $\mathcal{O}(\epsilon)$, and so when we take $\epsilon \rightarrow 0$ the two functions are the same.

Similarly we can calculate the shift in the microcanonical entropy, which will be important in the subsequent section for analyzing conjectured bounds on the Wilson coefficients. The actual expression is given in (\ref{entshiftads4}), and can be calculated straightforwardly using standard thermodynamic derivative identities 
\begin{equation}
\label{microSformula}
    (\Delta S)_{Q,E} = \lim_{\epsilon\rightarrow 0}\left[\left(\frac{\partial S}{\partial \epsilon}\right)_{q,\nu}- \left(\frac{\partial E}{\partial \epsilon}\right)_{q,\nu}\frac{\left(\frac{\partial S}{\partial \nu}\right)_{q}}{\left(\frac{\partial E}{\partial \nu}\right)_{q}}\right].
\end{equation}
The calculation for $\text{AdS}_5$ is similar, but in this case we have to be cautious about the Casimir energy. We calculate the free-energy in the physically motivated zero Casimir scheme.  To do so, we again fix the finite counterterms by evaluating the four-derivative on-shell action on pure $\text{AdS}_5$. Requiring the Casimir energy to vanish requires the following modification from the minimal subtraction counterterm action  
\begin{equation}
    I_{\text{CT}}^{(4)} \longrightarrow I_{\text{CT}}^{(4)}+\frac{1}{8 \pi } \int d^{4} x \sqrt{h} \left(-\frac{5 c_1 l^3}{48}\right) \mathcal{R}^2.
\end{equation}
Using this we calculate the four-derivative contribution to the renormalized free-energy 
\begin{align}
    F_{d=4}^{(4)} = &\frac{1}{256} \pi  c_1 \left(-\frac{43 q^4}{l^8 \nu ^8}+\frac{24 \left(5 \nu
   ^2+8\right) q^2}{l^4 \nu ^4}-32 \left(13 \nu ^4+41 \nu ^2+18\right)\right)\nonumber\\
   &    +\frac{3 \pi 
   c_2 \left(8 l^4 \nu ^4 q^2-3 q^4\right)}{32 l^8 \nu ^8}-\frac{9 \pi  c_3
   q^4}{16 l^8 \nu ^8}-\frac{9 \pi  c_4 q^4}{32 l^8 \nu ^8}.
\end{align}
We also obtain the entropy
\begin{align}
    S_{d=4} =&\frac{1}{2} \pi ^2 l^3 \nu ^3+\frac{\pi ^2 c_1 \epsilon  \left(8 l^8 \left(26 \nu ^2+41\right) \nu ^{10}+6 l^4
   \left(5 \nu ^2+16\right) \nu ^4 q^2-43 q^4\right)}{4 l^3 \nu ^3 \left(4 l^4 \left(2 \nu
   ^2-1\right) \nu ^4+5 q^2\right)}\nonumber\\
   &+\frac{6 \pi ^2 c_2 \epsilon  \left(4 l^4 \nu ^4
   q^2-3 q^4\right)}{l^7 \left(8 \nu ^9-4 \nu ^7\right)+5 l^3 \nu ^3 q^2}-\frac{36 \pi ^2
   c_3 q^4 \epsilon }{l^7 \left(8 \nu ^9-4 \nu ^7\right)+5 l^3 \nu ^3 q^2}\nonumber\\
   &    -\frac{18
   \pi ^2 c_4 q^4 \epsilon }{l^7 \left(8 \nu ^9-4 \nu ^7\right)+5 l^3 \nu ^3
   q^2}+\mathcal{O}\left(\epsilon^2\right),
\end{align}
and mass
\begin{align}
    M_{d=4} =&\frac{3 \pi  \left(4 l^4 \left(\nu ^2+1\right) \nu ^4+q^2\right)}{32 l^2
   \nu ^2}\nonumber\\
   &+ c_1\left[\frac{\pi  \epsilon  \left(384 l^{12} \left(\nu ^2+1\right) \left(26 \nu ^4+23
   \nu ^2+6\right) \nu ^{12}-32 l^8 \left(27 \nu ^4+32 \nu ^2+18\right) \nu ^8 q^2\right)}{256 l^8 \nu ^8 \left(4 l^4
   \left(2 \nu ^2-1\right) \nu ^4+5 q^2\right)}\right.\nonumber\\
   &\hspace{10mm}+\left.\frac{\pi  \epsilon  \left(-4 l^4
   \left(684 \nu ^2+253\right) \nu ^4 q^4+129 q^6\right)}{256 l^8 \nu ^8 \left(4 l^4
   \left(2 \nu ^2-1\right) \nu ^4+5 q^2\right)}\right]\nonumber\\
   &+\frac{3 \pi  c_2 q^2 \epsilon 
   \left(32 l^8 \left(10 \nu ^2+3\right) \nu ^8-4 l^4 \left(54 \nu ^2+19\right) \nu ^4
   q^2+9 q^4\right)}{32 l^8 \nu ^8 \left(4 l^4 \left(2 \nu ^2-1\right) \nu ^4+5
   q^2\right)}\nonumber\\
   &+\frac{9 \pi  c_3 q^4 \epsilon  \left(3 q^2-4 l^4 \nu ^4 \left(18 \nu
   ^2+7\right)\right)}{16 l^8 \nu ^8 \left(4 l^4 \left(2 \nu ^2-1\right) \nu ^4+5
   q^2\right)}\nonumber\\
   &+\frac{9 \pi  c_4 q^4 \epsilon  \left(3 q^2-4 l^4 \nu ^4 \left(18 \nu
   ^2+7\right)\right)}{32 l^8 \nu ^8 \left(4 l^4 \left(2 \nu ^2-1\right) \nu ^4+5
   q^2\right)} +\mathcal{O}\left(\epsilon^2\right).
\end{align}
The extremal mass shift is given by
\begin{align}
    (\Delta M_{d=4})_{Q,T=0} = &-\frac{1}{16} \pi  c_1 \left(138 \nu ^4+128 \nu ^2+31\right)-\frac{3}{2} \pi 
   c_2  \left(2 \nu ^2+1\right) \left(6 \nu ^2+1\right)\nonumber\\
   &    -9 \pi  c_3 \left(2 \nu
   ^2+1\right)^2-\frac{9}{2} \pi  c_4 \left(2 \nu ^2+1\right)^2 \, ,
\end{align}
which agrees exactly with the result (\ref{TdS5}). Likewise we can calculate the correction to the microcanonical entropy using (\ref{microSformula}), the explicit expression is given in (\ref{dSde5}).

\section{Constraints From Positivity of the Entropy Shift}
\label{sec:EntropyConstraints}

 Having derived the general entropy shift at fixed mass, we may now determine what constraints on the EFT coefficients are implied by the assumption that it is positive. 
Recall that the argument of \cite{Cheung:2018cwt}
for the positivity of the entropy
shift 
assumes the existence of a number of quantum fields $\phi$ with mass $m_{\phi}$, heavy enough so that they can be safely integrated out. 
In particular, such fields are assumed to couple to the graviton and photon in such a way that, after 
being integrated out, they generate \textit{at tree-level} the higher-dimension operators we are considering (with the corresponding operator coefficients scaling as $c_i \sim 1/m_{\phi}$). This assumption is essential to the proof; it may be that the entropy shift is universally positive (see \cite{Cheung:2019cwi} for a number of examples), but proving such a statement for non-tree-level completions would require a different argument from the one laid out here.

We revisit the logic of \cite{Cheung:2018cwt} in the context of flat space, before discussing how it may be extended to AdS asymptotics, and denote the Euclidean on-shell action of the theory that includes the heavy scalars $\phi$  by $I_{\text{UV}}[g, A, \phi]$.
First, note that when the scalars are set to zero and are non-dynamical, the action reduces to that of the pure Einstein-Maxwell theory, 
\begin{align}
    I_{\text{UV}}[g, A, 0] = I^{(2)}[g, A] \, .
\end{align}
This is a statement relating the value of the functionals $I_{\text{UV}}$ and $I^{(2)}$ (the two-derivative action) when we pick particular configurations for the fields. These fields may or may not be solutions to the equations of motion. 
Next, consider the corrected action, $I_C = I^{(2)} + I^{(4)}$, and note that it obeys  
\begin{align}
    I_C[g + \Delta g, A + \Delta A] \simeq I_{\text{UV}}[g, A, \phi] \, .
\end{align}
Here we have in mind that the fields are valid solutions of the respective theories, \emph{i.e.} 
$[g, A, \phi]$ is a solution of the $\text{UV}$ theory and  $[g + \Delta g, A + \Delta A]$ is a solution to the four-derivative corrected theory. The $\text{UV}$
theory and that with an infinite series of higher-derivative corrections should have exactly the same partition function; therefore, this expression is an equality up to quantum corrections and corrections that are $\mathcal{O}(\epsilon^2)$. 
Finally, let us choose $[g, A, \phi]$ to be solutions of the $\text{UV}$ theory with charge $Q$ and temperature $T$, and 
$[g_0, A_0]$ to be field configurations in the pure Einstein-Maxwell theory \emph{with the same charge and temperature} as those of the UV theory. One then finds the following inequality,
\begin{align}
    I_C[g + \Delta g, A + \Delta A]_{T, Q} \simeq I_{\text{UV}}[g, A, \phi]_{T, Q} < I_{\text{UV}}[g_0, A_0, 0]_{T, Q} = I^{(2)}[g_0, A_0]_{T, Q} \, .
    \label{action_ineqality}
\end{align}
Since $[g, A, \phi]$ is a solution of the UV theory,  it extremizes the action. To ensure the inequality that appears in (\ref{action_ineqality}), one must further require that this solution is a \textit{minimum} of the action. The inequality then follows because $[g_0, A_0,0]$ is \emph{not} a solution to the equations of motion, for the same charge and temperature.
Finally, as long as one works in the same ensemble, the boundary terms will be the same for both actions and thus do not affect the argument.

In general, different theories will have different relationships between mass, charge, and temperature. We are interested in the entropy shift at fixed mass and charge. Therefore we must compare the two action functionals at different temperatures. For simplicity, we use $T_4/T_2$ for the temperature that corresponds to mass $M$ and charge $Q$ for the theory with/without higher-derivative corrections, respectively. Then we have:
\begin{align}
    \begin{split}
        F_C(Q, T_4) & < F_2(Q, T_4) ,\\
        F_C(Q, T_4) & < F_2(Q, T_2) + (T_4 - T_2) \partial_T F_2(Q, T_2), \\
        F_C(Q, T_4) & < F_2(Q, T_2) - (T_4 - T_2) S_2, \\
        M - S_4 T_4 & < M - S_2 T_2 - (T_4 - T_2) S_2 ,\\
         - S_4 T_4 & <  - T_4  S_2, \\
         \Delta S & > 0,
    \end{split}
\end{align}
at fixed $M$ and $Q$ (and in the zero Casimir energy scheme).

Now that we have outlined the argument in flat space, we can ask whether it can be immediately extended to AdS. 
One subtle point in the derivation outlined above is that  the free-energy is only finite after the subtraction of the free-energy of a reference background. In the flat space context, the contributions of such terms to the two actions are identical because the asymptotic charges are the same. Thus, this issue does not affect the validity of the argument.

In AdS, the story is a little different-- the free-energy is computed using holographic renormalization. Different counterterms are required to render the two-derivative action $I^{(2)}$ and the corrected action $I_C$ finite. 
Moreover, $I_{UV}$ may also require a different set of counterterms involving contributions from the scalar, and unlike the bulk contribution, there is no reason to expect that their on-shell values are less than their off-shell values. 
This is a potential hole in the positivity argument in AdS. Apart from this issue, the rest of the argument can be immediately applied to AdS.

\subsection{Thermodynamic Stability }

As we have seen, the above proof requires that the uncorrected backgrounds are minima of the action. Thermodynamically, this amounts to the condition that the black holes are stable under thermal and electrical fluctuations. This translates to the following requirements on the free-energies,
\begin{align}
        \left( \frac{ \partial^2 F}{\partial T^2} \right)_Q \leq 0, \qquad  \left( \frac{ \partial^2 G}{\partial T^2} \right)_{\Phi} \leq 0, \qquad \epsilon_T = \left( \frac{ \partial^2 F}{\partial Q^2} \right)_T \geq 0 \, .
\end{align}
These conditions may be rewritten in terms of the specific heat and permittivity of the black hole,
which can be used to determine, respectively, the 
thermal stability and electrical stability of the black hole \cite{Chamblin:1999tk, Chamblin:1999hg}. We will ignore the specific heat at constant $\Phi$ now, as we are interested in the stability in the canonical ensemble, and consider 
\begin{align}
    C_Q = T \left( \frac{ \partial S}{\partial T} \right)_Q \geq 0, \qquad \epsilon_T = \left( \frac{ \partial Q}{\partial \Phi} \right)_T \geq 0 \, .
\end{align}
Positivity of the specific heat is equivalent to the statement that larger black holes should heat up and radiate more, while smaller ones should become colder and radiate less.
When the quantity $\epsilon_T$ is negative the 
black hole is unstable to electrical fluctuations, meaning that when more charge is placed into it, its chemical potential \textit{decreases}. We expect that it should instead increase, to make it more difficult to move a charge from outside to inside the black hole -- thus making it harder to move away from equilibrium \cite{Chamblin:1999hg}. We may compute these quantities using the results of the previous section. For AdS${}_4$, we find
\begin{align}
    C_Q =\frac{2 \pi l^2 \nu^2 (1 + 3 \nu^2) (2 - \xi) \xi}{2 - 6 \xi + 3 \xi^2 + 3 \nu^2 (4 - 6 \xi + 3 \xi^2)}, \qquad \epsilon_T =  \frac{(\xi - 2 ) \xi + 3 \nu^2 (2 - 2 \xi + \xi^2)}{\nu l \left( 2 - 6 \xi + 3 \xi^2 + 3 \nu^2 (4 - 6 \xi + 3 \xi^2)\right)} \, ,
    \label{specific_heats}
\end{align}
where we recall that $\nu = r_h / l$ and $Q = (1 - \xi) Q_{\text{ext}}$. These results have been obtained previously \emph{e.g.} in \cite{Shen:2005nu}. We find that both of these coefficients are positive when either 
\begin{equation}
\label{firstcondition}
    \nu< \nu^* =  \frac{1}{\sqrt{3}} \, , \qquad
    \xi< \xi^* = 1 - \sqrt{\frac{1 - 3 \nu^2}{1 + 3 \nu^2}}\, , 
\end{equation}
holds, or when
\begin{equation}
  \nu > \nu^* =  \frac{1}{\sqrt{3}} \, , 
  \qquad 0<\xi <1    \, ,
  \label{secondcondition}
\end{equation}
is satisfied.

Thus, for small black holes stability requires that the extremality parameter be less than some function of the radius, $\xi < \xi^*$. 
In particular, extremal black holes, for which $\xi \rightarrow 0$, are stable while neutral black holes, which correspond to $\xi \rightarrow 1$, are not. 
The implication of (\ref{secondcondition}) is that above a certain radius ($r_h >  l / \sqrt{3}$) all black holes are thermodynamically stable. 
\begin{figure}
    \includegraphics[scale=0.6]{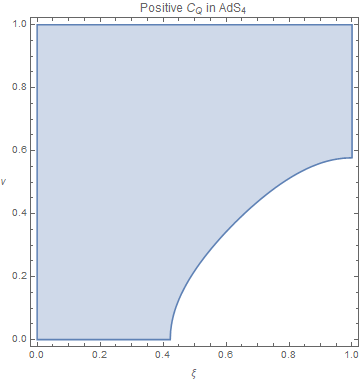}
    \includegraphics[scale=0.6]{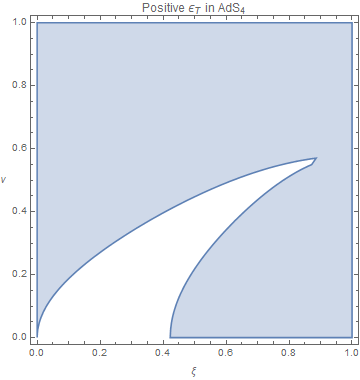}
    \caption{Blue represents the regions of parameter space where each quantity is positive.}
    \label{fig: stability} 
\end{figure}
This behavior is visible from Fig. \ref{fig: stability}, where we have plotted the allowed parameter space based on the $C_Q$ and $\epsilon_T$ conditions separately. This raises an interesting point in making contact with the flat space limit: if we require both parameters to be positive, there are no stable black holes at $\nu = 0$. Note that in \cite{Cheung:2018cwt} only $C_Q$ was considered. However, in applications involving AdS/CFT, we believe that both the specific heat and electrical permittivity should be taken into account.

Here we have only considered the leading-order stability. The higher-derivative corrections will shift the point where the specific heat crosses from positive to negative. However, in proving the extremality-entropy relation, we are only interested in the extremal limit, which is not affected by this consideration. In principal we could compute the order $\epsilon$ shifts to the stability conditions to obtain small corrections to the entropy bounds.

\subsection{Constraints on the EFT Coefficients}

 The entropy shift in AdS${}_4$ for a black hole with an \emph{arbitrary size and charge} takes the following form,
\begin{align}
\begin{split}
        & \left(\frac{\partial S}{\partial \epsilon} \right)_{Q, M} \, = \, \frac{l(1 + 3 \nu^2)}{5 \nu T} \Big(c_1 \left(   4 - 6 \xi + 19 \xi^2 - 16 \xi^3 + 4 \xi^4 +12 \nu^2 (\xi-1)^4  \right) \\
        & \ \  + c_2 (\xi-1)^2 \left(   2 - 14 \xi + 7 \xi^2 + 3 \nu^2 (12 - 14 \xi + 7 \xi^2)  \right) +  8 (2 c_3 + c_4) (1 + 3 \nu^2) (\xi -1)^4 \Big) \, ,
        \label{entshiftads4}
\end{split}
\end{align}
where the temperature is given by the expression
$$T(r_h, \xi)  = - \frac{(1 + 3 \nu^2) (\xi-2) \xi}{4 \pi \nu l} \, .$$
We can see from the $\xi$ dependence of the latter that in the $\xi \rightarrow 0$ limit the shift to the entropy blows up. If we examine the leading part in $1 / \xi$, we find that it is proportional to the mass shifts we have computed above. Thus, in the extremal limit we have
\begin{align}
     \left(\frac{\partial S}{\partial \epsilon} \right)_{\xi \rightarrow 0} \, = \,  \frac{l^2}{5 r_h T} \Big( 4 c_1 (1 + 3 \nu^2)^2 + 2 c_2 (1 + 3 \nu^2) (1 + 18 \nu^2) + 8 (2 c_3 + c4)  (1 +  3 \nu^2)^2  \Big) \, .
\end{align}
It is also interesting to note that in the chargeless limit $\xi \rightarrow 1$ the dependence of (\ref{entshiftads4}) on $c_2, c_3$ and $c_4$ drops out entirely, and  we are left with an entropy shift 
of the simple form
\begin{align}
     \left(\frac{\partial S}{\partial \epsilon} \right)_{\xi \rightarrow 1} \, = \,  \frac{l}{\nu T} c_1 \left( 1 + 3 \nu^2 \right) \,.
\end{align}
Our results above show that the large black holes are stable in the chargeless limit, which implies that the $c_1$ coefficient must be positive.

\begin{figure}
    \includegraphics[scale=0.6]{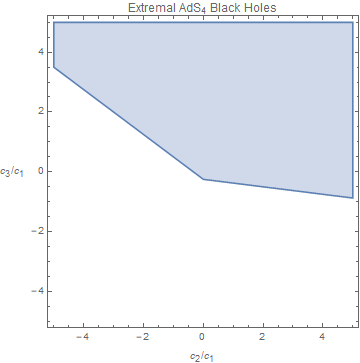}
    \includegraphics[scale=0.6]{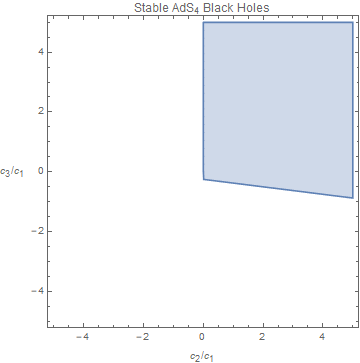}
    \caption{Blue regions are allowed after imposing that the entropy shift is positive. (Left): Allowed region after imposing that extremal black holes have positive entropy shift (Right):  Allowed region after imposing that all stable black holes have positive entropy shift}
    \label{AdS4_exclusion} 
\end{figure}

In Fig. \ref{AdS4_exclusion}, we have graphed the constraints on the coefficients that arise from demanding that the entropy shift is positive. We have included both the constraints from the extremal entropy shift and from considering the shift of all stable black holes. Considering only extremal black holes may be interesting because it is equivalent to the condition that the extremality shift, $\Delta (M / Q)$, is negative. Thus we may look at the constraints implied by positive entropy shift and by negative extremality shift independently. Note that we have divided by $c_1$, which we have already proven to be positive. We may write out the all the constraints obtained:
\begin{align}
    \begin{split}
        &c_1 \ \geq \ 0, \\
        &c_2 \ \geq \ 0, \\
        &c_3 \ \geq \ -\frac{1}{8}c_1 (2 + c_2).
    \end{split}
\end{align}

We have computed the corresponding bounds for AdS${}_5$ through AdS${}_7$. The results may be found in Appendix B. We would, however, like to comment on AdS${}_5$, where the positivity of the coefficient of the Riemann-squared term is of  particular interest. The stability analysis yields results that are qualitatively similar to (\ref{firstcondition}) and (\ref{secondcondition}), but with the following definitions
\begin{align}
    \xi^* = 1 - \sqrt{\frac{1 - 2 \nu^2}{1 + 2 \nu^2}}, \qquad \nu^* = \frac{1}{\sqrt{2}}\,.
\end{align}
Once again, we see that large black holes are stable for all values of the charge. 

When we examine the entropy shift in the neutral limit, we find
\begin{align}
    \frac{\pi l^2 }{32 T} c_1 \left( 87 + 164 \nu^2 + 52 \nu^4 \right) \, ,
\end{align}
whose overall sign is completely determined by that of $c_1$.
This means that there are stable black holes where the sign of the entropy shift is the same as the sign of the coefficient of $R_{abcd}^2$. Thus, a positive entropy shift for stable black holes implies that $c_1$ is positive. In fact, a positive value of $c_1$ was the necessary ingredient in \cite{Kats:2007mq} for obtaining the violation of the KSS bound\footnote{We have checked the calculation with a different basis, choosing to use Gauss-Bonnet instead of Riemann squared. As expected, we find that the coefficient of the Gauss-Bonnet term is positive.}. It is also interesting to note that in $d>3$, this sign constraint was shown to follow from the assumption of a unitary tree-level UV completion \cite{Cheung:2016wjt}. The entropy constraints given in this chapter are then strictly stronger since they also apply in $d=3$. 

In closing, we stress that we are not claiming that the entropy shift should be universally positive; the proof outlined above only applies when the higher-derivative corrections are generated by integrating out massive fields at tree-level (and relies on assuming that the corresponding solutions minimize the effective action). However, it is interesting that the conjecture that the entropy shift is universally positive appears to suggest that violations of the KSS bound are required to occur.
Our results extend and make more precise the earlier claim by some of us \cite{Cremonini:2009ih} of a link between the WGC and the violation of the KSS bound. We will come back to this point in the discussion section.

\subsection{Flat Space Limit}

\begin{figure}
    \includegraphics[scale=0.5]{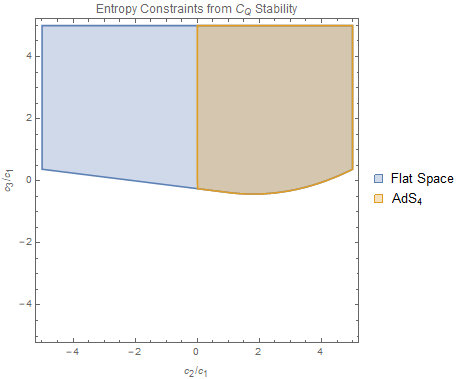}
    \includegraphics[scale=0.5]{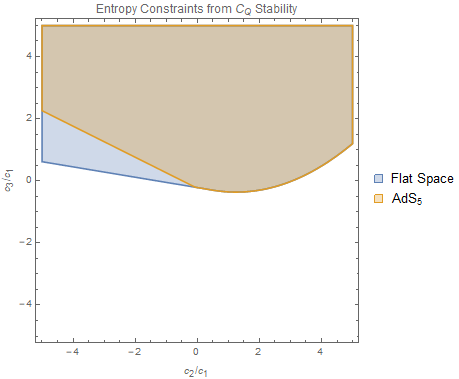}
    \caption{The blue regions are allowed in flat space and the orange in AdS-- note that the AdS regions are a subset of those from flat space.}
    \label{fig: flat space} 
\end{figure}

As we have pointed out above, we can not compare the results we have given above to the flat space limit. This is because if we impose both $C_Q > 0$ and $\epsilon_T> 0$, we find that there are no stable black holes in the flat space limit $\nu \rightarrow 0$ (as suggested by figure \ref{fig: stability}). In AdS/CFT, we expect that both conditions are necessary to ensure thermodynamic stability; nonethless, we may remove the condition $\epsilon_T> 0$ in order to compare with the flat space limit. In this case, we find that stability requires
\begin{align}
    \xi^* = 1 - \frac{1}{\sqrt{3}} \sqrt{\frac{1 - 3 \nu^2}{ 1 + 3 \nu^2}}, \qquad \nu^* = \frac{1}{\sqrt{3}},
\end{align}
for the AdS${}_4$ black holes, and 
\begin{align}
    \xi^* = 1 - \frac{1}{\sqrt{2}} \sqrt{\frac{1 - 2 \nu^2}{ 1 + 2 \nu^2}}, \qquad \nu^* = \frac{1}{\sqrt{2}},
\end{align}
for the AdS${}_5$ black holes. This allows for a more direct comparison between the two cases. In figure \ref{fig: flat space}, we contrast the bounds obtained in AdS and flat space. The bounds in AdS are stronger, as they should be given that there is an extra parameter's worth of stable black holes. Note also that $c_1> 0$ is implied by positivity in AdS, but not in flat space, because in flat space there are no stable neutral black holes.  

\section{Remarks}

We conclude this chapter with a few remarks on the results of the previous section.

\subsection{$c - a$ from the Entropy Shift}

As we have seen, for neutral black holes, the entropy shift is dominated by $c_1$, which is the coefficient of the Riemann squared term, so the positivity of the entropy shift implies the positivity of this coefficient. In AdS${}_5$, this coefficient may be related to the central charges of the dual field theory \cite{Henningson:1998gx, Nojiri:1999mh, Blau:1999vz} by 
\begin{align}
    c_1 = \frac{1}{8} \frac{c - a}{c} \, .
\end{align}
Thus, the positivity of the entropy shift appears to be violated in theories where $c - a < 0$. In \cite{Buchel:2008vz}, a number of superconformal field theories were examined, and all were found to satisfy $c - a > 0$. It is worth noting there are non-interacting theories where $c - a< 0$; for example, $\frac{a}{c} = \frac{31}{18}$ for a free theory of only vector fields \cite{Hofman:2008ar}. However, such theories do not have weakly curved gravity duals. If there are any bulk theories where $c_1 < 0$, we are not aware of them.
The question of whether holographic theories necessarily correspond to $c-a$ non-negative is interesting for a number of reasons -- both from a fundamental point of view and for phenomenological applications.

In particular, recall that the range of the Wilson coefficients and the sign of $c-a$ played an important role in the physics of the shear viscosity to entropy ratio $\eta/s$ and how it deviates from its universal    $1/4\pi$ result \cite{Policastro:2001yc,Buchel:2003tz}, as discussed extensively in the literature (see \cite{Cremonini:2011iq} for a review of the status of the shear viscosity to entropy bound).
Indeed, it is  interesting to compare our results to the higher-derivative corrections to $\eta/s$, which (for the $\text{AdS}_5$ case of interest to us here) were shown \cite{Myers:2009ij} to be given by 
\begin{align}
    \frac{\eta}{s} = \frac{1}{4 \pi} \left( 1 - 8 c_1 + 4 (c_1 + c_2) \frac{q^2}{r_0^6} \right) \, ,
\end{align}
where $r_0$ is a parameter of the solution defined in \cite{Myers:2009ij}; the factor ${q^2}/{r_0^6}$ goes from 0 (for neutral black holes) to 2 (at extremality). Our bounds on $c_1$ imply that neutral black holes will 
necessarily have a negative viscosity shift, violating the KSS bound. Models where this is realized are known to exist---the first UV complete counter-example to the KSS bound was given in \cite{Kats:2006xp}. 
For extremal black holes, the dependence on $c_1$ drops out and only the sign of $c_2$ matters, ${\eta}/{s} = \frac{1}{4 \pi} \left( 1 + 8 c_2\right)$. For AdS${}_5$, the $c_2$ coefficient may have both positive and negative values. However, imposing the null energy condition implies an additional constraint on the range of $c_2$, which in $\text{AdS}_5$ takes the form
\begin{align}
    \frac{13}{12} \, c_1 + c_2>0 \, .
    \label{eq:NECbound}
\end{align}
This may be seen by first noticing that the definition of the parameter $\gamma$ in equation (\ref{eq:gamma}) implies $\gamma > 0$ as long as the null energy condition holds. Then the bound in (\ref{eq:NECbound}) may be derived from the specific form of $\gamma$ given in (\ref{eq: gammaVal}). This alone is sufficient to bound $c_2$ from below, when $c_1$ is non-negative. 
Thus, one can see that utilizing such constraints it is at least in principle possible to bound $\eta/s$ \emph{from below}, in specific cases. 
To what extent this can be done generically is still an open question.

It might be interesting to try to relate the extremality bounds to the  transport coefficients of the boundary theory in a more concrete way. As the corrections to $\eta / s$ depend only on $c_1$ and $c_2$ in five dimensions, it is clear that the shift to extremality is not captured by the physics that controls $\eta / s$ alone. One might wonder, however, if some other linear combination of transport coefficients, such as the conductivity or susceptibility\footnote{These have been considered in \cite{Kovtun:2008kx}, which already in 2008 had an interesting comment about a possible relation to the WGC.}, might be related to the extremality shift. From a purely CFT point of view, this is certainly not that strange; the philosophy of conformal hydrodynamics is that scaling symmetry ties together ultraviolet quantities ($a, c$) that characterize the CFT to the transport coefficients, which characterize the IR, long-wavelength behavior of the theory. If we believe that EFT coefficients in the bulk are related to these UV quantities (as is known in the case of $c_1$), then a correspondence between higher-derivatives and hydrodynamics is very natural. 
The question is to what extent this can be used to efficiently constrain IR quantities. 
Finally, we should note that extending our analysis to holographic theories that couple gravity to scalars would be useful to make contact with the efforts to generate non-trivial temperature dependence for $\eta/s$ (see \emph{e.g.} the discussion in \cite{Cremonini:2011ej, Cremonini:2012ny}), which is expected to 
play a key role in understanding the dynamics of the strongly coupled quark gluon plasma. 

Our results also have potential to make contact with the work on CFTs at large global charge \cite{Hellerman:2015nra}. As we have seen above, the extremality curve for AdS-RN black holes is non-linear even at the two-derivative level. In an analysis of the minimum scaling dimension for highly charged 3D CFTs states of a given charge, it was found \cite{Loukas:2018zjh} that $\Delta \sim q^{3/2}$. This is in striking agreement with the extremality relationship $m \sim q^{3/2}$ that holds for large black holes. The large charge OPE may be powerful because it offers an expansion parameter, $1 / q$, which may be used even for CFTs which are strongly coupled. In principle, it should be possible to match our higher-derivative corrections to the extremality bound with corrections to the minimum scaling dimension that are subleading in $1 / q$. This might allow one to use the large charge OPE to compute the EFT coefficients of the bulk dual of specific theories where the minimum scaling dimensions are known.

\subsection{Weak Gravity Conjecture in AdS}

One of the motivations for this work is to address to question of to what extent the WGC is constraining in AdS space.
It is not obvious that it should be. In flat space, one looks for higher-derivative corrections to shift the extremality bound $m(q)$ to have a slope that is greater than one. In that case, a single nearly extremal black holes is (kinematically) allowed to decay to two smaller black holes, which can fly apart off to infinity and decay further if they wish.

In AdS, the extremality bound $m(q)$ has a slope that is greater than one at the two-derivative level. Therefore one might expect that large black holes are already able to decay without any new particles or higher-derivative corrections. This picture may be too naive, however; the AdS radius introduces a long range potential that is proportional to $\frac{r^2}{l^2}$. This causes all massive states emitted from the black hole to fall back in, contrary to the situation in flat space. 

A different decay path is provide by the dynamical instability \cite{Gubser:2008px, Hartnoll:2008kx, Hartnoll:2008vx, Denef:2009tp}, whereby charged black branes are unstable to formation of a scalar condensate. This occurs only if the theory also has a scalar with charge $q$ and dimension $\Delta$ that satisfies
\begin{align}
    (m_{\phi} l)^2 \leq \frac{1}{2}(q_{\phi}g M_{Pl} l)^2 - \frac{3}{2} \,.
\end{align}
Note that, even in the limit of large AdS-radius $l$, this does not approach the bound we have for small black holes, which is $m \leq q$. 
Numerical work in \cite{Hartnoll:2008kx} suggests that the endpoint of the instability is a state where all the charge is carried by the scalar condensate. 
Similar requirements appear for the superradiant instability of small black holes \cite{Bhattacharyya:2010yg, Dias:2010ma}. For a more thorough review, see \cite{Nakayama:2015hga}. In either case, it is curious that in AdS, a condition similar to the flat space WGC allows for black holes to decay through an entirely different mechanism.

Another remarkable hint of the WGC comes from its connection to cosmic censorship. In  \cite{Crisford:2017gsb, Horowitz:2019eum}, it is shown that a class of solutions of Einstein-Maxwell theory in AdS${}_4$ that appear to violate cosmic censorship \cite{Horowitz:2016ezu} are removed if the theory is modified to include a scalar whose charge is great enough to satisfy the weak gravity bound\footnote{The bound they consider is the bound for superradiance of small black holes, which requires $\Delta \leq q l$.}.

It may be possible to study these solutions in the presence of higher-derivative corrections. One might ask whether there is a choice of higher-derivative terms such that the singular solutions are removed. It would be interesting to check if this occurs when the higher-derivative terms are those that are obtained by integrating out a scalar of sufficient charge. It would also be interesting to compare constraints obtained by requiring cosmic censorship with constraints due to positivity of the entropy shift.

A more general proof of the WGC in AdS was given in \cite{Montero:2018fns}. In that paper, it was shown that, under mild assumptions, entanglement entropy for the boundary dual of an extremal black brane should go like the surface area of the entangling subregion, which is in tension with the volume law scaling predicted by the Ryu-Takayanagi formula. The contradiction is removed when one introduces a WGC-satisfying state. This violates one of the assumptions that imply the area law for the entropy-- that is, the assumption that correlations decay exponentially with distance.

This form of the WGC in particularly interesting to us because it makes no reference to whether or not the WGC-satisfying state is a particle, or a non-perturbative object like a black hole. Therefore, the contradiction pointed out in that paper may be lifted if the higher-derivative corrections allow for black holes with charge greater than mass. Heavy black holes in AdS have masses far greater than their charge-- therefore we expect that the WGC-satisfying states might be provided by small black holes whose higher-derivative corrections shift the extremality bound to allow slightly more charge.


%% file: Appendices/App1.tex
The Seeley-DeWitt coefficients $b_n(\Delta)$ depend on the field and the form of the second order operator $\Delta$.  In four dimensions, the appropriate operators for irreducible fields up to spin two are listed in \cite{Christensen:1978md}. Here we write down the analogous operators in six dimensions and compute the contribution of each to the anomaly.

We start with the basis of curvature invariants \cite{parker1987,Bastianelli:2000hi}
\begin{align}
& A_1 = \Box^2 R, \quad 
A_2 = \left(\nabla_a R \right)^2, \quad A_3 = \left(\nabla_a R_{m n} \right)^2, \quad 
A_4 = \nabla_a R_{b m} \nabla^b R^{a m}, \quad 
A_5 = \left(\nabla_a R_{m n i j} \right)^2,
\nn\\
& A_6 =  R \Box R, \quad
A_7 = R_{a b}\Box R^{a b}, \quad
A_8 = R_{a b} \nabla_m \nabla^b R^{a m}, \quad
A_9 = R_{a b m n}\Box R^{a b m n}, \quad
A_{10} = R^3
\nn\\
& A_{11} = R R_{ab}^2, \quad
A_{12} = R R_{abmn}^2, \quad
A_{13} = R_a^{\ m} R_m^{\ \, i} R_i^{\ a}, \quad
A_{14} = R_{ab} R_{mn} R^{ambn}, \quad
\nn\\
& A_{15}= R_{ab} R^{amnl} R^b_{\ \,mnl}, \quad
A_{16} = R_{ab}{}^{cd}R_{cd}{}^{ef}R_{ef}{}^{ab},\quad 
A_{17} = R_{aibj}R^{manb}R^i{}_m{}^j{}_n.
\end{align}
The $b_6$ coefficient may be computed from the expression (\ref{eq:b6coef}), where the $V_a$'s are given by
\begin{align}
    & 
    V_1 = \nabla_k F_{ij} \nabla^k F^{ij}, \quad
    V_2 = \nabla_j F_{ij} \nabla^k F^{ik}, \quad
    V_3 =  F_{ij} \Box F^{ij}, \quad
    V_4 = F_{ij} F^{jk} F_{k}{}^i, \quad
    \nn\\
    &
    V_5 = R_{mnij} F^{mn} F^{ij}, \quad
    V_6 = R_{jk} F^{jn} F^k{}_n, \quad
    V_7 = R F_{ij} F^{ij}, \quad
    V_8 = \Box^2 E, \quad
    V_9 = E \Box E, \quad
    \nn\\
    &
    V_{10} = \nabla_k E \nabla^k E, \quad
    V_{11} = E^3, \quad
    V_{12} = E F_{ij}^2, \quad
    V_{13} = R \Box E, \quad
    V_{14} = R_{i j} \nabla^i \nabla^j E, \quad
    \nn\\
    &
    V_{15} = \nabla_k R \nabla^k E, \quad
    V_{16} = E^2 R, \quad
    V_{17} = E \Box R, \quad
    V_{18} = E R^2, \quad
    \nn\\
    &
    V_{19} = E R_{ij}^2, \quad
    V_{20} = E R_{ijkl}^2.
\end{align}
Here $\Delta=-\nabla^2-E$ and $F_{ij}$ is the curvature of the connection, $[\nabla_i,\nabla_i]=F_{ij}$. Below we present the $V$-terms for each field after tracing over the representation.

\section{Conformally Coupled Scalar}

The conformally coupled scalar has $E = -\frac{1}{5} R$ and $F_{ij} = 0$, so the $V$-terms are:
\begin{center}
$\begin{array}{| c | c | c | c | c | c | c | c | c | c |}
    \hline
   V_1 & V_2 & V_3 & V_4 & V_5 & V_6 & V_7 & V_8 & V_9 & V_{10}\\
   \hline
   0 & 0 & 0 & 0 & 0 & 0 & 0 &  -\frac{1}{5} A_1 & \frac{1}{25} A_6 & \frac{1}{25} A_2 \\
   \hline
   \hline
   V_{11} & V_{12} & V_{13} & V_{14} & V_{15} & V_{16} & V_{17} & V_{18} & V_{19} & V_{20} \\
   \hline 
    -\frac{1}{125} A_{10} & 0 & -\frac{1}{5} A_6 & \frac{2}{5} (-A_8 + A_{13} - A_{14}) & -\frac{1}{5} A_2 & \frac{1}{25} A_{10} & -\frac{1}{5} A_6 & -\frac{1}{5} A_{10} & -\frac{1}{5} A_{11} & -\frac{1}{5} A_{12}  \\
   \hline
\end{array}$
\end{center}
The $b_6$ coefficient is
\begin{align}
b_6(\mathcal{O}) &= \frac{1}{(4 \pi)^3 7!} \bigg[ \frac{6}{5} A_1 + \frac{1}{5} A_2 - 2 A_3 -4 A_4 + 9 A_5 - 8  A_7 + \frac{8}{5} A_8 + 12 A_9 \nn\\
    &\quad-\frac{7}{225} A_{10} + \frac{14}{15} A_{11} - \frac{14}{15} A_{12} - \frac{32}{45} A_{13} -\frac{16}{15} A_{14} -\frac{16}{3} A_{15} + \frac{44}{9} A_{16} +\frac{80}{9} A_{17} \bigg].
\end{align}

\section{Weyl Fermion}
The appropriate second order operator for the Dirac fermion may be obtained as the square of the Dirac operator:
\begin{align}
    \mathcal{O} \psi = -\Box \psi + \frac{1}{4} R  \psi.
\end{align}
The endomorphism and curvature of the connection coincide with the result obtained in \cite{Bastianelli:2000hi}. 
\begin{align}
    E = -\frac{1}{4} R , \qquad F_{i j} = \frac{1}{4} R_{i j a b} \gamma^{a  b} .
\end{align}
Then the $V$-terms contributing to the anomaly are (after tracing):
\begin{center}
$\begin{array}{| c | c | c | c | c | c | c | c | c | c |}
    \hline
   V_1 & V_2 & V_3 & V_4 & V_5 & V_6 & V_7 & V_8 & V_9 & V_{10}\\
   \hline
   -\frac{1}{2} A_5 & A_4 - A_3 & -\frac{1}{2} A_9 & \frac{1}{2} A_{17} & -\frac{1}{2} A_{16} & -\frac{1}{2} A_{15} & -\frac{1}{2} A_{12} & -A_1 & \frac{1}{4} A_6 & \frac{1}{4} A_2 \\
   \hline
    \hline
   V_{11} & V_{12} & V_{13} & V_{14} & V_{15} & V_{16} & V_{17} & V_{18} & V_{19} & V_{20} \\
   \hline 
   -\frac{1}{16} A_{10} & \frac{1}{8} A_{12} & -A_6 & -2(A_8 -A_{13} + A_{14}) & -A_2 & \frac{1}{4} A_{10} & -A_6 & -A_{10} & -A_{11} & -A_{12} \\
   \hline
\end{array}$
\end{center}
The $b_6$ coefficient is
\begin{align}
b_6(\mathcal{O}) &= \frac{4}{(4 \pi)^3 7!} \bigg[ -3 A_1 + \frac{5}{4} A_2 - 9A_3 +3 A_4 -5 A_5 + \frac{7}{2} A_6 - 8 A_7 -4 A_8 -9 A_9 \nn\\
    &\quad- \frac{35}{72} A_{10} +\frac{7}{3} A_{11} + \frac{49}{24} A_{12} + \frac{44}{9} A_{13} -\frac{20}{3} A_{14} +\frac{5}{3} A_{15} - \frac{101}{18} A_{16} - \frac{109}{9} A_{17} \bigg].
\label{eq:b6-100}
\end{align}

\section{Vector}
The $(0, 1, 0)$ vector representation of $SU(4)$ is a one form, so the correct Laplacian may be obtained by computing the Hodge-deRham operator $d \delta + \delta d$. We get 
\begin{align}
    \mathcal{O} A_{\mu} = -\Box A_{\mu} + R^{\nu}_{\ \, \mu} A_{\nu}.
\end{align}
The endomorphism and curvature of the connection here are: 
\begin{align}
    E^{a}{}_{b} = -R^{a}{}_{b}, \qquad (F_{i j})^{a}{}_{b} = R^{a}{}_{b i j},
\end{align}
so that (after tracing)
\begin{center}
$\begin{array}{| c | c | c | c | c | c | c | c | c | c |}
    \hline
   V_1 & V_2 & V_3 & V_4 & V_5 & V_6 & V_7 & V_8 & V_9 & V_{10}\\
   \hline
   - A_5 & 2(A_4 - A_3) & - A_9 & A_{17} & - A_{16} & - A_{15} & - A_{12} & -A_1 & A_7 &  A_3 \\
   \hline
    \hline
   V_{11} & V_{12} & V_{13} & V_{14} & V_{15} & V_{16} & V_{17} & V_{18} & V_{19} & V_{20} \\
   \hline 
   - A_{13} & A_{15} & -A_6 & -2(A_8 -A_{13} + A_{14}) & -A_2 & A_{11} & -A_6 & -A_{10} & -A_{11} & -A_{12} \\
   \hline
\end{array}$
\end{center}
and
\begin{align}
b_6(\mathcal{O}) &= \frac{1}{(4 \pi)^3 7!} \bigg[ 24 A_1 - 66 A_2 + 352 9A_3 + 32 A_4 -58 A_5 -140 A_6 + 792 A_7 - 32 A_8 - 96 A_9 \nn\\
    &\quad- \frac{140}{3} A_{10} + 420 A_{11} -70 A_{12} - \frac{2600}{3} A_{13} + 16 A_{14} + 444 A_{15} - \frac{164}{3} A_{16} - \frac{344}{3} A_{17} \bigg].
\end{align}

\section{Self-Dual Three-Form}

The field which transforms under the $(2, 0, 0)$ representation is the 10-component self-dual three-form. A three-index antisymmetric tensor has 20 components and the self-duality condition removes half of these. The operator acting on this field is
\begin{align}
    \mathcal{O} C_{\mu \nu \rho} &= -\Box C_{\mu \nu \rho} + R_{\mu}{}^{\lambda}  C_{\lambda \nu \rho} + R_{\nu}{}^{\lambda}  C_{\mu \lambda \rho} + R_{\rho}{}^{\lambda}  C_{\mu \nu \lambda} \nn\\
    &\quad- R_{\mu \nu}{}^{\lambda \sigma} C_{\lambda \sigma \rho} - R_{\nu \rho}{}^{\lambda \sigma} C_{\mu \lambda \sigma}- R_{\rho \mu}{}^{\lambda \sigma} C_{ \lambda \nu \sigma}.
\end{align}
This means that the endomorphism and connection curvature are given by 
\begin{align}
    E^{\ \ \ def}_{abc} = -3 R^{\ [d}_{[a}\delta^{e}_{b} \, \delta^{f]}_{c]} + 3 R_{[a b}{}^{[d e} \, \delta^{f]}_{c]} ,
    \qquad (F_{i j})_{abc}{}^{def} = -3 R_{i j [a}{}^{[d}\delta^{e}_{b} \delta^{f]}_{c]}.
\end{align}
Then we can compute the relevant terms:
\begin{center}
$\begin{array}{| c | c | c | c | c |}
    \hline
    V_1 & V_2 & V_3 & V_4 & V_5 \\ 
    \hline 
    -6 A_5 & 12(A_4 - A_3) & -6 A_9 & 6 A_{17} & -6 A_{16} \\ 
    \hline
    \hline
    V_6 & V_7 & V_8 & V_9 & V_{10}\\
    \hline
    -6 A_{15} & -6 A_{12} & -6 A_1 & 2 A_6 -2 A_7 + 2 A_9 & 2 A_2 -2 A_3 + 2 A_5\\
    \hline
    \hline
    V_{11} & V_{12} & V_{13} & V_{14} & V_{15}\\ 
    \hline 
    * & 2 A_{12} -2 A_{15} + 2 A_{16} & - 6 A_6 & -12 (A_8 -A_{13} + A_{14}) & - 6 A_2\\ 
    \hline
    \hline
    V_{16} & V_{17} & V_{18} & V_{19} & V_{20}\\
    \hline
    2 (A_{10} -A_{11} + A_{12}) & - 6 A_6 & - 6 A_{10} & - 6 A_{11} & - 6 A_{12}\\ 
    \hline
\end{array}$
\end{center}
where $V_{11}=- A_{10} +6 A_{11} -3 A_{12} -6A_{13} - 12A_{14} + 12 A_{15} - 2 A_{16} + 8 A_{17}$.  (Again, all $V$-terms are given after tracing over the representation.)  So the $b_6$ coefficient is given by
\begin{align}
b_6(\mathcal{O}) = \frac{1}{(4 \pi)^3 7!} \bigg[& -144 A_1 +172 A_2 -1216 A_3 + 256 A_4 + 348 A_5 + 392 A_6 - 1840 A_7 \nn\\
&- 192 A_8 + 912 A_9 
    + \frac{3080}{9} A_{10} +\frac{12824}{3} A_{11} - \frac{4004}{3} A_{12} - \frac{43472}{9} A_{13} \nn\\ &-\frac{30976}{3} A_{14} +\frac{28408}{3} A_{15} - \frac{11216}{9} A_{16} + \frac{53008}{9} A_{17} \bigg].
\end{align}
The self-duality condition reduces each of these terms by a factor of two, reproducing the $A_{16}$ and $A_{17}$ terms found in table~\ref{tbl:b6coef}.

\section{Gravitino}
The gravitino with the gauge condition $\gamma^{\mu} \psi_{\mu}=0$ corresponds to the $(1,1,0)$ representation. In this case the operator $\mathcal{O}$ is the square of the Rarita-Schwinger operator: 
\begin{align}
    \mathcal{O} \psi_{\mu} = -\Box \psi_{\mu} + \frac{1}{4} R  \psi_{\mu} - \frac{1}{2}\gamma^{\rho}\gamma^{\sigma} R_{\rho \sigma \mu \nu}  \psi^{\nu}.
\end{align}
The endomorphism and connection curvature are given by 
\begin{align}
    E_{b}{}^{a} = -\frac{1}{4} R \delta^{a}_{b} + \frac{1}{2} R_{c d b}{}^{a} \gamma^{c d},\qquad (F_{i j})^{\ \ a}_{b} = \frac{1}{4} R_{i j c d} \gamma^{c d} \delta^{a}_{b} +  R_{i j b}{}^{a} ,
\end{align}
so (after tracing)
\begin{center}
$\begin{array}{| c | c | c | c | c |}
    \hline
   V_1 & V_2 & V_3 & V_4 & V_5\\ 
   \hline 
   -7 A_{15} & -7 A_{12} & -6 A_1 & \frac{3}{2} A_6 + 2A_9 & \frac{3}{2} A_2 + 2A_5 \\ 
   \hline
   \hline
   V_6 & V_7 & V_8 & V_9 & V_{10}\\
   \hline
   -7 A_5 & 14(A_4 - A_3) & -7 A_9 & 7 A_{17} & -7 A_{16}\\
   \hline
   \hline
   V_{11} & V_{12} & V_{13} & V_{14} & V_{15} \\ 
   \hline 
   -\frac{3}{8} A_{10} - \frac{3}{2} A_{12} + 4A_{17} & \frac{7}{4} A_{12} + 2 A_{16} & - 6 A_6 & -12 (A_8 -A_{13} + A_{14}) & - 6 A_2\\ 
   \hline\hline
   V_{16} & V_{17} & V_{18} & V_{19} & V_{20} \\
   \hline 
   \frac{3}{2} A_{10} + 2A_{12} & - 6 A_6 & - 6 A_{10} & - 6 A_{11} & - 6 A_{12} \\
   \hline
\end{array}$
\end{center}
and
\begin{align}
b_6(\mathcal{O}) = \frac{1}{(4 \pi)^3 7!} \bigg[ & -60 A_1 + 25 A_2 - 404 A_3 + 284 A_4 + 292 A_5 + 70 A_6 - 160 A_7 \nn\\ 
    &- 80 A_8 +828 A_9 
    - \frac{175}{18} A_{10} +\frac{140}{3} A_{11} - \frac{1435}{6} A_{12} - \frac{880}{9} A_{13} \nn\\ & -\frac{400}{3} A_{14} +\frac{772}{3} A_{15} + \frac{3526}{9} A_{16} + \frac{22012}{9} A_{17} \bigg].
\label{eq:b6-110}
\end{align}

\section{Two-Form}
The adjoint representation $(1, 0, 1)$ corresponds to the two-form computed in \cite{Bastianelli:2000hi}. 
\begin{align}
    \mathcal{O} B_{\mu \nu} = -\Box B_{\mu \nu} + R^{\lambda}_{\mu}  B_{\lambda \nu} - R^{\lambda}_{\nu}  B_{\lambda \mu} - R^{\ \ \ \ \rho \sigma}_{\mu \nu} B_{\rho \sigma} .
\end{align}
This means that the endomorphism and connection curvature are given by 
\begin{align}
    E_{ab}{}^{cd} = -2 R^{[c}_{[a}\delta^{d]}_{b]} + R_{ab}{}^{cd} , \qquad (F_{i j})_{ab}{}^{cd} = 2R_{ij[a}{}^{[c} \delta^{d]}_{b]},
\end{align}
so (after tracicng)
\begin{center}
$\begin{array}{| c | c | c | c | c |}
    \hline
    V_1 & V_2 & V_3 & V_4 & V_5 \\
    \hline
    -4 A_5 & 8(A_4 - A_3) & -4 A_9 & 4 A_{17} & -4 A_{16} \\
    \hline
    \hline
    V_6 & V_7 & V_8 & V_9 & V_{10}\\
    \hline
    -4 A_{15} & -4 A_{12} & -4 A_1 & A_6 + A_9 & A_2 + A_5 \\
    \hline
    \hline
    V_{11} & V_{12} & V_{13} & V_{14} & V_{15}\\
    \hline 
    -3 A_{11} + 4 A_{13} + 6 A_{14} -6 A_{15} + A_{16} & A_{12} + A_{16} & - 4 A_6 & -8 (A_8 -A_{13} + A_{14}) & - 4 A_2 \\
    \hline 
    \hline
    V_{16} & V_{17} & V_{18} & V_{19} & V_{20} \\
   \hline A_{10} + A_{12} & - 4 A_6 & - 4 A_{10} & - 4 A_{11} & - 4 A_{12} \\
   \hline
\end{array}$
\end{center}
and
\begin{align}
b_6(\mathcal{O}) &= \frac{1}{(4 \pi)^3 7!} \bigg[ -66 A_1 + 3 A_2 - 254 A_3 + 164 A_4 + 107 A_5 + 28 A_6 - 120 A_7 - 88 A_8 + 348 A_9 \nn\\
    &\kern-1em+ \frac{595}{3} A_{10} -2478 A_{11} + 518 A_{12} + \frac{10384}{3} A_{13} -4912 A_{14} - 4896 A_{15} + \frac{2992}{3} A_{16} - \frac{1616}{3} A_{17} \bigg].
\end{align}

\section{Graviton}

The symmetric spin-two field is the $(0, 2, 0)$ representation. The appropriate kinetic operator is the Lichnerowicz operator \cite{Lich}:
\begin{align}
    \mathcal{O} h_{\mu \nu} = -\Box h_{\mu \nu} + R_{\mu}{}^{\lambda} h_{\lambda \nu} + R_{\nu}{}^{\lambda} h_{\lambda \mu} - 2 R_{\mu \rho \nu \sigma} h^{\rho \sigma}.
\end{align}
The endomorphism and connection are given by
\begin{align}
    E_{\mu \nu}^{ \rho \sigma} = -2 R_{ \{ \mu}^{\ \{ \rho} \delta_{\nu \} }^{\ \sigma \} } + R_{\mu \ \ \nu}^{\ \ \rho \ \ \sigma} + R_{\mu \ \ \nu}^{\ \ \sigma \ \ \rho}, \qquad (F_{a b})_{\mu \nu}^{ \rho \sigma} = 2 R_{a b \{ \mu}^{\ \ \ \ \{ \rho} \delta_{\nu \}}^{\ \sigma \}}.
\end{align}
Then we can compute the relevant terms:
\begin{center}
$\begin{array}{| c | c | c | c | c |}
        \hline
   V_1 & V_2 & V_3 & V_4 & V_5 \\
   \hline
   -8 A_5 & 16(A_4 - A_3) & -8 A_9 & 8 A_{17} & -8 A_{16} \\
   \hline
   \hline
   V_6 & V_7 & V_8 & V_9 & V_{10}\\
   \hline
   -8 A_{15} & -8 A_{12} & -8 A_1 & A_6 + 12 A_7 + 3 A_9 & A_2 + 12 A_3 + 3 A_5 \\
   \hline
   \hline
   V_{11} & V_{12} & V_{13} & V_{14} & V_{15} \\
    \hline 
   * &  A_{12} + 12 A_{15} +3 A_{16}  & - 8 A_6  & -16 (A_8 -A_{13} + A_{14})  & - 8 A_2  \\
   \hline
   \hline
   V_{16} & V_{17} & V_{18} & V_{19} & V_{20} \\
   \hline
   A_{10} +12 A_{11} + 3A_{12}  & - 8 A_6  & - 8 A_{10}  & - 8 A_{11}  & - 8 A_{12} \\
   \hline
\end{array}$
\end{center}
where $V_{11}=-3A_{11} -16 A_{13} -6A_{14}  -18 A_{15} -A_{16} + 8A_{17}$, and these terms are given after tracing over the representation. The $b_6$ coefficient is
\begin{align}
b_6(\mathcal{O}) = \frac{1}{(4 \pi)^3 7!} \bigg[ & -312 A_1 -584 A_2 - 4552 A_3 + 368 A_4 + 544 A_5 - 1064 A_6 + 9920 A_7 \nn\\ &- 416 A_8 + 1416  A_9 
    - \frac{560}{9} A_{10} +\frac{7952}{3} A_{11} + \frac{2968}{3} A_{12} - \frac{117056}{9} A_{13} \nn\\ &-\frac{16528}{3} A_{14} -\frac{29216}{3} A_{15} - \frac{1388}{9} A_{16} + \frac{49984}{9} A_{17} \bigg].
\end{align}


%% file: Appendices/App2.tex
We are interested in a general formula to compute the heat kernel coefficients for spins higher than two, analogous to the algorithm \cite{Christensen:1978md} in four dimensions. We consider fields transforming in an irreducible representation of the spacetime symmetry group that are acted on by a generalized second-order operator $\Delta = -\Box - E$. In four dimensions, the method of computing the heat kernels for general representations assumes that the endomorphism term $E$ for fields transforming as $(A,B)$ of $SO(4)\simeq SU(2)_L\times SU(2)_R$ is given by:
\begin{align}
    E = \Sigma_{ab}R^{abcd}\Sigma_{cd}\qquad\mbox{or} \qquad E = \frac{1}{A}\Sigma_{ab}R^{abcd}_{+}\Sigma_{cd},
\end{align}
for bosonic ($A + B =$ integer) or fermionic ($A + B =$ half-integer, $A>B$) representations, respectively. Here $R^{abcd}_{+} = \frac{1}{2}(R^{abcd} + R^{* \ abcd})$. This prescription is shown to be valid for fields up to spin two in four dimensions, and is conjectured to be the appropriate operator for general spins. In six dimensions, it appears that this prescription is reasonable for bosonic representations, but straightforward generalizations for fermions fail to reproduce the conventional endomorphism terms for the Weyl fermion and gravitino. So it remains unclear what endomorphism term is appropriate for general fermions. Below we use this method for bosonic representations to compute all the $V$ terms, which are built out of the endomorphism $E$ and the connection $F_{ij}$. 
\section{Tracing Over Generators}

Computing the heat kernel using this method requires computing the trace of a number of generators; the most we will need is six, as $E^3 \sim \Sigma^6$. We perform these traces using the algorithm presented in \cite{GroupTheory}, which requires expanding the trace into a sum of symmetric traces, and then writing each symmetric trace in a basis of orthogonal tensors and higher order Dynkin indices. For example, the trace of two generators of an irreducible representation is 
\begin{align}
    \text{Tr}[T_R^A T_R^B] = I_2(R) g^{AB}.
\end{align}
Here $R$ refers to the representation, and the capital Roman letters $A, B, \ldots=1,2,\ldots,15$ label the generators of $SU(4)$. Each $SU(4)$ index is interchangeable with a pair of antisymmetrized six-dimensional spacetime indices $\{\mu, \nu\}$.

If the number of generators is greater than two, we will first need to break the trace into a sum of symmetrized traces. For a trace of $n$ generators, this is accomplished by writing out each of the $n!$ terms in the symmetrized trace, and then using commutation relations to return each term to the original order, plus a number of traces of lower numbers of generators. For example, we may look at the trace of six generators. First consider the symmetrized trace
\begin{align}
    \text{STr} & [ T_{A}  T_{B} T_{C}  T_{D} T_{E}  T_{F}] \nn\\
    &= \frac{1}{6!} \left(    \text{Tr}[ T_{A}  T_{B} T_{C}  T_{D} T_{E}  T_{F}] +     \text{Tr}[ T_{B}  T_{A} T_{C}  T_{D} T_{E}  T_{F}] + 718 \text{ more terms} \right).
\end{align}
Using the fact that $T_B T_A = [T_B, T_A] + T_A T_B$ and the algebra, we may rewrite this  trace as
\begin{align}
        \text{STr} & [ T_{A}  T_{B} T_{C}  T_{D} T_{E}  T_{F}]
    = \frac{1}{6!} (    \text{Tr}[ T_{A}  T_{B} T_{C}  T_{D} T_{E}  T_{F}] 
    \nn\\
    & \quad +     \text{Tr}[ T_{A}  T_{B} T_{C}  T_{D} T_{E}  T_{F}] +     \text{Tr}[ f_{BAX} T^X T_{C}  T_{D} T_{E}  T_{F}] + 718 \text{ more terms}).
\end{align}
This gives two factors of the non-symmetrized trace plus a term which has a trace over only five generators. Each of the other 718 terms may be dealt with in the same way: commute the generators to put them in the order $(ABCDEF)$ and keep track of all of the traces over five generators which are picked up along the way. This adds $5\cdot5!$ terms with five generators. Using this and rearranging the trace and symmetric trace, we get the schematic relation
\begin{equation}
    \text{Tr} [ T_{A}  T_{B} T_{C}  T_{D} T_{E}  T_{F}]
    = \text{STr}[ T_{A}  T_{B} T_{C}  T_{D} T_{E}  T_{F}] -  \frac{1}{6!}\cdot600 \, \text{Tr}[TTTTT].
\end{equation}
Each of these five-generator traces may be treated the same way-- they each yield a symmetric trace with five generators plus $4\cdot4!$ terms with a trace over four generators. Schematically, the trace may be expanded as
\begin{align}
    \text{Tr} [ T_{A}  T_{B} T_{C}  T_{D} T_{E}  T_{F}] &\nn\\
    &\kern-4em= \text{STr}[ T_{A}  T_{B} T_{C}  T_{D} T_{E}  T_{F}] -  \frac{1}{6!} \left( 600 \Big( \text{STr}[TTTTT] - \frac{1}{5!}\cdot 96 \, \text{Tr}[TTTT] \Big) \right),
\end{align}
and so on, until the result is a sum of symmetric traces of 2, 3, 4, 5, and 6 generators. Clearly this computation is not tractable by hand. Using the XACT package for Mathematica, we calculated all the necessary terms. The symmetric traces over an odd number of generators cancel each other out (which appears to be a sort of generalization of Furry's theorem).  The result of this procedure includes a symmetric trace over six generators and a large number of symmetric traces over four generators and two generators.

\section{Orthogonal Tensors}

The symmetrized traces may be expanded in a set of orthogonal symmetric tensors. The two needed for this calculation are
\begin{align}
    \text{STr}[T^A T^B T^C T^D] =& I_4 (R) d_{\bot}^{ABCD} + I_{2,2}(R) (\delta^{AB} \delta^{CD} + \delta^{AC} \delta^{BD} +\delta^{AD} \delta^{BC})/3,
\end{align}
and
\begin{align}
    \text{STr} [T^A T^B T^C T^D T^E T^F] &= I_6 (R) d_{\bot}^{ABCDEF} + I_{4,2}(R) (d_{\bot}^{ABCD} \delta^{EF} + d_{\bot}^{ABCE} \delta^{DF} + \cdots)/15 \nn\\
    &\kern-6em+ I_{3,3}(R) (d_{\bot}^{ABC}d_{\bot}^{DEF} + d_{\bot}^{ABD}d_{\bot}^{CEF} +\cdots)/10 + I_{2,2,2}(R) (\delta^{AB} \delta^{CD}\delta^{EF} + \cdots)/15.
\end{align}
Note that $I_6 = 0$ for all representations of $SU(4)$. The tensors $d_{\bot}^{ABCD}$ and $d_{\bot}^{ABC}$ are fixed by the condition of orthogonality; $d_{\bot}^{ABC}$ is the six-dimensional epsilon tensor (recalling that $A = \{\mu_1 \nu_1\}$, etc.) The fourth order $d_{\bot}^{ABCD}$ may be expressed in terms of the six-dimensional metric --- its terms include $g^{\mu_1 \nu_4 } g^{\mu_2 \nu_3} g^{\mu_3 \nu_2} g^{\mu_4 \nu_1}$ and the other 47 ways of arranging the indices. 
The indices
$I_{4,2}$, $I_{3,3}$, and $I_{2,2,2}$ are not unique; imposing orthogonality and other group-theoretic relations yields the system of equations (158)--(160) in \cite{GroupTheory}. Solving these allows $I_{4,2}$, $I_{3,3}$, and $I_{2,2,2}$ to be expressed in terms of the Dynkin indices $I_4$, $I_3$, and $I_2$.

\section{Dynkin Indices}
A representation $R$ with Dynkin labels $(a,b,c)$ has dimension
\begin{align}
    &\text{Dim}_R (a, b, c) = \frac{1}{12} (a+1) (b+1) (c+1) (a+b+2) (b+c+2) (a+b+c+3) .
\end{align}
The Weyl character formula may be used to show that
\begin{align}
    I_2(a,b,c) = \frac{\text{Dim}_R}{60} \left(3 a^2+2 a (2 b+c+6)+4 b^2+4 b (c+4)+3 c
   (c+4)\right).
\end{align}
The third and fourth order generalization to this index were computed in \cite{Okubo:1981td}, which finds
\begin{align}
   I_3(a,b,c) &= \frac{\text{Dim}_R}{120} (a-c) (a+c+2) (a+2 b+c+4)
   \nn\\
   I_4(a,b,c) &= \frac{\text{Dim}_R}{3360}
   \big( 3 a^4+8 a^3 b+4 a^3 c+24 a^3+2 a^2 b^2+2 a^2 b c+30 a^2 b
   \nn\\
   &\quad -4 a^2 c^2+6 a^2 c +54 a^2 -12 a b^3-18 a b^2 c-50 a b^2+2 a b c^2-28 a b c
   \nn\\
   &\quad -34 a b+4 a c^3+6 a c^2-2 a c+24 a -6 b^4-12 b^3 c-48 b^3+2 b^2 c^2
   \nn \\
   &\quad-50 b^2 c-122 b^2+8 b c^3+30 b c^2-34 b c-104 b+3 c^4+24 c^3+54 c^2+24  c\big).
\end{align}


\section{Results}
As the trace of each of the $V_a$ coefficients may be reduced to a trace of generators variously contracted with the Riemann tensor, this method will allow each of them to be computed. The entire list of traced coefficients is presented here:
\begin{align}
    V_1 &= - \frac{A_5}{2} I_2, \qquad
    V_2 = (A_4- A_3) I_2, \qquad
    V_3 = - \frac{A_9}{2} I_2, \qquad
    V_4 =  \frac{A_{17}}{2}  I_2\nn\\
    V_5 &= - \frac{A_{16}}{2}  I_2,\qquad
    V_6 = - \frac{A_{15}}{2}  I_2, \qquad
    V_7 = - \frac{A_{12}}{2}  I_2, \qquad
    V_8 = - \frac{A_1}{2}  I_2,\nn\\
    V_9 &= \left(-\frac{A_6}{51} + \frac{A_7}{6} - \frac{25 A_9}{204} \right) I_2 + \left( \frac{15 A_6}{68} + \frac{15 A_9}{34} \right) \frac{I_2^2}{\text{Dim}_R} + \left( \frac{11 A_6}{51} - \frac{4 A_7}{3} + \frac{5 A_9}{51}\right) I_4,\nn \\
    V_{10} &= \left(-\frac{A_2}{51} + \frac{A_3}{6} - \frac{25 A_5}{204} \right) I_2 + \left( \frac{15 A_2}{68} + \frac{15 A_5}{34} \right) \frac{I_2^2}{\text{Dim}_R} + \left( \frac{11 A_2}{51} - \frac{4 A_3}{3} + \frac{5 A_5}{51}\right) I_4, \nn\\
    V_{11} &= 
    \left( \frac{A_{10}}{612} - \frac{11 A_{11}}{357}-\frac{3A_{12}}{238} - \frac{55 A_{13}}{2142} + \frac{151 A_{14}}{714} + \frac{3 A_{15}}{34} - \frac{383 A_{16}}{4284} - \frac{338 A_{17}}{1071}
    \right) I_2\nn\\
    & \quad +\left( \frac{5 A_{10}}{136}-\frac{375
   A_{11}}{952}+\frac{1095
   A_{12}}{3808}+\frac{115
   A_{13}}{476}+\frac{345
   A_{14}}{952}-\frac{165
   A_{15}}{136}+\frac{325
   A_{16}}{476}+\frac{725
   A_{17}}{952}
    \right) \frac{I_2^2}{\text{Dim}_R}\nn\\
     & \quad+\left( \frac{10 A_{10}}{153}-\frac{41
   A_{11}}{51}+\frac{6
   A_{12}}{17}+\frac{280
   A_{13}}{153}+\frac{38
   A_{14}}{51}-\frac{42
   A_{15}}{17}+\frac{43
   A_{16}}{153}-\frac{8 A_{17}}{153}
    \right) I_4\nn\\
    & \quad +\left( -\frac{5 A_{10}}{68}-\frac{165
   A_{11}}{952}-\frac{1845
   A_{12}}{3808}+\frac{115
   A_{13}}{476}+\frac{345
   A_{14}}{952}-\frac{45
   A_{15}}{136}-\frac{305
   A_{16}}{476}-\frac{115
   A_{17}}{952}
    \right)\frac{I_2^3}{\text{Dim}_R^2}\nn
    \\
    & \quad +\left( -\frac{7 A_{10}}{24}+\frac{209
   A_{11}}{56}-\frac{183
   A_{12}}{224}-\frac{437
   A_{13}}{84}-\frac{437
   A_{14}}{56}+\frac{57
   A_{15}}{8}-\frac{101
   A_{16}}{84}+\frac{437 A_{17}}{168}
    \right)\frac{I_3^2}{\text{Dim}_R}
    \nn\\
    & \quad +\left( -\frac{13 A_{10}}{102}+\frac{4
   A_{11}}{17}-\frac{12
   A_{12}}{17}+\frac{76
   A_{13}}{51}+\frac{38
   A_{14}}{17}+\frac{54
   A_{15}}{17}-\frac{2
   A_{16}}{51}-\frac{38 A_{17}}{51}
    \right)\frac{I_2 I_4}{\text{Dim}_R},\nn\\
    V_{12} &= \left(-\frac{A_{12}}{51} + \frac{A_{15}}{6} - \frac{25 A_{16}}{204} \right) I_2 + \left( \frac{15 A_{12}}{68} + \frac{15 A_{16}}{34} \right) \frac{I_2^2}{\text{Dim}_R} + \left( \frac{11 A_{12}}{51} - \frac{4 A_{15}}{3} + \frac{5 A_{16}}{51}\right) I_4,\nn\\
    V_{13} &= -\frac{A_6}{2} I_2, \qquad
    V_{14} = -\left( A_8 -A_{13} + A_{14} \right) I_2, \qquad
    V_{15} = -\frac{A_2}{2} I_2,\nn\\
    V_{16} &= \left(-\frac{A_{10}}{51} + \frac{A_{11}}{6} - \frac{25 A_{12}}{204} \right) I_2 + \left( \frac{15 A_{10}}{68} + \frac{15 A_{12}}{34} \right) \frac{I_2^2}{\text{Dim}_R} + \left( \frac{11 A_{10}}{51} - \frac{4 A_{11}}{3} + \frac{5 A_{12}}{51}\right) I_4,\nn\\
    V_{17} &= -\frac{A_6}{2} I_2, \qquad
    V_{18} = -\frac{A_{10}}{2} I_2, \qquad
    V_{19} = -\frac{A_{11}}{2} I_2, \qquad
    V_{20} = -\frac{A_{12}}{2} I_2.
\end{align}
Since these expressions pertain to an endomorphism of the form $E=\Sigma_{ab}R^{abcd}\Sigma_{cd}$, where $\Sigma_{ab}$ are $SU(4)$ generators in an arbitrary representation specified by Dynkin labels $(a,b,c)$, we refer to this as the ``group theory method'' for determining the heat kernel coefficients.

Now that the $V_a$'s are known, we may compute the $b_6$ coefficient using the group theory method. We present the coefficient for a representation $R$ on Ricci-flat backgrounds:
\begin{align}
    b_6(R)\Big|_{R_{ab} = 0} &= \frac{1}{(4 \pi)^3 7!} \Bigg[ A_5 \left(\frac{3150
   I_2^2}{17
   \text{Dim}_R}+9
   \text{Dim}_R-\frac{1827
   I_2}{17}+\frac{700
   I_4}{17}\right)
   \nn\\
   &\kern5em + A_9 \left(\frac{6300
   I_2^2}{17
   \text{Dim}_R}+12
   \text{Dim}_R-\frac{3178
   I_2}{17}+\frac{1400
   I_4}{17}\right) \nonumber
    \\
    &\kern-4.5em + A_{16} \left(-\frac{9150
   I_2^3}{17
   \text{Dim}_R^2}+\frac{12900
   I_2^2}{17
   \text{Dim}_R}-\frac{560
   I_2 I_4}{17
   \text{Dim}_R}-\frac{1010
   I_3^2}{\text{Dim}_R}+\frac{44
   \text{Dim}_R}{9}-\frac{8597
   I_2}{51}+\frac{14140
   I_4}{51}\right) \nonumber
   \\
   &\kern-4.5em+A_{17}\left(-\frac{1725
   I_2^3}{17
   \text{Dim}_R^2}+\frac{10875
   I_2^2}{17
   \text{Dim}_R}-\frac{10640
   I_2 I_4}{17
   \text{Dim}_R}+\frac{2185
   I_3^2}{\text{Dim}_R}+\frac{80
   \text{Dim}_R}{9}-\frac{17804
   I_2}{51}-\frac{2240
   I_4}{51}\right) \Bigg].
\label{eq:b6gen}
\end{align}
In general, the full $b_6$ coefficients obtained by the group theory method do not match the expressions (\ref{eq:b6-100}) and (\ref{eq:b6-110}), for the fermion and gravitino, respectively, as the group theory method does not correspond to the square of the Dirac operator when acting on fermions. This indicates that some modification may be necessary for fermionic representations, as was already noted in the four-dimensional case \cite{Christensen:1978md}.
Curiously, however, this mismatch disappears when restricted to Ricci-flat backgrounds.  This suggests that (\ref{eq:b6gen}) may potentially be valid for fermions as well as bosons.  If this were true, we could then derive a general expression for $\delta(c-a)$ for arbitrary higher spin supermultiplets.

Finally, we find that the expression $\delta\mathcal A$ in (\ref{eq:deltaA}) vanishes on arbitrary (ie not just Ricci-flat) backgrounds for long multiplets using the group theory method for the heat kernel. This is in contrast to the conventional method where the fermions are treated by squaring the Dirac operator.  There, $\delta\mathcal A$ for long multiplets only vanished on Ricci-flat backgrounds, but was otherwise non-vanishing on more general backgrounds.  This complete vanishing of $\delta\mathcal A$ for long multiplets is consistent with expectations from AdS$_5$/CFT$_4$ \cite{Ardehali:2013gra,Beccaria:2014xda}, and lends credibility to the idea that the group theory method may yield the correct expression for $\delta\mathcal A$ for general spins.

%% file: Appendices/App4.tex
Operator redundancies in EFTs arise due to the field reparametrization invariance of physical observables \cite{Arzt:1993gz}. For example, in Einstein-Maxwell we consider redefinitions of the metric of the form
\begin{equation}
g'_{\mu\nu} \equiv g_{\mu\nu} + c_1R_{\mu\nu} + c_2R g_{\mu\nu}+c_3 F_{\mu\rho}{F^\rho}_\nu+...
\end{equation}
where $c_i$ are independent coefficients. In the complete effective action (including all possible terms of all mass dimensions consistent with the assumed symmetries) the effect of such a field redefinition is to shift the Wilson coefficients. By choosing $c_i$ in a particular way, certain operators can be removed from the effective action entirely; these are the so-called redundant operators. One approach to constructing a non-redundant basis of operators is to first enumerate all local operators, then use the most general field reparametrization to remove redundant operators. In this appendix we describe an alternative approach that makes use of on-shell scattering amplitudes methods. 

The S-matrix corresponding to the effective action is likewise a physical observable, and independent of the choice of field parametrization. In the tree approximation, gauge invariant effective operators generate Lorentz invariant on-shell matrix elements without kinematic singularities. The on-shell method begins with the observation that there is a one-to-one correspondence between non-redundant gauge invariant local operators and Lorentz invariant local matrix elements \cite{Henning:2017fpj}. By making use of the spinor-helicity formalism for massless on-shell states \cite{Elvang:2015rqa}, it is sometimes more efficient to construct an independent set of the latter. Below we use this correspondence to construct a complete basis for operators coupling gravity to $N$ $U(1)$ gauge fields with up to four derivatives.

The on-shell matrix elements we construct are in the helicity basis. Lorentz invariance is encoded in the requirement that the expressions we construct are rational functions of spinor brackets
\begin{equation}
	\langle ij \rangle = \epsilon^{\dot{\alpha}\dot{\beta}} \tilde{\lambda}_{i\dot{\alpha}} \tilde{\lambda}_{j\dot{\beta}}, \hspace{5mm} [ ij ] = \epsilon_{\alpha\beta} \lambda_i^\alpha \lambda_j^\beta.
\end{equation}
On-shell matrix elements corresponding to gauge invariant local operators are given by polynomials of spinor brackets; we first construct a basis of monomials satisfying certain physical conditions. The first condition we impose is consistency with the action of the massless little group. Such monomials must scale homogeneously with the correct little group weight determined by the helicities $h_i$ of each of the external states
\begin{equation}
M\left(t\lambda_i, t^{-1}\tilde{\lambda}_i\right) = t^{2h_i}M\left(\lambda_i, \tilde{\lambda}_i\right) \, .
\end{equation}
Here we are scaling the spinors of particle $i$ separately, leaving the remaining spinors unchanged. Since the expressions we are constructing are simply strings of $\tilde{\lambda}$s and $\lambda$s, this constraint is equivalent to the following 
\begin{equation}
  2 h_i = (\text{\# of }\lambda_i)-(\text{\# of }\tilde{\lambda}_i).
\end{equation}
This constraint places a lower bound on the mass dimension of the monomial. The minimal dimension monomial we could construct with the correct little group weight for each state contains no anti-holomorphic spinors ($\tilde{\lambda}$) for positive helicity states, no holomorphic spinors ($\lambda$) for negative helicity states and no spinors of either chirality for helicity zero states. As an example, the schematic form of such a minimal dimension monomial
\begin{equation}
  M_4\left(1^{+2},2^{+1},3^{-2},4^0\right) \sim \lambda_1^4 \lambda_2^2 \tilde{\lambda}_3^4. 
\end{equation}
As described above, we need to contract the implicit spinor indices in all inequivalent ways to form a basis of such monomials. The mass dimension of such a string is given simply by $[\lambda]=[\tilde{\lambda}]=1/2$. In this example the minimal dimension is 5. Non-minimal monomials may be generated by introducing further pairs of spinors $\lambda_i \tilde{\lambda}_i \sim p_i$, which have zero little group weight. In general, for a monomial with $k$ photon states and $m$ graviton states the dimension of the monomial is bounded below as:
\begin{equation}
  [M_n] \geq k + 2m.
\end{equation}
To connect this to the EFT basis, such a monomial must correspond to the Feynman vertex rule derived from a gauge invariant local operator. Since polarization vectors for Bosonic states are dimensionless, $[\epsilon]=0$, the mass dimension of the monomial can only arise from powers of momenta generated from derivative interactions. For a local operator with $D$ derivatives the matrix element of $k$ photons and $m$ gravitons has the schematic form
\begin{equation} \label{mnfeyn}
	M_n\left(\{\epsilon,p\}\right) \sim  \epsilon_\gamma^k \epsilon_h^m p^D,
\end{equation}
and so the dimension of the monomial is simply 
\begin{equation}
  [M_n]=D.
\end{equation}
Putting these results together we find that the number of photons and gravitons in a local matrix element is bounded above by the number of derivatives in the corresponding local operator
\begin{equation} \label{Dbound}
D \geq k + 2m.
\end{equation}
This also bounds the total number of states $n=k+m$ (since both $k$ and $m$ are non-negative) as $D\geq n$. Our task is now to enumerate all inequivalent monomials for photon and gravitons with $D=3$ and $D=4$ and identify the corresponding local operators. Here inequivalent means constructing a basis of monomials that are not related to each other by momentum conservation
\begin{equation}
  \sum_{j=1}^n \langle i j\rangle [jk] = 0, 
\end{equation}
 or Schouten identities
\begin{equation}
  \langle ij \rangle \langle kl\rangle + \langle ik\rangle \langle lj\rangle + \langle il\rangle \langle jk\rangle = 0, \hspace{10mm} [ij][kl]+[ik][lj]+[il][jk]=0. 
\end{equation}
A straightforward (though certainly not optimal) approach to this is to first generate a complete basis of monomials, and then numerically evaluate on sets of randomly generated spinors to find a linearly independent subset.

To construct local operators corresponding to the monomials we can make use of the following replacement rules, for photons:
\begin{align}
  \lambda_\alpha\lambda_\beta \rightarrow F^+_{\alpha\beta}\equiv \sigma^{\mu\nu}_{\alpha\beta}F_{\mu\nu}, \hspace{10mm} \tilde{\lambda}_{\dot{\alpha}}\tilde{\lambda}_{\dot{\beta}} \rightarrow F^-_{\dot{\alpha}\dot{\beta}} \equiv \overline{\sigma}^{\mu\nu}_{\dot{\alpha}\dot{\beta}}F_{\mu\nu},
\end{align}
and for gravitons\footnote{Here we are defining $\sigma^{\mu\nu}_{\alpha\beta} \equiv \frac{i}{4}\epsilon^{\dot{\alpha}\dot{\beta}}
\left(\sigma^\mu_{\alpha\dot{\alpha}} \sigma^\nu_{\beta\dot{\beta}}-\sigma^\nu_{\alpha\dot{\alpha}} \sigma^\mu_{\beta\dot{\beta}}\right)$ and $\overline{\sigma}^{\mu\nu}_{\dot{\alpha}\dot{\beta}} \equiv \frac{i}{4}\epsilon^{\alpha\beta}
\left(\sigma^\mu_{\alpha\dot{\alpha}} \sigma^\nu_{\beta\dot{\beta}}-\sigma^\nu_{\alpha\dot{\alpha}} \sigma^\mu_{\beta\dot{\beta}}\right)$. Using standard trace identities, we can rewrite the local operators we construct in the more familiar (though less compact) Lorentz vector notation.}:
\begin{align}
  \lambda_\alpha\lambda_\beta\lambda_\gamma\lambda_\delta \rightarrow W^+_{\alpha\beta\gamma\delta} \equiv \sigma^{\mu\nu}_{\alpha\beta}\sigma^{\rho\sigma}_{\gamma\delta}W_{\mu\nu\rho\sigma}, \hspace{5mm} \tilde{\lambda}_{\dot{\alpha}}\tilde{\lambda}_{\dot{\beta}}\tilde{\lambda}_{\dot{\gamma}}\tilde{\lambda}_{\dot{\delta}} \rightarrow  W^-_{\dot{\alpha}\dot{\beta}\dot{\gamma}\dot{\delta}} \equiv\overline{\sigma}^{\mu\nu}_{\dot{\alpha}\dot{\beta}}\overline{\sigma}^{\rho\sigma}_{\dot{\gamma}\dot{\delta}}W_{\mu\nu\rho\sigma},
\end{align}
where $F^\pm$ and $W^\pm$ are the (anti-)self-dual field strength and Weyl tensors respectively. For non-minimal operators there are additional helicity spinors; these must come in pairs with zero net little group weight and so we can replace:
\begin{equation}
  \lambda^i_\alpha \tilde{\lambda}^i_{\dot{\alpha}} \rightarrow \sigma^\mu_{\alpha\dot{\alpha}}\nabla_\mu \, ,
\end{equation}
where the derivative acts on the local operator creating state $i$. As an illustrative example, consider the following matrix element
\begin{align} \label{matrixex}
  &M_4\left(1^{+1},2^{+1},3^{-1},4^{-2}\right)\nonumber\\
  &= [12]^3\langle 34 \rangle^2 \langle 14 \rangle \langle 24 \rangle \nonumber\\
  &= (\lambda_{1}^{\alpha_1} \lambda_{1}^{\alpha_2})(\lambda_{2\alpha_1}\lambda_{2\alpha_2})(\tilde{\lambda}_{3\dot{\alpha}_1}\tilde{\lambda}_{3\dot{\alpha}_2})(\tilde{\lambda}_{4}^{\dot{\alpha}_1}\tilde{\lambda}_{4}^{\dot{\alpha}_2}\tilde{\lambda}_{4}^{\dot{\alpha}_3}\tilde{\lambda}_{4}^{\dot{\alpha}_4})(\tilde{\lambda}_{1\dot{\alpha}_3}\lambda_1^{\alpha_3})(\tilde{\lambda}_{2\dot{\alpha}_4}\lambda_{2\alpha_3}) \, .
\end{align}
Using the replacement rules given above, this can be generated from the following local operator
\begin{equation}
  [12]^3\langle 34 \rangle^2 \langle 14 \rangle \langle 24 \rangle  \rightarrow \epsilon^{\dot{\alpha}_3\dot{\alpha}_4}\sigma^\mu_{\alpha_3\dot{\alpha}_3}\sigma^\nu_{\alpha_4\dot{\alpha}_4}(\nabla_\mu F^{1+\alpha_1 \alpha_2})(\nabla_\nu F^{2+}_{\alpha_1\alpha_2})F^{3-}_{\dot{\alpha}_1\dot{\alpha}_2}W^{-\dot{\alpha}_1\dot{\alpha}_2\dot{\alpha}_3\dot{\alpha}_4}
\end{equation}
Here we have used a superscript $F^{i}$ to indicate that the spin-1 states correspond to distinct $U(1)$ gauge groups. If two or more states with the same helicity correspond to the same $U(1)$ factor, then we must Bose symmetrize over the particle labels in the matrix elements before applying the replacement rules. This generically reduces the number of independent local operators at a given order in the derivative expansion. 

Finally we must discuss the constraints of parity conservation. In the spinor-helicity formalism, parity $\mathcal{P}$ acts by interchanging the chirality of the spinors $\lambda_{i\alpha}\leftrightarrow \tilde{\lambda}_{i\dot{\alpha}}$, or equivalently interchanging angle and square spinor brackets\footnote{This definition of parity makes sense only if we write the entire matrix element in terms of spinor brackets. For example, to see that local matrix elements containing a single instance of the Levi-Civita symbol are parity odd we must use the identity $\epsilon^{\mu\nu\rho\sigma}p_{1\mu}p_{2\nu}p_{3\rho}p_{4\sigma} \propto [12]\langle 23 \rangle [34] \langle 41\rangle - \langle 12 \rangle [23] \langle 34 \rangle [41]$. }. A local operator is called parity conserving if it generates local matrix elements that satisfy
\begin{equation}
 \mathcal{P}\cdot M_n\left(1^{h_1},2^{h_2},..., n^{h_n}\right) = M_n\left(1^{-h_1},2^{-h_2},..., n^{-h_n}\right).
\end{equation}
This means that when constructing a basis of local operators using the method described above, in a parity conserving model the matrix elements $M_n\left(1^{h_1},2^{h_2},..., n^{h_n}\right)$ and $M_n\left(1^{-h_1},2^{-h_2},..., n^{-h_n}\right)$ should not be counted separately, while in a parity non-conserving model they should be. \section{Three-Derivative Operators}
In accord with the constraint (\ref{Dbound}) the possible, non-redundant, three-derivative operators that generate on-shell matrix elements with $k$-photons and $m$-gravitons have
\begin{equation}
  (k,m) \in \{(3,0)\}.
\end{equation}
The list of possible matrix elements modulo Schouten and momentum conservation, and the corresponding local operators is:\\
$(+1,+1,+1):$
\begin{align}
  [12][23][31] \rightarrow F^{1+}_{\alpha\beta}F^{2+\beta\gamma}{{F^{3+}}_\gamma}^\alpha.
\end{align}
$(-1,-1,-1):$
\begin{align}
  \langle 12 \rangle \langle 23\rangle \langle 31\rangle \rightarrow  F^{1-}_{\dot{\alpha}\dot{\beta}}F^{2-\dot{\beta}\dot{\gamma}}{{F^{3-}}_{\dot{\gamma}}}^{\dot{\alpha}}.
\end{align}
There are two independent, three-derivative local operators. Imposing parity conservation there is only a single independent local operator. Such operators vanish unless all field strength tensors are from distinct $U(1)$ factors. To preserve Bose symmetry of the matrix element we see that the associated Wilson coefficients must be totally antisymmetric in flavor indices.

An equivalent form of the three-derivative effective Lagrangian is
\begin{equation}
  \mathcal{L}^{(3)} = a_{ijk}F^{i}_{\mu\nu}F^{j\nu\rho}{F^{k}_\rho}^\mu + b_{ijk}F^{i}_{\mu\nu}F^{j\nu\rho} \tilde{F}_\rho^{k\mu},
\end{equation}
where both $a_{ijk}$ and $b_{ijk}$ are totally antisymmetric. The first operator ($a$) is parity even while the second ($b$) is parity odd. 
\section{Four-Derivative Operators}
The possible, non-redundant, four-derivative operators generate on-shell matrix elements with $k$-photons and $m$-gravitons with
\begin{equation}
  (k,m) \in \{(2,1),(4,0)\}.
\end{equation}
The list of possible matrix elements modulo Schouten and momentum conservation, and the corresponding local operators is :\\
$(+1,+1,+2):$
\begin{align}
  [13]^2[23]^2 \rightarrow F^{1+}_{\alpha_1\alpha_2}F^{2+}_{\alpha_3\alpha_4}W^{+\alpha_1\alpha_2\alpha_3\alpha_4}.
\end{align}
$(-1,-1,-2):$
\begin{align}
  \langle 13\rangle^2\langle 23\rangle^2 \rightarrow F^{1-}_{\dot{\alpha}_1\dot{\alpha}_2}F^{2-}_{\dot{\alpha}_3\dot{\alpha}_4}W^{-\dot{\alpha}_1\dot{\alpha}_2\dot{\alpha}_3\dot{\alpha}_4}.
\end{align}
$(+1,+1,+1,+1):$
\begin{align}
  [13]^2[24]^2 &\rightarrow F^{1+}_{\alpha_1\alpha_2}F^{3+\alpha_1\alpha_2}F^{2+}_{\alpha_3\alpha_4}F^{4+\alpha_3\alpha_4}\nonumber\\
  [12][23][34][41] &\rightarrow F^{1+}_{\alpha_1\alpha_2}F^{2+\alpha_2\alpha_3}F^{3+}_{\alpha_3\alpha_4}F^{4+\alpha_4\alpha_1}\nonumber\\
  [12]^2[34]^2 &\rightarrow F^{1+}_{\alpha_1\alpha_2}F^{2+\alpha_1\alpha_2}F^{3+}_{\alpha_3\alpha_4}F^{4+\alpha_3\alpha_4}.
\end{align}
$(-1,-1,-1,-1):$
\begin{align}
  \langle 13\rangle^2\langle 24\rangle^2 &\rightarrow F^{1-}_{\dot{\alpha}_1\dot{\alpha}_2}F^{3-\dot{\alpha}_1\dot{\alpha}_2}F^{2-}_{\dot{\alpha}_3\dot{\alpha}_4}F^{4-\dot{\alpha}_3\dot{\alpha}_4}\nonumber\\
  \langle 12\rangle \langle 23\rangle \langle 34\rangle \langle 41\rangle &\rightarrow F^{1-}_{\dot{\alpha}_1\dot{\alpha}_2}F^{2-\dot{\alpha}_2\dot{\alpha}_3}F^{3-}_{\dot{\alpha}_3\dot{\alpha}_4}F^{4-\dot{\alpha}_4\dot{\alpha}_1}\nonumber\\
  \langle 12\rangle^2\langle 34\rangle^2 &\rightarrow F^{1-}_{\dot{\alpha}_1\dot{\alpha}_2}F^{2-\dot{\alpha}_1\dot{\alpha}_2}F^{3-}_{\dot{\alpha}_3\dot{\alpha}_4}F^{4-\dot{\alpha}_3\dot{\alpha}_4}.
\end{align}
$(+1,+1,-1,-1):$
\begin{align}
  [12]^2\langle 34 \rangle^2 \rightarrow F^{1+}_{\alpha_1\alpha_2}F^{2+\alpha_1\alpha_2}F^{3-}_{\dot{\alpha}_1\dot{\alpha}_2}F^{4-\dot{\alpha}_1\dot{\alpha}_2}.
\end{align}
There are five independent, four-derivative local operators. Imposing parity conservation there are only three independent local operators. An equivalent form of the four-derivative effective Lagrangian is
\begin{align}
  \mathcal{L}^{(4)} &= \alpha_{ijkl}F^{i}_{\mu\nu}F^{j\mu\nu}F^{k}_{\rho\sigma}F^{l\rho\sigma} + \beta_{ijkl}F^{i}_{\mu\nu}\tilde{F}^{j\mu\nu}F^{k}_{\rho\sigma}\tilde{F}^{l\rho\sigma} + \gamma_{ij}F^{i}_{\mu\nu}F^{j}_{\rho\sigma}W^{\mu\nu\rho\sigma} \nonumber\\
                    &\hspace{10mm}+ \chi_{ijkl}F^{i}_{\mu\nu}F^{j\mu\nu}F^{k}_{\rho\sigma}\tilde{F}^{l\rho\sigma} + \omega_{ij}F^{i}_{\mu\nu}\tilde{F}^{j}_{\rho\sigma}W^{\mu\nu\rho\sigma}.
\end{align}
The first three operators ($\alpha$, $\beta$ and $\gamma$) are parity even, while the remaining two ($\chi$ and $\omega$) are parity odd.

%% file: Appendices/App5.tex
In this appendix we shall review the derivation of (\ref{Maxwell Corrections}). Recall the corrected equation of motion for the gauge field:
\begin{align}
    \begin{split}
    \nabla_{\mu} F^{i \mu \nu} =& \, \nabla_{\mu} \Big( 8 \, \alpha_{ijkl}  F^{j \mu \nu} F^k_{\alpha \beta} F^{l \alpha \beta} + 8 \, \beta_{ijkl}  \tilde{F}^{j \mu \nu} F^k_{\alpha \beta} \tilde{F}^{l \alpha \beta} + 4 \, \gamma_{ij}  F^j_{\alpha \beta} W^{\mu \nu \alpha \beta}  \\
    &  \qquad \qquad + 4 \, \left( \chi_{ijkl} \tilde{F}^{j \mu \nu} F^k_{\alpha \beta} F^{l \alpha \beta} + \chi_{klij} F^{j \mu \nu} \tilde{F}^k_{\alpha \beta} F^{l \alpha \beta} \right) +  4 \, \omega_{ij}  \tilde{F}^j_{\alpha \beta} W^{\mu \nu \alpha \beta} \Big) \, .
    \end{split}
\end{align}
For simplicity we label the term in the parentheses on the right-hand side of (\ref{Maxwell}) by $G^{i \, \mu \nu}$. First note that the anti-symmetry of $F^{\mu \nu}$ allows us to rewrite the equation of motion as
\begin{align}
    \begin{split}
    \frac{1}{\sqrt{-g}} \partial_{\mu} \left[ \sqrt{-g} \, F^{i \mu \nu} \right] =     \frac{1}{\sqrt{-g}}  \partial_{\mu} \left[ \sqrt{-g} \, G^{i \, \mu \nu} \right] \, .
    \end{split}
\end{align}
We expand this equation in power of the coefficients $\alpha, \ ... \ \omega$. The zeroth- and first-order equations are:
\begin{subequations}
\begin{align}
    & \partial_{\mu} \left[ \sqrt{-g} \, F{}^{i \mu \nu} \right]^{(0)} = 0 \label{EOM0}
    \\
    & \partial_{\mu} \left[ \sqrt{-g} \, F{}^{i \mu \nu} \right]^{(1)} = \partial_{\mu} \left[ \sqrt{-
    g} \, G^{i \, \mu \nu}  \right]^{(1)} \, . \label{EOM1}
\end{align}
\end{subequations}
The solution to the zeroth-order equation is the uncorrected Reissner-Nordstr{\"o}m solution. We are interested in obtaining the first-order part, which represents the corrections to the background. The derivative may be removed from (\ref{EOM1}) because an additive constant has the same fall-off in $r$ as the solution to (\ref{EOM0}), so we may absorb it into the definition of integration constant in the zeroth-order solution, which is $q$. As a result, we have 
\begin{align}
    \begin{split}
        \left[ \sqrt{-g} \, F{}^{i \mu \nu} \right]^{(1)} = \left[ \sqrt{-
    g} \, G^{i \, \mu \nu}  \right]^{(1)} \, .
    \end{split}
\end{align}
 Note that $G^{\mu \nu}$ depends explicitly on $(\, \alpha, ..., \omega \, ) $, so $(G^{\mu \nu})^{(1)}$, which is first-order in the coefficients, depends only on the zeroth-order value of the fields $F^{\mu \nu}$ and $W^{\mu \nu \rho \sigma}$. 
 
 In addition to the Maxwell equation, the gauge fields must satisfy the Bianchi identity
\begin{equation}
    \partial_\mu F^{i}_{\nu\rho}+\partial_\nu F^{i}_{\rho\mu}+\partial_\rho F^{i}_{\mu\nu}=0.
\end{equation}
Together with the assumed spherically symmetry, which imposes that only $F^{i}_{t r}$ and $F^{i}_{\theta \phi}$ are non-zero, this gives the following constraint on the \textit{magnetic} component of the gauge field
\begin{equation} \label{magBI}
    \partial_{r}F^{i}_{\theta \phi}=0.
\end{equation}
Since the leading order magnetic field (\ref{zerosol}) is the unique spherically symmetric field with magnetic monopole moment $p^i$, and by (\ref{magBI}) there can be no subleading $1/r$ corrections, it remains the exact solution even with the addition of higher-derivative interactions. Therefore we are only interested in the corrections to the electric fields $F^{(i)}_{t r}$. Using that $g^0_{tt} = - g^0_{rr}$, we have
\begin{align}
    \begin{split}
    & \left[ \sqrt{-g} F{}^{i \, t r} \right]^{(1)} \ = \  \sqrt{-g}^{(0)} \left(  8 \alpha_{ijkl}  F^{(0) j}{}_{t r} F^{(0) k}{}_{tr} F^{(0)l}{}_{tr} + ... \right).
    \end{split}
\end{align}
Now we may use this to compute the first contribution to the stress tensor corrections. This relies on the non-trivial fact that this combination of $\sqrt{-g}$ and $F$ is the only combination that appears in the corrections to the stress tensor. To see this consider the stress tensor for a Maxwell field, 
\begin{align}
\begin{split}
    T_{\mu \nu} = F^i_{\mu \alpha} F^i_{\nu}{}^{\alpha} - \frac{1}{4} F^i_{\alpha \beta} F^{i \alpha \beta} g_{\mu \nu}.
\end{split}
\end{align}
We are interested only in the corrections to 
\begin{align}
\begin{split}
    T_t {}^t = F^i{}_{t \alpha} F^{i t \alpha} - \frac{1}{4} F^i{}_{\alpha \beta} F^i{}^{\alpha \beta} \delta_t{}^t \, .
\end{split}
\end{align}
We use the fact that only $F_{t r}$ and $F_{\theta \phi}$ are non-zero, and only the former is corrected, to write 
\begin{align}
\begin{split}
    T_t{}^t \ =& \ \frac{1}{2} F^i{}_{t r} F^{i t r} - \frac{1}{2} F^i{}_{\theta \phi} F^{i \theta \phi} \\
=& \ (T^{(0)})_t{}^t - \left[  \sqrt{-g} F^{i t r} \right]^{(1)} \left[  \sqrt{-g} F^{i t r} \right]^{(0)} / (g_{\theta \theta} g_{\phi \phi}) + \mathcal{O} \left[ (\alpha, ...)^2 \right] \, .  
\end{split}
\end{align}
So we have found that 
\begin{align}
\begin{split}
    (T^{(1)}_{Max})_t{}^t =& - \left[  \sqrt{-g} F^{i t r} \right]^{(1)} \left[  \sqrt{-g} F^{i t r}     \right]^{(0)} / (g_{ \theta \theta } g_{\phi \phi})  \\
    = & -  \sqrt{-g}^{(0)} \left(  8 \alpha_{ijkl}  F^{(0) j}{}_{t r} F^{(0) k}{}_{t r} F^{(0)l}{}_{t r}  + ... \right) \sqrt{-g}^{(0)} F^{i t r}{}^{(0)} / (g_{\theta \theta } g_{\phi \phi}) \\
    = & \left(  8 \alpha_{ijkl}  F^{(0) j}{}_{t r} F^{(0) k}{}_{t r} F^{(0)l}{}_{t r}  + ... \right)  F^i{}_{t r}{}^{(0)} \, .
\end{split}
\end{align}
Evaluating this expression gives the result obtained in (\ref{Maxwell Corrections}).

%% file: Appendices/App6.tex
In chapter II, we computed the shift to the geometry by first computing the shift to the stress tensor due to the presence of higher-derivative operators. One source of stress tensor corrections comes from varying the four-derivative operators with respect to the metric. The variations of each of these terms are recorded here for reference.
\begin{align}
    \begin{split}
        (F^i F^j) (F^k F^l): \qquad \qquad & g_{\alpha \beta} (F^i F^j) (F^k \cdot F^l) - 4 \left( F^i_{\mu \alpha} F^{j \mu}{}_{\beta} (F^k  F^l) + (F^i  F^j) F^k_{\mu \alpha} F^{l \mu}{}_{\beta}  \right)  \\[5pt]
        (F^i  \tilde{F}^j) (F^k  \tilde{F}^l) : \qquad \qquad & - g_{\alpha \beta} (F^i  \tilde{F}^j) (F^k  \tilde{F}^l) \\[5pt]
        W F^i F^j :  \qquad \qquad & g_{\alpha \beta} W F^i F^j - 3 R^{\mu}{}_{\alpha \rho \sigma} (F^i_{\mu \beta} F^{j \rho \sigma}+ F^{i \rho \sigma} F^j_{\mu \beta} ) + 4 R_{\alpha \mu} (F^i_{\beta \nu} F^{j \mu \nu} + F^{i \mu \nu} F^j_{\beta \nu} )\\
            & \qquad + 4 R_{\mu \nu} F^{i \mu}{}_{\alpha} F^{j \nu}{}_{\beta}  - \frac{4}{3} R F^i_{\alpha \mu} F^j_{\beta}{}^{\mu} - \frac{2}{3} R_{\alpha \beta} (F^i  F^j) \\
            & \qquad - 4 \nabla_{\mu} \nabla_{\nu} (F^{i \mu}{}_{\alpha}  F^{j \nu}{}_{\beta}) - 4 \nabla_{\mu} \nabla_{\alpha} (F^{i \mu}{}_{\nu} F^j_{\beta}{}^{\nu}) + 2 g_{\alpha \beta} \nabla_{\mu} \nabla_{\nu} (F^{i \mu}{}_{\rho}  F^{j \nu \rho}) \\
            & \qquad + 2 \Box (F^i_{\alpha \mu} F^j_{\beta}{}^{\mu}) + \frac{2}{3} \nabla_{\alpha} \nabla_{\beta} (F^i  F^j) - \frac{2}{3} g_{\alpha \beta} \Box (F^i  F^j) \\[5pt]
        (F^i  \tilde{F}^j) (F^k  F^l) :  \qquad \qquad & -4 (F^i  \tilde{F}^j) F^k_{\mu \alpha} F^{l \mu}{}_{\beta} \\[5pt]
        W F^i \tilde{F}^j:   \qquad \qquad & - 2 R^{\mu}{}_{\alpha \rho \sigma} F^i_{\mu \beta} \tilde{F}^{j \rho \sigma} + 4 R_{\alpha \mu} F^i_{\beta \nu} \tilde{F}^{j \mu \nu}  - \frac{2}{3} R_{\alpha \beta} (F^i  \tilde{F}^j) \\
            & \qquad - 4 \nabla_{\mu} \nabla_{\nu} (F^{i \mu}{}_{\alpha}  \tilde{F}^{j \nu}{}_{\beta}) - 4 \nabla_{\mu} \nabla_{\alpha} (F^{i \mu}{}_{\nu} \tilde{F}^j_{\beta}{}^{\nu}) + 2 g_{\alpha \beta} \nabla_{\mu} \nabla_{\nu} (F^{i \mu}{}_{\rho}  \tilde{F}^{j \nu \rho} )\\
            & \qquad + 2 \Box (F^i_{\alpha \mu} \tilde{F}^j_{\beta}{}^{\mu}) + \frac{2}{3} \nabla_{\alpha} \nabla_{\beta} (F^i  \tilde{F}^j) - \frac{2}{3} g_{\alpha \beta} \Box (F^i  \tilde{F}^j)
    \end{split}
\end{align}
Each of the terms on the left-hand side are multiplied by $\sqrt{-g}$ in the action. Note that we use the shorthand $( F^i  F^j) $ to denote $F^i_{\mu \nu} F^{j \mu \nu}$, and $W A B$ to denote $W_{\mu \nu \rho \sigma} A^{\mu \nu} B^{\rho \sigma}$.

%% file: Appendices/App7.tex
In this appendix we give a short proof of the claim made in section \ref{sec:decay}, that in the perturbative regime, $Q^2\gg 1$, the extremality surface bounds a convex region. Though convexity is a global property, we can reduce the problem to a local one through the \textit{Tietze-Nakajima theorem} \cite{bjorndahl_karshon_2010}: if $X\subset \mathds{R}^n$ is closed, connected and \textit{locally convex}, then $X$ is convex. Here local convexity means that for each $x\in X$, for some $\delta>0$ the set $B_\delta(x)\cap X$ is convex. 
    
Since the requirements of closure and connectedness are trivial for the kinds of regions we are considering, it remains to show that the extremality surface is the boundary of a locally convex set. The key idea of the argument is to show that on a sufficiently small neighborhood of any point, the surface is well approximated by an inverted paraboloid up to $\mathcal{O}(1/Q^2)$ corrections. Local convexity is then a consequence of the convexity of the paraboloid hypograph.
    
Consider a general co-dimension-1 hypersurface $X$ embedded in $\mathds{R}^n$, defined by an equation of the form
    \begin{equation}
        \label{defS}
        \sum_{i=1}^n x_i^2 = 1 + T(x_i),
    \end{equation}
    where $T(x_i)$ is \textit{small} in the sense that 
    \begin{equation}
        \label{smallness}
        \biggr|\sum_{i=1}^n x_i^2-1\biggr| < \epsilon, 
    \end{equation}
for all points $x_i\in X$, for some arbitrarily small $\epsilon >0$. Since this condition is preserved under orthogonal rotations, every point on $X$ can be mapped to $x_i=0$ for $i>1$ up to a redefinition of the function $T(x_i)$. Without loss of generality then we will study the local neighbourhood of such a point. We use the fact that we are interested in functions of the form
    \begin{equation}
        T(x_i) = \sum_{ijkl}T_{ijkl}x_ix_jx_kx_l \, .
    \end{equation}
Here the smallness condition (\ref{smallness}) is equivalent to the statement that $|T_{ijkl}|\sim \epsilon$. To begin with we can rewrite the equation (\ref{defS}) in a useful form
    \begin{align}
        x_1^2 = 1-\sum_{i\neq 1} x_i^2 &+T_{1111}x_1^4+ 4x_1^3\sum_{i}T_{111i}x_i+ 6x_1^2 \sum_{ij\neq 1}T_{11ij}x_i x_j \nonumber\\ 
        &+4x_1\sum_{ijk\neq 1}T_{1ijk}x_i x_j x_k   +\sum_{ijkl\neq 1}T_{ijkl}x_i x_j x_k x_l.
    \end{align}
At $x_i=0$, $i>1$, for small $\epsilon$ there is a single value of $x_1>0$ on $X$. Since we are interested in the surface on an arbitrarily small convex neighbourhood $D$ of $x_i=0$, $i>1$, we can construct a local parametrization of the surface as a function $x_1:D\rightarrow \mathds{R}$
    \begin{equation}
    \label{local}
        x_1(x_2,...,x_n) = 1-\frac{1}{2}\sum_{i\neq 1}x_i^2+\frac{1}{2}T_{1111} +\frac{1}{2}T_{1111}\sum_{i\neq 1}x_i^2+ 3\sum_{i}T_{111i}x_i+ 3 \sum_{i,j\neq 1}T_{11ij}x_i x_j + \mathcal{O}(x_i^3).
    \end{equation}
It is an elementary theorem that the \textit{hypograph} of a function $f:D\rightarrow \mathds{R}$, with $D$ a convex set in $\mathds{R}^{n-1}$, is a convex set in $\mathds{R}^n$ if the Hessian of $f$ is negative definite on the interior of $D$. From (\ref{local}) we can read off the eigenvalues of the Hessian matrix at this point as $-1 + \mathcal{O}(\epsilon)$. Since the eigenvalues of the Hessian are continuous on $X$ they must all be negative on some neighbourhood of this point. This completes the proof that $X$ is locally convex.

%% file: Appendices/App8.tex
\noindent In chapter IV, we computed the constraints on the coefficients in AdS${}_4$. Here we will present the results of this calculation for AdS${}_5$ through AdS${}_7$ using the entropy shifts, which corresponds to working in the zero Casimir energy scheme.  For completeness, we also present the Casimir energies for AdS$_5$ and AdS$_7$.

\section{AdS${}_5$}

\noindent In AdS${}_5$ we find that the stability condition obtained by demanding positive specific heat and permittivity is given by $\xi < \xi^*$ for $\nu<\nu^*$, with
\begin{align}
    \xi^* = 1 - \sqrt{\frac{1 - 2 \nu^2}{1 + 2 \nu^2}}, \qquad \nu^* = \frac{1}{\sqrt{2}} \, ,
\end{align}
and that all black holes with $\nu>\nu^*$ are stable for all values of the charge. The full entropy shift is simpler to express as a function of charge $q$ than extremality parameter $\xi$. We find 
\begin{align}
\label{dSde5}
    \begin{split}
      &\left( \frac{\partial S }{\partial \epsilon} \right)_{M, Q} \ = \ \frac{\pi}{256 l^6 \nu^8 T} \Big( \ c_1 \left( 43 q^4 - 24 l^4 q^2 \nu^4 (8 + 5 \nu^2) + 32 l^8 \nu^8(18 + 41 \nu^2 + 13 \nu^4) \right) \\
      & \qquad  \qquad \qquad \qquad + 24 c_2 q^2 \left( 3 q^2 - 8 l^4 \nu^4  \right) +72 ( 2 c_3 + c_4)q^4 \Big)\,.
    \end{split}
\end{align}
Note that holographic renormalization in AdS$_5$ with a Riemann-squared correction yields a Casimir energy
\begin{equation}
    E_c=\fft{\omega_3}{16\pi}\left(\fft34l^2-\fft{15}4c_1l^2\right),
\end{equation}
where $\omega_3=2\pi^2$.  This Casimir energy must be removed from the thermodynamic energy in order to obtain the mass $M$ of the black hole.  Alternatively, it can be cancelled right from the beginning by adding an appropriate finite counterterm to the action, in which case the thermodynamic energy would then correspond directly to the mass.  If the Casimir energy is not removed, then the thermodynamic energy shift becomes a combination of mass shift and Casimir energy shift since $E_c$ depends explicitly on the $c_1$ Wilson coefficient.

We find the following expression for the extremal limit,
\begin{align}
    &\left( \frac{\partial S }{\partial \epsilon} \right)_{M, Q} \nonumber\\
    & =\frac{\pi l^2 }{16 T} \left( c_1 (31 + 128 \nu^2 + 138 \nu^4) + 24 c_2 (1 + 2 \nu^2)(1 + 6 \nu^2) + 72(2c_3 + c_4) (1 + 2 \nu^2)^2 \right)  ,
\end{align}
while in the neutral limit we have
\begin{align}
    \left( \frac{\partial S }{\partial \epsilon} \right)_{M, Q} \ = \ \frac{\pi l^2 }{16 T} c_1 \left( 18 + 41 \nu^2 + 13 \nu^4 \right) \, .
\end{align}
Once again, the entropy shift is proportional to $c_1$ in this limit.
\begin{figure}
    \includegraphics[scale=0.6]{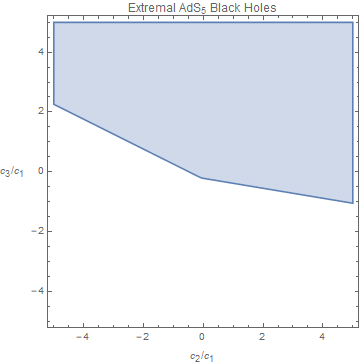}
    \includegraphics[scale=0.6]{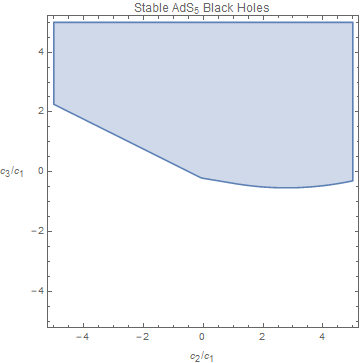}
    \caption{Allowed regions for AdS${}_5$ EFT coefficients.}
    \label{AdS5_exclusion} 
\end{figure}
It is interesting that we do not find a positivity constraint on $c_2$, as we did in AdS${}_4$. There is a lower bound on $c_3 / c_1$ of about $-$0.5339. The general constraints obtained by the Reduce function of Mathematica are extremely complicated and probably of little interest.

\section{AdS${}_6$}

\begin{figure}
    \includegraphics[scale=0.6]{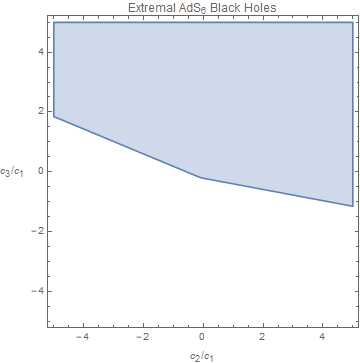}
    \includegraphics[scale=0.6]{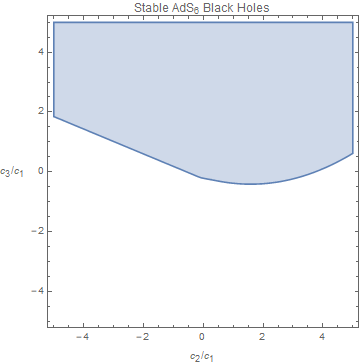}
    \caption{Allowed regions for AdS${}_6$ EFT coefficients.}
    \label{AdS6_exclusion} 
\end{figure}

\noindent In AdS${}_6$ the stability condition obtained by demanding positive specific heat and permittivity is of the same general structure as in AdS${}_5$, 
but with the following identifications:
\begin{align}
    \xi^* = 1 - \sqrt{\frac{3 - 5 \nu^2}{3 + 5 \nu^2}}, \qquad \nu^* = \sqrt{\frac{3}{5}}\,.
\end{align}
The entropy shift is given by:
\begin{align}
    \begin{split}
      &\left( \frac{\partial S }{\partial \epsilon} \right)_{M, Q} \ = \ \frac{\pi}{264 l^9 \nu^{11} T} \Big( \ c_1 \left( 189 q^4 - 22 l^6 q^2 \nu^6 (36 + 29 \nu^2) + 264 l^{12} \nu^{12} (8 + 17 \nu^2 + 7 \nu^4) \right) \\
      & \qquad  \qquad \qquad \qquad + 2 c_2 q^2 \left( 153 q^2 - 44 l^6 \nu^6 (9 + 5 \nu^2) \right) + 288 ( 2 c_3 + c_4)q^4 \Big) \, ,
    \end{split}
\end{align}
and in the extremal limit takes the form:
\begin{align}
    &\left( \frac{\partial S }{\partial \epsilon} \right)_{M, Q} \nonumber\\
    &= \ \frac{2 \nu \pi l^3 }{99 T} \left( c_1 (369 + 1263 \nu^2 + 1124 \nu^4) + 4 c_2 (3 + 5 \nu^2)(27 + 100 \nu^2) + 96 (2c_3 + c_4) (3 + 5 \nu^2)^2 \right) \, .
\end{align}
Finally, in the neutral limit we find
\begin{align}
    \left( \frac{\partial S }{\partial \epsilon} \right)_{M, Q} \ = \ \frac{ \nu \pi l^3 }{ T} c_1 \left( 8 + 17 \nu^2 + 7 \nu^4 \right) \, .
\end{align}
Note that no Casimir energy subtraction is needed in AdS$_6$.
We again find that $c_1$ is positive. The other bounds are displayed in figure \ref{AdS6_exclusion}. In AdS${}_6$ and AdS${}_7$, the Reduce function of Mathematica was not able to find the general constraints over all stable values of $\xi$ and $\nu$. However, we believe that the strongest constraints will come from the boundaries of the region of stable black holes. Specifically, we imposed positivity at the neutral $\xi \rightarrow 1$ limit, the extremal $\xi \rightarrow 0$ limit, the planar limit $\nu \rightarrow \infty$ limit, and at $\xi = \xi^*$. We believe this method should give the same answer, and we have checked explicitly that it does in the case for AdS${}_4$ and AdS${}_5$.

\section{AdS${}_7$}

\noindent In AdS${}_7$ the stability window is determined by 
\begin{align}
    \xi^* = 1 - \sqrt{\frac{2 - 3 \nu^2}{2 + 3 \nu^2}}, \qquad \nu^* = \sqrt{\frac{2}{3}} \, ,
\end{align}
and the entropy shift is:
\begin{align}
    \begin{split}
      &\left( \frac{\partial S }{\partial \epsilon} \right)_{M, Q} \nonumber\\
      &= \ \frac{\pi^2}{896 l^{12} \nu^{14} T} \Big( \ c_1 \left( 556 q^4 - 14 q^2 l^8 \nu^8 (160 + 141 \nu^2) + 56 l^{16} \nu^{16} ( 100 + 207 \nu^2 + 8 \nu^4) \right)  \\
      & \qquad  \qquad \qquad \qquad + 80 c_2 q^2 \left( 11 q^2 - 7 l^8 \nu^8 (4 + 3 \nu^2) \right) + 800 ( 2 c_3 + c_4) q^4 \Big)\,.
    \end{split}
\end{align}
The Casimir energy that must be removed from the thermodynamic energy in AdS$_7$ is
\begin{equation}
    E_c=\fft{\omega_5}{16\pi}\left(-\fft58l^4+\fft{35}8c_1l^4\right),
\end{equation}
where $\omega_5=\pi^3$.

\begin{figure}
    \includegraphics[scale=0.6]{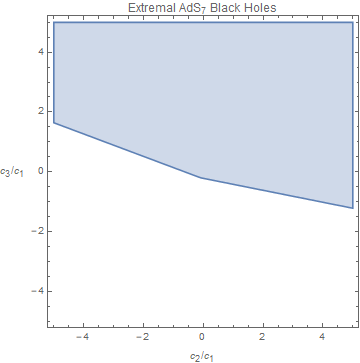}
    \includegraphics[scale=0.6]{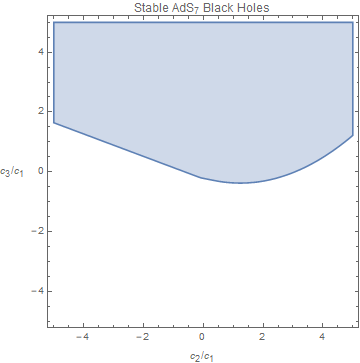}
    \caption{Allowed regions for AdS${}_7$ EFT coefficients.}
    \label{AdS7_exclusion} 
\end{figure}

We find the following expression for the extremal limit,
\begin{align}
\begin{split}
        & \left( \frac{\partial S }{\partial \epsilon} \right)_{M, Q} \ = \ \frac{\pi^2 \nu^2 l^4}{224 T} \Bigg( c_1  \left( 1384 + 4236 \nu^2 + 3345 \nu^4 \right) \\ 
    &\qquad \qquad \qquad \qquad \qquad + 40 c_2 (2 + 3 \nu^2)(16 + 45 \nu^2) + 800 (2c_3 + c_4)(2 + 3 \nu^2)^2 \Bigg)\, ,
\end{split}
\end{align}
while in  the neutral limit we find
\begin{align}
    \left( \frac{\partial S }{\partial \epsilon} \right)_{M, Q} \ = \ \frac{ \pi^2 l^2 \nu^2 }{16 T} c_1 \left( 100 + 207 \nu^2 + 93 \nu^4 \right) \, .
\end{align}
Once again, $c_1$ is positive. The other bounds are displayed in figure \ref{AdS7_exclusion}. Again, we used the method of extremizing over the boundaries of the space of stable black holes. 

